\documentclass[a4paper, twoside, openright]{Style/discothesis}

\usepackage[disable]{todonotes}

\usepackage{multicol}

\newcounter{countitems}
\newcounter{nextitemizecount}
\newcommand{\setupcountitems}{%
  \stepcounter{nextitemizecount}%
  \setcounter{countitems}{0}%
  \preto\item{\stepcounter{countitems}}%
}
\makeatletter
\newcommand{\computecountitems}{%
  \edef\@currentlabel{\number\c@countitems}%
  \label{countitems@\number\numexpr\value{nextitemizecount}-1\relax}%
}
\newcommand{\nextitemizecount}{%
  \getrefnumber{countitems@\number\c@nextitemizecount}%
}
\newcommand{\previtemizecount}{%
  \getrefnumber{countitems@\number\numexpr\value{nextitemizecount}-1\relax}%
}
\makeatother    
\newenvironment{AutoMultiColItemize}{%
\ifnumcomp{\nextitemizecount}{>}{3}{\begin{multicols}{2}}{}%
\setupcountitems\begin{itemize}}%
{\end{itemize}%
\unskip\computecountitems\ifnumcomp{\previtemizecount}{>}{3}{\end{multicols}}{}}

\usepackage{wrapfig}

\usepackage{lmodern}
\usepackage{lipsum}

\usepackage{paralist}
\usepackage{enumitem}
\setlist{nosep}
\usepackage{esvect}

\usepackage{geometry}
\usepackage{floatrow}
\newfloatcommand{capbtabbox}{table}[][\FBwidth]

\usepackage{mathptmx}
\usepackage{anyfontsize}
\usepackage{t1enc}
 
\usepackage{cancel}
 
\usepackage{mathrsfs}
\usepackage{tcolorbox}
 
\usepackage{color} 
\usepackage{colortbl}

\usepackage[utf8]{inputenc}
\usepackage[T1]{fontenc}

\usepackage{xcolor}
\usepackage{etoolbox}
\usepackage[printonlyused]{acronym}
\usepackage{setspace}
\makeatletter
\AtBeginDocument{%
  \renewcommand*{\AC@hyperlink}[2]{%
    \begingroup
      \hypersetup{hidelinks}%
      \hyperlink{#1}{#2}%
    \endgroup
  }%
}

\AtBeginEnvironment{acronym}{%
  
  \renewcommand*{\descriptionlabel}[1]{\color{red}#1}
  \renewcommand*{\AC@hyperlink}[2]{%
    \begingroup
      \hypersetup{hidelinks}%
      \hyperlink{#1}{#2}%
    \endgroup
  }%
}
\makeatother

\usepackage{multirow}
\usepackage{adjustbox}
\newcommand{\tabitem}{~~\llap{\textbullet}~~}
\usepackage{subcaption}
\usepackage{placeins}
\usepackage{tabularx}

\usepackage{cite}

\usepackage[toc,page]{appendix}

\newcommand{\mypictures}[1]{\textcolor{gray}{Image published by the author of this thesis in \cite{#1} $\dagger$.}}

\newcommand{\ccbypictures}[1]{\textcolor{gray}{Image published in \cite{#1}, released under a CC-BY license $\ddagger$.}}

\newcommand{\revigopictures}[0]{\textcolor{gray}{The scatter plots have been generated with Revigo \cite{supek2011revigo}.}}

\thesistype{PhD Thesis} 
\title{Convergent transcriptomic and neuroimaging signature of Autism Spectrum Disorder}

\author{Elisa Ferrari}
\email{elisa.ferrari@sns.it}

\institute{\textbf{PhD in Data Science} \\[4pt]
Joint program offered by:\\[2pt]
\centering
\begin{tcolorbox}[text width=5cm,arc=0mm,boxrule=0.2mm,colback=gray!10]
\centering
\footnotesize
Scuola Normale Superiore\\
Scuola IMT Alti Studi Lucca\\
Università di Pisa\\
Consiglio Nazionale delle Ricerche\\
Scuola Superiore Sant'Anna
\end{tcolorbox}
}

\supervisors{Davide Bacciu, Alessandro Cellerino and Alessandra Retico.}

\date{Academic year 2021-2022}

\begin{document}

\frontmatter 
\maketitle

{\pagestyle{empty}
\null\newpage{\pagestyle{empty}}
\null\newpage
\null\newpage
}
\begin{abstract}
Autism Spectrum Disorder (ASD) is a multi-factorial neurodevelopmental disorder, whose causes are still poorly understood. 
Effective therapies to reduce all the heterogeneous symptoms of the disorder do not exists yet, but behavioural programs started at a very young age may improve the quality of life of the patients. For this reason, many efforts have been dedicated to the research of a reliable biomarker for early diagnosis. Machine learning approaches to distinguish ASDs from healthy controls based on their brain Magnetic Resonance Images (MRIs) have been plagued by the problem of confounders, showing poor classification performance and inconsistency in the biomarker definition. Brain transcriptomics studies, instead, showed some converging results, but being based on data that can be acquired only post-mortem they are not useful for diagnosis.\\
In this work, using an imaging transcriptomics approach, the following results have been obtained.
\begin{itemize}
    \item A deep learning based classifier resilient to confounders and able to exploit the temporal dimension of resting state functional MRIs has been developed, reaching an AUC of 0.89 on an independent test set.
    \item Five gene network modules involved in ASD have been identified, by analyzing brain transcriptomics data of subjects with ASD and healthy controls.
    \item By comparing the brain regions relevant for the classifier obtained in the first step and the brain-wide gene expression profiles of the modules of interest obtained in the second step, it has been proved that the regions that characterize ASD brain at the neuroimaging level are those in which four out of the five gene modules take a significantly high absolute value of expression.
\end{itemize}
These results prove that, despite the heterogeneity of the disorder, it is possible to identify a neuroimaging-based biomarker of ASD, confirmed by transcriptomics.
\end{abstract}

\clearpage
\null\newpage
\begingroup
\hypersetup{linkcolor=black}
\chapter*{Acronyms}
\addcontentsline{toc}{chapter}{Acronyms}
\begin{acronym}[MPCAAAA] 

\acro{2S-CNN}{two-stream CNN}
\acro{AAL}{Automated Anatomical Labelling}
\acro{ABIDE}{Autism Brain Imaging Data Exchange}
\acro{Acc}{Accuracy}
\acro{ADOS}{Autism Diagnostic Observation Schedule}
\acro{AE}{Autoencoder}
\acro{AHBA}{Allen Human Brain Atlas}
\acro{AI}{Artificial Intelligence}
\acro{AS}{Alternative Splicing}
\acro{ASD}{Autism Spectrum Disorder}
\acro{ATP}{Autism Tissue Program}
\acro{AUC}{Area Under the ROC Curve}
\acro{BH}{Benjamini–Hochberg}
\acro{BOLD}{Blood Oxygen Level Dependent}
\acro{BR-NN}{Bias-Resilient Neural Network}
\acro{C}{Cerebellum}
\acro{cDNA}{Complementary DNA}
\acro{CE}{Confounder Effect}
\acro{CI}{Confounding Index}
\acro{CNN}{Convolutional Neural Network}
\acro{DE}{Differentially Expressed}
\acro{DL}{Deep Learning}
\acro{DL}{Deep Learning}
\acro{ds-cDNA}{Double Stranded Complementary DNA}
\acro{DS}{Differential Stability}
\acro{DSM}{Diagnostic and Statistical Manual of Mental Disorders}
\acro{EA}{Enrichment Analysis}
\acro{EEG}{Electroencephalography}
\acro{F}{Frontal Cortex}
\acro{FDR}{False Discovery Rate}
\acro{FET}{Fisher's Exact Test}
\acro{fMRI}{Functional MRI}
\acro{FN}{False Negative}
\acro{FNR}{False Negative Rate}
\acro{FP}{False Positive}
\acro{FPR}{False Positive Rate}
\acro{FV}{Fixed Variables}
\acro{GE}{Gene Expression}
\acro{GO}{Gene Ontology}
\acro{IG}{Imaging Genetics}
\acro{iPSC}{Induced Pluripotent Stem Cell}
\acro{IT}{Imaging Transcriptomics}
\acro{LMM}{Linear Mixed Model}
\acro{LMMs}{Linear Mixed Models}
\acro{lncRNA}{Long Non-Coding RNA}
\acro{LR}{Logistic Regression}
\acro{MDS}{Multidimensional Scaling}
\acro{miRNA}{Micro RNA}
\acro{ML}{Machine Learning}
\acro{MNI}{Montreal Neurological Institute}
\acro{MRI}{Magnetic Resonance Imaging}
\acro{mRNA}{messenger RNA}
\acro{NICHD}{National Institute of Child Health and Human Development}
\acro{NN}{Neural Network}
\acro{NPCs}{Neuronal Progenitor Cells}
\acro{Prec}{Precision}
\acro{PWACR}{Prader-Willi/Angelman critical region}
\acro{ReLU}{Rectified Linear Units}
\acro{RIN}{RNA Integrity Number}
\acro{RMSE}{Root Mean Squared Error}
\acro{RNN}{Recurrent NN}
\acro{ROC}{Receiver Operating Characteristic}
\acro{rRNA}{ribosomal RNA}
\acro{rsfMRI}{Resting State fMRI}
\acro{RTT}{Rett's syndrome}
\acro{RV}{Random Variables}
\acro{Sens}{Sensitivity}
\acro{SFT}{Scale-free Topology}
\acro{SML}{Supervised Machine Learning}
\acro{sMRI}{Structural MRI}
\acro{sncRNA}{Small Non-Coding RNA}
\acro{snRNA}{Small Nuclear RNA}
\acro{Spec}{Specificity}
\acro{SVM}{Support Vector Machine}
\acro{T}{Temporal Cortex}
\acro{TDC}{Typically Developing Control}
\acro{TMM}{Trimmed Mean of M-values}
\acro{TN}{True Negative}
\acro{TNR}{True Negative Rate}
\acro{TOM}{Topological Overlap Matrix}
\acro{TP}{True Positive}
\acro{TPR}{True Positive Rate}
\acro{tRNA}{transfer RNA}
\acro{WGCNA}{Weighted Gene Co-Expression Network Analysis}
\acro{XAI}{Explainable Artificial Intelligence}


\end{acronym}

\endgroup
\clearpage
\null\newpage
\tableofcontents


\clearpage
{
\pagestyle{empty}
\null\newpage
\null\newpage
\null\newpage
}

\mainmatter 

\chapter{Introduction}
\subsubsection*{Motivations}
\todo[inline]{tabella soggetti negli studi di trascrittomica, aggiungere il paper bellissimo alle motivazioni per cui vale la pena fare uno studio di imaging genetics nell'ASD, leggere meglio MM e risultati, controllare figure (per estetica e per quanto riguarda le autocitazioni), controllare gli acronimi.}
The research on Autism Spectrum Disorder (ASD) is still in its infancy with respect to that on other psychiatric disorders \cite{thurm2012importance}. This delay can be attributed partly to the late identification of the disorder, described for the first time in 1943 \cite{harris2018leo}, to the constant changes in its definition (current diagnostic criteria have been fixed only in 2013 \cite{american2013diagnostic}) and to flawed initial etiological theories, such as the speculation that the condition was caused by a lack of parental affection \cite{cohmer2014early}. However, the main hurdle in the research is represented by the heterogeneity of ASD and by its definition exclusively defined on behavioural traits, which can be difficult to objectively evaluate and measure, especially during early childhood.\\
Current understanding of the disorder indicates that ASD is a lifelong neurodevelopmental disorder, mainly affecting social and communicative abilities, with a clear inheritance component \cite{folstein1977infantile,rutter2000genetic,bailey1995autism,sandin2014familial,spiker2002behavioral}, probably caused by the interaction of genetic abnormalities and environmental factors \cite{ivanov2015autism}, in which the impact of the latter ones is suggested to depend on the timing of the exposure \cite{hornig2002infectious,rodier2011environmental}.\\
Although the aforementioned findings contributed to reach a better understanding of ASD, the exact causes behind this condition are not clear and many fundamental aspects require further research; first among all, the identification of a quantitative biomarker to facilitate the diagnosis and anticipate it. In fact, to date, the diagnosis of ASD is a complex procedure, requiring a psychiatric assessment by trained evaluators \cite{eaves2006screening,zander2016objectivity}. As a consequence, subjects with ASD are often diagnosed after the fourth year of life \cite{apaaverage}, or even later, because of an initial misdiagnosis \cite{kentrou2019delayed}, preventing the possibility to benefit from behavioural programs, which proved to alleviate some ASD symptoms only if promptly started \cite{estes2015long,reichow2012early}.
The research of a biomarker of ASD ranges across various subfields of biology and medicine, but being it a neurodevelopmental disorder with an inherited susceptibility, genetics and neuroimaging have been extensively explored.
\subsubsection*{State of the art}
From a genetic point of view, causative DNA variants have been detected in a small fraction of ASDs, but they may occur also in non-spectrum subjects \cite{fernandez2017syndromic}.  Currently, hundreds of genes, with diverse functions, have been associated to ASD \cite{rylaarsdam2019genetic}. 
Although this genetic variety may explain the phenotypical heterogeneity observed in ASD, it is commonly agreed that shared downstream mechanisms may characterize the disease, which can be captured by transcriptomic analysis. Reproducible findings in transcriptome studies performed on post-mortem brain samples highlight mainly a dysregulation of neuronal genes \cite{giovedi2014involvement} and an upregulation of the immune response \cite{gupta2014transcriptome}, as well as other genetic pathways, whose relevance for the ASD is instead less consensual and requires further investigations. Unfortunately, brain transcriptomic signatures can not be used as a biomarker useful for diagnosis because they can be measured only post-mortem.

On the morphological and functional brain characteristics of ASD, instead, there is less consensus. Although various abnormalities possibly implicated in the neurodevelopmental trajectory of ASD have been reported,
the search for a useful biomarker remains obscured by inconsistent or incompatible findings \cite{pua2017autism}. 
In fact, in addition to the unavoidable heterogeneous nature of ASD, the research for neuroimaging based biomarkers is confronted with: 
\begin{itemize}
    \item the general lack of labelled dataset of consistent size and good quality, freely available for medical studies;
    \item the high number of features composing neuroimaging data;
    \item the high variance in the data arising from different demographics characteristics of the sample and acquisition modalities adopted;
\end{itemize}
Statistical analyses made under these conditions may be strongly influenced by the presence of biases in the dataset, particularly when using \ac{DL} \cite{goodfellow2016deep} algorithms, that can capture even the most subtle spurious correlations. In addition, the definition of a biomarker is further complicated by the current poor understanding of the disorder, that does not allow to validate data-driven findings.\\
Structural information has been thoroughly explored in literature, not only with in-vivo neuroimaging studies but even analyzing post-mortem brains, without finding a clear shared pattern \cite{donovan2017neuroanatomy}.
Thus, even supposing to find new solutions to deal with the problems of the neuroimaging-based research mentioned above, the existence of an univocal structural biomarker of ASD seems unlikely. Instead, large margins of improvement in the combined study of the temporal and spatial dimensions of functional data seem to be available because of new advancements in DL-based video analysis, making current research mainly focused on this type of information.
\subsubsection*{Imaging transcriptomic approach}
Summarizing the previous paragraph, post-mortem transcriptomic analyses have given some insights of a shared pattern among ASDs, but can not be used for diagnosis, while neuroimaging data could, but present the aforementioned issues hampering reproducibility. Thus, this work proposes an imaging transcriptomic study that combines the strengths of the two techniques. 
The adopted approach is based on the comparison of the spatial pattern extracted from a classifier trained on resting state functional magnetic resonance imaging (rsfMRI) data to distinguishing ASDs from Typically Developing Controls (TDCs) with the brain-wide expression distribution of genes found to be relevant for ASD. 
In this way, the neuroimaging-based pattern, useful for diagnosis, can be validated by transcriptomic characteristics of ASD, that are supposedly supported by literature.\\
The possibility to compare neuroimaging and brain-wide transcriptomic data is grounded on many recent experimental evidences obtained both in human and animal models \cite{richiardi2015correlated,hawrylycz2012anatomically}. For instance, the gene expression profiles of non-adjacent regions belonging to the same functional network have been found to be more correlated with respect to those belonging to different networks despite being spatially closer \cite{richiardi2015correlated}. This result means that the synchronous activity of brain regions, observable with functional neuroimaging, reflects the correlations found at the transcriptome level.
The observed enrichment in ASD-associated genes among those whose transcriptome correlates with brain activity in regions involved in the default mode network (i.e. a functional network activated also during rest) \cite{wang2015correspondence}, further suggests that an imaging transcriptomic analysis of ASD is worth pursuing.
\subsubsection*{The proposed work}
The work of this thesis can be divided into three main analyses: 
\begin{itemize}
    \item \textit{Neuroimaging}\\
    To address the presence of possible confounding biases in the dataset and the challenging extraction of spatio-temporal features from rsfMRI, a bias resilient classifier (based on adversarial learning \cite{lowd2005adversarial}) able to jointly process the 4D rsfMRI data and its optical flow has been developed and trained to distinguish ASDs and TDCs.
    Then, using an explainability framework \cite{burkart2021survey} that allows to understand the decisional pattern learned by classifier, the most important brain regions for classification have been identified.
    \item \textit{Trascriptomics}\\
    The genes showing a significantly different expression pattern have been identified, by comparing transcriptome data of post-mortem brain samples of ASDs and TDCs.
    Then, through a weighted gene co-expression network analysis, the modules enriched for the genes detected have been selected as potentially involved to the disorder. 
    \item \textit{Imaging transcriptomics}\\
    The expression profiles of the network modules defined at the previous step are analyzed across the brain regions in a transcriptomic dataset with high spatial resolution, trying to understand whether the modules identified as associated to ASD have a significantly higher average absolute expression value in the regions most important for the neuroimaging-based classification.
\end{itemize}
Note that the convergence of the first two analyses mutually validates the results of each of them and proves that, despite its heterogeneity, the disorder can be characterized as a whole at the transcriptome and neuroimaging levels.
\subsubsection*{Structure of the thesis}
The thesis follows a standard structure and is articulated in 5 parts: background, contribution, materials and methods, results, discussion and conclusions. Most of these parts, with the exception of the contribution and the conclusion, contain three main chapters: neuroimaging, genetics and imaging genetics, so that readers with different backgrounds can easily identify the sections meeting their interests.\\
The background part describes the theoretical topics necessary to understand the thesis and a summary of the state-of-the-art to which this work should be compared. The contribution part, in light of what is reported in the background, details the objectives of this thesis and the advancements with respect to the current knowledge that this work intends to bring. Starting from this section, everything described has been exclusively developed by the author. The materials and methods and the results parts follow the same structure, reporting in detail the design, the methodological choices made in this work and the consequent results, respectively. Finally, the results are critically analyzed, linked together and compared with the current state of the art in the discussion, while the conclusions summarize the main thesis discoveries and presents possible avenues for future works.

The main manuscript does not report any information on the computational resources and time of execution necessary for each analysis. This choice was made to keep the writing flow focused on the scientific contents of this work. However, key computational details about the performed analyses can be found in Appendix \ref{Appendix_computational}.


\part{Background}
\chapter{Autism Spectrum Disorder}

\section[Basic notion on ASD]{Basic notion on ASD\footnote{Many of the concepts mentioned in this section have already been published by the author of this thesis in a book chapter on artificial intelligence in medicine \cite{Ferrari2020autismChapter}.}}
\label{sec:bkg_asd_notions}
\ac{ASD} is a lifelong and extremely heterogeneous neurodevelopmental disorder characterized by impaired communication, social interaction, restricted and repetitive interests and activities \cite{american2013diagnostic}.
It can manifest with various degrees of severity and many different comorbidities and its symptomatology may vary widely across the lifespan \cite{georgiades2017editorial}.\\
According to the Autism and Developmental Disabilities Monitoring Network, the latest estimate of ASD prevalence in the USA is 1 in 59 \cite{baio2018prevalence}, with a male to female ratio of approximately 3:1 \cite{loomes2017male}.\\
Symptoms may be visible already at 6 months of age  \cite{maestro2002attentional}, becoming more evident around the 2nd or 3rd year \cite{rogers2009infant}, when the behavior of the children affected by ASD appears clearly different from their peers.

ASD is considered a permanent condition and as of today, no therapies are known, to effectively relieve all the symptoms of the disorder \cite{myers2007management,mandell2009nih}.
However, intensive behavioral treatments can improve long-term developmental outcomes and the overall quality of life \cite{estes2015long,reichow2012early}. Such treatments proved to be more effective when promptly started in the early childhood \cite{fernell2013early,bryson2004early, koegel2014importance}.
Few children with milder forms of ASD can reach a clinical profile that is no longer consistent with the ASD diagnostic criteria, the so-called “optimal outcome” \cite{orinstein2014intervention}.\\
For these reasons, an early diagnosis would be fundamental. However, no reliable quantitative biomarkers of ASD exist \cite{frye2019emerging}, and the diagnosis remains a long and complex process, requiring a psychiatric evaluation based on behavioral observation, clinical interviews and questionnaires to the child and their parents \cite{eaves2006screening}. 

\section{Pathophysiology of ASD}\label{causes_of_ASD}
The exact causes and the pathophysiology of ASD are still unclear. It could be argued that the ASD definition itself is somewhat "uncertain". In fact, the triad of symptoms used to characterize ASD (impaired communication, impaired social interaction, and restricted and repetitive interests and activities) occurs also in the general population with no clear distinction between pathological and common traits \cite{london2007role}.
This suggests that the gross behavioral descriptors of ASD may be poorly related to the biological mechanisms causing it, further complicating the research on this disorder.

\begin{figure}
    \centering
    \includegraphics[width=0.5\textwidth]{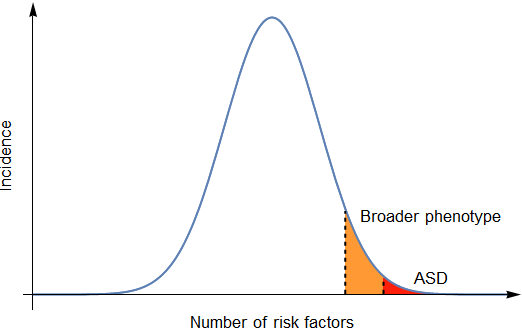}
    \caption{Multi-factorial threshold model applied to ASD. According to this model, ASD is caused by the accumulation of certain amount of risk factors, which are naturally present in the population, but are more numerous in ASD subjects. The broader phenotype can be explained by the presence of a lower but still supranormal number of risk factors.\\ 
    \mypictures{Ferrari2020autismChapter}
    }
    \label{fig:gaussian_asd}
\end{figure}

Several lines of evidence indicate a genetic susceptibility to ASD. For instance, the individual risk of ASD is much higher in families with one member already diagnosed with it \cite{folstein1977infantile,rutter2000genetic,bailey1995autism}: approximately 10 times greater in subjects with a sibling with ASD \cite{sandin2014familial}. Furthermore, in these families even undiagnosed members can show autism-related traits, termed "broader autism phenotype" \cite{spiker2002behavioral}.
Despite substantial progress has been made in understanding ASD genetic underpinnings, most ASD-associated DNA variations account for no more than 1\% of ASD cases and very few of them lead to ASD with high penetrance \cite{fernandez2017syndromic}.
All these considerations suggest a complex (or multi-factorial) nature of ASD \cite{ivanov2015autism}. According to various researchers, in fact, the ASD etiology may be explained in terms of a multi-factorial threshold model: ASD is caused by the interplay of genetic and environmental risk factors that contribute to reach a threshold that determines a diagnosable phenotype \cite{werling2013sex}.
In such a model, the broader phenotype can be explained by the presence of a lower, but not negligible, number of risk factors (see Fig. \ref{fig:gaussian_asd}).

To date, hundreds of genes linked to ASD were identified, that may contribute to its etiology \cite{rylaarsdam2019genetic}.
It remains elusive whether the heterogeneity of ASD may be explained by different genetic mutations converging on few shared molecular pathways or by different disorders with diverse mechanisms (similarly to intellectual disabilities) \cite{geschwind2008autism}.\\
Reproducible findings highlight the involvement of genes coding for synaptic proteins \cite{giovedi2014involvement} and modulators of apoptosis \cite{wei2014apoptotic} in the nervous system, together with upregulation of genes related to the immune system in ASD brain and peripheral tissues \cite{gupta2014transcriptome,estes2015immune}.
Altered apoptosis and synaptogenesis are thought to be involved in disrupted brain connectivity and early brain overgrowth \cite{wei2014apoptotic,piven2017toward}, the latter being repeatedly reported in neuroimaging studies \cite{amaral2008neuroanatomy,bonnet2018autism}. However, the works describing these structural brain features display several inconsistencies and often analyzed sub-optimal cohorts of subjects \cite{vasa2016disrupted, raznahan2013compared}.
Inflammation, combined with gastrointestinal comorbidities \cite{penzol2019functional} leads to hypothesize that gut abnormalities may cause neuroinflammation and brain dysfunction. However, gastrointestinal diseases do not occur in all the ASDs and neuroinflammation has also been observed in depression \cite{lotrich2009risk,dowlati2010meta}, bipolar disorder \cite{goldstein2009inflammation}, neurodegenerative diseases \cite{nuzzo2014inflammatory} and schizophrenia \cite{potvin2008inflammatory}.\\
Along with gastrointestinal comorbidities, also motor difficulties, which have not been considered core features of ASD so far, are recently gaining attention \cite{whyatt2013sensory,gowen2013motor}.
In fact, behavioral, neurophysiological, neuroimaging and histopathological studies observed abnormal motor system development in the majority of ASDs \cite{mosconi2015sensorimotor}.
It has been estimated that around 87\% of ASDs are affected to some extent by motor difficulty \cite{bhat2020motor}, ranging from an atypical gait to problems with handwriting, but again these phenomena are also not specific of ASD.\\
Another ASD feature coming to the fore is represented by hypo- and hyper-responsivity to sensory stimuli \cite{robertson2017sensory,balasco2020sensory}. This was historically believed to represent a secondary consequence of ASD social-cognitive problems. However, the high occurrence (more than 90\% \cite{marco2011sensory}) of atypical sensory experiences among ASDs and the correlation observed between measures of sensory sensitivity and autistic traits \cite{tavassoli2014sensory} in the general population represent strong arguments to reconsider the role of this feature in the pathophysiology of the disorders.\\
An attempt to explain the onset of ASD based on its canonical core symptoms, hypothesizes that abnormal development of the mirror neuron system interferes with imitation and leads to social and communication difficulties \cite{williams2008self}.
While several studies show structural and functional anomalies in the ASD brain regions involved in the mirror neuron system \cite{iacoboni2006mirror}, abnormalities have been found also in other regions and this model does not explain the normal performance of ASDs in imitation tasks involving objects \cite{hamilton2008emulation}.

In summary, from these ASD pathophysiology hypotheses (that should not be intended as an exhaustive list) it emerges that:
\begin{itemize}
    \item none of the proposed theories alone fully describes the heterogeneity of the spectrum;
    \item no biomarker has been found to clearly distinguish ASD from \ac{TDC} or from others disorders;
    \item the current definition of ASD provides gross descriptors of the disorder, but to understand its etiology it may be important to consider also common comorbidities and side traits of the disorder.
\end{itemize}

\section{Rett and Dup15q syndromes: two examples of difficult research related to ASD.}
In this section, two syndromes closely related to ASD are briefly described: the \ac{RTT}, that before the release of the V edition of the \ac{DSM} was included in the "Pervasive Developmental Disorders" along with "Autism Disorder" \cite{american2000diagnostic}, and the Dup15q syndrome, which is considered the most common syndromic form of ASD. Despite these being syndromes, i.e. with a clear and unambiguous genetic cause, they display a phenotypic heterogeneity that may shed light on the complexity of ASD and for this reason are described in this section.

RTT is a monogenic X-linked\footnote{Linked to the X chromosome} disorder caused by mutations of the MECP2 gene. Male carriers usually die within the second year of life from severe encephalopathy, while the clinical profile of females ranges from asymptomatic carriers to classical RTT. The latter is characterized by neurological deterioration after an initial period of apparently normal life, resulting in severe physical, social and cognitive impairments. Similarly to males, also females display a reduced life expectancy, but with a higher age of death: on average around 20 years \cite{hagberg2002clinical}.\\
RTT extreme gender bias is due to the fact that the responsible gene is located on the X chromosome, so male carriers can only express its mutated allele, while in females some cells express the mutated allele from one chromosome and the others from the opposite chromosome, thanks to a process called X-inactivation. It has been proved that asymptomatic female carriers and the ones with very mild symptoms show skewed X-inactivation that favors the inactivation of the mutated allele \cite{huppke2006very}, so non-uniform X-inactivation is one demonstrated source of clinical heterogeneity.\\ Another certain source of variability is given by the exact location of the mutation in the MECP2 gene. In fact, different locations of the mutation lead to different levels of impairment of the protein products coded by the gene, from no functional consequences, to binding instability, abnormal folding or truncation \cite{bienvenu2006molecular}.\\
Despite its phenotypical heterogeneity, in general RTT's symptomatology is much more severe than that of ASD, and this might seem in contradiction with the monogenic nature of RTT versus the alleged complex etiology of ASD. However, MECP2  proved to be an essential epigenetic regulator in human brain development, playing a role in the expression of thousands of genes involved in neural maturation and signalling \cite{della2016mecp2}; some of which seem to be associated also to ASD, probably accounting for the common traits of the two diseases \cite{lasalle2009evolving}.

Dup15q syndrome accounts for approximately 0.2-0.4\% of the ASD cases \cite{rarediseases} and it is caused by the partial multiplication of the 15th chromosome, in particular of the 15q11.2-q13.1 portion also referred to as the \ac{PWACR}, which comprises around one hundred genes.\\
The extra copies of that region may be more than one (besides duplication also triplication, tetrasomy and hexasomy have been observed), and they can be located in different positions, more often within the 15th chromosome or in a supernumerary chromosome \cite{finucane201615q}. The replicated portion of DNA can have different breakpoints (i.e. start and end positions), but must contain the PWACR to be causative of Dup15q syndrome. All these differences contribute to the heterogeneity of the disorder \cite{ortiz202115q} that in its mild manifestations may be barely recognizable, but it usually includes typical ASD communication impairment, seizure, intellectual disabilities, hypotonia, motor difficulties and sometimes also abnormal facial features. \\
In addiction, phenotypes resulting from maternal or paternal replications can differ significantly, because the PWACR region is subjected to genomic imprinting: an epigenetic process wherein a gene is differentially expressed depending on whether it has been inherited from the mother or from the father \cite{mao2000characteristics}.\\
Besides displaying a clear genetic cause, detectable through genetic tests, Dup15q is also the only subgroup of ASD to have a clear and explainable neuroimaging biomarker \cite{frohlich2016quantitative}. Its distinctive \ac{EEG} signature similarly appears also in subjects treated with drugs enhancing the effect of the GABA neurotransmitter, which is in fact regulated by genes of the multi-copied PWACR region.  
This promising finding suggests that functional brain imaging may provide biomarkers to diagnose non-syndromic forms of ASD too.

To summarize, two syndromes closely related to ASD and well defined from a genetic point of view show a wide phenotypical heterogeneity. Considering that the sources of variability here reported (and also others \cite{pizzo2019rare}) may exist for each single gene or cluster of genes involved in the multifactorial model of ASD etiology, it becomes clear why the research on ASD is so riddled with difficulties. Furthermore, for RTT and Dup15q it has been shown that genetic traits apparently opposed (such as mutation causing a complete loss of function of MECP2 and duplication of that gene, or duplication of the PWACR and its deletion) actually induce some common symptoms \cite{na2013impact,cassidy2000prader}. Thus, the ASD category, solely defined by gross behavioural descriptors, may include subtypes caused by different and opposed mechanisms, which again complicates the understanding of the disorder through classical ASD-vs-TDC comparison.

\section{Research objectives and challenges}\label{cap1_research_objectives}
Current research is far from a cure for ASD \cite{zaky2017autism}. As of today, the main efforts are focused on the identification of a biomarker to anticipate the diagnosis and understand the nature of the disorder. However, the research of a biomarker is hampered, in the first place, by the uncertainty in defining the boundaries of ASD. In fact, the ASD population identified by current diagnostic tools may include subjects affected by other neurodevelopmental disorders or healthy controls with a clinical profile similar to mild forms of the disorder. For instance, it has been shown that the percent agreement for diagnostic classification (i.e. ASD/non-spectrum) among clinically trained examiners using the \ac{ADOS} is around 64–82\% \cite{zander2016objectivity}.\\
As it emerges from Sec. \ref{causes_of_ASD}, another important concern regards the possibility that different disorders fall under the current definition of ASD, which for this reason may not be explained by a single biomarker. The fact that the Dup15q EEG signature does not characterize non-syndromic forms of ASD seems to support this hypothesis. In fact, there is an increasing interest in decomposing the heterogeneity of ASD, identifying more uniform subtypes that can individually be distinguished from TDCs \cite{lombardo2019big}.
However, finding a single biomarker, complex at will,  would help to facilitate the diagnosis of all the ASDs at once and to understand what the biological causes shared by all the ASDs are. 

\chapter{Neuroimaging}
\label{chap:bkg_neuro}
\section{Introduction}
Neuroimaging refers to several techniques that allow to investigate brain structure and functioning. They can be divided into:
\begin{itemize}
\item \textit{Techniques based on the use of ionizing radiation.}\\
They include computed tomography, single photon emission computed tomography and positron emission tomography. The first one differentiates the tissues by measuring how much radiation passes through them. The other ones provide functional and metabolic information by scanning how a radioactive tracer, injected into the patient, is distributed into the body.
\vspace{0.2cm}
\item \textit{Techniques based on the direct or indirect detection of brain electrical activity.}\\
They include EEG and magnetoencephalography which both provide functional information on the brain.
\vspace{0.2cm}
\item \textit{Techniques based on the nuclear magnetic resonance.}\\
They are collectively called \ac{MRI} and can be of three types: \ac{sMRI}, \ac{fMRI} and \ac{rsfMRI}. The first one provides morphological information allowing a differentiation of the tissues based on how they interact with magnetic fields. The others, instead, give functional information studying how the magnetic properties of the tissues changes over time during brain activity.
\end{itemize}
These techniques are employed for diagnosis, disease monitoring, treatment planning and during post treatment follow up.
However, with respect to the others, the MRI presents various advantages that makes it more suitable for open-ended research, such as:
\begin{itemize}
     \item its high spatial ($\sim$ $mm^3$) and temporal resolution ($\sim$ second);
     \item it does not use ionizing radiations, thus it is a safe exam that can be repeated for longitudinal studies;
     \item within a single MRI acquisition session and with the same instrumentation it is possible to acquire both structural and functional information;
     \item usually it does not require any particular preparation for the patient\footnote{MRI requires a preparation of the patient only when: (1) the subject needs to be sedated to maintain a fixed position during the exam; (2) the effects of a certain stimulus or condition (such as sleep deprivation) are analyzed. Instead, positron emission tomography requires the injection of a radioactive tracer and EEG needs a preparation of the scalp area to avoid signal degradation.}
 \end{itemize}
For these reasons, collections of MRIs for research purposes are usually larger than other types of neuroimaging data. Besides the availability of data and their high resolution, it must be considered that finding a biomarker based on a safe and preparation-free medical exam is highly desirable for diagnostic practice. As a consequence, neuroimaging research on ASD, whose goal is to simplify diagnosis, has been mainly based on the analysis of MRI signals.\\
This chapter provides a description of MRI and how it is usually analyzed to study ASD.
\section{Magnetic resonance imaging}
\subsection*{Signal generation process}
\begin{figure}
     \centering
     \begin{subfigure}[b]{0.3\textwidth}
         \centering
         \includegraphics[width=\textwidth]{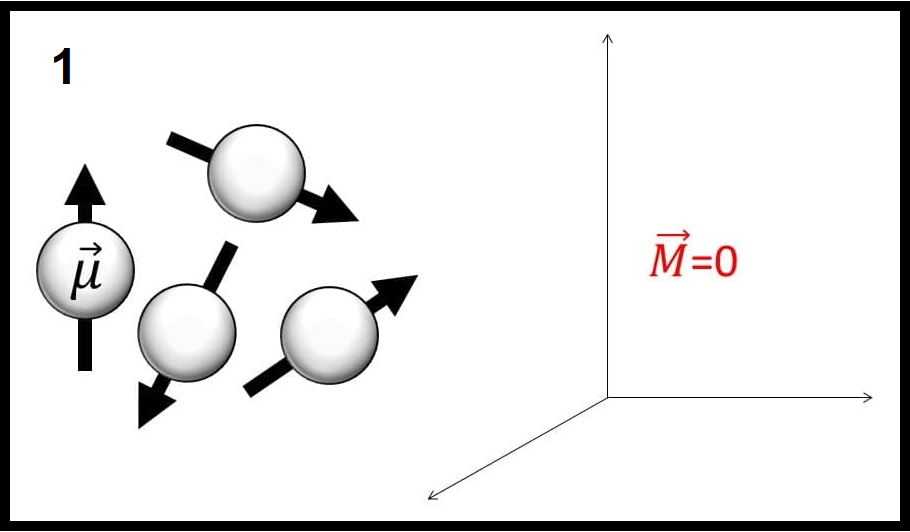}
         \label{fig:mri_spins_1}
     \end{subfigure}
     \hspace{-2.3mm}
     \begin{subfigure}[b]{0.3\textwidth}
         \centering
         \includegraphics[width=\textwidth]{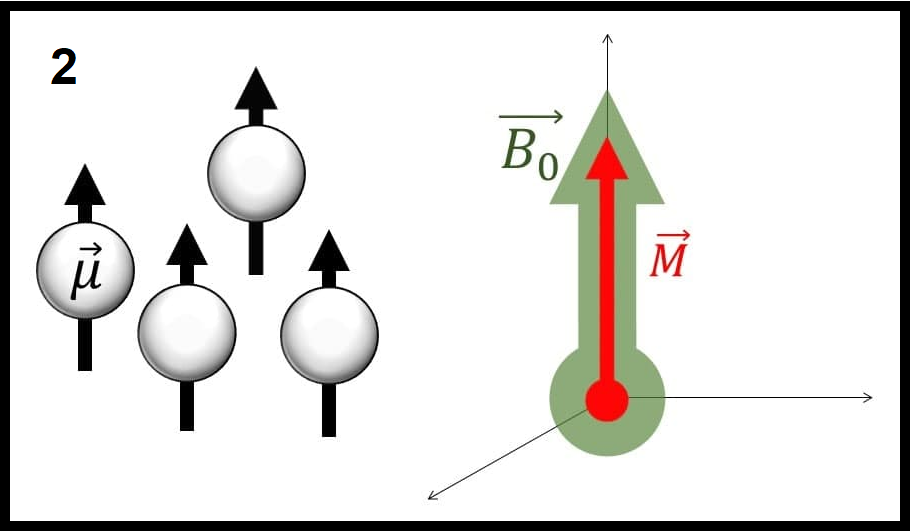}
         \label{fig:mri_spins_2}
     \end{subfigure}
     \hspace{-2.3mm}
     \begin{subfigure}[b]{0.3\textwidth}
         \centering
         \includegraphics[width=\textwidth]{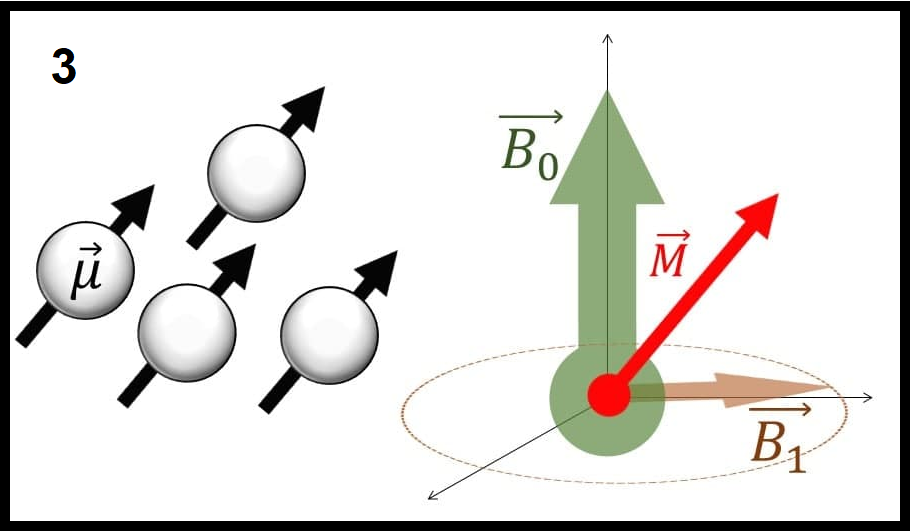}
         \label{fig:mri_spins_3}
     \end{subfigure}\\
     \vspace{-5.75mm}
     \begin{subfigure}[b]{0.3\textwidth}
         \centering
         \includegraphics[width=\textwidth]{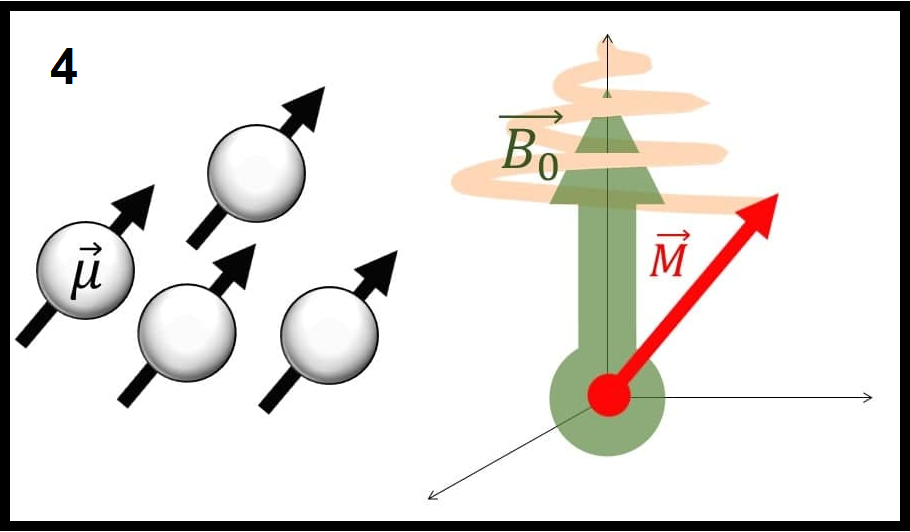}
         \label{fig:mri_spins_4}
     \end{subfigure}
     \hspace{-2.3mm}
     \begin{subfigure}[b]{0.3\textwidth}
         \centering
         \includegraphics[width=\textwidth]{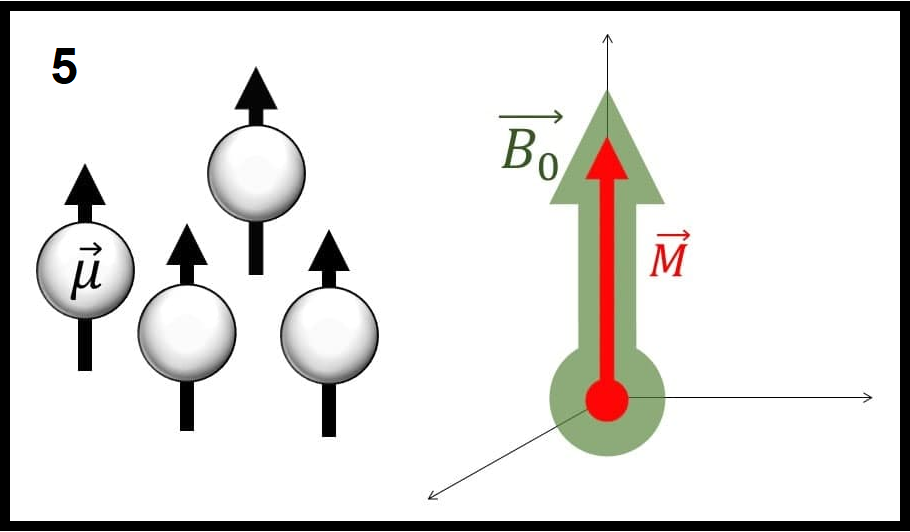}
         \label{fig:mri_spins_5}
     \end{subfigure}     
        \caption{Steps of MRI signal generation. Step 1 illustrates the randomly oriented magnetic moments in a tissue. In step 2 a constant magnetic field $\vec{B_0}$ is applied to the tissue, thus its $\vec{\mu}$ align to the field, giving rise to $\vec{M}$. In step 3 an electromagnetic radiofrequency field $\vec{B_1}$ is further applied, perturbing the equilibrium state of step 2. Step 4 shows in light red the typical motion of the magnetization $\vec{M}$ when the perturbation has ended. Finally, at step 5 the system comes back to the equilibrium state.}
        \label{fig:mri_spins}
\end{figure}

To understand MRI signal generation, the human body has to be considered at a nuclear level, simply as a distribution of nuclei with randomly oriented magnetic moments $\vv{\mu}$ (see Fig. \ref{fig:mri_spins}). \\
During an MRI exam, a uniform and constant magnetic field $\vv{B_0}$ is applied to the portion of the body under examination, aligning its $\vv{\mu}$s along the line of force of the field to reach an equilibrium state. In this configuration, the aligned $\vv{\mu}$ will jointly produce a magnetization $\vv{M}$, whose intensity depends on the density $\rho$ of the nuclei.\\
After the magnetic moments have been aligned to the static field, an electromagnetic radiofrequency field  $\vv{B_1}$ is applied, altering the value of $\vv{M}$. When the perturbation $\vv{B_1}$ ends, the nuclei magnetic moments relax and return to their equilibrium state, bringing $\vv{M}$ back to its equilibrium value reached when only $\vv{B_0}$ was applied to the system.
The observable and measurable relaxation of $\vv{M}$ happens on two distinct time scales, characterized by the temporal constants $T_1$ and $T_2$, that depend on the chemical composition of the tissues under examination.

Thus, the MRI signal measures the variations of $\vv{M}$ during the relaxation phase. These variations depend on three parameters: $\rho$, $T_1$ and $T_2$, on the basis of which the various tissues can be differentiated.\\
By carefully choosing the MRI acquisition parameters, it is possible to make the detected signal dependent mainly on one of the three parameters; in this case the signal is said to be "weighted" in one parameter. 

\subsection*{Structural and functional imaging}
\begin{figure}
    \centering
    \includegraphics[width=0.45\textwidth]{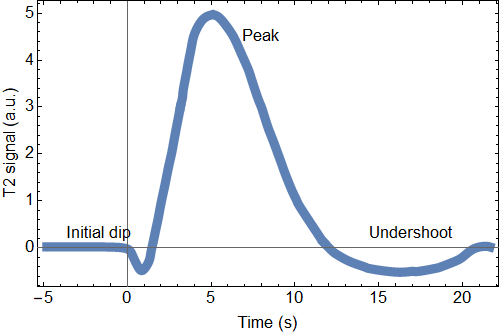}
    \caption{Visualization of the temporal evolution of the BOLD signal.}
    \label{fig:bold}
\end{figure}

Structural MRIs are 3D images, that, depending on the spatial resolution adopted, can contain by up to tens of millions of voxels (i.e. 3D pixels) and occupy tens of MBs \cite{Ferrari2020autismChapter}, when decompressed. They are commonly $T_1$-weighted since this allows to better differentiate the brain tissues: meninges, cerebrospinal fluid, white matter and grey matter.\\
Functional MRIs are 4D images (three spatial dimensions and a temporal one) and depending on the protocol adopted, they can contain between 10 and 100 millions of data points, corresponding to 20-200 MBs \cite{Ferrari2020autismChapter}. Functional MRI indirectly measures brain activity, thanks to the opposite magnetic properties of oxygenated\footnote{Oxygenated hemoglobin is diamagnetic.} and deoxygenated hemoglobin\footnote{Deoxygenated hemoglobin is paramagnetic.} whose ratio changes during neural activity. Variations of this ratio, called \ac{BOLD} signal, can be detected acquiring temporal sequences of $T_2$-weighted images. The BOLD signal is characterized by three elements \cite{faro2006functional} (see Fig. \ref{fig:bold}):
\begin{itemize}
    \item \textit{Initial dip.}\\
    Increased neural activity in a certain region of the brain induces a higher consumption of glucose and oxygen from the local capillary bed, which causes a decrease in the BOLD signal.
    \vspace{0.2cm}
    \item \textit{Asymmetric peak with fast rise and slower fall.}\\
    After the initial reduction of oxygen and glucose, local cerebral blood flow increases to carry more of these molecules to the area. The oxygen brought in typically exceeds the local rate of oxygen consumption in burning glucose and this causes a rapid increase of the BOLD signal followed by a slower fading, due to gradual oxygen consumption, in the active area.
    \vspace{0.2cm}
    \item \textit{Final undershoot.}\\
    When neural activity comes back to normal, cerebral blood flux drops down drastically but the local vasodilation decreases more slowly, leading to an undershoot of the BOLD signal.
\end{itemize}

The acronym fMRI refers to the $T_2$-weighted temporal signal acquired during a functional MRI independently from the experimental condition. Conventionally, the term rsfMRI is used when the acquisition is performed while the subject is at rest and as a consequence, when no extra information is given, fMRI refers to a task-based acquisition.
\subsection*{Non-quantitative nature of MRI images}\label{MRI-non-quantitative}
To correctly interpret MRI data, it is important to note that:
\begin{compactitem}
\item even in the images that are said to be weighted on one parameter, it is impossible to fully eliminate the dependency of the signal from the other two parameters;
\item in addition, the MRI images do not exclusively depend on the magnetic/chemical properties of the tissues, but also on the acquisition settings in a complex and until now undetermined way.
\end{compactitem}
Thus the MRI signal depends on a mixture of $\rho$, $T_1$ and $T_2$, as well as extraneous factors. These factors include hardware characteristics, acquisition parameters and unavoidable and undetectable defects in $\vv{B_0}$ and $\vv{B_1}$ \cite{faro2006functional,jezzard1999sources}. This nonlinear blend of signal sources makes the physical interpretation of the measured signal challenging and precludes any direct comparison of intensity values across subjects, time-points or imaging centers \cite{deoni2010quantitative}.\\
For these reasons, MRI is commonly said to be a non quantitative technique \cite{basser2011microstructural}, which should not be used as a scientific measuring instrument but as a camera that allows to see brain morphology and functioning.
\section{Machine learning studies based on MRI images}
\label{sec:bkg_studies_on_mri}
The research of a MRI-based biomarker has been initially addressed using standard univariate analyses, comparing single brain morphological or functional properties between ASDs and TDCs. However, the inconsistency of results prompted a move towards the adoption of multivariate approaches and in particular of \ac{ML} \cite{mosconi2006structural}. \\
In this section, first an introduction on ML is provided, then the typical workflow analysis of an MRI-based classification study is presented. Finally, the state-of-the-art on MRI classification studies of ASD is described, followed by a discussion on the challenges of this area of research and some possible solutions to address them.
\subsection{Basic notions on machine learning}
The following paragraphs introduce the terminology and some important ML practices that are needed to understand the review of the literature and the methodological choices made in this work. 

A mathematical description of ML and \ac{DL} algorithms is out of the scope of this thesis, but can be found in dedicated textbooks, such as \cite{goodfellow2016machine,goodfellow2016deep}.\\
However, to facilitate the comprehension of some technical details described in the Materials and Methods part, Appendix \ref{app:neural_networks} provides a brief description of the \ac{NN} paradigm. 
\subsubsection*{Basic definitions}
ML refers to a class of computational models able to discover patterns from data without being explicitly programmed to do so. Among the various ML approaches, \ac{SML} is particularly suited for diagnosis applications. SML is characterized by a training phase, in which the algorithm tries to find a pattern that maps an input to an output based on example input-output pairs, called training set. This is achieved through an optimization process that searches the best set of parameters of the ML model that minimizes a cost-function related to the mismatch between the predicted output and the true output on the training set. \\
Contrary to standard regressors, SML does not necessarily need any assumption on the functional form of the mapping function, which is instead learnt automatically during the optimization process. Thanks to this feature, SML has been widely adopted for open-ended applications.\\
For research on ASD diagnosis, SML is trained to distinguish between ASD and TDC subjects, using for instance their MRI derived features (or raw images) as inputs and their corresponding diagnosis labels as outputs. Given that the output labels are classes (i.e. ASD or TDC), such a model is called a classifier and for each input instance it returns a probability of that instance to belong to one of the two classes, which is complementary to the probability of belonging to the other class. 

Within the ML context, some confusion may occur in the distinction between ML and DL. DL is a subcategory of ML, composed by deep NN algorithms. NNs are a family of ML models that learn to map the inputs to their outputs through a sequence of mathematical operations which define multiple and progressively abstract representations of the input information, called layers. NNs with more layers are said to be deeper, but a minimum number of layers characterizing DNN has not been defined \cite{deng2014deep}.
A more thorough description of NNs is provided in Appendix \ref{app:neural_networks}\\
From a practical point of view, the main difference between DL and other ML algorithms is that the former ones have a much larger number of internal parameters that are optimized in the training phase. 
As a consequence, DL can discover much more complex patterns, which means that it can work with raw or minimally pre-processed MRI data, at the cost of requiring a larger number of examples in the training set.
\subsubsection*{Performance evaluation}
The goodness of a trained classifier, being either a ML or a DL algorithm, is usually evaluated on a set of input-output pairs, called validation set, with no intersection with the training set. Different metrics exist to evaluate the performances of a binary classifier, which are described in Box \ref{test}.\\

\newtcolorbox[auto counter]{mytable}[2][]{float=htb, halign=left,  title=Box ~\thetcbcounter: {#2}, every float=\centering, #1}
\begin{mytable}[label=test,float=ht!]{Performance metrics for binary classifiers}
\small
When evaluating the results of a classification problem, the two classes are generically defined with the terms positive and negative. The positive class is used by convention to refer to the class that represents the condition to be detected/recognized, if there is any.\\
As already said, in a classification task the model returns a probability and not a categorical class of membership. Thus, to decide whether a certain instance has been correctly or wrongly predicted it is necessary to define a threshold above/below which the predicted probability is assumed to indicate the membership to the positive/negative class. Since this probability is in the range $[0,1]$, an obvious threshold is 0.5.\\
Once the classification threshold has been defined, it is possible to draw the confusion matrix:\\
\vspace{0.1cm}
{
\centering
    \begin{tabular}{|c|c|cc|}
         \hline
         \multicolumn{2}{|c|}{}&\multicolumn{2}{c|}{\cellcolor{white!20!gray!40}{True classes}}  \\
         \cline{3-4}
         \multicolumn{2}{|c|}{}&\cellcolor{blue!50!green!30}{Positive}&\cellcolor{blue!50!green!30}{Negative}\\
         \hline
         \cellcolor{white!20!gray!40}{}& \cellcolor{blue!50!green!30}{Positive}& \ac{TP}& \ac{FP}\\
         \multirow{-2}{*}{\cellcolor{white!20!gray!40}{Predicted classes}}& \cellcolor{blue!50!green!30}{Negative}& \ac{FN}&\ac{TN}\\
         \hline
    \end{tabular}\\
\vspace{0.12cm}
}

From the confusion matrix the most used metrics to evaluate the classification performance can be defined:
\begin{itemize}
    \item \textbf{\ac{Acc}} = $\frac{TP+TN}{TP+FP+FN+TN}$
    \item \textbf{\ac{Prec}} = $\frac{TP}{TP+FP}$
    \item \textbf{\ac{TPR}} = \textbf{\ac{Sens}} = $\frac{TP}{TP+FN}$
    \item \textbf{\ac{TNR}} = \textbf{\ac{Spec}} = $\frac{TN}{FP+TN}$
    \item \textbf{\ac{FPR}} =$\frac{FP}{FP+TN}$
    \item \textbf{\ac{FNR}} =$\frac{FN}{TP+FN}$
\end{itemize}

Each of them represents the percentage of a certain group of predictions (i.e. all the correctly classified data, TP, FP, etc.) over a certain category of data (i.e. all the data, all the predicted positives, all the true positives, etc.), ignoring everything else. Thus, one indicator is not sufficient to fully represent the performance of a classifier. For instance, a predictor with a low number of TP and high number of TN may reach a good accuracy but a very low precision. Furthermore, the value of all these metrics depends on the chosen classification threshold.\\ 
A more comprehensive metric can be computed by plotting the \ac{ROC} curve, i.e. the TPR vs FPR obtained at different classification thresholds. The \ac{AUC} is independent from the classification threshold and takes into account both the TPR and FPR, which are complementary to the FNR and TNR, respectively, so the information provided by this metric is complete. 
The Fig. below shows an example of a ROC curve and its associated AUC.\\
{
\centering
\includegraphics[width=0.3\textwidth]{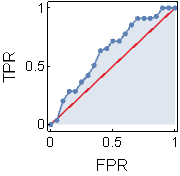}\\
}
Being both the TPR and FPR two percentages, the AUC has values in the range [0,1].\\
The red line indicates the ROC curve of a classifier guessing randomly, whose AUC would be 0.5. Thus this value sets the baseline to which the AUCs obtained must be compared: a classifier that has learned a true pattern should have an AUC significantly greater than 0.5.

\end{mytable}

Depending on the application of interest, one metric may represent the utility of the classifier better than the others. For instance, for a screening, capturing most of the positive cases is essential, which means that sensitivity is more important than specificity. \\
In the ASD context, considering that research is very far from the definition of a classifier of clinical utility and that the main aim of this kind of studies is to understand whether the input data actually contains the information necessary for an objective diagnosis, errors on both the categories should be equally considered. Thus, the AUC seems to be the most suited metric, being also threshold-invariant.\\

To guarantee that the performance measured on the validation set really reflects the overall performance of the classifier, a method called K-fold cross-validation is often employed. Basically, according to this strategy, the available data are divided into K-subsets and the model is trained on all but one of the subsets and validated on the remaining one. This process is repeated for all the K possible configuration of training and validation.\\ The mean and standard deviation of the performance measured on all the validation sets (that all together form the entire dataset) are considered a good measure of the classifier's overall performance and of its uncertainty, respectively. 
\subsubsection*{Good practices in ML applications}
In this paragraph, some important, but sometimes overlooked, practices in the design of a ML study are described \cite{scheinost2019ten, quaak2021deep}:
\begin{enumerate}
    \item \textit{Validation-independent pre-processing.}\\
   Whenever the data properties play a role in the definition of a possible pre-processing operation, only the training set must be used to define the pre-processing function to be applied to all the instances.
    If this rule is broken, the training set can carry some information of the entire dataset, which means that the performance estimated on the validation set would be biased and thus not reliable.\\
    Note that also filtering the input data is considered a pre-processing operation, thus, feature selection should be performed only considering the training set. 
    \vspace{0.2cm}
    \item \textit{Class-independent pre-processing.}\\
    Data should be pre-processed in the same way, independently from their class membership. Otherwise, treating differently data belonging to distinct classes may artificially introduce group differences. As a result, the classification performances will be overestimated. Furthermore, this approach is unfeasible out of an exploratory analysis and can produce only useless classifiers. In fact, even if an ML trained in such a way would display high performances, it could be used only on data pre-processed in the same way, which means that their true membership class should be known before classification.
    \vspace{0.2cm}
    \item \textit{Performance evaluation on truly unseen data.}\\
    When multiple classifiers (that may differ in their underlying algorithms, parametrization or input data, etc.) are compared and the best one is chosen based on the performance on the validation set, this value does not represent anymore a reliable estimate of the true classification performance. In fact, given that the validation set has been used for model selection, its role has now become more similar to that of the training set. 
    Thus, to fairly evaluate the performance of the chosen classifier, a new, completely unseen dataset must be used, which is usually called test set.
    \vspace{0.2cm}
    \item \textit{Generalizability assessment on data drawn from a distribution different from the one of the training set.}\\
    The data used in a study are always somehow biased, because of the selection criteria or the acquisition modalities adopted. As a consequence, the performance estimated on a validation (or test) set, composed by instances chosen according to the same data generation/selection process of the training set, does not provide an estimate of the generalizability of the classifier and of its clinical utility. For this reason, when possible, it is recommended to validate or test the performance on a set of data slightly different from the one used for model training and selection.
    For instance, given that as already explained the MRI signal depends also on acquisition settings that are difficult to regress out, in an MRI study it is recommended to test the performance of the classifier on a dataset acquired from a different scanner than the one (or ones) used for the training set. 
\end{enumerate}
Overlooking one of the first three rules often implies that the reported results are overestimated and in any case not reliable, while the last one is just a good practice that may help in understanding the true impact of a study. Unfortunately, many neuroimaging works on ASD mentioned in the literature review reported in Sec. \ref{soa_neuroimaging} do not comply with the third rule and in some cases the description of the methodology does not clarify whether the first two have been thoroughly followed. The fourth recommendation is instead rarely applied \cite{quaak2021deep}.\\ 
The ML part of the present study instead has been designed in order to adhere to the mentioned principles.
\subsection{MRI-based classification studies on ASD}\label{MRI_classification_ASD}
In the last 25 years, many works studied MRI-based characteristics of the ASD brain through ML and DL techniques. Only in the last 5 years, over 2000 papers on predictive models for different neurological disorders have been published, with a substantial number of them focusing on ASD \cite{biswas2020prediction}.
\subsubsection*{Workflow of an MRI-based classification study}
Despite the great variety of workflows adopted in MRI-based classification studies, they can be essentially divided in two strategies \cite{khodatars2021deep}:
\begin{enumerate}
    \item \textit{Heavy pre-processing and data manipulation followed by a ML analysis.}\\
    Typically the raw MRI images are pre-processed using dedicated software such as Freesurfer \cite{fischl2012freesurfer}, SPM \cite{penny2011statistical} and FSL \cite{jenkinson2012fsl} which provide various pipelines both for sMRI and fMRI (or rsfMRI) comprising low and high level pre-processing steps that can be applied integrally, partially or even customized (see Tab. \ref{tab:mri_preproc}). 
    Either using the mentioned pipelines or with any other technique, a common step characterizing this strategy is the reduction of the MRI data to a very condensed form of information. For sMRI, a vector of morphological features such as volume, surface area and thickness of various brain regions is usually extracted \cite{ferrari2020dealing,hrdlicka2005subtypes,van2018cortical}. For fMRI and rsfMRI instead, a set of temporal signals characterizing each brain region is obtained. These data are often further processed and transformed into functional correlation matrices, showing the signal correlation between each pair of regions \cite{monk2009abnormalities,rausch2016altered,alaerts2016sex}.
    After the pre-processing step, given the low dimensionality of the resulting data, any ML algorithm can be applied to perform the final classification task.
    \vspace{0.2cm}
    \item \textit{Minimal or no pre-processing followed by a DL analysis.}\\
    In this approach the process of feature extraction and classification is accomplished intelligently and integrally by DL. Despite various DL algorithms proved to be insensitive to noise and different resolution and orientations during machine vision contests with a large number of available data \cite{goodfellow2013challenges,mundhenk2016large}; the scarcity of MRI images often motivates the adoption of some low level pre-processing to denoise and standardize the inputs. In case of both minimal and no pre-processing at all, the core of this type of analysis is the design of the DL-based classifier that is tasked to discover discriminating spatial patterns in 3D sMRIs and spatio-temporal ones in 4D fMRIs or rsfMRIs.  
\end{enumerate}
The first strategy is the most explored one in literature. In fact, the high dimensionality of the data, the strong non-biological noise (i.e. head movements, acquisition artifacts, etc.) and the non-quantitative nature of the MRI suggest that more digested data could isolate better the biological information of interest. Furthermore, given that medical studies usually have access to a small number of instances, reducing their dimensionality has also the objective to prevent from overfitting the data during the ML training.\\
However, the main drawbacks are that: information useful for classification may be lost during the pre-processing, each operation potentially introduces new errors and finally the large number of arbitrary choices in the pre-processing reduces confidence in the final result. In addition, the expected removal of the data dependency on external factors such as acquisition modalities is not assured, as shown in Appendix \ref{site_dependency}.\\
Among the ML algorithms, the one that has been adopted the most by works following the first strategy is the \ac{SVM} \cite{hyde2019applications,biswas2020prediction,rashid2020use}, that searches in the feature space (or in a higher-dimensional space based on a user-defined function) the hyperplane that maximizes the separation between the two classes \cite{suthaharan2016support}.

\begin{table}\centering
\begin{tabular}{|l|l|l|l|}

\hline                   
& \textbf{Operation} & \textbf{MRI data type} & \textbf{Description}  \\
\hline
\multirow{16}{1.8cm}{\textbf{Low-level}} &
\multirow{4}{2.2cm}{Upsampling and downsampling} & \multirow{4}{2cm}{structural functional} & \multirow{4}{6cm}{Standardizing the dimensionality of a dataset by increasing/decreasing the spatial and/or temporal resolution of its images.}\\
& & &\\
& & &\\
& & &\\
\cline{2-4}
& \multirow{2}{2.2cm}{Skull stripping} & \multirow{2}{2cm}{structural functional} & \multirow{2}{6cm}{Removal of the skull.}\\
& & &\\
\cline{2-4}
 & \multirow{2}{2cm}{Spatial smoothing} & \multirow{2}{2cm}{structural functional} & \multirow{2}{6cm}{Application of a gaussian filter to smooth the image.}\\
& & &\\
\cline{2-4} 
 & \multirow{2}{2cm}{Motion correction} & \multirow{2}{2cm}{structural functional} & \multirow{2}{6cm}{Alignment of all the slices (and frames) to a reference one.}\\
& & &\\
\cline{2-4} 
 & \multirow{3}{2cm}{Intensity normalization} & \multirow{3}{2cm}{structural functional} & \multirow{3}{6cm}{The image gray levels are brought to a common scale across subjects and scanners.}\\
& & &\\
& & &\\
\cline{2-4} 
 & \multirow{2}{2cm}{Registration} & \multirow{2}{2cm}{structural functional} & \multirow{2}{6cm}{The image is aligned to and morphed into a reference space.}\\
& & &\\
\cline{2-4} 
 & \multirow{4}{2cm}{Temporal filtering} & \multirow{4}{2cm}{functional} & \multirow{4}{6cm}{Removal of unwanted frequencies, resulting from technical sources and physiological fluctuations like breathing and heartbeat.}\\
& & &\\
& & &\\
& & &\\
\cline{2-4} 
 & \multirow{2}{2cm}{Slice timing correction} & \multirow{2}{2cm}{functional} & \multirow{2}{6cm}{Correction for slice-dependent delays.}\\
& & &\\
\cline{2-4} 
\hline
\multirow{10}{1.8cm}{\textbf{High-level}} 
& \multirow{2}{2cm}{Parcellation} & \multirow{2}{2cm}{structural functional} & \multirow{2}{6cm}{Images are parcellated according to a certain atlas.}\\
& & &\\
\cline{2-4} 
& \multirow{4}{2cm}{Feature extraction} & \multirow{4}{2cm}{structural functional} & \multirow{4}{6cm}{Estimation of high-level features such as brain volume or temporal properties of the functional MRI in the whole brain and in its regions.}\\
& & &\\
& & &\\
& & &\\
\cline{2-4}
 & \multirow{2}{2cm}{Signal averaging} & \multirow{2}{2cm}{functional} & \multirow{2}{6cm}{Average estimation of the MRI signals within brain regions.}\\
& & &\\
\cline{2-4} 
& \multirow{3}{2cm}{Sliding window} & \multirow{3}{2cm}{functional} & \multirow{3}{6cm}{Feature extraction of temporal properties calculated in a time-moving window.}\\
& & &\\
& & &\\
\cline{2-4} 
 & \multirow{3}{2cm}{Fourier feature extraction} & \multirow{3}{2cm}{functional} & \multirow{3}{6cm}{Extraction of temporal features based on Fourier analysis.}\\
& & &\\
& & &\\
\cline{2-4} 
 & \multirow{2}{2cm}{Functional connectivity} & \multirow{2}{2cm}{functional} & \multirow{2}{6cm}{Computation of the connectivity matrix.}\\
& & &\\
\hline
\end{tabular}
\caption{Summary of the most commonly employed pre-processing techniques for MRI data.}
\label{tab:mri_preproc}
\end{table}

\begin{figure}
    \centering
    \includegraphics[width=0.6\textwidth]{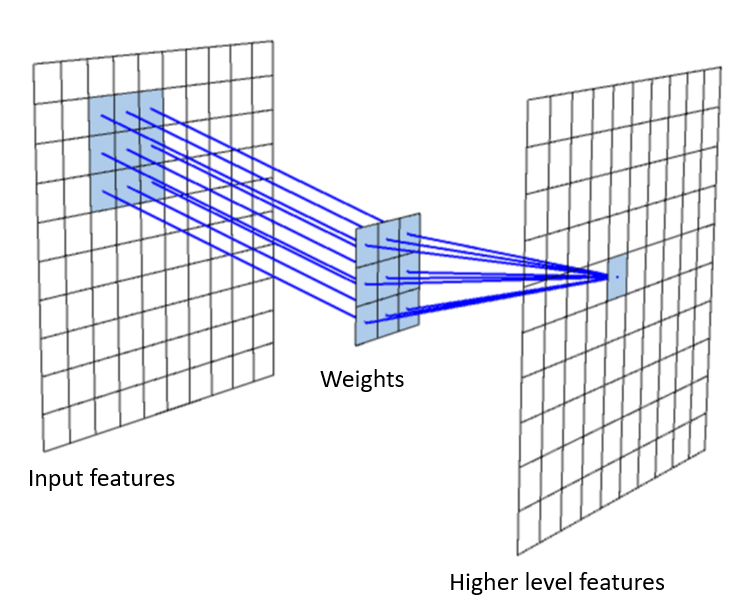}
    \caption{Representation of a convolutional layer. Parts of the input, usually an image, are processed by a set of weights that form the receptive field of the layer, and condensed into a single high-level feature. This process is repeated scanning the whole input with the same receptive field, which is thus able to extract spatially invariant features.}
    \label{fig:cnn}
\end{figure}

The second workflow is less explored in the history of MRI-based ASD research and it is mainly adopted in more recent works \cite{ingalhalikar2021functional}. In fact, starting from 2012 \cite{krizhevsky2012imagenet} machine vision research has seen a huge raise in performances, showing that for certain problems of image classification DL can match \cite{ciregan2012multi,esteva2017dermatologist,taigman2014deepface} and overcome \cite{ciregan2012multi,he2015delving,krizhevsky2012imagenet} human abilities. These milestones have been achieved mainly thanks to the use of a specific type of NN, called \ac{CNN} \cite{lecun1998gradient}, that contains convolution operations able to extract spatial features with tunable size (see Fig. \ref{fig:cnn}).\\

\begin{figure}
    \centering
    \includegraphics[width=0.5\textwidth]{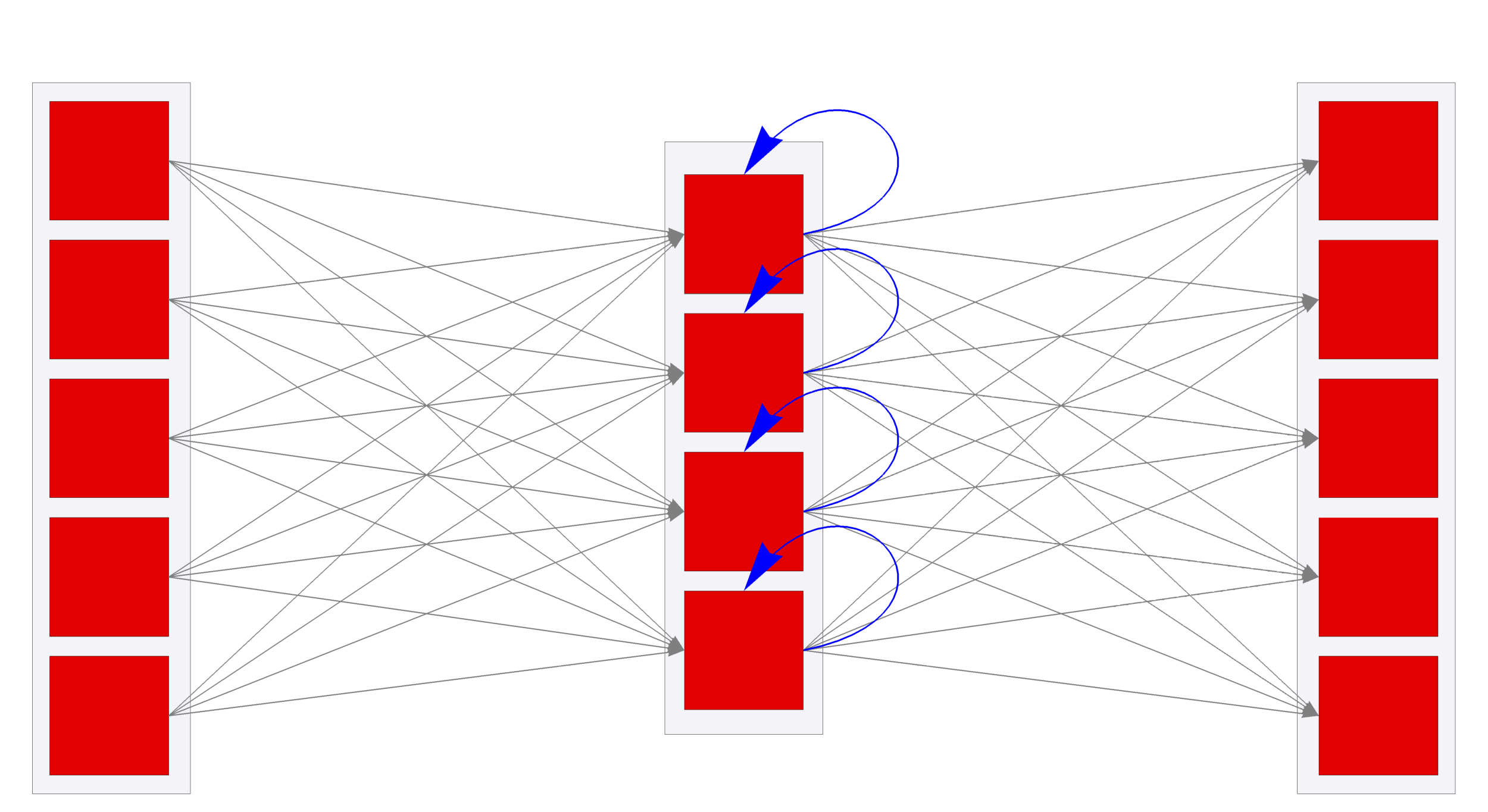}
    \caption{Structure of an RNN. An RNN is based on the concept of memory. Each layer retains memory of what has previously passed through it, thanks to connections to itself (the blue arrows in the picture). Temporal data are treated as a stream of information and these connections allow the analysis and extraction of temporal features from values that would not simultaneously available to the layer. \mypictures{Ferrari2020autismChapter}}
    \label{fig:rnn}
\end{figure}

\begin{figure}
    \centering
    \includegraphics[width=0.75\textwidth]{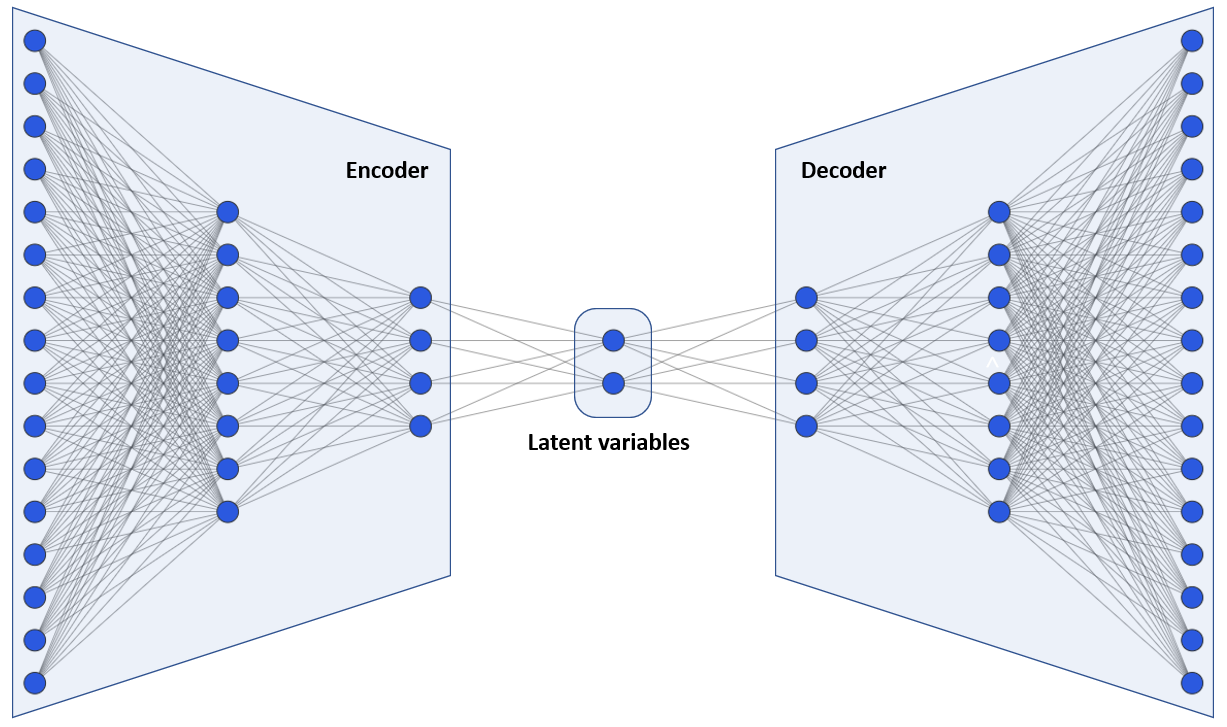}
    \caption{Schematic of the functioning of an auto-encoder. The encoder block compresses the input into a latent variable representation, eliminating noise and superfluous information. The decoder is then used to reconstruct the input from the latent variable space.}
    \label{fig:latent_variables}
\end{figure}

To date, most of the neuroimaging studies adopting the second workflow use DL architectures based on 3D-CNN (i.e. CNN adapted to process 3D images) and 4D-CNN to analyze sMRI and fMRI/rsfMRI respectively. The 4D-CNN basically treats the temporal dimension exactly as the other spatial ones, which is clearly a simplistic approach. For instance, usually an algorithm should look for rotationally invariant spatial patterns, but in the temporal dimension there is a preferred direction. Another type of DL commonly adopted for the analysis of fMRI or rsfMRI is the \ac{RNN}\cite{sherstinsky2020fundamentals}, which processes the temporal signal as a stream of data (in this case 3D images) and contains memory retaining layers that simultaneously process the current input and the previous ones (See Fig. \ref{fig:rnn}). Despite treating the temporal dimension differently from the spatial ones, its ability to find temporal patterns is limited by the memory retained by its recurrent layers.
A slightly different solution, falling under the definition of the second strategy, despite sharing some commonalities with the first one, consists in the use of an \ac{AE} \cite{meng2017relational,wang2016auto} followed by a standard ML classifier. An AE is a special type of DNN that tries to reproduce the inputs through the use of an encoder that first compresses the data and a decoder that expands it back to the input dimensionality (See Fig. \ref{fig:latent_variables}). AEs are used as an intelligent pre-processing operation, that finds a compressed and denoised representation of the inputs that preserves the information necessary to reproduce them. CNNs, RNNs and AEs are the most adopted forms of DL in the study of MRI data \cite{khodatars2021deep,rashid2020use}.

Summarizing, independently on the specific DL algorithm used, the adoption of the second strategy allows to analyze the entire information encoded in the MRIs and it has thus the potential to extract the complex pattern expected to characterize a heterogeneous disorder such as ASD. However, the ability to find complex relationships in high dimensional images can be also dangerous as it allows the algorithm to fit noise or spurious correlations due to other sources of variations such as sex and age related differences.

\subsubsection*{Small or large datasets?}
Across the vast literature of MRI-based classification studies of ASD, it has been noted that studies using small samples reported promising classification performances, that however drop significantly on larger populations \cite{khosla2019ensemble,heinsfeld2018identification,ghiassian2016using,quaak2021deep}. For instance, a review made in 2017 \cite{arbabshirani2017single} showed that good accuracies have been obtained only involving fewer than 100 participants, with accuracies above 90\% obtained exclusively on dozens of subjects. This phenomenon has been attributed to multiple causes:
\begin{itemize}
    \item \textit{Participants selection \cite{ghiassian2016using}}\\
    Participants to small studies are usually recruited according to specific selection criteria (including age, sex, severity of the ASDs and cognitive abilities of the TDCs), that do not represent the complexity of both ASD and TDC brain characteristics. Thus, the performance obtained on restricted cohorts are not easily replicable on a more large and heterogeneous dataset.
    \vspace{0.2cm}
    \item \textit{Data acquisition homogeneity \cite{ghiassian2016using,auzias2016influence}}\\
    Small samples are usually acquired in a single site (i.e. hospital or research center), thus, using the same scanner and acquisition modalities for all the patients, while larger datasets usually consist in multi-site collections of samples. Given that, as already said in Sec. \ref{MRI-non-quantitative}, the MRI signal unavoidably depends on acquisition settings, the variability introduced by aggregating multi-site data makes it harder to reach high classification performance.
    \vspace{0.2cm}
    \item \textit{Overfitting \cite{ghiassian2016using,rashid2020use}}\\
    The low number of subjects enrolled and the high number of features usually analyzed in this kind of studies increase the probability to overfit the data. Furthermore, studies analyzing small datasets may try to maximize the dimension of the training set, sacrificing the validation and, if present, the test set \cite{xiaoxiao2018channel}. Assessing the performance on few subjects very similar to the training set, either for recruitment criteria or for acquisition modalities, may cause an overestimation of the classification abilities, without revealing the underlying overfitting problem. This is less probable on larger cohorts.
\end{itemize}
Despite the greater difficulty in reaching high performances, only by using large and multi-site datasets that include a greater diversity of subjects and acquisition modalities it is possible to find a generalizable biomarker, useful for clinical practice \cite{ghiassian2016using}. However discovering a complex pattern generalizable to a heterogeneous set of data, requires a high number of model parameters, which increases the risk to learn from subtle biases in the training set, the so called \ac{CE} \cite{ferrari2020measuring}.\\
The CE occurs because supervised learning tries to minimize the prediction error made on the training set simply looking for correlation relationships between the inputs and their paired outputs; no causal reasoning is included in the algorithm.
As a consequence, if the distribution of the target classes is correlated to a variable that affects the data but is not causally related to the task of interest, the confounder, the classifier may learn a pattern based on how it affects the data. As an intuitive example, if in the training set most of the ASDs are males and most of the TDCs are females, learning to systematically assign the labels ASD and TDC to all the males and females respectively, may be "convenient" because the easier task of recognizing sex differences allows to reduce the prediction error. Another well known confounding effect occurs when input data obtained from different sources are not equally represented. In this case the predictor may learn to perfectly fit the data coming from the most represented source, almost ignoring the remaining ones, because this allows to better minimize the mean error across the entire dataset.\\
The CE is notoriously one of the most deceitful problem in supervised learning. In fact, if the biases present in the training set are equally present in the validation (or test) set the performance may be overestimated without giving a clue on the presence of this phenomenon. Clearly this is an important issue for neuroimaging applications, because MRI data depend on a large number of biological and technical variables, so the probability that one or more of them are biased in the dataset is high and it increases when analyzing heterogeneous multi-site datasets.

Summarizing, searching a pattern to distinguish ASDs and TDCs using large and heterogeneous datasets is more difficult and classification performance are generally lower \cite{stanfield2008towards,ghiassian2016using}, but it is the only way towards the development of a diagnostic tool of clinical utility. However, when dealing with heterogeneous data it is also possible to obtain unrealistically high performance, due to the effect of one or more confounders that misled the learning process.
\subsubsection*{Structural or functional MRI?}
Structural images have been thoroughly analyzed in literature, much more than functional ones \cite{cody2002structural}. In fact they are easier to interpret, analyze and can be acquired for various reasons (e.g. suspected trauma, to check for pathologies characterized by an abnormal brain structure, etc.), not exclusively for research purposes. Despite considerable effort has been devoted to the identification of anatomical abnormalities in individuals with ASD \cite{amaral2008neuroanatomy,courchesne2007mapping}, the biomarkers found in this kind of studies are very heterogeneous and often not significant, since they allow to distinguish ASDs from TDCs with an accuracy lower than 60\% \cite{haar2014anatomical}, while the few higher-performing results are never replied in subsequent studies or in larger samples \cite{haar2014anatomical,anagnostou2011review,abraham2015learning}. \\
These results do not automatically imply that ASDs and TDCs do not differ from a structural point of view. In fact, anatomical abnormalities could be detectable only at a certain point in life \cite{courchesne2011brain} or for distinct subtypes of ASD \cite{haar2014anatomical}.
For instance, a longitudinal study \cite{courchesne2011brain} shows that the debated brain overgrowth in ASD is detectable only during infancy, followed instead by an accelerated decline in size. Another study \cite{shen2018extra} reports that the increase of extra-axial cerebrospinal fluid volume in early infancy is a high-sensitivity predictive biomarker of a possible subtype of ASD involving sleep disturbances and low non-verbal ability. \\
However the lack of a common and generalizable result turned many researchers to the analysis of fMRI/rsfMRI data or to the joint analysis of functional and structural information. 

RsfMRI is usually preferred with respect to task-based fMRI, because the latter requires both a careful experimental setup to control for psychological confounds in the fulfillment of the task, and the active and focused participation of the subject, which may be difficult to obtain depending on the severity of the disorder. Despite the results obtained with rsfMRI until now seem to suffer from the same problems described for sMRI \cite{quaak2021deep}, recent progress in machine learning video analysis may provide new methods to properly treat the temporal and spatial dimensions, overcoming the limitations of the commonly employed 4D-CNNs and RNNs. 
\subsubsection*{State-of-the-art on large datasets}\label{soa_neuroimaging}
To provide a meaningful representation of current state-of-the-art across the vast and heterogeneous literature described above, the studies involving at least 500 ASDs and 500 TDCs mentioned by one of four \cite{biswas2020prediction,hyde2019applications,khodatars2021deep,quaak2021deep} very recent reviews on ML/DL for ASD research are summarized in Table \ref{tab:asd_studies}.
In confirmation of what has been said in the previous paragraphs, in this selection of recent studies, most of them analyze rsfMRIs, use DL algorithms and the only ML one mentioned is an SVM. \\
All the listed papers, except the two highlighted in yellow, report an accuracy below 71\%, which is not enough to reach diagnostic utility.
Both the papers \cite{wang2019identification, ahmed2020single} showing higher performance, analyzed heavily pre-processed data. Despite the algorithms used in these two studies require parametrization, none of them reports its results on a test set, breaking the third rule listed in the ML good practices listed before. The second paper, also made a comparison among different pre-processing operations, which further aggravate the lack of a test set. The pre-processing adopted by both the papers is based on group analyses and their methodology description do not clarify if the first rule has been respected\footnote{This problem has been highlighted also by a recent review \cite{quaak2021deep} when discussing the results of \cite{wang2019identification}}. None of them proved the resilience of their approaches to overfitting or confounding effects. Finally, the authors did not identify the features of interest for classification.\\ 
All these elements raise doubts on the generalizability and reliability of the reported performance, which are significantly different from the other ones found in literature. \\
Despite these shortcomings, the 16 papers reported in the table obtained classification accuracies between 62\% and 97\% analyzing MRI data (mostly rsfMRI), with an average value of 70\%, suggesting that some information to distinguish ASDs and TDCs is encoded in rsfMRI data.

\begin{table}
\centering
\noindent\adjustbox{max width=\columnwidth}
{
\begin{tabular}{|c|c|c|c|c|c|c|c|}

\hline
\textbf{Review} &    \textbf{Paper}                &   \textbf{Year}&  \textbf{ASDs} & \textbf{TDCs} &   \textbf{Data analyzed}& \textbf{ML/DL}&   \textbf{Performance}  \\
\hline
\multirow{2}{*}{\cite{biswas2020prediction}}    & \multirow{2}{*}{\cite{ghiassian2016using}}     &   \multirow{2}{*}{2016}&  \multirow{2}{*}{538}  &  \multirow{2}{*}{573} &  rsfMRI + p.char &    SVM&    Acc 65\%, Sens 71\%, Spec 58\%  \\
                          &                          &       &       &    &   sMRI + p.char& SVM& Acc 64\%, Sens 76\%, Spec 52\%   \\
\cite{biswas2020prediction,quaak2021deep}    &    \cite{sen2018general} &   2018&   538 &    573&    rsfMRI + sMRI& AE + SVM& Acc 64\%, Sens 60\%, Spec 68\%\\
\cite{biswas2020prediction}    &    \cite{eslami2019asd}  &   2019&   505 & 530& rsfMRI& AE + SLP&  Acc 70\%, Sens 68\%, Spec 72\%  \\
\cite{biswas2020prediction}    &    \cite{sharif2019novel} &   2019&   538 & 573& sMRI&    CNN&    Acc 66\%  \\
\multirow{2}{*}{\cite{quaak2021deep}}& \multirow{2}{*}{\cite{xing2018convolutional}} & \multirow{2}{*}{2018} & \multirow{2}{*}{527} & \multirow{2}{*}{569} & \multirow{2}{*}{rsfMRI} & SVM & Acc 63\%, Sens 62\%, Spec 64\% \\
& & & & & & CNN &  Acc 67\%, Sens 66\%, Spec 70\% \\
\cite{biswas2020prediction,hyde2019applications,khodatars2021deep,quaak2021deep} &  \cite{heinsfeld2018identification} &   2017&   505 &    530&    rsfMRI &    AE + MLP&    Acc 70\%, Sens 74\%, Spec 63\%  \\
\cite{hyde2019applications} &    \cite{rane2017developing}          &   2017&   538 &   573&   rsfMRI&   SVM & Acc 62\%  \\
\cite{khodatars2021deep}    &    \cite{leming2020ensemble}   &   2020&   $\approx 1700$ & $\approx 2300$&  rsfMRI or fMRI&  CNN&  AUC 68\% Acc 67\%  \\
\cite{khodatars2021deep,quaak2021deep}    &    \cite{dvornek2017identifying} &   2017&   538 &  573&  rsfMRI&  RNN & Acc 69\%  \\
\cite{khodatars2021deep}    &    \cite{el2019simple}         &   2019&    993 &    1092&   rsfMRI&   CNN&   Acc 65\%  \\
\cellcolor{yellow}{\cite{khodatars2021deep}}    &\cellcolor{yellow}{\cite{ahmed2020single}}     &\cellcolor{yellow}{2020}
&\cellcolor{yellow}{529}
&\cellcolor{yellow}{573}
&\cellcolor{yellow}{rsfMRI}
&\cellcolor{yellow}{CNN}
&\cellcolor{yellow}{Acc 87\%, Rec 85\%, Prec 87\%}\\
\cite{khodatars2021deep}    &    \cite{sherkatghanad2020automated} &   2019& 505& 530& rsfMRI& CNN& Acc 70\%, Sens 77\%, Spec 62\%  \\
\cite{khodatars2021deep}    &    \cite{sairam2019computer}   &   2019&    505&    530&    rsfMRI&    AE + CNN&    Acc 70\%, Sens 74\%, Spec 63\%\\
\cellcolor{yellow}{\cite{khodatars2021deep,quaak2021deep}}
&\cellcolor{yellow}{\cite{wang2019identification}}
&\cellcolor{yellow}{2019}
&\cellcolor{yellow}{501}
&\cellcolor{yellow}{553}
&\cellcolor{yellow}{rsfMRI}
&\cellcolor{yellow}{SVM-RFE + AE + softmax}
&\cellcolor{yellow}{AUC 97\%, Acc 94\%, Sens 93\%, Spec 95\%}\\
\cite{khodatars2021deep}    &    \cite{shahamat2020brain}    &   2020&  500&  500&  sMRI& CNN& Acc 70\%  \\
\multirow{2}{*}{
\cite{khodatars2021deep}}    &
\multirow{2}{*}{\cite{thomas2020classifying}}&  \multirow{2}{*}{2020}&\multirow{2}{*}{620}&    \multirow{2}{*}{542}&    \multirow{2}{*}{rsfMRI}&    SVM&    Acc 66\%  \\
&&&&&& CNN & Acc 64\%\\
\cite{khodatars2021deep}    &    \cite{jiao2020improving}    &   2020&  505&  530&  rsfMRI&  CapsNet&  Acc 71\%, Sens 73\%, Spec 66\%  \\
\hline
\hline
SLP & \multicolumn{7}{l|}{Single Layer Perceptron}\\
MLP & \multicolumn{7}{l|}{Multi Layer Perceptron}\\
SVM-RFE & \multicolumn{7}{l|}{Support Vector Machine - Recursive Feature Elimination}\\
CapsNet & \multicolumn{7}{l|}{Capsule Neural Network}\\
\hline
\end{tabular}
}
\caption{Recent studies involving at least 500s ASD and 500 TDCs. The listed works have been selected based on this size criteria from the ones mentioned in the four most recent reviews of ML/DL applied to MRI for the diagnosis of ASD \cite{biswas2020prediction,hyde2019applications,khodatars2021deep,quaak2021deep}. The studies that reached an accuracy above 0.85 are highlighted in yellow.}
\label{tab:asd_studies}
\end{table}

\subsubsection*{Challenges}
Summarizing the previous sections, as of today, rsfMRI seems a more promising source of information to find an ASD biomarker with respect to sMRI. The main challenge in the pursuit of this objective is the development of a DL algorithm that has simultaneously the following properties:
\begin{enumerate}
    \item \textit{The ability to analyze raw or minimally pre-processed high-dimensional rsfMRIs and automatically find meaningful spatio-temporal patterns.}\\
    Avoiding arbitrary pre-proceessing operations of data reduction and feature extraction would eliminate the risk to remove useful information and introduce new noise, and would increase the trustability of the results. In addition, improving the analysis of temporal and spatial dimensions by using more advanced DL techniques could help in finding a diagnosing pattern based on the complex information of BOLD flux across the brain.
    \vspace{0.2cm}
    \item \textit{Resilience to confounding effects.}\\ 
    This means that the algorithm should be able to learn a pattern that does not depend on the presence of biases among the classes to predict. An algorithm resilient to CEs has the same classification performance on all the data subcategories, defined for example by subject related characteristics such as sex and age or by acquisition modalities. This property increases both the generalizability and trustability of the learned pattern.
    \vspace{0.2cm}
    \item \textit{Explainability.}\\ If the brain regions important for the classification can be extracted from the model, it is possible to discuss the medical validity of the pattern learned and increase current knowledge on the disorder. Furthermore an explainable algorithm allows to better evaluate the performance obtained. For instance, a classifier may not have perfect predicting capabilities on the overall dataset but be able to correctly identify a specific subtype of ASD with a well defined neuroimaging profile. 
\end{enumerate}
Despite some papers sought to achieve one of these properties individually \cite{eslami2021explainable,bengs20204d,abbas2018machine}, an algorithm addressing all these aspects and reaching performance of diagnostic utility has not been developed yet. The development and test of the classifier should clearly be performed on a large multi-centric dataset, which ensures the heterogeneity of the subjects and data cohort. Furthermore, it is fundamental to adopt an evaluation scheme able to spot overfitting and verify the generalizability of the model, i.e. its applicability to new unseen data, possibly obtained with different selection criteria or acquisition modalities.  

\section{New ML/DL tools useful for MRI analysis}
\label{sec:bkg_ml_tools}
The challenges listed before represent issues affecting also a great variety of other DL applications and new methodologies to address them are continuously developed. 
Making a comprehensive review of all these techniques is out of the scope of this thesis. Therefore, this section describes exclusively the methodologies taken from literature that have been adopted in this work to tackle the aforementioned problems:
\begin{itemize}
    \item The issue of learning from spatial and temporal dimensions without pre-reducing the information is addressed by using a DL architecture, called the \ac{2S-CNN} \cite{carreira2017quo}, developed for video action recognition applications, that aims at recognizing object and movement features, separately.  
    \item The resilience to confounding effects has been addressed thanks to the use of two tools:
    \begin{itemize}
        \item[--] the \ac{CI} \cite{ferrari2020measuring}, specifically developed for MRI applications, that helps to find and rank the CEs of different variables and evaluate the efficacy of correction methodologies.
        \item[--] a DL architecture based on adversarial learning, called \ac{BR-NN} \cite{adeli2021representation}, developed for machine vision applications, that avoids learning data dependency from a list of confounder variables. 
    \end{itemize}
    \item The explainability has been accomplished using a well known gradient-based algorithm, called Smoothgrad \cite{smilkov2017smoothgrad}, applicable to each instance to understand which of its features contributed the most to determine the final outcome.
\end{itemize}

\subsection{Learning from high-dimensional 4D rsfMRIs with a two-stream CNN}
As already said in Sec. \ref{MRI_classification_ASD}, the study of rsfMRIs can benefit from recent advancements in video action recognition methods as standard approaches to video analysis such as RNNs and CNNs suffer from poor representation of the temporal dimension.
In fact, the challenge in analyzing a video is that its classification usually depends both on spatial features, such as objects present in the scene, and spatio-temporal features in which the temporal characterization is intrinsically tied to the spatial one, because the information to identify is not simply a time length but a \textit{movement}.\\
The importance of recognizing movements is well known in video action recognition problems, and intuitively this is also the focus of rsfMRI information, which is related to the blood flux in the brain.

Recently, several evidences \cite{tanberk2020hybrid,carreira2017quo} suggest that the optical flow stream, containing the distribution of apparent velocities of brightness patterns in a video, is a representation from which it is easier to identify motion features that are meaningful for action recognition. For instance a work comparing the performances of different DL algorithms, some including optical flow into their input and some based exclusively on the raw original video, such as CNNs or RNNs, shows that all the methods including optical flow perform better than the others \cite{carreira2017quo}.

In the following paragraphs, the optical flow is described from a mathematical point of view and the DL architecture used in this thesis to analyze this information, i.e. 2S-CNN, is presented.  

\subsubsection{Optical flow}
\begin{figure}
    \centering
    \includegraphics[width=0.5\textwidth]{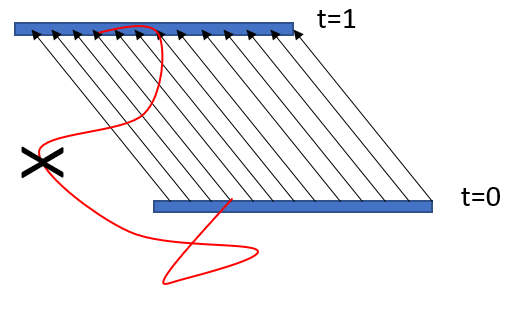}
    \caption{Trying to guess the movement of the blue bar from its configuration at t=0 to the one at t=1, the regularization term in Eq. \ref{eq:optical_flow_2} basically says that the most likely path of the bar's pixels are the ones represented with the black arrows, because they ensure that at each intermediate time, the points are close to each other. Paths such as the one depicted with the red line would bring to the same outcome at t=1, but are very unlikely.}
    \label{fig:optical_flow_2}
\end{figure}

The mathematical derivation of the optical flow of a video relies on the initial hypothesis that the image brightness $I$ at a certain point in space and time $\{\vec{p},t\}$ remains constant for small displacements:
\begin{equation}
    I(\vec{p},t) = I(\vec{p}+\vec{\Delta p}, t+\Delta t )
    \label{eq:brightness_constancy}
\end{equation}
where $\vec{p}$ is a point in the $N$-dimensional space of a single frame. In the case of the rsfMRI it is a 3D point, identified by the triplet $\{x,y,z\}$. 
Thus, expanding the right term with a first-order Taylor series in $\vec{\Delta p}$ and $\Delta t$ and isolating the temporal element, the following equation is obtained:
\begin{equation}
    \frac{\partial I}{\partial x}\Delta x +  \frac{\partial I}{\partial y}\Delta y +  \frac{\partial I}{\partial z}\Delta z = -  \frac{\partial I}{\partial t}\Delta t  
\end{equation}
Dividing both the terms by $\Delta t$, and observing that $\Delta x/ \Delta t$ is the x component of the velocity $V_x$ (and similarly for the other spatial coordinates), the optical flow equation takes form:
\begin{align}
        &\vec{\nabla}I \cdot \vec{V} = -  \frac{\partial I}{\partial t}\\
   & \vec{\nabla}I \cdot \vec{V} +  \frac{\partial I}{\partial t} = 0
   \label{eq:optical_flow_1}
\end{align}
The vector $\vec{V}$, evaluated at every spatial and temporal point of a video, constitutes its optical flow stream.
However, finding $\vec{V}$, that has 3 components, from this scalar equation is an ill-posed problem. \\
Therefore, several methods have been developed to compute an approximation of $\vec{V}$ from the video alone.
The approach followed in this thesis is the so-called Dual TV-L1 optical flow method, described in \cite{zach2007duality}.
According to it, the best approximation of $\vec{V}$ can be found by minimizing the absolute value of the left-hand side of Eq. \ref{eq:optical_flow_1} together with a constraint on the absolute value of the gradient of the optical flow itself:
\begin{equation}
    \min_{\vec{V}} \left\{\int_{\Omega} \left|\vec{\nabla}I \cdot \vec{V} +  \frac{\partial I}{\partial t}\right| + \lambda |\vec{\nabla}\vec{V}| \right\}
    \label{eq:optical_flow_2}
\end{equation}
Where $\Omega$ is the set of all the points of the frame for which the optical flow is being computed and $\lambda$ is a hyperparameter that regulates the weight of the regularization. The constraint on the norm of $\vec{\nabla}\vec{V}$ basically imposes that points that are close to each other in a frame have similar velocities, i.e., they end up close to each other in the subsequent frame (see Fig. \ref{fig:optical_flow_2}).

\begin{figure}
    \centering
    \includegraphics[width=0.6\textwidth]{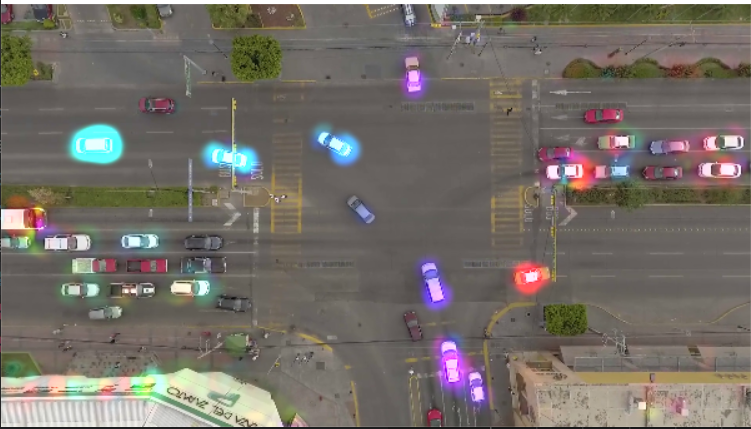}
    \caption{Superposition of a traffic scene and its optical flow (in module). Pixels of the optical flow are color coded according to their estimated velocity.}
    \label{fig:optical_flow}
\end{figure}

Summarizing, the optical flow can be viewed as a video that has the same temporal dimension\footnote{Strictly speaking, if the video is composed by N frames, its optical flow has N-1 frames, because its is computed on each pair of frames.} of the input video, but in which every point of every frame is a vector that represents the supposed displacement of that point in the original video between that frame and the next one (see an example of optical flow in Fig. \ref{fig:optical_flow}).\\
The main limitations in the computation of the optical flow are that:
\begin{itemize}
    \item The initial brightness constancy hypothesis (Eq. \ref{eq:brightness_constancy}) may not be respected in every video. For example, a video may not have a sufficient temporal resolution to allow to record smooth movements of all objects, or again, there may be sudden luminosity variation in the video, that are not related to objects moving. 
    Luckily, both these situations are almost impossible to occur in a rsfMRI video.
    \item The additional constraint imposing that the velocity gradients must be small is always false at the edges of objects. However, in rsfMRIs there is not a clear interface between the moving parts (oxygenated blood) and the static ones (the brain), because the blood flows through arteries, veins and capillaries that form a complex network in the whole brain, making difficult to identify their edges.
    \item The optimization process used to find $\vec{V}$ does not guarantee to find the real optical flow (and not even a good approximation of it), because it may end up in a local minimum.
\end{itemize}
Despite these limitations, the optical flow computation here described is widely adopted \cite{geiger2012we,szeliski2010computer,carreira2017quo,yue2015beyond}, showing to be a good source of information for DL video classification. According to the type of information encoded in the rsfMRI, it may be useful as well. 
\subsubsection{The two-stream CNN}
\label{subsec:bkg-2s-cnn}
\begin{figure}
    \centering
    \includegraphics[width=0.5\textwidth]{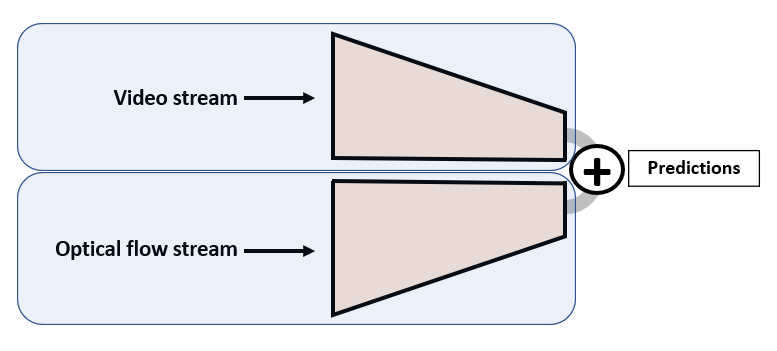}
    \caption{Schematic of the two-stream CNN architecture. The optical flow and the unprocessed video are analyzed separately by two standard convolutional networks. Their outputs are then summed together at the end to give the prediction of the whole network.}
    \label{fig:2s-cnn}
\end{figure}

The DL algorithms that are recently becoming popular in action recognition tasks are mainly based on a two-streams architecture that starts by processing the video (integrally or just a selection of frames) and its optical flow (again integrally or just a selection of frames) separately and then joins their extracted features to compute the final prediction. This architecture has been inspired by the two-streams hypothesis \cite{goodale1992separate}, according to which in the human brain two separate visual pathways are responsible for object recognition (the ventral stream) and visual guidance of movements towards the objects (the dorsal stream) \cite{simonyan2014two}.\\
A recent paper on action recognition \cite{carreira2017quo} shows that, despite the greater data dimensionality, analyzing in the two streams the entire video and its entire optical flow, it is possible to achieve better results with respect to exclusively analyze the video or a selection of frames for both the video and the optical flow. The architecture presented by this paper is illustrated in Fig. \ref{fig:2s-cnn}. \\
Note that in almost all the works adopting a two stream approach, the optical flow analyzed contains only the information of $|\vec{V}|$. Despite this choice implies removing the directional information, it significantly reduces the number of parameters to process the optical flow stream. The good results obtained in literature \cite{carreira2017quo} with this approach suggest that the module of the velocity is sufficiently informative to extract the motion features of interest.

\subsection{Resilience to confounding effects, through the combined use of the Confounding Index and the Bias-Resilient NN.}
\label{sec:confounders_ci}
\begin{figure}
    \centering
    \includegraphics[width=0.5\textwidth]{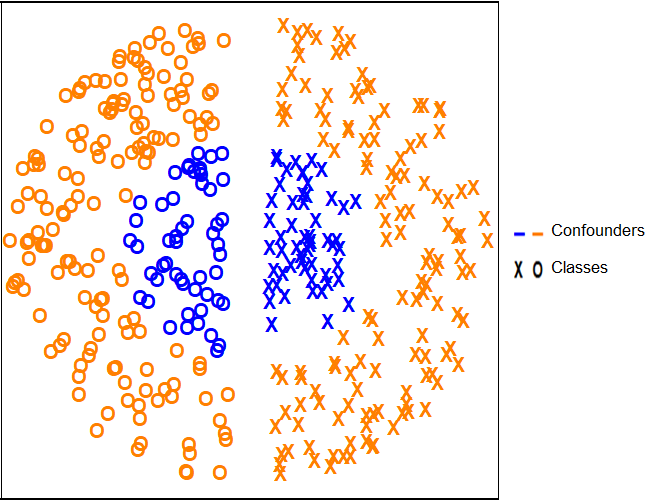}
    \caption{Representation of the case in which a confounder variable affects the input data in a highly non linear way, while they are linearly separable with respect to the target classes in the feature space.}
    \label{fig:confounders_classes}
\end{figure}

Addressing the problem of the confounding effects requires to:
\begin{enumerate}
    \item \textit{Identify the confounding variables for the problem of interest.}\\
    A variable to be confounding should affect the input data in a way that is detectable by the classification algorithm used. The strength of its CE depends on the relative complexity of the real task and the one of recognizing the confounder.
    In an extreme case, such as the one depicted in Fig. \ref{fig:confounders_classes}, if a variable affects the input data in a highly non linear way, while they are linearly separable with respect to the target classes in the feature space and the classification algorithm is linear; that variable is not confounding for the problem of interest, despite it affects the data.\\
    Only two tools for assessing the confounding effect of a variable in a ML/DL analysis have been developed. One of them provides an estimate of the confounding effect that depends on the training set bias \cite{neto2019detecting}, while the other one, called Confounding Index \cite{ferrari2020measuring}, is independent from the dataset composition and basically measures whether it is easier for a specific classifier to learn the confounder or the classes under examination. This second approach makes easier to rank, in absolute terms, the effects of multiple variables and correct for the most important ones independently from the specific training set composition, that can significantly vary in a cross-validation scheme. 
    \vspace{0.2cm}
    \item \textit{Discourage the DL algorithm to learn from possible biases related to the confounding variables.}\\
    Various strategies have been explored in literature to pursue this objective. Approaches developed in the context of fair AI \cite{mehrabi2021survey} penalize the algorithm when its errors are not equally distributed among instances with different confounder values, by adding a fairness constraint to the cost-function \cite{ghassami2018fairness,singh2018fairness,blum2019recovering}. \\
    Approaches based on cost-sensitive learning \cite{kukar1998cost,mandal2020ensuring,ferrari2021addressing}, instead, try to avoid learning from the bias by weighing differently the error committed on the less represented clusters of data, defined by their confounder and class values. \\
    Finally a third type of approach tries to avoid learning information related to how the confounders affect the data. For instance, this can be achieved with an adversarial learning framework, that tries to maximize the classification performances while minimizing the ability of the model to predict each confounders \cite{adeli2021representation}, or with a variational autoencoder that searches a feature representation that preserves the information on the task, while removing the ones due to the confounders. While the first two types of approaches have proved to be a critical a trade-off between prediction accuracy and fairness fulfilment, all the others showed good resilient abilities if well calibrated \cite{ferrari2021addressing}. However, adversarial learning has the great advantage to be applicable, almost limitless, to a high number of confounders regardless whether they are categorical or continuous, which is fundamental for a medical study.
\end{enumerate}

In the following paragraphs the CI and the BR-NN used in this work to address the two steps described above, respectively, are briefly described.

\subsubsection*{The confounding index \footnote{Note that this index has been developed by the author of this thesis within the PhD program, but being a content quite separated from the present work, it is presented as a tool taken from literature.
}}
\begin{figure}
     \centering
     \begin{subfigure}[b]{0.417\textwidth}
         \centering
         \includegraphics[width=\textwidth]{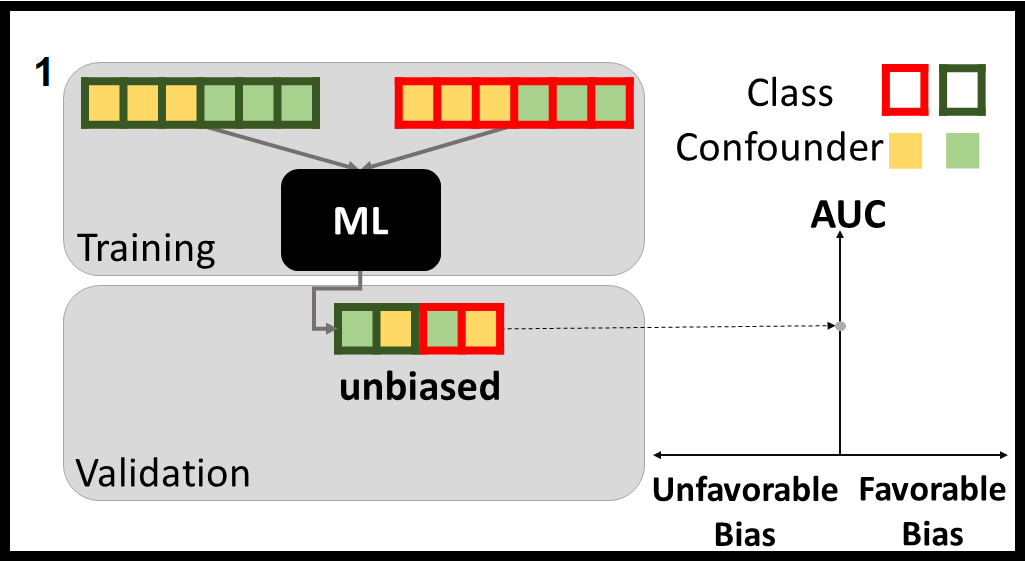}
         \label{fig:ci1}
     \end{subfigure}
     \hspace{-2.3mm}
     \begin{subfigure}[b]{0.417\textwidth}
         \centering
         \includegraphics[width=\textwidth]{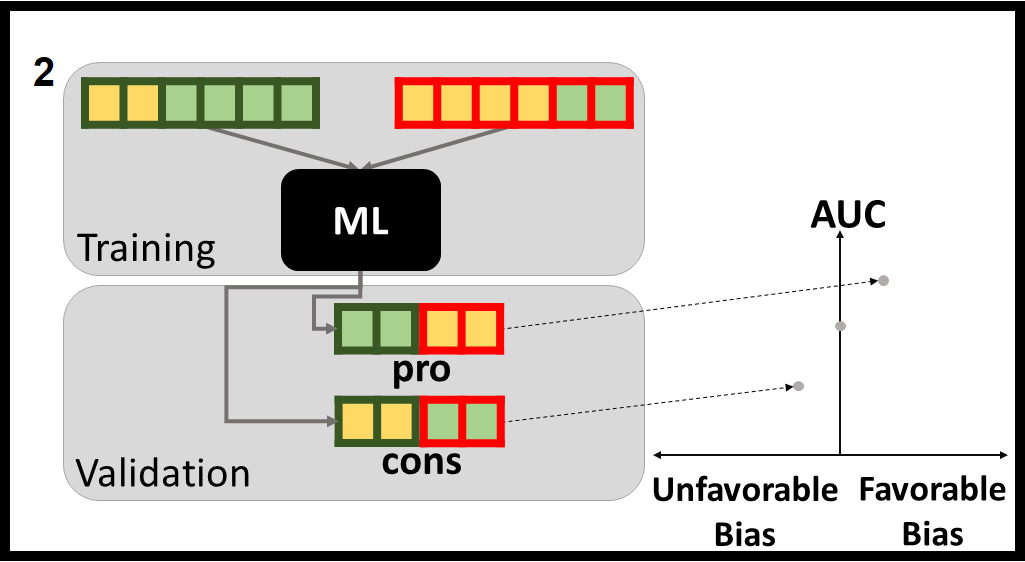}
         \label{fig:ci2}
     \end{subfigure}
     \hspace{-2.3mm}\\
     \vspace{-5.6mm}
     \begin{subfigure}[b]{0.417\textwidth}
         \centering
         \includegraphics[width=\textwidth]{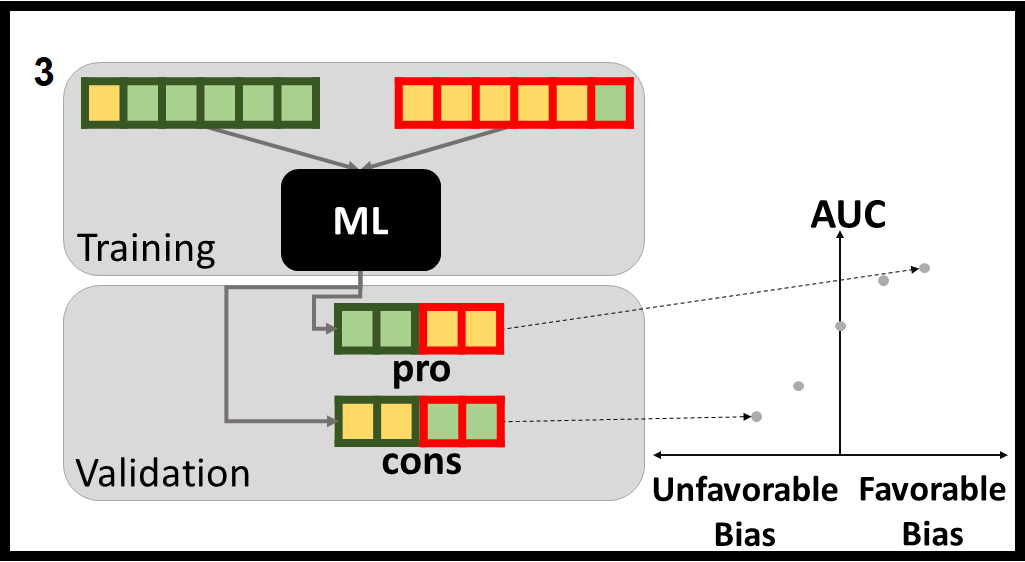}
         \label{fig:ci3}
     \end{subfigure}
     \hspace{-2.3mm}
     \begin{subfigure}[b]{0.417\textwidth}
         \centering
         \includegraphics[width=\textwidth]{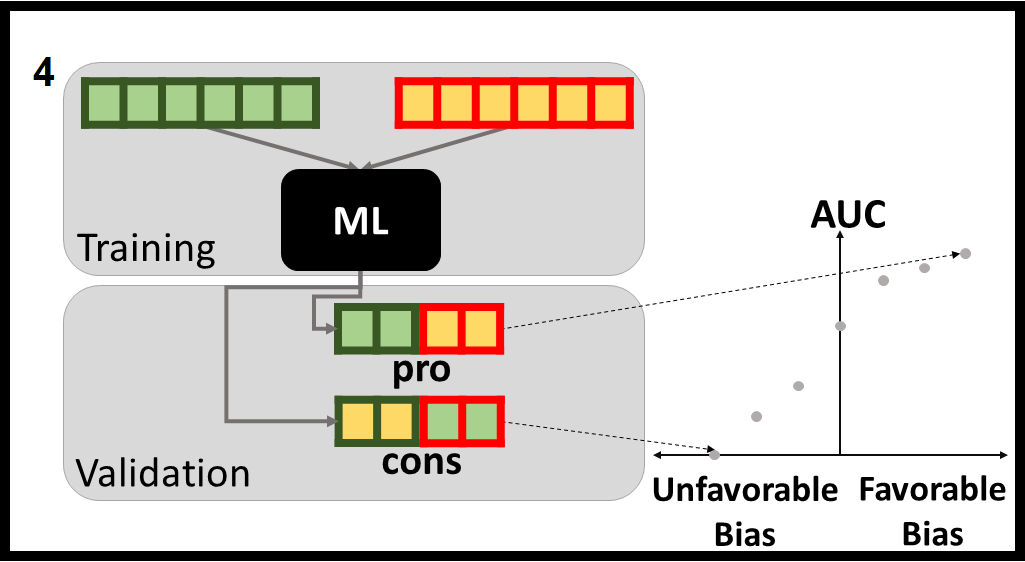}
         \label{fig:ci4}
     \end{subfigure}
     \hspace{-2.3mm}
     \begin{subfigure}[b]{0.16\textwidth}
         \centering
         \includegraphics[width=\textwidth]{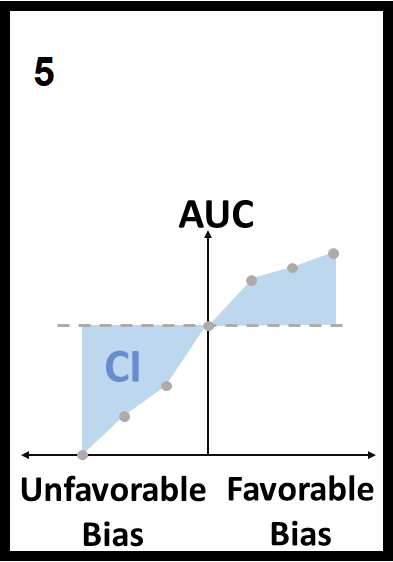}
         \label{fig:ci5}
     \end{subfigure}     
        \caption{Steps of the calculation of the CI.}
        \label{fig:ci_steps}
\end{figure}

The CI measures the confounding effect of a categorical data attribute in a binary classification problem in a bias-agnostic way.
It can be also used for continuously distributed variables, by binning their values.\\
The main steps for computing the CI are schematically illustrated in Fig. \ref{fig:ci_steps}. First, the algorithm under examination is trained for the classification task of interest using various training sets with an increasing percentage of bias with respect to the supposed confounder variable (Fig \ref{fig:ci_steps}-1). Then, the AUCs of these models are evaluated on two different validation sets called \textit{pro} and \textit{cons}, which are composed by data coherently and oppositely 100\% biased with respect to the training set configuration (Fig \ref{fig:ci_steps}-2).\\ 


If the tested variable is not a confounder, the performances obtained should not depend on the dataset composition with respect to it, so all the AUCs should be equivalent within a certain uncertainty. On the contrary, if the tested variable has a confounding effect, the AUCs obtained on the \textit{pro} validation set should monotonically increase with the percentage of training set bias, because the pattern learned on the biased training set is still valid on a validation set coherently biased. Analogously, a monotonic decrease is instead detectable on the \textit{cons} set, because the learned pattern is based on an association (the one between the confounder and the classes under examination) that is oppositely present in the validation set. 
The speed of performance increase and decrease with respect to the bias percentage in the training set varies between variables with different confounding effects. Clearly, when small biases cause a high variation in performances the variable under examination has to be considered a strong confounder for the classification problem of interest. So, the CI is based on the area determined by the AUCs increase and decrease with respect to the not-biased configuration and returns a value within 0 and 1, where 0 is a non-confounding variable and 1 refers to the strongest CE measurable, meaning that the smallest training set bias tested brings simultaneously to an $AUC=1$ on the \textit{pro} set and an $AUC=0$ on the \textit{cons} set.
Clearly, this occurs only on synthetic data, with a null classification task and a confounding variable that heavily affect the data; in real applications the CI value usually never exceeds 0.7.

The CI has been widely tested on heavily pre-processed neuroimaging structural data \cite{ferrari2020measuring,ferrari2020dealing} showing that site effect is an important confounder even for these kind of data in a ML based study for ASD diagnosis (see Appendix \ref{site_dependency}). Another interesting result is that also single-site data collected in different "samples", probably because acquired in different periods or with different physical scanners (despite being the same model with the same acquisition setting), display differences that can be easily detected by a ML algorithm, thus, constituting a potential source of CEs.
This suggests that heavily pre-processed features extracted from MRI still depend on scanner software updates, replacement of hardware components, or even time wear of the machine. This result confirms what has been previously said in Sec.
\ref{MRI-non-quantitative}: MRI is a non-quantitative measure whose signal is highly entangled with fine non traceable machine characteristics. These considerations further motivate the importance of detecting and measuring the confounding effect of all the potential sources of variation in a MRI dataset and to adopt a DL framework resilient to confounders. In this sense, given that the CI provides a model-specific confounding effect estimation, it can be also used to test the resilience of an algorithm to CEs.

\subsubsection*{The Bias-resilient Neural Network}
\label{subsec:bkg-br-nn}
\begin{figure}
    \centering
    \includegraphics[width=0.7\textwidth]{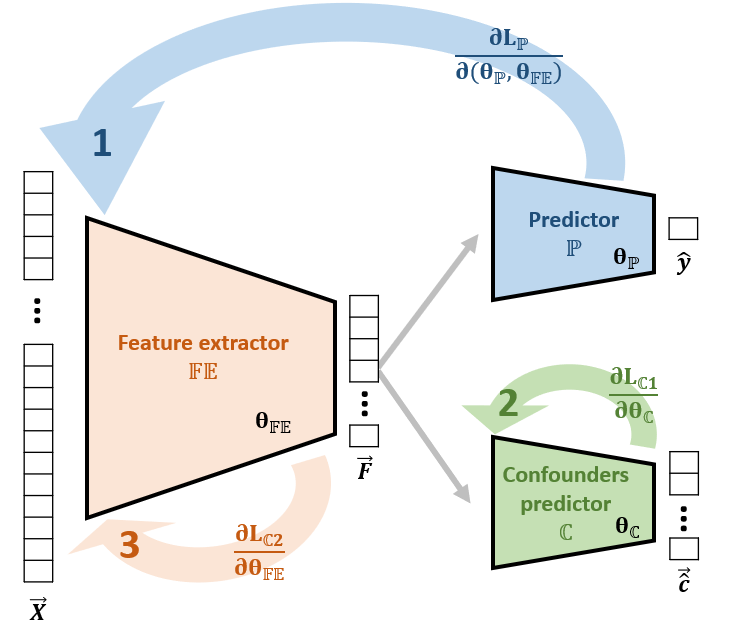}
    \caption{Architecture of the BR-NN\protect\footnotemark.}
    \label{fig:BR-Net}
\end{figure}

Consider a classification problem in which each of the $N$ instances $i$ is a triplet, composed by: a vector of input features, its binary target output (0 or 1) and a vector of confounders (containing both categorical and continuous variables): $\{(\vec{X}_i,y_i,\vec{c}_i)\}_{i=1}^N$. The BR-NN allows to avoid CEs by predicting $y_i$ from a set of features $\vec{F}_i$, derived from the input $\vec{X}_i$, that does not contain information about the confounders $\vec{c}_i$. This is achieved with an adversarial learning approach implemented using a tripartite architecture (see Fig. \ref{fig:BR-Net}), whose elements are:
\begin{itemize}
    \item \textit{The Feature extractor $\mathbb{FE}$}: it processes the $\vec{X}_i$ to extract $\vec{F}_i$.
    \vspace{0.2cm}
    \item \textit{The Predictor $\mathbb{P}$}: it takes in input the $\vec{F}_i$, generated by the $\mathbb{FE}$, and tries to predict the target class $y_i$, returning the output $\hat{y}_i$, which is by convention the estimated probability of the $i$-th instance of belonging to class 1: $\hat{p}_{1i}$. Clearly, in a binary problem $\hat{p}_{1i} = (1-\hat{p}_{0i})$ and $\hat{y}_i \in [0,1]$. 
    \vspace{0.2cm}
    \item \textit{The confounders predictor $\mathbb{C}$}: it tries to predict the counfounders $\vec{c}_i$ from $\vec{F}_i$.
\end{itemize}
Note that in picture \ref{fig:BR-Net} the symbols $\theta_\mathbb{FE}$, $\theta_\mathbb{P}$ and $\theta_\mathbb{C}$ indicate the weights of the three parts of the net that are optimized during the training, trying to minimize a set of cost-functions computed on a subset, called batch $B$, of the $N$ input triplets. 
Intuitively, these cost-functions are combined to adversarially force the $\mathbb{FE}$ to process the inputs into a set of features that maximizes the prediction of $\mathbb{P}$ while minimizing the performances of $\mathbb{C}$. In practice, this is achieved by updating the weights of the net in three iterative rounds, involving three different losses (see Fig \ref{fig:BR-Net}):
\begin{enumerate}
    \item \textit{Training $\mathbb{FE}$ and $\mathbb{P}$ to predict the classes of interest.}\\
    Both the set of parameters $\theta_\mathbb{FE}$ and  $\theta_\mathbb{P}$ are optimized with the cost-function:
    \begin{equation}
        L_{\mathbb{P}} = - \sum_{i}^{B}\sum_{k \in \{0,1\}} \delta_{k,y_{i}}\log(\hat{p}_{ki}) =  - \sum_{i}^{B}  \left( y\log(\hat{y}_{i}) + (1-y)\log(1-\hat{y}_{i}) \right) 
    \end{equation}
    where $k$ is a categorical index that spans over all the possible classes and $\delta_{k,y_{i}}$ is the Kronecker delta, which takes value 1 when $k = y_{i}$ and 0 otherwise. This loss is the standard cross-entropy function, which is the most widely used cost-function to train categorical classifiers. Considering its extended form on the right, for the instances belonging to class 1, only the first addendum contributes to $L_{\mathbb{P}}$ and the error is estimated as the inverse of the logarithm of the predicted probability that those instances actually belonging to class 1. The same calculation is done in the second addendum for the instances belonging to class 0.
    The characteristic log-based error of the cross-entropy non-linearly takes into account the predictions, weighing more the cases in which the classifier assigns an instance to its true class with a low probability. 
    \vspace{0.2cm}
    \item \textit{Training $\mathbb{C}$ to predict the confounders of the problem.}\\
    In the second round the $\theta_\mathbb{C}$ are optimized trying to minimize the following loss:
    \begin{equation}
     L_{\mathbb{C}1} = \sum_j L_{\mathbb{C}1,j} \; \; \rightarrow \;\; L_{\mathbb{C}1,j} =  
     \begin{cases} 
      \sqrt{ \frac{1}{B}\sum_{i}^{B} \left(\hat{c}_{ij} - c_{ij}\right)^2} & \text{if } c_{ij} \in \mathbb{R}\\
      - \frac{1}{B} \sum_i^B\sum_m^M \delta_{m,c_{ij}}\log(\hat{p}_{ijm}) & \text{if } c_{ij} \in \mathbb{Z}
     \end{cases}
    \end{equation}
    where $i$ and $j$ are the indexes of the specific instance and confounder, respectively, and $m$ is a categorical index that spans over all the possible values of the categorical confounders. As it can be noted, the loss is computed differently for continuous confounders, using the \ac{RMSE}, and categorical ones, using the cross-entropy. Despite this choice may seem awkward, it should be noted that errors on categorical and continuous confounders are unavoidably incomparable. However the good results obtained by the authors, testing it on both simulated and real data, suggest that the overall loss is able to correct both the types of confounders.
    \vspace{0.2cm}
    \item \textit{Training $\mathbb{FE}$ to extract features  $\vec{F}_i$ that minimizes th ability of $\mathbb{C}$ to predict the confounders.}\\
    This is achieved by optimizing the $\theta_\mathbb{FE}$ using the following loss:
     \begin{equation}
    L_{\mathbb{C}2} = - \lambda \sum_j corr^2(\vec{c}_j,\vec{\hat{c}}_j)   
    \label{eq:lc2}
    \end{equation}
    which is simply the inverse of the Pearson Correlation between the confounder values predicted by $\mathbb{C}$ and the true ones, weighted by an hyper-parameter $\lambda$ that defines the strength of the adversarial correction.
    Note that $L_{\mathbb{C}1}$ and $L_{\mathbb{C}2}$ performs opposite tasks, using different metrics. This choice has been made by purpose. In fact, $L_{\mathbb{C}2}$ reduces the Pearson correlation between the true and predicted confounders, which are obtained from $\vec{F}_i$. This indirectly forces $\vec{F}_i$ to be linearly uncorrelated to $\vec{c}_i$, but does not remove from it non-linear dependencies. However, thanks to $L_{\mathbb{C}1}$ which ensures that the outputs of $\mathbb{C}$ are as close as possible to the confounder values, both linear and non-linear dependencies between $\vec{F}_i$ and $\vec{c}_i$ are destroyed.
\end{enumerate}
These three steps, repeated for every batch, are equivalent to optimizing the net with a saddle-point loss of the form:
\begin{equation}
    L = L_\mathbb{P} + L_{\mathbb{C}2}
\end{equation}
Interestingly, besides preventing learning from possible confounders, this sort of 'tug of war' indirectly reduces overfitting, because it forces the algorithm to find alternative patterns with respect to the one that would simply maximize the performances alone.

As already mentioned, among the various strategies explored in literature to deal with CEs, an approach based on adversarial learning is easily applicable to an almost limitless number of continuous and categorical confounders. Furthermore, according to a study made by the author of this thesis on both synthetic and real structural MRI data, \cite{ferrari2021addressing}, an adversarial approach also obtains superior performance compared to other algorithms.
Among the different approaches based on adversarial learning \cite{zhang2018mitigating,elazar2018adversarial,wang2019balanced,sadeghi2019global} addressing the confounding problem, the BR-NN presents the unique feature of using two diverse quantities $L_{\mathbb{C}1}$ and $L_{\mathbb{C}2}$ to learn the confounders and adversarially discourage learning from them, respectively. Despite no experimental tests have been made to assess the efficacy of this choice, it is the opinion of the author that this elegant solution may make the difference when compared to other more straightforward implementations of the adversarial approach.

\FloatBarrier

\footnotetext{The description of the BR-NN depicted in this figure may seem very different from its counterpart in the original paper \cite{adeli2021representation} presenting this algorithm. However, it is the opinion of the author of the present thesis that this figure provides a more faithful representation of the code implementation of the BR-NN made available by the developers on GitHub \cite{adeli_git}. Instead, the description and illustration present in the original paper \cite{adeli2021representation} are quite liberal and with respect to the complexity of the actual algorithm.}

\subsection{Explainability with Smoothgrad}
\label{subsec:bkg_smoothgrad}
A wide number of model-agnostic explainability algorithms have been developed to unbox the decisional pattern learned by any DL algorithm. They are usually categorized in local and global \ac{XAI} approaches. The first ones provide individual explanation for each instance, while global ones try to explain the learning algorithm as a whole. Local explanations are usually preferred in medical applications because:
\begin{itemize}
    \item they show the most relevant features for case by case predictions, which is a fundamental for a possible clinical adoption;
    \item they can shed light on the heterogeneity of a disorder and help in defining groups of patients with shared features, which is useful for research purposes;
    \item they allow to study individually misclassified instances and spotting possible confounding effects, that, as already said, often occur when dealing with medical data. 
\end{itemize}
To briefly describe a XAI algorithm for local explanations, consider a classifier $\mathbb{C}$ already trained to predict the probability of input $\vec{X}_i$ to belong to the classes $c_1$, $c_2$ and $c_3$. An intuitive way to understand how important a specific value $X_{ij}$ of the $j$-th feature is in the prediction of the instance $\vec{X}_i$ as belonging to $c_1$ consists in computing $\frac{\partial p_{c_1}}{\partial X_{ij}}$.
Where $p_{c_1}$ is the predicted probability of $\vec{X}_i \in c_1$.

\begin{figure}
    \centering
    \includegraphics[width=0.5\textwidth]{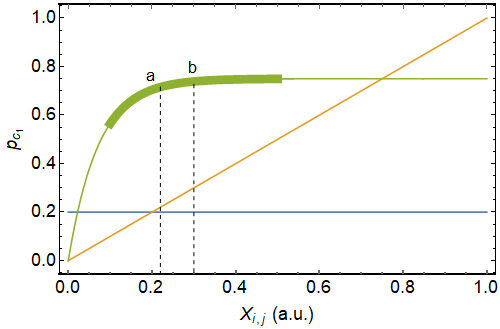}
    \caption{Plot of different kinds of relationship between $p_{c_1}$ and $X_{i,j}$. Linear (orange), no relationship (blue) and saturating relationship (green)}
    \label{fig:figura_smoothgrad}
\end{figure}

\begin{figure}
    \centering
    \includegraphics[width=1\textwidth]{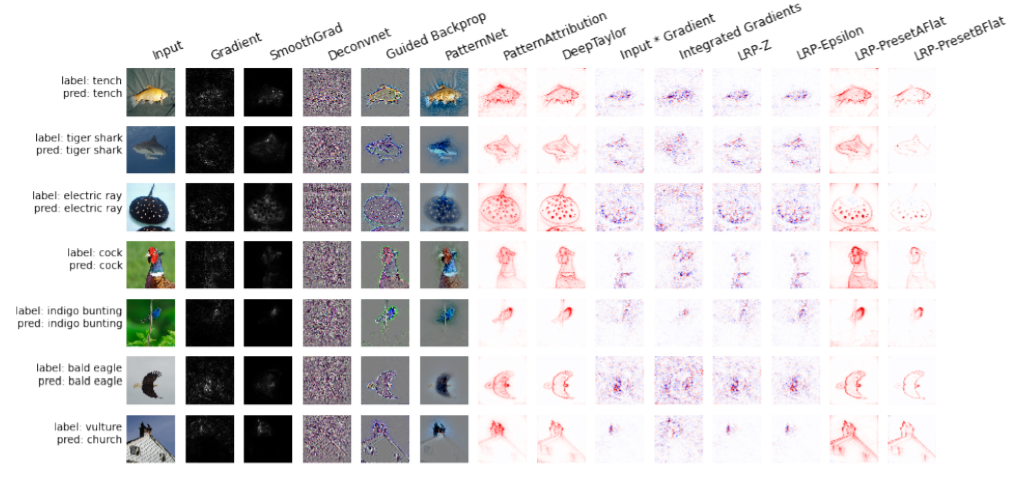}
    \caption{Comparison of the local explanations provided by different XAI algorithms on a sample of images portraying single objects. \ccbypictures{linardatos2021explainable}}
    \label{fig:comparison_xai}
\end{figure}

In fact, supposing that the value of $X_{ij}$ is linearly correlated to $p_{c_1}$, as illustrated by the orange line Fig. \ref{fig:figura_smoothgrad}, the partial derivative would be sensibly different from zero for any value of $X_{ij}$. Instead, if the value of $X_{ij}$ has no effect on the prediction of $c_1$, as depicted by the blue line in Fig. \ref{fig:figura_smoothgrad}, $\frac{\partial p_{c_1}}{\partial X_{ij}}$ would be equal to zero.\\
However, these are very unrealistic examples and $p_{c_1}$ usually depend on $X_{ij}$ in much more complex ways. For instance, the green line in Fig. \ref{fig:figura_smoothgrad} shows a situation in which the relevance of $X_{ij}$ has a saturation point for $X_{ij} > a$. So, the partial derivative $\frac{\partial p_{c_1}}{\partial X_{ij}}$ computed for example in $X_{ij} = b$ would be zero, contrary to the real relevance of that value for the $c_1$ prediction, that also depends on the first part of the curve. An easy solution consists in computing $\frac{\partial p_{c_1}}{\partial X_{ij}}$ in a large neighbourhood of $b$ and averaging the results obtained. This also guarantees to have more reliable results in case the relationship between $p_{c_1}$ and $X_{ij}$ shows some fluctuations. \\
Averaging the partial derivatives of the predicted class with respect to each input feature in a neighbourhood of its value is the approach at the basis of many XAI algorithms with local explanations, among which the widely used SmoothGrad.
Fig. \ref{fig:comparison_xai} shows a comparison of the local explanations obtained by different XAI algorithms \cite{linardatos2021explainable}, when classifying images portraying objects that are easy to recognize at a glance.
As it can be seen SmoothGrad provides a very rough explanation: just useful to identify the location of the object of interest. While more advanced algorithms highlight its edges, which is considered a representation easier to understand. However, neuroimaging data are very different from the classical images used to compare these algorithms. Considering the spatial 3D nature of a rsfMRI, the subject variability of brain characteristics and as a consequence the difficulty to recognize brain features at a glance, a rough explanation that allows to identify a fuzzy region of interest may be a good option, that could also facilitate comparison among different instances.   
  
\chapter{Genetics}
\section{Introduction}\label{intro_gen}
\begin{figure}
    \centering
    \includegraphics[trim={0 7cm 0 0},clip,width=0.99\textwidth]{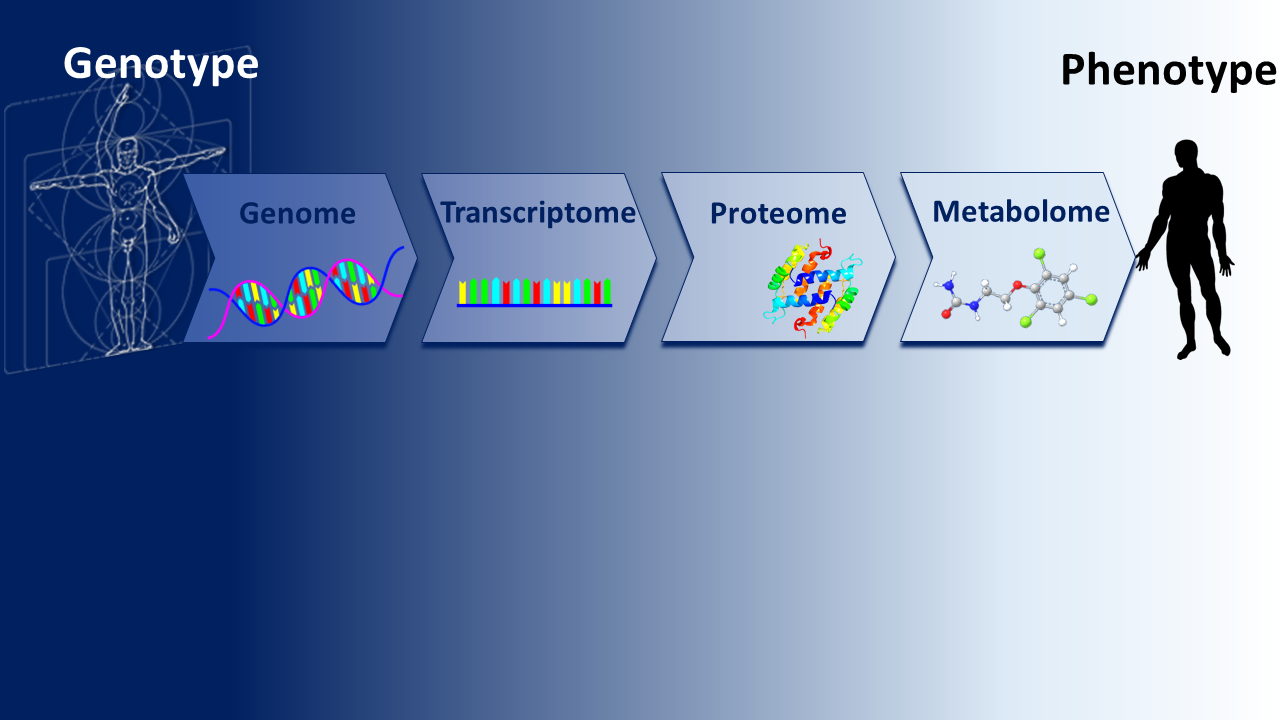}
    \caption{Steps linking the DNA information to the phenotype}
    \label{fig:genotype-phenotype}
\end{figure}

Genetics studies the basis for the heritability of phenotypes, which include the study of how DNA variations are transmitted from parents to offspring and how they affect the individual phenotype \cite{snustad2015principles}. This last aspect is particularly interesting in the study of complex traits and disorders, such as ASD, that by definition exhibit a large distribution of phenotypes derived from the effect of multiple genetic and environmental factors \cite{lander1994genetic}.\\
The process that translates the information contained in the DNA in its phenotypic expression, including the interaction with the environment, is called \ac{GE} and it is articulated in various steps, as shown in Fig. \ref{fig:genotype-phenotype}  \cite{josephs2021gene}, each one characterized by different biological molecules whose abundance is measured by the so called "omics" technologies.

This chapter is dedicated to the study of the branch of genetics called transcriptomics. First, the biological meaning of the transcriptome is described, followed by an illustration of the two key modern techniques to measure it and of the typical pipeline of analysis to study it. Finally, after a brief overview of the state-of-the-art on ASD transcriptomic studies, some important challenges of this field of research are presented.
\begin{figure}
    \centering
    \includegraphics[width=0.99\textwidth]{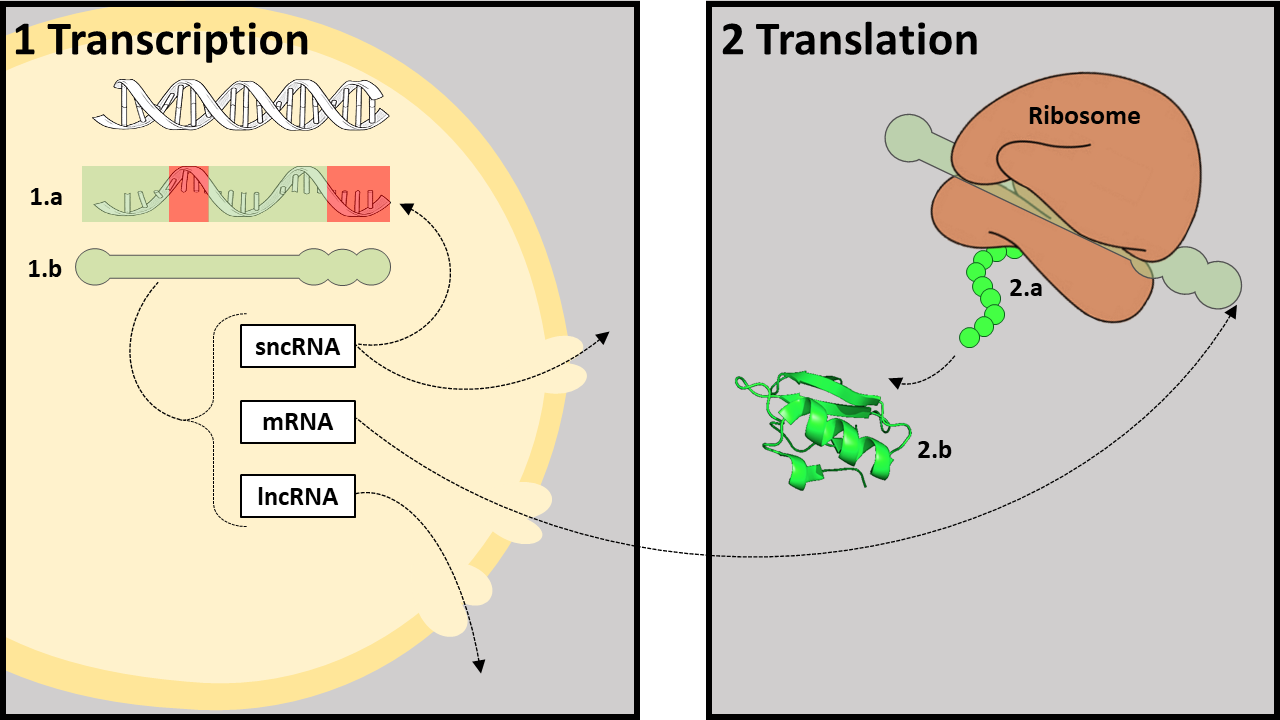}
    \caption{Schematic representation of the protein synthesis process.\\
    The transcription phase takes place inside the nucleus, represented in yellow. The DNA is represented by the double stranded white helix. Its primary transcript is depicted in 1.a by a single strand with green and red parts that represent the exons and introns, respectively. In 1.b there is a schematic representation of a mature RNA. The translation phase, instead, takes place in the cytoplasm, where the mRNAs are translated by the ribosome, the pink structure, into a polypeptide chain, represented by the green figure 2.a. Finally, the polypeptide chain folds into a protein, the green structure 2.b. }
    \label{fig:GE}
\end{figure}
\section{Transcriptome}
As briefly introduced, the GE is the process with which the genes in the DNA of an organism are differentially expressed across the spatial and temporal dimension, generating a phenotype composed by different cell types organized in tissues and organs, that continuously change in their function and structure with age. The regulation of which genes are expressed in a cell and the intensity of their expression is at the basis of the physiological complexity of eukaryotic multicellular organisms, such as humans.\\
According to the central dogma of molecular biology \cite{crick1970central} (often referred as the "one gene, one protein" model) the genes expressed by a cell act as templates for the synthesis of \ac{mRNA} molecules, which in turn provide the information for the synthesis of the proteins that the cell has to assemble to accomplish its functions. Thus, the flow of genetic information within a cell goes from the DNA to the mRNA, with a process called \textit{transcription}, and subsequently from the mRNA to the proteins, with a process called \textit{translation}.
\newcommand*\mystrut[1]{\vrule width0pt height0pt depth#1\relax}
\begin{equation}
\underbrace{\mystrut{1.5ex}\textbf{DNA} \Longrightarrow \textbf{mR}}_\text{Transcription}\!\underbrace{\mystrut{1.5ex}\textbf{NA} \Longrightarrow \textbf{Proteins}}_\text{Translation}
\end{equation}
The transcriptome is the result of the transcription process and in a simplistic way it can be defined as a measure of which genes are expressed in a cell or in a tissue and of the intensity of this expression. This intensity is estimated by counting the copies of the mRNAs associated to each gene, on the basis that they will be later translated into proteins.\\
However, with modern research it is becoming clear that some aspects of the central dogma are not entirely accurate and that the GE process is more complex.
A more thorough, although not exhaustive, explanation of the GE process is illustrated in Fig. \ref{fig:GE}, which depicts the steps described below.
\begin{enumerate}
    \item \textbf{Transcription}
    \begin{enumerate}
    \item \textit{Primary transcription.}\\
    The DNA present in the cell nucleus serves as template for the synthesis of its corresponding RNA molecule copy, which is called the primary transcript.
    \item \textit{Maturation of the RNA, through splicing, capping and tailing.}\\
    The primary transcript undergoes several changes. Some parts of its sequence, called introns, are removed and the remaining parts, called exons, are joined. This phenomenon is called splicing and it is often \textit{alternative}, meaning that each gene can be expressed by producing different sequences of RNA composed by different combination of exons. So, \ac{AS} is a mechanism that increases the diversity of RNAs expressed from the genome and thanks to it the $\sim 25,000$ human protein-coding genes can produce up to $90,000$ different proteins \cite{collins2004finishing,wang2015mechanism}.
    Note that the discovery of AS contributed in revealing one limitation of the central dogma of molecular biology according to which one gene codes for one protein.\\  
    Besides splicing, during the RNA maturation both the extremities of the RNA portions coding for proteins are marked with the addition of a cap at the start and a tail at the end\footnote{Strictly speaking, all the starting extremities of protein coding regions are marked with a cap, while most of the final ones (but not all) are marked with a tail.} characterized by a sequence of A nucleotides, thus called poly(A) tail. These markers are of help during the translation phase.\\
    At the end of the maturation stage 
    Different types of RNAs are produced:
    \begin{itemize}
        \item mRNA: it is the only type of protein-coding RNA, its length usually exceed 200 nucleotides and it is characterized by its cap and tail. In humans, mRNA accounts for only 1-4\% of the total transcriptome \cite{berg2002biochemistry,mattick2006non}. 
        \item \ac{lncRNA}: it is a type of transcript with a length over 200 nucleotides that is not translated into any protein. It regulates GE at multiple levels, during transcription, translation and also in epigenetic processes \cite{statello2021gene,saxena2011long}.
        One abundant lncRNA is the \ac{rRNA}, which comprises > 80\% of the transcriptome and it is a constituent of the molecular machines responsible for protein synthesis, the ribosomes.
        \item \ac{sncRNA}: this category includes all the non-coding RNAs with a length inferior to 200 nucleotides.\\
        Among sncRNAs some are essential to make GE possible, such as: \ac{snRNA}, that is involved in the splicing process and \ac{tRNA} that carries the amminoacids necessary to build the proteins.
        Other types of sncRNA have regulatory roles, such as \ac{miRNA} that post-transcriptionally regulates GE by degradation of mRNAs \cite{bartel2018metazoan}.
    \end{itemize}
    Note that the presence of lncRNAs and sncRNAs is another important point missed in the central dogma of molecular biology, according to which the transcriptome encompasses only protein-coding mRNA transcripts. 
    \end{enumerate}
    \vspace{0.2cm}
     \item \textbf{Translation}
        \begin{enumerate}
        \item \textit{Protein synthesis.}\\
        The mature mRNA transcripts are transported to the cytoplasm, where the protein synthesis takes place.
        The ribosomes read the mRNAs, starting from their cap, in informational units of three nucleotides, that code for a specific amino acid. The tRNAs bring the amino acids to the ribosomes which chemically join them following the correct sequence coded in the mRNAs and thus producing polypeptide chains.
        \item Protein folding.\\
        Thanks to the chemical interaction among the amino acids within the polypeptide chains, they fold into a 3D structures giving rise to functional proteins. 
        \end{enumerate}
    \end{enumerate}
From this explanation, it clearly emerges that the transcriptome is the set of all RNA transcripts, in a cell or group of cells of the same tissue at any given time. So, the transcriptome provides an insight into the GE process of a tissue and represents an intermediate characterization between the genotype of the organism and the protein production of that tissue, which is closely related to its phenotipycal aspect.\\
As already mentioned throughout this chapter, mechanisms of regulation to increase, decrease or forbid, in the end, the production of certain proteins can occur at any of the GE steps.
However, the analysis of the transcriptome is technologically more advanced and it is currently possible to obtain a measurement of expression of all genes in the genome.
The regulatory mechanisms that take place in a tissue can be inherited, influenced by the environment, such as by the exposition to chemical products, by the experience of the organism with external factors, as in the case of motor activity, or even by other signals within the organism, due for example to changes in the hormone levels due to age. So, in conclusion, the transcriptome qualitatively and quantitatively reports the composition of the RNA transcripts of the cells that in a tissue at any given time, depending on genetic, exogenous and endogenous factors.
\section{Transcriptomics technologies}
\label{sec:bkg_transcriptomics}
Transcriptome profiling is a very hard task, that requires to associate each transcript to its related gene and to quantify it. Considering that transcripts are unequivocally identified by their base sequence and that the total RNA present in a single mammalian cell is composed by around 10 gigabases \cite{scott2000transcripts}, the complexity arises mainly in the time required to analyze this huge quantity of biological information.
Large-scale transcriptome profiling has become accessible only recently with the advent of \textit{microarray} technology and subsequently with the more performant \textit{RNA-Seq}, that thanks to a wide combination of different techniques are able to parallelize the task and speed up the analysis. Both of them consist in multi-stage processes that go from the total RNA sample isolation to the quantification of how many sample transcripts are related to each gene. Each step of the process can introduce additional noise to the final gene expression estimation, so both the techniques generate noisy data that can be interpreted only after extensive normalization procedures.\\
In this thesis, data acquired with both microarrays and RNA-seq are analyzed. In the following paragraphs, the two technologies and their necessary corrections are briefly described.
\subsection{Microarray}
Despite the complexity of its realization, the idea behind microarray analysis is simple. A typical microarray consists in a glass slide on top of which several short sequences of DNA (oligonucleotides), complementary to specific DNA fragments, are attached. On the slide surface, the molecules complementary to different regions of the same gene are grouped in clusters, called probes. Thus, by spreading a solution containing the transcripts to analyze over the microarray, they will hybridize with their complementary probes. Knowing the exact location of each probe on the slide and through a fluorescence reaction that illuminates the hybridized transcripts, it is possible to quantify the expression of each gene.
Although the relationship is not linear, the intensity of the fluorescence reflects the amount of gene-specific transcripts present in the sample of interest. This indirect measure based on fluorescence allows to quantify the expression of thousand of genes, basically at once.
\subsubsection*{Microarray analysis steps}
\begin{figure}
    \centering
    \includegraphics[width=0.80 \textwidth]{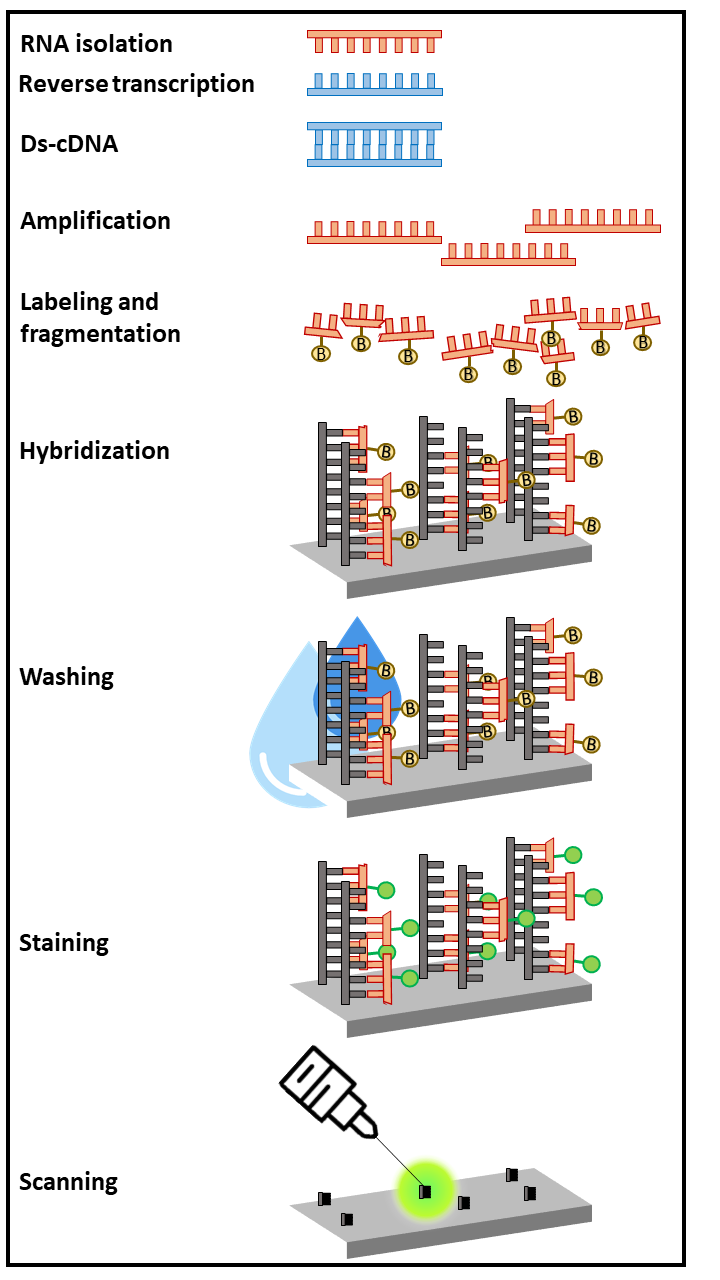}
    \caption{Steps of the microarray data acquisition process.}
    \label{fig:microarray}
\end{figure}

To understand why microarray data are considered very noisy and difficult to compare, the most common steps of their complex generation process are represented in Fig \ref{fig:microarray} and explained below\footnote{Note that there are different microarray technologies implementing slightly different procedures. The explanation reported here is quite generic and has the only aim to list the main passages where measurement noise can be introduced.}.
\begin{enumerate}
    \item \textit{RNA isolation from the sample.}\\
    The total RNA is isolated from the sample. Usually its quality is evaluated using the \ac{RIN} to decide whether to proceed or repeat the RNA isolation from another sample. This is important because during the GE process the RNAs that are no longer necessary to the cell are digested by some enzymes that continue to be active even after the cell is dead. As a consequence, the RNA extracted from a sample often includes some partially digested molecules, which are not of interest and can potentially affect the microarray analysis \cite{schroeder2006rin}. The RIN helps to discard highly degraded samples, however a certain degree of noise due to this phenomenon is always introduced already at this stage, especially in post-mortem human samples.
    \vspace{0.2cm}
    \item \textit{Reverse transcription.}\\
    The \ac{cDNA} strands of the RNAs in the sample are synthesized with a process called reverse transcription, made possible by the introduction of some primers. 
    The most often used primers are oligo-dT molecules, that allow the reverse transcription of only ploy-A tailed RNAs, so basically mRNAs. However, in some cases rRNA can have poly-A too \cite{slomovic2006polyadenylation}, and conversely, not all the mRNAs have a poly-A tail and sometimes they even exist both with and without the tail \cite{yang2011genomewide}.
    Furthermore, the use of oligo-dT alone prevents to analyze other interesting RNAs without the tail, such as lncRNA and miRNA. For these reasons, some technologies uses a combination of oligo-dT and random primers, that allow the reverse transcription of any RNAs, including rRNA which however is very abundant and not very informative. In both the cases the measure is somehow imperfect, either because of missing information or because of noisiness.
    \vspace{0.2cm}
    \item \textit{Ds-cDNA generation.}\\
    After the generation of the cDNAs, the RNA cleavage is performed with the addition of ribonuclease. Then the second strand of each cDNA is synthesized by the polymerase enzyme, producing the \ac{ds-cDNA}.
    \vspace{0.2cm}
    \item \textit{Amplification.}\\
    The newly-synthesized ds-cDNA is replicated (amplified), obtaining a large quantity of cRNAs.
    The linearity and reproducibility of this reaction, meaning the capability to maintain the proportionality of each RNA species present in the original sample and to equally amplify each sample, are essential to correctly quantify the gene expression levels in a sample and compare them among different ones. However, the efficiency of this reaction depends on many factors, ranging from experimental conditions to transcript structure, such as the amount of GC nucleotides in the ds-cDNAs \cite{arezi2003amplification,degrelle2008amplification}.
    The high level of noise introduced in this step motivates the development of microarray protocols that do not require amplification \cite{sudo2012use}.
    \vspace{0.2cm}
    \item \textit{Labeling and Fragmentation.}\\
    The cRNAs are labeled with biotin and cut into 50–100 nucleotides long fragments.
    \vspace{0.2cm}
    \item \textit{Hybridization.}\\
    The labeled and fragmented cRNAs are transferred on the microarray to start the hybridization, that is the process in which the cRNAs bind to their complementary (or partially complementary) oligonucleotides placed on the slide surface. Hybridization is a delicate chemical process that lasts approximately 16 h and depends on many factors \cite{sykacek2011impact,koltai2008specificity}. For instance, sample drying, uneven distribution of the material on the microarray, changes in salt concentration due to evaporation and the GC content of the oligonucleotides can significantly affect the efficiency of the process. Furthermore, the hybridization is never a perfect match: the cRNAs and other molecules in the sample may bind to non-complementary but similar oligonucleotides. To detect and reduce the measurement errors due to these and other factors, the oligonucleotides representing the same gene are grouped in clusters called probes and more clusters are placed in different positions in the microarray. With this trick, it is possible to detect and correct for macroscopic uneven hybridation effects. In addition, the microarray contains some background probes, that are non-complementary to any gene sequence, whose function is to evaluate background intensity levels due to spurious bindings.
    \vspace{0.2cm}
    \item \textit{Washing.}\\
    After hybridization the microarray is washed, to remove the molecules that did not strongly bind to any probe. This step requires a perfect balance between avoiding to break weak bounds and leaving untied molecules, that may increase the background level of the entire microarray.
    \vspace{0.2cm}
    \item \textit{Staining.}\\
    A molecular complex containing a fluorophore, that binds to the biotin labels on each cRNA, is added. Also in this step some variability is introduced. For instance, the fluorophore quality and stocking can affect the final fluorescence intensity \cite{munier2014physicochemical}.
    \vspace{0.2cm}
    \item \textit{Scanning.}\\
    Finally, the microarray is put into a scanner where a laser excites the fluorophores bound to the cRNAs. The level of fluorescence is measured by a detector and it depends, in a non-linear manner, to the amount of cRNAs bound to the corresponding probe. The fluorescence of some fluorophores may depend on laboratory conditions, that are time- and location-dependent.
\end{enumerate}
At the end of all these steps, the light signal coming from each probe is usually measured as a voltage or current produced by the photosensor. However, this signal is often amplified by the read-out circuitry and may be further processed with non-linear transformations via software (as it will better explained later) so the unit of measurement is rarely interesting and often cannot be defined better than "arbitrary units of fluorescence intensity". For this reason, in many papers it is not even indicated. \\
The measured value of light intensity related to a certain probe should always be compared to the one measured on its related background probe. When the latter is higher than the first one, the gene related to the probe is not necessarily unexpressed in the sample, it is simply not detectable under the experimental conditions in which the sample was analyzed. This remains a critical point of any microarray analysis. In addition, a microarray can detect only the genes whose probes are present on the chip and these may represent a subset of the entire genome.  

\subsubsection*{Data normalization}
As it emerges from the previous paragraph, several sources of noise and bias are present in the microarray analysis pipeline. To account for them, usually two macrocategories of corrections are applied to microarray data:
\begin{itemize}
    \item\textit{Within-array corrections}.\\
    Systematic within-array measurement errors mainly occur for the presence of bubbles of air or hybridization inhomogeneities that cause the fluorescence intensities measured to depend on their location on the slide. This can be quantified by comparing the intensities of the replicated probes across the array. Another systematic bias is due to the GC content of the probes analyzed, that affect both the efficacy of the amplification and hybridization process.
    Finally, given that the fluorescence intensity is non linearly proportional to the number of RNAs bound to the probes, usually an intensity-based correction is performed to rescale the values into a more linear scale.
    All these measurement errors, and possibly others depending on the specific experiment and microarray pipeline adopted, may be difficult to quantify because of their covariation. So they are usually detected using a multivariate regression and corrected by removing their model-fitted effects.
    \vspace{0.2cm}
    \item \textit{Cross-batches corrections.}\\
    As already mentioned, the outputs of a microarray analysis depend on many laboratory conditions, that are difficult to control. As a consequence, microarrays analyzed in different places or at different times may be impossible to compare. Sources of variability occur also before the microarray analysis, that can be related to the conservation, treatment and dissection of the samples. So the term batch is used to refer to the set of data whose data generation process is considered consistent among them. Homogenizing cross-batch data is a hard task because of the impossibility of keeping track of all the variables that may have affected the outcome and thus modeling them. To address this issue, various tools have been developed such as ComBat \cite{johnson2007adjusting}, which uses an Empirical Bayes method to adjust for hidden variables. More conservative approaches mainly rely in operations that modify the data distribution of the batches in order to make them more similar, such as mean or k-percentile alignment and quantile normalization. The hypothesis at the basis of these approaches is that the gene levels measured in different batches, may be individually very different, but the shape and centering of their distribution should be almost equal. 
    When the experiment allows it, one or more control samples can be included in each batch, making it possible to regress out the batch effects.
\end{itemize}
In addition to these corrections, depending on the experiment and on its purpose, other sources of variation may be corrected or included as covariates in the subsequent analysis, such as the RIN of the samples or, in case of human samples, cause of death and post-mortem interval.\\
Despite the large number of corrections often employed, microarray data remain highly noisy. For this reason, the microarray technology has been almost replaced by the RNA-Seq, described in next section.
\subsection{RNA-Seq}
\label{sec:bkg_genetics_rnaseq}
The technical revolution between the microarray technology and the RNA-Seq is that the first one identifies the transcripts by means of their hybridization to probes, which is an indirect measure, and quantifies their abundance by analyzing the intensity of their emitted fluorescence, which is another indirect measure. With RNA-seq, instead, the transcripts are literally \textit{read} nucleotide by nucleotide, thanks to a procedure called \textit{high-throughput sequencing}. 
Despite being a much more direct type of measurement, RNA-Seq is not free from noise and biases and, thus, requires various corrections too. In the next paragraphs a typical pipeline of RNA-Seq analysis\footnote{As already said for microarrays, there are various RNA-Seq technologies following different pipelines. Specifically, in this paragraph the Illumina sequencing technology is described, but with the only scope to illustrate with a practical example the complexity of the technique and the number of steps in which the errors can be accumulated.} is described as well as the corrections commonly applied to its outputs.
\subsubsection*{RNA-Seq analysis steps}
\begin{figure}
    \centering
    \includegraphics[width=0.90\textwidth]{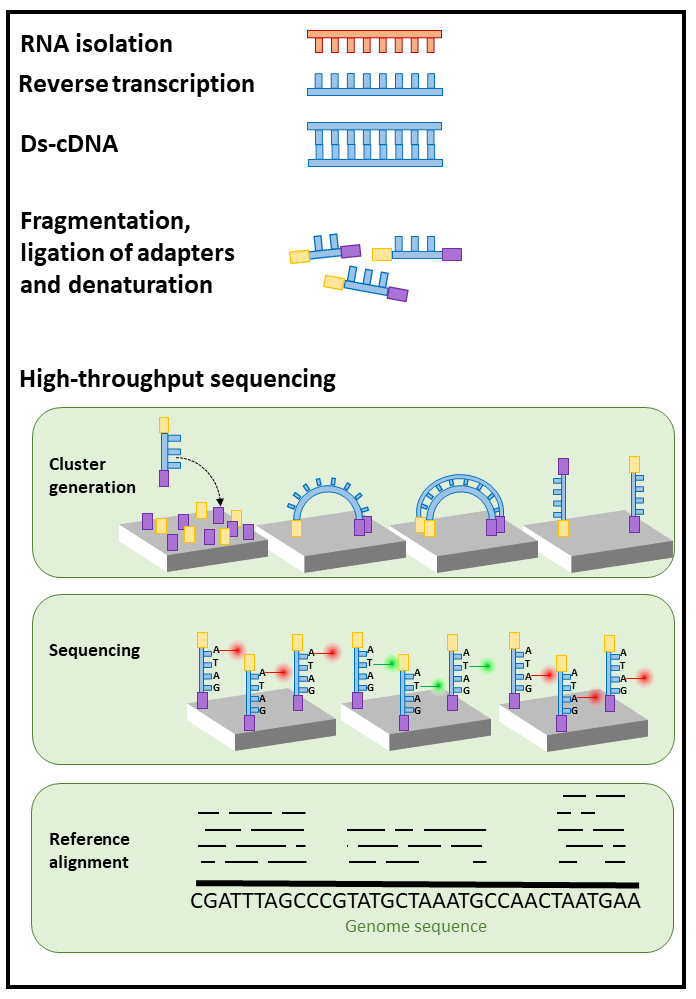}
    \caption{Steps of the RNA-seq data acquisition process.}
    \label{fig:rnaseq}
\end{figure}

The RNA-Seq starts with the same three preliminary steps of the microarray analysis, consisting in the isolation of RNA, reverse transcription and synthesis of the ds-cDNA complementary to the RNAs isolated. With respect to the microarray analysis in which it is possible to detect only the molecules that hybridize to the ones placed on the array, in the RNA-Seq all the RNAs molecules sampled are processed and read. Considering that almost the 80\% of the total RNA is constitute by rRNA, which is not of interest in transcriptome analysis, this is usually removed by rRNA-depletion or filtered out by polyA-selection. The RNA-seq steps that follows the ds-cDNA synthesis are shown in Fig. \ref{fig:rnaseq} and described below:
\begin{enumerate}
  \setcounter{enumi}{3}
  \item \textit{Fragmentation, ligation of adapters and denaturation.}\\
  The ds-cDNA are fragmented and size filtered in order to select the molecules with the appropriate length for the sequencing protocol (next step).
  In this step, miRNAs are usually filtered out, but they can be analyzed separately with other techniques. Then, the selected fragments are ligated at either end with two specific oligonucleotides, called right and left \textit{adapters}, that bind to the end and the beginning of each strand of the fragment, respectively. Note that the efficiency of this ligation step is not 100\% \cite{kozarewa2009amplification}. Finally each ds-cDNA is denaturated.
  \vspace{0.2cm}
  \item \textit{High-throughput sequencing.}
  \begin{itemize}
      \item Cluster generation: is a process where each fragment cDNA is amplified, forming clusters of its copies. The cDNAs are placed on a glass slide covered by the two oligonucleotides used as adapters. By repeating a sequence of (1) bridge binding, in which the cDNA binds its adapters to their closer complementary oligonucleotides on the slide forming a bridge, (2) polymerization, in which the second strand of the cDNAs is synthesized and finally (3) denaturation, in which the two strands are separated, the original cDNAs are amplified.
      This articulated process of amplification proved to be more linear with respect to the amplification techniques commonly employed in microarray analyses, particularly towards GC biases \cite{kozarewa2009amplification}. 
      \item Sequencing: this is the process by which the cluster are read. A set of primers that binds to one of the two adapters is placed on the slide. Then, fluorescently tagged nucleotides with a diverse emission wavelength for each of the four base types (A, T, C, G) are added to the sample. These nucleotides bind \textit{in sequence} to the amplified cDNAs, synthesizing a complementary strand. During this operation, progressive fluorescence images are taken, making it possible to reconstruct the exact sequence of the DNA copies forming a cluster from the sequence of emission wavelengths captured. Finally, in some cases the same operation is repeated starting from the opposite adapter.
      \item Reference alignment: The sequences of the fragments obtained with the last step are digitized and called "reads". An algorithm maps the reads to genes, by comparing the fragment sequence and a reference genome, and finally quantifies gene expression. This is the most delicate operation since shorts fragments and chimeric sequences can match multiple loci. Furthermore, mutations, alternative splice variants and base substitutions by inaccurate polymerases make the comparison with the reference harder. For these reasons various alignment algorithms exists based on different assumptions. 
  \end{itemize}
\end{enumerate}
As it emerges from this description, RNA-seq outputs consist in transcript counts for each gene, which are much simpler to interpret with respect to the fluorescent intensity emitted by molecules hybridized to pre-defined probes, with all the related issues of probe redundancy and background noise. Furthermore, a comparison between RNA-seq and microarray technologies showed that the first one is superior in detecting low abundance transcripts and has broader dynamic range, thus it can detect more differentially expressed genes with higher fold-change \cite{zhao2014comparison}.

\subsubsection*{Data normalization}
Despite all the advantages of RNA-Seq over the microarray technology, RNA-Seq remains a multi-step process in which various errors can occur at different stages and despite initially it was believed that its outputs did not need sophisticated normalization \cite{wang2009rna}, it is now well known that multiple corrections are necessary to fairly compare gene expression levels within and between samples.
There is not a consensus on which normalization procedures and algorithms should be adopted, but the most used ones are listed below.
\begin{itemize}
\item \textit{Gene length}\\
The transcripts of longer genes are fragmented in a higher number of pieces than those of shorter ones, which means that longer genes have a higher number of reads even if they are expressed as much as shorter genes. Thus, it is important to correct the number of counts of each gene by its length, in order to fairly compare gene expression levels within a sample. However, correcting for gene length is an approximation, because due to alternative splicing multiple transcripts with different length can originate from the same gene. For this reason, this correction is often skipped when the expression levels are not compared across genes, but across samples.
\vspace{0.2cm}
\item \textit{Sequencing depth}\\
The total number of RNA fragments extracted from each sample depends on many factors, often difficult to control; this makes the RNA-Seq outputs not comparable between samples.
For instance, a sample with half of fragments will give, on average, half the number of reads mapping to each gene. As a consequence, an intuitive correction consists in dividing the number of counts for each gene by the total number of reads in the sample, also called sequencing depth.
\vspace{0.2cm}
\item \textit{RNA composition}\\
Sequencing depth correction alone does not guarantee comparability among different samples. In fact, the number of reads mapping to a gene also depends on the RNA population that is being sampled. Let's consider, for instance, two sample A and B, if some genes are uniquely or highly expressed in A but not in B, the sequencing 'real estate' available for the remaining genes in A is less than the one of B. Correcting for such type of RNA composition artifacts is fundamental to properly compare inter-sample gene expression levels.
This normalization can be performed only under the hypothesis that the major part of the genes are equally expressed across different samples, which makes it possible to infer RNA composition from the data themselves.
In fact, this is the main hypothesis at the basis of the two most reliable methods to correct for RNA composition \cite{dillies2013comprehensive}: the \ac{TMM} \cite{robinson2010scaling} implemented in the edgeR Bioconductor package and the one implemented adopted in the DESeq Bioconductor package \cite{anders2010differential}. The main difference between the two is that the TMM estimates the normalization factor for the RNA composition correction excluding the genes with a higher variability among the dataset, the so called "trimming" operation. This step improves the robustness of the normalization process and it is especially important when the number of samples is limited. A brief description of the TMM algorithm, which is the one used for the analysis of the RNA-Seq data of this thesis, can be found in Box \ref{test2}.
\end{itemize}
As already said for microarray data, also for RNA-Seq data, depending on the specific experiment, further normalizations may be necessary to correct for other sources of variablity.

\begin{mytable}[label=test2,float=ht!]{Trimmed Mean of M-values}
\small
According to the TMM the expected level of expression $E[Y_{gs}]$ of a gene $g$ in a sample $s$ measured by the RNA-Seq, depends on the length of the gene $L_g$, the total number of fragments in the sample $N_s$ and the total RNA output of the sample $R_s$, as follows:
\begin{equation*}
    E[Y_{gs}] = \mu_{gs} \cdot L_g \cdot \frac{N_s}{R_s}  \text{\;\;\;where\;\;\; $R_s=\sum_{g}  \mu_{gs} \cdot L_g$}
\end{equation*}
where $\mu_{gs}$ is the true gene expression level of $g$ in $s$. Correcting for the diverse RNA composition of each sample would require to remove the dependence of $E[Y_{gs}]$ from $R_s$, but clearly its value is not known nor measurable. However, according to the TMM algorithm, comparing the gene expression of two samples $s$ and $s'$ it is possible to estimate the $\frac{R_{s'}}{R_s}$ ratio, which is called $TMM^{s'}_{s}$ and is calculated with:   
\begin{equation*}
    log_2 (TMM_{s}^{s'}) = \frac{\sum_{g \in G^*} w_{gs}^{s'} \cdot M_{gs}^{s'}}{\sum_{g \in G^*} w_{gs}^{s'}}  \text{\;\;\;where\;\;\;}  \begin{cases}
      M_{gs}^{s'}=log_2 \left(\frac{Y_{gs}/N_s}{Y_{gs'}/N_{s'}} \right) =log_2 \left(\frac{\mu_{gs} \bcancel{L_g} R_{s'}}{\mu_{gs'} \bcancel{L_g} R_{s}} \right)\\
      w_{gs}^{s'}=\left( \frac{N_s-Y_{gs}}{N_s \cdot Y_{gs}} + \frac{N_{s'}-Y_{gs'}}{N_{s'} \cdot Y_{gs'}} \right)^{-1}
    \end{cases}       
\end{equation*}
In the formula above the $log_2 (TMM_{s}^{s'})$ is computed as the inverse-variance weighted average of the $M_{gs}^{s'}$. In fact, the weights $w_{gs}^{s'}$ are the inverse variance of the $M_{gs}^{s'}$ according to the delta method \cite{cox2005delta}. Note that all the sums span over the $G^*$ ensemble, which is defined as the set of genes whose measured expression value is similar across the two samples $s$ and $s'$. 
So the normalization simply consists in taking one sample $s'$ as a reference, computing the $TMM_{s}^{s'}$ for all the other sample with respect to the reference. Finally, by dividing the measured gene expression levels $E[Y_{gs}]$ of a sample $s$ by the $TMM_{s}^{s'}$, which provides the best estimation of the $\frac{R_{s'}}{R_s}$ ratio, it is possible to correct for differences in RNA composition across different samples.
\end{mytable}

\section{Typical gene expression analysis}
\label{sec:bkg_gene_expr_analysis}
There are countless different ways to analyze the huge amount of information encoded in transcriptome data. However, the studies in which two groups of data are compared, such as data belonging to TDCs and subjects affected by a disease under examination, usually include the following steps:
\begin{itemize}
    \item \textit{Pre-processing.}\\
    The pre-processing includes all the corrections described in the previous section. In addition, depending on the specs of the technology adopted, it may be necessary to filter out genes whose transcripts length would be barely detectable by the method. Other filtering procedures may be done to select genes that are reliably expressed in most of the samples, thus cutting out the most noisy data. Finally, gene expression levels are often Log2 transformed. 
    This operation, in fact, brings many advantages:
    \begin{enumerate}
        \item the typical skewed distribution of gene expression data is brought it closer to a normal distribution, which is an essential prerequisite of many statistical tests;
        \item the high variability of the data decreases, benefiting commonly employed correlation coefficients\cite{jaskowiak2018clustering};
        \item the original value range is stretched to a scale that better represents biologically relevant changes. For instance, a doubling of a value, which is considered a relevant change, on the Log2 scale corresponds to one unit variation.
    \end{enumerate}
    \vspace{0.2cm}
    \item \textit{Differential expression analysis.}\\
    This is the most important step of gene expression analysis and consists in identifying the genes that are significantly \ac{DE} among two groups of data, usually cases and healthy controls. Gene disregulation, with respect to controls, may be either a cause or an effect of the disorder under examination. In both the cases, DE genes are of interest to understand the pathophysiology of the disorder.\\
    Various softwares exist to identify the DE genes. The Bioconductor package alone lists 351 software components dedicated to this purpose \cite{bioconductor}. Some are dedicated specifically either to microarray or RNA-Seq data, some differ in the in the type of hypothesis testing (i.e. null hypothesis or Bayesian inference) performed, others in the statistical assumptions made on the distribution of the data to evaluate the significance of the DE genes. About the latter aspect, an interesting deepening on the distribution of the RNA-Seq data can be found in Box \ref{test3}.\\
    A large part of these packages is based on the use of mixed models, which perform robust multivariate regressions allowing to include all the variables of interest for the experiment at hand. Basically, the expression of each gene across the various samples is modeled with a regression that includes various sources of variation, such as sex and age, and of course the presence/absence of disease. The genes for which the mixed models attributed a significant importance to the disease are considered DE. For their level of customizability, approaches based on mixed models are often preferred when the data available requires to take into account the effects of confounding variables that may mislead the results such as phenotypical information or quality metrics of the sequencing process. 
    \vspace{0.2cm}
    \item \textit{Co-expression network analysis.}\\
    The differential expression analysis provides a list of genes possibly implicated in the disease under examination. However, the typical limited size of transcriptomic dataset makes it unlikely to capture all the susceptibility genes of very heterogeneous disorders such as ASD. For this reason, it is common to include also a co-expression network analysis. In practice, a network is constructed depending on the degree of correlation of the expression levels of each pair of genes across samples.
    Several evidences show that, in this type of network, the genes with shared molecular or biological functions are highly correlated and thus are grouped into modules. 
    Hypothesizing that the genes implicated in a disorder, but not detectable with the DE analysis because of the limited sample size, have similar functions to those detected, it is possible to expand the gene list found at the previous step to all the genes in the modules enriched in DE genes.
    Although this step may appear redundant or remotely related to the objective of DE studies, the importance of network analysis is documented by many works.
    For instance, it has been proved that selection criteria based on network connectivity find gene lists biologically more informative than methods that ignore gene-to-gene relationships \cite{langfelder2013hub}.
    Furthermore, several empirical evidences commonly agree that the most connected genes within disease related modules are of clinical importance and can lead to relevant biological insights \cite{horvath2006analysis,carlson2006gene,oldham2006conservation,keller2008gene,ivliev2010coexpression,gargalovic2006identification,ghazalpour2006integrating,fuller2007weighted,presson2008integrated,langfelder2012systems,miller2010divergence,dawson2012r}.
    \vspace{0.2cm}
    \item \textit{Validation analyses: permutation test, false discovery rate and enrichment analysis.}\\
    All the gene expression analyses suffer from the high degree of noise characterizing their data and from the discrepancy between the high dimensionality of the transcriptome ($\sim$ 25,000 genes) and the typical low number of biological samples available. Both these conditions increase the chances to get spurious results. Furthermore there are countless different ways to analyze the huge amount of information encoded in transcriptome data, which increases the arbitrariness of the methodological choices. For these reasons, the validity of each intermediate result should be always assessed by a permutation test and corrected for false discovery rate. In addition to these precautions, it is common to perform some sub-analyses to check whether the results obtained are confirmed by literature. A typical example consists in \ac{EA}, that tests if the intersection between two lists is random or not. It is usually used to compare lists of genes known to be implicated in a disorder or in a biological condition/function of interest with the gene list found during the analysis. When the list under examination is compared to all the known gene lists that have common molecular function, or are involved in the same biological process or play a role in the development and maintenance of the same cellular component, the analysis is called \ac{GO} enrichment analysis.
    \vspace{0.2cm}
    \item \textit{Visualization of results.}\\
    Due to the high dimensionality of the data, gene expression studies often adopt some visualization tools, specifically developed to condense such information, to ease the comprehension of the results.
    Commonly adopted approaches to visualize how distant multidimensional data (with a dimension higher then 3) are include dendrograms and \ac{MDS}. Various tools have also been developed to ease the interpretation of the GO analysis results. In fact, the large number of biological functions to tests and their cross-hierarchical structure may represent an obstacle to the biological interpretation of transcriptomic data.
\end{itemize}
Except for the pre-processing step, which has been already illustrated in Sec. \ref{sec:bkg_transcriptomics}, the algorithms and tools used in this thesis to perform the above points are described in the following subsections. 
\begin{mytable}[label=test3,float=ht!]{RNA-Seq data distribution}
\small
The process of counting the transcript's fragments with RNA-Seq, as described in Sec. \ref{sec:bkg_genetics_rnaseq}, can be easily viewed as that of a detector, for which each event (in this case 'fragment') has a certain probability $p$ to be detected (in this case 'read'). Thus, the RNA-Seq represents a Poissonian process \cite{cellerino2018transcriptome}. \\
Nevertheless, it can be experimentally observed that the transcript counts associated to a gene across a set of samples, that can be considered copies of each other\footnote{Note that in biology, samples taken from the same tissue in subjects with the same condition are considered replicates of each other.} presents an overdispersion due to biological and technical variations that makes the distribution more similar to a negative binomial \cite{cellerino2018transcriptome}.
For this reason some popular packages to identify DE genes assume the data to follow a negative binomial, such as DESeq2 \cite{love2014moderated}.\\
However, the parameters of a negative binomial distribution fitting RNA-Seq counts can not be derived from the data generation process; they can only be estimated from data. In addition, it has been shown that the negative binomial assumption is violated in many RNA-Seq datasets and this may explain the high number of false positive findings in DE analyses based on this assumption \cite{hawinkel2020sequence}.\\
Despite the shape of the counts distribution is often skewed, many other packages for DE analysis are based on the assumption of normality, such as LIMMA \cite{ritchie2015limma} and Sleuth \cite{pimentel2017differential}. 
This choice is often motivated by the following reasons:
\begin{itemize}
    \item the Log2 transformation of the raw counts often reduces the skewness of the distribution enough to make the normality assumption acceptable (an example of this phenomenon, using real data from this thesis, is reported below);
    \begin{center}
    \includegraphics[width=0.9\textwidth]{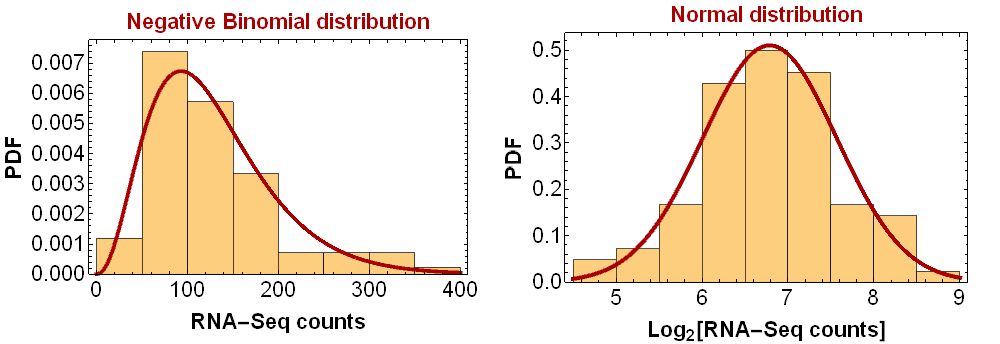}
    \end{center}
    \item the central limit theorem ensures that the sum of a large number of independent variables is normally distributed, regardless their individual distribution. This theorem applies to the RNA-Seq counts for those genes that have a sufficiently high number of counts and for which their overdispersion is explained by the summation of a sufficiently high number of variables;
    \item given that no distribution is theoretically justified to model RNA-Seq counts, normality is a reasonable and conservative assumption to make that paves the way to the adoption of many statistical methods.
\end{itemize}
In conclusion, the problem of modeling RNA-Seq data is controversial. Good practices to reduce false positive findings include the adoption of strict selection criteria on both the size effect and the p-value and comparisons with current literature.
\end{mytable}
\subsection{DE analysis with linear mixed models.}
\label{sec:bkg_lmm}
As already mentioned, a good option to study DE genes consists in using mixed models. Mixed models are a particular type of regression in which the independent variables can be of two kinds: \ac{FV} and \ac{RV}.
They differ in how their weights are estimated from data. The coefficients of FVs are assumed to be estimated without error, while the ones of RVs are considered to have a prior distribution.
Mixed models can be mainly divided into:
\begin{itemize}
    \item \ac{LMMs}: they are based on a linear regression and the random variables, as well as the regression error, are assumed to follow a Gaussian distribution.
    \item Non-Linear Mixed Models: with respect to LMM they can model also non-linear dependencies.
    \item Generalized Mixed Models: they are a generalization of the LMMs that allow the response variable to have an error distribution other than the Gaussian distribution.
\end{itemize}
The functional form of transcriptome data dependency is mainly unknown. The LMM has clearly less degrees of freedom with respect to the other two models described above and, being based on the standard hypotheses of linearity and normality, it is considered more conservative and less arbitrary, so it is usually preferred. This paragraph provides an introduction to LMMs, focusing on how to properly use them in a DE analysis and how to interpret their results. A thorough explanation of the math behind LMMs can be found in \cite{jiang2007linear}.

\subsubsection*{How to use LMM in a DE analysis.}
A linear mixed model can be represented as:
\begin{equation}
    \vec{y} = X \vec{\alpha} + Z \vec{\beta} + \vec{\epsilon}
    \label{eq:lmm}
\end{equation}
Where $\vec{y}$ contains the values that the dependent variable assumes in $N$ instances; $X$ and $Z$ are the design matrices that report the values that the fixed and random variables, respectively, taken in the $N$ instances.
These variables can be either continuous or categorical.
The $\vec{\alpha}$ and $\vec{\beta}$ terms are vectors of coefficients to be estimated by the LMM, representing the strength with which fixed and random variables influences the dependent variable, while $\vec{\epsilon}$ is the fit error. 
In a typical DE analysis, $\vec{y}$ is the gene expression level of one gene across all the available samples and the independent variables put into $X$ and $Z$ can be either technical, demographic or medical/biological information of the samples that are supposed to affect $\vec{y}$. 
Among the fixed effects, a binary variable is usually employed to indicate whether the data belong to the group affected by the disease under examination or to the control one. The LMM jointly estimates the coefficients of each variable, including the case/control membership. The coefficient assigned to the latter provides an estimate of how strongly the disease affects the gene expression of the gene under examination. By repeating this analysis for each gene, it is possible to find a list of genes possibly related to the disorder under examination.\\
How to choose FVs and RVs is a common critical point of the LMM analysis, due to the profusion of conflicting definitions that makes the distinction between the two very confusing \cite{gelman2005analysis}. 
As briefly mentioned before, the only difference between random and fixed effects is in how their coefficients are estimated: $\vec{\alpha}$ is calculated with least squares methods (or, more generally, maximum likelihood), while $\vec{\beta}$ is estimated with partial pooling (often called shrinkage). In practice, the coefficients of fixed variables are supposed to have no intrinsic errors, meaning that the errors depend only on the modeled random noise $\vec{\epsilon}$ and on the goodness of the fit. Instead, the dependency of $\vec{y}$ on RVs are supposed to be intrinsically variable, and their coefficients are supposed to follow a Gaussian distribution, whose variance is estimated by the LMM. Furthermore, when a RV has multiple categorical values and the data are clustered with respect to its values, computing the coefficients by partial pooling allows to include into the model the fact that these clusters have some commonalities. One practical advantage of this approach is that if one cluster has few data points, its group's effect estimate will be based partially on the more abundant data from other groups. This is a nice compromise between estimating an effect by completely pooling all clusters, which masks group-level variation, and estimating an effect for all clusters completely separately, which could give poor estimates for low-sample clusters.\\
In conclusion, no variable can be considered intrinsically FV or RV, but under certain conditions it may be convenient to compute their coefficients with partial pooling, and in others it may be better to use least squares which is computationally less expensive. 
The criteria for which a variable may benefit from being defined as RV are:
\begin{itemize}
    \item the variable affects the data in a non-completely deterministic way, however, the data that share the same variable value are affected by it in a similar way;
    \item the variable is categorical and has many different values \cite{fox2015ecological};
    \item each value of the variable is represented by a relatively low number of data points, but multiple data point are available for most of them \cite{fox2015ecological};
    \item the different values of the variable are unevenly sampled \cite{fox2015ecological}.
\end{itemize}
Consistently with these rules, a typical RV in gene expression studies is the batch \cite{espin2018comparison}.

\subsubsection*{How to interpret LMM results.}
The outputs of the LMM are:
\begin{itemize}
    \item the coefficients estimated for each variable, that show how strongly the data depends on a variable.
    \item the p-values associated to each coefficient, that represent the probability of the latter to be different from 0, taking into consideration the fitting error.
    \item the variances of the coefficients associated to each random variable.
\end{itemize}
The coefficient of the case/control variable is usually converted into an effect size, by dividing it for the standard deviation of the data distribution. To estimate which genes are significantly DE with respect to the control and case group both the size effect and the p-value (appropriately corrected for the false discovery rate) are relevant, as good fits, with low p-values, may arise even for biologically insignificant effects. On the other hand, large effect sizes are not reliable when the fit is bad, with large errors and thus high p-values. 
While it is important to consider both these figures of merit when evaluating DE genes, there is not a consensus on how to do it, with some studies attributing more importance to p-values \cite{zhou2014efficient,jo2013statistical}, others to effect size \cite{sullivan2012using,lee2016alternatives}, and others advocating a more balanced approach \cite{marot2009moderated,goodman2019proposed}.
\subsection{Co-expression network}\label{sec:bkg_network}
\begin{figure}
    \centering
    \includegraphics[width=0.99 \textwidth]{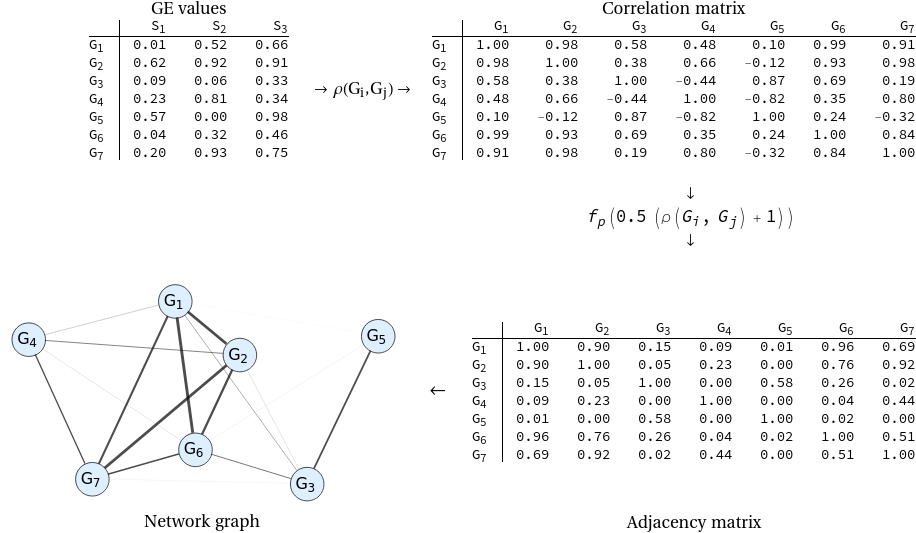}
    \caption{Graphical representation of the steps required for constructing a signed gene co-expression network.}
    \label{fig:network-construction}
\end{figure}

Several evidences show that genes grouped within modules of a co-expression network have similar functions. Thus the identification of the network modules enriched by a list of genes implicated in a disease, is a good way to expand the latter to other genes of possible interest.\\
The construction of a co-expression network (graphically summarized in Fig. \ref{fig:network-construction}) articulates in three steps \cite{zhang2005general}:
\begin{enumerate}
    \item Definition of the similarity matrix $s_{ij}$, that quantifies the similarity between pairs of genes, usually through correlation.
    \item Definition of a family of adjacency functions $f_p(s_{ij})$, used to transform the similarity matrix by suppressing small correlations and enhancing large ones.
    \item Selection of the best adjacency function $f^*$ and construction of the adjacency matrix $a_{ij}=f^*(s_{ij})$.
\end{enumerate}
After the computation of the network, the following steps are performed to identify its modules:
\begin{enumerate}
    \setcounter{enumi}{3}
    \item Definition of a measure of gene dissimilarity based on the adjacency matrix, used to calculate which genes are more similar to each other with respect to the rest of the network. Note that this is a different concept from the similarity matrix, because the latter is defined pairwise without considering the rest of the genes in the network.
    \item Identification of network modules using a clustering algorithm on the dissimilarity matrix.
\end{enumerate}
The following sections provide an in-depth description of all the steps reported above. In addition, paragraph \ref{subsec:bkg-deltacon} describes a method to compare different networks, which is a difficult but useful operation in many gene expression studies, including the one presented in this thesis.    
\subsubsection{Similarity matrix}
Each element of the similarity matrix weighs the similarity between pair of genes with a number between 0 and 1, where 1 indicates perfect similarity.
The similarity is commonly defined starting from the Pearson's correlation between pairs of genes across the available samples: $\rho(g_i,g_j)$.
The similarity matrix is then derived applying a transformation on the correlation values, depending on whether one wants to preserve the sign of the correlation or not, building a \textit{signed} or \textit{unsigned} network, respectively.\\
For an \textit{unsigned} network, the absolute value of the correlation is taken: $s_{ij}=|\rho(g_i,g_j)|$, while for a \textit{signed} network, the correlation is transformed in the following way: $s_{ij}=(1+\rho(g_i,g_j))/2$. If the first (second) transformation is applied, genes perfectly anti-correlated to each other will have a transformed similarity value of 1 (0).
Note that obviously $s_{ii}=1$, since every node is perfectly similar to itself, however, for the purpose of calculating all the network concepts described in the following section, it is assumed by convention that the diagonal elements of the similarity matrix are set to 0, as it simplifies many formulas.

\subsubsection{Definition of the family of adjacency functions}
After constructing the similarity matrix and appropriately transforming it, the adjacency matrix must be derived from it. In order to do this, several adjacency functions are tested and the best one is selected. Usually, functions belonging to the same family all have a common base form, with one or two parameters that distinguish them from each other. Common examples are:
\begin{itemize}
    \item Powers: $f_p(s_{ij})=s^p_{ij}$.
    \item Sigmoids: $f_{ab}(s_{ij}=\frac{1}{1+e^{-a(s_{i,j}-b)}}$
    \item Hard thresholds: $ f_{t}(s_{ij}) = \begin{cases}
      0 &  \text{if }x < t \\
      1 &  \text{if }x \geq t
    \end{cases}\, $
\end{itemize}
The first two families of functions can be described as soft thresholds and generate weighted networks, i.e., networks in which the connection between two genes is represented by a weight in the interval $[0,1]$, with 1 being the strongest connection and 0 being the weakest one. Many other families of soft thresholds have been proposed in literature, but they all behave very similarly \cite{zhang2005general}. The key characteristics that they all share is that they are monotonic in $s_{ij}$ and that they enhance the difference between small similarity values and large ones.\\
The last family instead generates binary, un-weighted networks, in which the connection between two genes is either present (1) or absent (0). Obviously, the use of hard thresholds generates simpler networks, for which concepts such as node connectivity are easier to define and have straightforward interpretations, however they are also highly sensitive to the specific threshold used and lose a substantial amount of information, while technically no information is lost when applying soft thresholds.\\

\subsubsection{Selection of the adjacency function}
The best adjacency function is usually selected from the chosen family using the scale-free topology criterion. \ac{SFT} is a condition in which the number $P$ of nodes (i.e., genes) with connectivity $k$ is proportional to a (negative) power of said connectivity: $P(k) \sim k^{-\gamma}$.
The emergence of SFT follows directly from the growth of the network (larger networks are more commonly SFT) and is a property that is also thought to characterize the evolution of biological systems \cite{barabasi1999emergence,albert2000topology}. An important mathematical feature of SFT networks is that they exhibit great tolerance towards errors, making them suitable for the analysis of inherently noisy data such as GE values.
To understand how to verify the SFT criterion, first the concept of node connectivity must be introduced. In an un-weighted (i.e., binary) network, the connectivity of the $i$-th node is the number of nodes it is connected to. This can be generalized for a weighted network as follows:
\begin{equation}
    c_i=\sum_j{a_{ij}}
\end{equation}
A common way to proceed for the evaluation of SFT, is then to calculate $P(k)$ for various values of $k$, fit the function $\log(P(k)) = -\gamma \log(k)$ with free parameter $\gamma$ and calculate the $R^2$ of the fit.\\
It has been empirically observed that, for example for the power functions, a SFT is more easily reached at higher power values. However, since $s_{ij}<1$, high power values also mean a contraction of the connectivity and thus a compression of the information towards the lower end of the $[0,1]$ interval. This is often undesirable and therefore a common way to choose the power function in such cases is to plot the $R^2$ of the SFT fit against the mean connectivity of the network and find a compromise between a good $R^2$ (above 0.8) and a non-zero connectivity.

\subsubsection{Definition of a measure of gene dissimilarity}
Now that the network is fully constructed, it is necessary to define a measure of dissimilarity between pairs of genes, and hence the dissimilarity matrix, that can be used to derive gene modules.
This is usually done through the calculation of the \ac{TOM} \cite{ravasz2002hierarchical}. The elements $t_{ij}$ of the TOM describe how similarly connected the nodes $i$ and $j$ are, and can be defined from the adjacency matrix as follows:
\begin{equation}
    t_{ij} = \frac{\sum_u{a_{iu} \cdot a_{uj}} + a_{ij}  }{ \min{(k_i,k_j)} + 1 - a_{ij} }
\end{equation}
Where $k_i$ and $k_j$ are the connectivities of node $i$ and $j$ respectively. This formula has an easy interpretation in case of binary networks. Consider as an example two nodes that are connected to all other nodes in the same way: $a_{iu} = a_{uj} \ \forall u$. Then, the numerator of $t_{ij}$ becomes $\min (k_i,k_j)+a_{ij}$ (remember $a_{ii}=0$). If also $a_{ij}=1$ then $t_{ij}=1$, showing that two nodes that are connected to the same other nodes, and that are connected to each other, have a topological overlap value of 1.\\
Similarly, if two nodes do not share any connection and are not connected to each other, then $t_{ij}=0$.\\
The interpretation is more difficult for weighted networks, but it can be shown that $t_{ij} \in [0,1]$ and increases when the two nodes share many connections.\\
From the TOM values, it is easy to derive a dissimilarity measure as $d_{ij} = 1 - t_{ij}$.

\subsubsection{Identification of network modules}
There are many ways to identify groups of genes that are well connected to each other. The work described in this thesis uses an average linkage clustering algorithm on the dissimilarity matrix as commonly applied in many other studies \cite{d2005does,yip2007gene,kerr2008techniques}.
Average linkage clustering is a particular case of hierarchical clustering, that is a process that starts by treating each element (network nodes in the case at hand) as a separate cluster. Then, it repeatedly merges the two clusters that are closest to each other until all elements are in a single cluster. 

From the definition above, it can immediately be seen that hierarchical clustering needs two things to function:
\begin{itemize}
    \item a function that measures distances between two elements $x$ and $y$: $f(x,y)$. This is straightforward in the network case: the dissimilarity matrix $f(x,y) = d_{xy}$ is used.
    \item a linkage criterion which measures the dissimilarity between two clusters as a function of the pairwise distances of their elements. There are several ways of defining this, as outlined in \cite{murtagh2012algorithms}. The approach adopted in this work is the average linkage, which defines the dissimilarity between two clusters $X$ and $Y$ with cardinality $|X|$ and $|Y|$ as:
    \begin{equation}
        \frac{\sum\limits_{\substack{x\in X, y \in Y }}{f(x,y)}}{|X||Y|}
    \end{equation}
\end{itemize}
Note that these criteria define the hierarchy of clusters, but if one is interested in extracting network modules, a further decision regarding the level at which the hierarchy is split must be made. In fact, at the top level, all the elements of the network belong to a single cluster, while at the bottom level, every element is its own cluster. The researcher must decide at which intermediate level to observe the network, considering cluster number and size. This is an arbitrary choice, as no hard criteria exists.

The identified modules are often represented through their eigengenes and/or their hubs as they provide a compact information of regulation of the whole module \cite{langfelder2007eigengene,liu2016weighted}.\\
Eigengenes are linear combinations of the genes of the module, identified with the principal component analysis \cite{ringner2008principal}. Usually only the first eigengene is used, corresponding to the first component. In fact, due to the high correlation between the members of a module, the first principal component explains a large fraction of the variance in gene expression.\\
Hubs instead are defined as the genes with the highest intra-modular connectivity, calculated as $c^M_{i}=\sum_{m\in M}{a_{im}}$, where $M$ is a module and $m$ is an index running through all the genes in $M$.

\subsubsection{Evaluating similarity between networks}
\label{subsec:bkg-deltacon}
Sometime it is necessary to evaluate the similarity between two weighted networks, for instance between different species, tissues or the case and control groups. This is not a simple task, because the two networks may have a different number of connections. There are various different approaches to compare networks, as outlined in \cite{tantardini2019comparing}, but the description of all of them is out of the scope of this thesis. This paragraph only describes the one used in this work, called DeltaCON \cite{koutra2013deltacon}.\\
Its is based on comparing the matrix $b_{ij}$ of the pairwise node affinities of both networks. Node affinity is a concept that aims at quantifying the influence that a certain node $i$ has on another node $j$.
The authors of \cite{koutra2013deltacon} propose to calculate it as follows:
\begin{equation}
    S = (I + \epsilon^2 C - \epsilon A)^{-1} 
\end{equation}
Where $\mathbf{I}$ is the identity matrix, $C$ is the diagonal matrix of the connectivity values $c_i$ and $A$ is the adjacency matrix. The parameter $\epsilon$ is calculated as $1/(1+\max_i(c_i))$.
The idea behind this formula stems from the fact that, ignoring the $\epsilon^2$-weighted term, $S$ can be written as:
\begin{equation}
    S \approx (1-\epsilon A)^{-1} \approx 1 + \epsilon A + \epsilon^2 A^2 + ...
\end{equation}
Considering that $k$-th powers of $A$ encode information on the interaction between $k$-neighbors (just as the matrix $A$ encodes the interaction between first neighbors), it can be said that $S$ takes into account all the interactions at different levels, by appropriately discounting them with a decreasing weight (remember $\epsilon < 1$). In addition to this, the $\epsilon^2 D$ term also considers the total connectivity of each node.\\
The DeltaCON measure then contains a measure of the similarity of the two matrices $S_1$ and $S_2$ computed for the two networks:
\begin{equation}
    \delta_c =\sqrt{ \sum_{i,j} ( \sqrt{S_{1_{ij}}} - \sqrt{S_{2_{ij}}} )^2 }
\end{equation}
This measure of distance is known as Matusita distance \cite{matusita1953estimation} and the authors argue that it is better than other more standard distance functions such as euclidean distance because the square root boosts the values of $S_{ij}$ (because they are smaller than 1) and thus enhances the sensitivity of $\delta_c$ to small differences between $S_1$ and $S_2$.
\subsection{Validation analyses.}
In this paragraph four common analysis methods employed to validate results obtained with gene expression data are briefly described.

\subsubsection{Permutation test}
The result of any scientific analysis made to validate an hypothesis $H$ is measurable with a number.
Without an in-depth prior knowledge of the phenomenon under examination and of the type and distribution of the data involved in the analysis, it is really difficult to understand how significant the numerical value of the result is. The permutation test solves this issue, being a type of hypothesis test that does not require any previous knowledge and can be applied to any type of analysis. It consists in evaluating what is the probability that the result obtained belongs to the null distribution, i.e. the distribution of results that would be obtained under the hypothesis that $H$ is not true, the so called null-hypothesis $H_0$.
What makes the permutation test so versatile is that the null distribution is user-generated by defining an appropriate data generation process respecting the null hypothesis.\\
For instance, consider two lists:
\begin{itemize}
    \item A: the list of $n_1$ genes possibly involved in a disease according to a certain analysis.
    \item B: the list of $n_2$ genes known to be involved in the same disease, according to literature.
\end{itemize}
Both these lists have been obtained by analyzing a larger list $C$ of $N$ genes.
Suppose that $A\cap B$ has length $l$ and that the hypothesis  $H$ to test is that $A$ has a significant intersection with $B$. As a consequence, $H_0$ would state that the intersection between $A$ and $B$ is random. To generate a null distribution compatible with this hypothesis the intersection length $l_i$ of pairs of randomly generated lists of length $n_1$ and $n_2$, independently drawn from $C$, is  calculated. The permutation test requires to repeat this operation for a number of times $P$ in order to get a list of lengths $D={l_1,l_2,...,l_P}$.
To test the null hypothesis, the probability that $l$ belongs to the distribution that generated $D$ is calculated as:
\begin{equation}
    p = \frac{x+1}{P+1}
\label{eq:pvalue}
\end{equation}
Where $x$ is the minimum between the number of times in which $l\leq l_i$ and the number of times in which $l \geq l_i$. The addition of 1 to the numerator and the denominator is meant to include $l$ in the list $D$, as supposed by $H_0$.
At this point, $H_0$ should be considered accepted/rejected if $p$ is greater/smaller than a certain threshold of significance. This threshold is conventionally set to 0.05, which means that the probability of erroneously rejecting the null hypothesis, i.e a false positive, should be approximately less than 5\%.
Because of the way in which it is calculated, $p$ cannot assume a value below $1/(P+1)$, thus $P$ must be set considering the chosen threshold.
Note that Eq. \ref{eq:pvalue} is by definition the way to empirically calculate the p-value testing whether a certain value belongs to a distribution.
Therefore, this equation applies to any permutation test analysis, not limited to the case at hand of the intersection of two lists.
With a different analysis, and thus a different starting hypothesis, the only thing that would change would be the data generation process of $D$, that should follow the new $H_0$.
In some cases, generating enough data in $D$ may be computationally very expensive. However, if the analytical form of the null distribution is known or can be fitted from the data available in $D$, one may use it to calculate the probability of a certain value to belong to it instead of using Eq. \ref{eq:pvalue}.
On a final note, it is important to underline that the permutation test is an approximate test, because the estimation of the p-values is based on randomization.  However, approximate tests are often computationally less expensive than exact tests, that instead explore all the possible configurations of results.  

\subsubsection{False discovery rate}
In the previous paragraph it has been explained that a p-value below a threshold of $\tau$\% implicates that the probability of false positives is approximately below $\tau$\%. However, the probability of false positives increases with the number of hypotheses, referred to the same problem, that are being tested. 
This problem is particularly exacerbated in genomewide studies, where each analysis is applied to each gene, meaning that thousands of hypothesis tests are conducted simultaneously. In these cases, it is necessary to correct the obtained p-values, being them obtained with a permutation test or with other techniques, to take into account the \ac{FDR} expected when a certain number $\nu$ of independent hypotheses has been tested. \\
There exists different methods to implement FDR correction.
The most straightforward approach is the Bonferroni method \cite{bland1995multiple}, which simply consists in multiplying each p-value for $\nu$. However, the Bonferroni method is considered too stringent, since conservatively guarding against the occurrence of false positives it often overlooks many false negatives, i.e. missing findings \cite{narum2006beyond} when $\nu$ is large. Another broadly applied FDR correction is the \ac{BH} \cite{benjamini1995controlling} method that requires to:
\begin{itemize}
\item sort in ascending order the p-values of the $\nu$ tests;
\item choose a significance threshold $\tau_{fdr}$ to use to select the significant results after the FDR correction;
\item compute the correction: $C(p)= p_i\frac{\nu}{i}$, where $i$ refers to the index of the sorted list obtained at the first step;
\item consider significant only those analyses for which $C(p_i) \leq \tau_{fdr}$. 
\end{itemize}
In practice, by correcting the p-values according to their ranking, the BH method represents a weaker correction with respect to the Bonferroni one but it constitutes a better compromise between the probability to get false positives and false negatives \cite{white2019beyond}.

\subsubsection{Enrichment analysis}
In a gene expression study, it is often useful to assess whether a certain gene list obtained with an analysis is enriched by genes known to be implicated in a disease or biological function of interest. The objective is thus to determine if the intersection between two lists is random or significant. This can be approached with a permutation test, as explained before, or with an exact test, such as the widely used  \ac{FET}. 
To explain FET, let us consider the contingency table in tab \ref{tab:FET}, in which $Q_1$ and $Q_2$ are the research questions that are true for the elements of the two lists under examination, and $n$ is the total number of elements analyzed with the two questions.
The numbers $a$, $b$, $c$ and $d$ in the table represent the length of the various sublists of elements depending on whether they satisfy $Q_1$ and $Q_2$ or not.
According to FET, the probability of observing the set $\{a,b,c,d\}$ can be calculated as:
\begin{equation}
    \pi = \frac{(a+b)!(c+d)!(a+c)!(b+d)!}{a!b!c!d!n!} 
    \label{eq:fet_prob}
\end{equation}

\newcolumntype{P}[1]{>{\centering\arraybackslash}p{#1}}
\newcolumntype{N}{@{}m{0pt}@{}}
\begin{table}\centering
\begin{tabular}{|P{1.62cm}|P{1.62cm}|P{1.62cm}|P{1.62cm}|P{1.62cm}|N}
\cline{3-5}         
\multicolumn{2}{c}{} & \multicolumn{3}{|c|}{\textcolor{red}{\textbf{$Q_1$}}}&\\
\cline{3-5}     
\multicolumn{1}{c}{}& & \textbf{T} & \textbf{F} & \textbf{T+F}&\\
\hline
 \multirow{3}{*}{\textcolor{red}{\textbf{$Q_2$}}} & \textbf{T} & a & b & a+b&\\
 & \textbf{F} & c & d & c+d&\\
 & \textbf{T+F} & a+c & b+d & a+b+c+d=n&\\
\hline
\end{tabular}
\caption{Contingency table for the computation of the Fisher's exact test. Note that \textbf{T} and \textbf{F} stands for True and False.}
\label{tab:FET}
\end{table}

Thus, the probability of any arrangement of $\{a,b,c,d\}$ can be calculated in this way. However, Fisher showed that to understand the significance of a certain intersection length $a$, one needs to calculate only the probabilities of those arrangements with the same marginal totals (showed in the last column and row), and among those, only the ones with an intersection length equal or greater than $a$.
The sum of the probabilities of these events (calculated with eq. \ref{eq:fet_prob}), and the probability of the observed event is the exact p-value measuring the significance of the observation of an intersection length $a$.

\subsubsection{Gene ontology enrichment analysis}
In the previous paragraph, it has been shown how to conduct an EA when comparing two lists. In many applications, it is interesting to study the intersection between a gene list obtained from an analysis and all the lists of genes (called GO terms) that share some biological/molecular/cellular functions, according to current knowledge. However, the GO enrichment analysis presents many additional issues with respect to a standard EA, due to the structure of the GO database:
\begin{itemize}
    \item The GO terms are not univocally defined and are continuously changing following new discoveries. 
    \item The GO terms are partially overlapping and organized in a hierarchical structure. As a consequence, the multiple tests can not be considered independent, a prerequisite for the application of the permutation and the Fisher tests.
    \item Not all the genes have been equally analyzed in literature, some of them do not belong to any GO term while others belong to many.
\end{itemize}
These aspects make the GO analysis open to a wide number of arbitrary choices in its implementation. For this reason, various free tools have been developed to standardize the way in which these analyses are conducted and favor the comparability of results.
Examples of these tools are: ToppGene, GSEA, Enrichr, DAVID, ENCODE, Revigo, WebGestalt. They all provide the FDR corrected p-values of the EAs performed and most of them also add some visualization tools to ease the comprehension of the GO analysis results. 
\subsection{Data visualization methods}\label{sec:bkg_visualization}
The results of a gene expression study have often very high dimension and are thus difficult to visualize. In this thesis, three types of data visualizations techniques and tools have been used: 
\begin{itemize}
    \item multidimensional scaling, that performs a data dimensionality reduction allowing to plot the information encoded in the data in intelligible 2D or 3D plots.
    \item Revigo \cite{supek2011revigo}, which is a data visualization tool that helps to graphically represent the results of a GO enrichment analysis by reducing redundancy.
    \item hierarchical clustering, that helps in understanding distances among multidimensional data when they are not clearly grouped in clusters. The output of hierarchical clustering algorithms is usually a dendrogram, which depicts the relationships among data with a tree.
\end{itemize}
Given that the last one has been already described in Section \ref{sec:bkg_network}, this paragraph provides a brief description of the remaining two.

\subsubsection{Multi-dimensional scaling}
The objective of MDS is to find a representation of the data, at a lower dimensionality, that preserves their distances\footnote{This is possible because the definition of "distance" is invariant for space dimensionality. For instance it is possible to apply euclidean distance in 2D, 3D and others.}. In order to do that, it is first necessary to select a distance function $\delta$ and construct a matrix $D$ whose elements $d_{i,j}$ are the distances between pairs of data points $\delta(\vec{p_i},\vec{p_j})$. Note that the vector notation is necessary because the data are multidimensional. \\
As stated before the objective of the MDS is to find a new representation $x_i$ of each $p_i$ with a lower dimensionality, such as the equation   $\delta(\vec{x_i},\vec{x_j})=d_{i,j}=\delta(\vec{p_i},\vec{p_j})$ holds for every other data point $\vec{p_j}$.\\
Obviously, this is an ill-posed problem because there are less equations than unknowns\footnote{In fact, $d_{i,j}$ is a scalar while $\vec{x_i}$ is a vector.}. 
Typically, this is addressed by transforming this equation into an optimization problem, whose solution is searched by minimizing a certain error function, such as the one below: 
\begin{equation}
    \Omega(\vec{x_i}) = \frac{\sum_{i,j}{(\delta(\vec{x_i},\vec{x_j})-d_{i,j})^2}}{\sum_{i,j}{d_{i,j}^2}}
\end{equation}
By minimizing $\Omega$, it is possible to reduce the mean squared error between the distances in the new representation $\delta(\vec{x_i},\vec{x_j})$ and the ones computed in the original space $d_{i,j}$.
Commonly adopted definitions of $\delta$ are euclidean distance, standardized euclidean distance (euclidean distance applied to the Z-scored data, so that the results are not influenced by large scale variations between coordinates), Mahalanobis distance and the inverse of the Spearman correlation.\\
The dimensionality of the vector $\vec{x_i}$ can be chosen at will, but common choices are clearly 2 and 3 so that the vectors can be visualized in a plot.

\subsubsection{Revigo} \label{revigo}
According to an estimate of 2018 \cite{gene2019gene}, approximately 45000 GO terms have been defined, divided into: 29,698 biological processes, 11,147 molecular functions and 4,201 cellular components. All the terms have parent-child relationships for example, "glucose transmembrane transport" (GO:1904659) is a child of "monosaccharide transport" (GO:0015749), forming a big tree of terms. Clearly, in a GO analysis it is highly probable that the gene list under examination results to be enriched by multiple GO terms belonging to the same branch. As a consequence the results of a GO analysis are highly redundant, and thus difficult to interpret.
The Revigo ("\underline{Re}duce and \underline{vi}sualize \underline{g}ene \underline{o}ntology") web service helps in that interpretation, by reducing the redundancy and applying different visualization tools.
The reduction step consists in a modified version of the hierarchical clustering that relies on semantic similarity between pairs of GO terms. The elements within the same clusters, defined by a certain cutting point, are reduced to only one GO term, which is selected trying to avoid the more general terms and taking in considerations the p-values obtained from the EA and the parent-child relationships among the terms. Then the reduced list of GO terms can be visually represented in Revigo either with a scatterplot, a graph or a tree, starting from the matrix of the GO terms semantic similarities. At several steps the user can customize the analysis.
\section{Transcriptome studies on ASD}
\label{sec:bkg_transcriptome_soa}
\subsection{Premise}
\label{premise}
A review of transcriptomic studies on ASD is very complicated because:
\begin{enumerate}
    \item Transcriptomic studies may differ for the tissue under examination and for the technique used to measure gene expression.
    \item Usually the number of available samples is very limited and this increases the probability to get spurious results, clearly not reproducible. Furthermore, given the heterogeneity of ASD, the few subjects included in the analysis often do not allow to draw any generalizable conclusion on the whole spectrum of the disorder. 
    \item Some works may analyze only a certain group of genes. Furthermore, even the studies that declare to analyze all the gene transcripts may overlook the expression of some of them due to technical limitations of the measurement method adopted. 
    \item The pipelines of analysis can be very different and often include multiple arbitrary choices.
    \item The list of genes found to be of interest by a certain study not only depends on all the previous points, but also on the level of confidence desired by the author. For instance, on the effect size, p-value threshold and FDR correction type chosen by the author.
    \item As a consequence of the last point, also the GO terms associated to a gene list of interest depend on many factors. In addition, the results of a GO analysis are often open to a wide range of interpretations. Thus, depending on the focus of the study and on the personal taste of the authors some possible interpretations are emphasized, while others are omitted.
\end{enumerate}
For all these reasons, comparing either the composition, the length or the enriched GO terms of lists of genes found to be associated with ASD does not constitute alone a proper literature review. It can be said that hypotheses supported by converging results from multiple studies are likely to be true, while nothing can be said about hypotheses based on a single study or still unconfirmed by any work.

With these premises in mind, the following paragraphs illustrate the results of the largest review \cite{ansel2017variation} on ASD transcriptomic studies analyzing different tissues, a selection of works comparing ASD and TDC brain samples and some hypothesis-driven analyses based on brain transcriptome data.
Finally, the challenges of the study of ASD through transcriptomic data are illustrated.

\subsection{Summary of the largest review on ASD transcriptomic studies}\label{sec:bkg_summary_review_GE_ASD}
\todo[inline]{Devo aggiungere una tabella con il numero di soggetti inclusi negli studi menzionati nei paragrafi 3.5.2 e 3.5.3}
The largest available review on ASD transcriptomic studies \cite{ansel2017variation} summarizes the results found by 26 works analyzing one out of five different biological sources: brain, gastrointestinal tissue, peripheral blood, adult olfactory stem cells, and scalp hair follicles. The first two constitute tissues of interest for ASD, but are difficult to sample; in fact, brain can be only examined post-mortem, while invasive procedures are necessary to obtain gastrointestinal samples. The other biological sources, instead, can be harmlessly sampled thus making it possible to collect more data and possibly find a transcriptomic biomarker useful for screening programs. 
The main observations made in the review are the following:
\begin{enumerate}
    \item Around $\sim 12,000$ genes were found DE between ASDs and TDCs when considering all the 26 studies reviewed.
    \item Around 100 genes have been reported by more than one study.
    \item Most of these 100 genes were consistently down or up-regulated across different sources.
    \item The pathways that were detected three times are: cell cycle, cell death, gastro-intestinal disease, immune function, and neurogenesis.
    \item The pathways that were detected twice are: alternative splicing, arrhythmogenic right ventricular cardiomyopathy, cellular assembly and organization, cell-to-cell signaling and interaction, gap junction, inflammation, small molecule biochemistry, and ubiquitin mediated proteolysis. 
\end{enumerate}
Despite the aforementioned difficulties in commenting the total number and the intersection of genes found by different transcriptomic studies, looking at the first two points the following observations can be made:
\begin{itemize}
    \item Given that the human genes are about twice as many as the total number of genes found DE, two scenarios are possible:
    \begin{itemize}
        \item ASD is actually characterized by the disregulation of half of the human genes.
        \item A large portion of the gene lists found comes from spurious effects due either to the small size of the datasets or to the noisiness of the data.
    \end{itemize}
    In any case, this data combined with the fact that only 100 genes have been found relevant in more than one study, suggests that ASD may be characterized by the dysregulation of different genes in different tissues.
    \item Randomly selected gene lists with a length similar to that of the reviewed ones should generate a much larger set of elements present in more than one list: around 50 times greater.
    This suggests that the list of genes found to be relevant strongly depends on the methodology adopted, including the tissue analyzed. So different studies find different genes, much more different than randomly sampling them.
\end{itemize}
The third observation highlighted by the review suggests that the $\sim 100$ genes reported at least twice, might actually be active players in the underlying pathogenesis of the disorder.\\
Finally, regarding the pathways identified (points 4 and 5), the ones found to be relevant by three studies already constitute areas of extensive research in ASD, while some of the ones found twice are less explored and deserve further investigations. 

In conclusion, despite this review shows some convergent results, it clearly emerges a difficulty in drawing unquestionable conclusions by comparing different transcriptomic studies, especially if they analyze different tissues. This last point has been outlined also by the authors of the review, pointing out that due to the high variance that characterize transcriptomic data and the absence of a study comparing different tissues from the same subjects, it remains unclear if the heterogeneity of the results obtained is due to the different subjects or the different biological sources analyzed.

\subsection{ASD vs TDC brain sample analyses}\label{brainASD-s-o-a}
Despite all the difficulties in obtaining brain samples, the results are clearly of major interest for a neurodevelopmental disorder such as ASD. Furthermore, given that brain regions with different functions have proved to be characterized by different transcriptome profiles \cite{richiardi2015correlated} (as it is better described in Sec. \ref{successful_link}), analyzing different brain locations can shed light on the spatial, functional and transcriptomic characteristics of the ASD brain.\\
One important study in this sense has been published in 2011 \cite{voineagu2011transcriptomic}, using microarray technology. In this work, the authors analyzed 3 different brain tissues (cerebellum and temporal and frontal cortex) and observed that group differences among ASDs and TDCs are much more pronounced in the cortex. Deepening on this result, they also performed a DE analysis between temporal and frontal samples within the ASD and TDC groups, separately, observing a reduced differentiation of the cortical transcriptome in the ASD brain.\\
In addition, the authors identified two modules of co-expressed genes associated with the disorder: a module enriched for neuronal functions and one enriched for immune genes and glial markers.

The dataset used for this study has been later expanded including more samples and the gene expression has been re-quantified with RNA-Seq in 2016 \cite{parikshak2016genome}. A second analysis on this renewed dataset corroborated previous findings, including the attenuation of cortical differences in gene expression. New insights emerging from this larger cohort, analyzed with a more advanced technique, include ASD-associated dysregulation of primate-specific lncRNAs and the observation that a subsample of Dup15q ASDs shares the core transcriptomic signature observed in idiopathic ASDs.\\
Even though being analyzed by the same research group and sharing a portion of data somewhat weakens their reciprocal validation, both the aforementioned works \cite{voineagu2011transcriptomic,parikshak2016genome} find that ASDs and TDCs show different gene expression patterning across the brain and that the transcriptome of some regions may not differ significantly between the two groups.

This observation emerges also by other works analyzing multiple regions. For instance, a study \cite{anitha2012brain} focusing on the expression of a reduced number of genes known to be related to mitochondrial functions across three brain locations (anterior cingulate gyrus, motor cortex and thalamus), observed region-specific alterations in the expression of a good portion of the genes analyzed in the ASD group as compared to the TDC one.

Similarly, another work \cite{khan2014disrupted} aimed at discussing the role of thyroid hormone in ASD, analyzed the levels of oxidative stress markers and the expression of a limited number of thyroid hormone dependent genes across 6 locations (cerebellum, brain stem, cingulated gyrus, orbitofrontal cortex, putamen, and Wernicke's area). The results obtained suggest that ASD brain is characterized by region-specific disruption of thyroid hormone homeostasis and gene expression.

Finally, multiple evidences indicate that upregulation of immune system genes may characterize the whole brain. In fact, this phenomenon has been observed in temporal \cite{garbett2008immune,parikshak2016genome,voineagu2011transcriptomic}, frontal and occipital cortex \cite{voineagu2011transcriptomic, parikshak2016genome,gupta2014transcriptome} in humans and in the hippocampus of two robust mouse models of ASD \cite{provenzano2016comparative}. Furthermore, postmortem measurements of cytokines in the cerebrospinal fluid of autistic children \cite{vargas2005neuroglial} also confirmed the presence of inflammation.

As already said in the previous paragraphs, depending on the methodological choices and on the focus of the study, different transcriptomic features of the ASD brain have been observed. The problem of defining which are the core ones remains. Reproducible findings highlight mainly two broad classes of genes: synapse-related genes \cite{voineagu2013converging, voineagu2011transcriptomic, parikshak2016genome} and those involved in the immune system \cite{gupta2014transcriptome,voineagu2013converging}.
Another phenomenon observed multiple times, despite not thoroughly analyzed, is the presence of region-specific gene expression abnormalities in ASD brain, which may deserve further attention.

\subsection{Hypothesis driven studies}\label{hp_driven_studies}
Hypothesis driven studies do not always necessitate to compare ASDs and TDCs data, but can nonetheless improve current understanding of ASD transcriptome, and are thus worth to be mentioned. An important work \cite{werling2016gene} of such type describes an analysis to explore two separate hypotheses to explain why males are at higher risk for ASD, which are:
\begin{itemize}
    \item ASD risk genes are expressed at different levels in males and females. 
    \item ASD risk genes are expressed at the same level in males and females, but mainly interact in molecular pathways and/or cellular processes that are differentially expressed by sex. Thus, the downstream impact of ASD risk genes is modulated by their interactions with sexually dimorphic processes.
\end{itemize}
The authors found that sex-differentially expressed genes among TDCs are not enriched by known ASD risk genes. However, they observed that genes expressed at higher levels in TDC males are significantly enriched for co-expression modules that in literature have been found upregulated in post-mortem autistic brain, including astrocyte and microglia markers. These results seem to validate the second hypothesis, stating that the sex-differential risk of ASD should be considered a naturally occurring sexually dimorphic process.

Another very interesting work \cite{herrero2020identification} investigates the role of the amygdala, known to be involved in fear response and memory of emotional events, in ASD. Given that some signs of ASD can be detectable at as early as 12 months \cite{young2020video,mckinnon2019restricted} of age, and that postmortem studies revealed cortical deficits explainable by alterations in prenatal development \cite{donovan2017neuroanatomy}, the authors analyzed the gene expression of the amygdala during fetal to early postnatal stages of development, considering it a critical point for ASD onset. The results showed that most of the known ASD susceptibility genes are expressed during at least one of the developmental stages examined and that in the temporal window explored these genes have much greater variance with respect to all the other genes expressed in the amygdala, proving that they are dynamically regulated during the development of the amygdala.

Finally, a new hypothesis on the possible role of ribosomal protein genes in ASD has been recently formulated, after the publication of a study \cite{griesi2021transcriptome} comparing the ASD and TDC gene expression of \ac{iPSC}-derived \ac{NPCs}, that have been estimated to belong to the mid-fetal temporal window of interest for the disease onset \cite{willsey2013coexpression}. In fact, this study found a module of co-expressed genes involved in protein synthesis upregulated in ASD NPCs.
The authors claimed that in brain post-mortem studies there was no prior evidence of ASD-related translation modules. However, as already mentioned in point 6 of Sec. \ref{premise}, one difficulty in making a review of transcriptomic studies is that, due to the complex nature of the results, each author emphasizes some elements of interest according to his/her taste and objective, omitting others. So, another work \cite{lombardo2017hierarchical} re-analyzed the data of other brain post-mortem studies finding two works in which a module enriched by ribosomal protein genes was detected as consistently upregulated in ASDs. As noted by one of the authors of this re-analysis in another paper \cite{lombardo2021ribosomal}, the convergence of results on the role of ribosomal protein genes in ASD is not insignificant, taking also in consideration that many known single gene mutations with high penetrance for ASD are involved in translational control of GE (e.g., FMR1, TSC2, PTEN) \cite{kelleher2008autistic}.

\subsection{Challenges}
Summarizing, in paragraph \ref{sec:bkg_summary_review_GE_ASD} it has been shown that the number of genes found to be DE in ASD in at least one transcriptomic study (independently from the tissue analyzed) is disproportionately larger than the one obtained when considering genes found at least twice. This disproportion raises doubts on the reliability of the results reviewed and may suggest that the transcriptomic signature of ASD changes across tissues. Similarly, paragraph \ref{brainASD-s-o-a} shows that multiple studies focusing on ASD brain observed region-specific dysregulation of gene expression. \\
Reviewing studies both based on different biological sources and on the brain alone, convergent results highlight the involvement of genes with synaptic, glial and immune functions. However, a lot of other molecular/biological functions have been observed, but their significance remains elusive. According to the opinion of the author of this thesis, future works should focus on the identification of the core pathways involved in ASD, intended as those that are specific of the disorder and drives all the other downstream consequences. This would facilitate the research for a biomarker and elucidate the upstream mechanism of ASD that should be addressed with therapy.

In conclusion, future challenges for transcriptomic research on ASD are:
\begin{itemize}
    \item \textit{Comparing individual tissue/regional differences between ASDs and TDCs.}\\
    This would allow to understand whether the consistently reported phenomenon of tissue/region-specific transcriptomic signature of ASD is an important trait of this disorder. Furthermore, as explained in Appendix \ref{app:differences}, there are several other good reasons for which studying brain regional differences would be interesting. For instance, individual regional differences may have a reduced variability across subjects, due to the attenuation of subject-specific dependencies (see the simulation in Appendix \ref{app:differences}).
    \vspace{0.2cm}
    \item \textit{Developing experiments and analyses that cross-validate the significance of the results with another source of information.}\\
    The inclusion of a second validation, possibly based on data obtained with different techniques, would allow to refine the significant results or corroborate them, increasing the reliability of the study and avoiding to generate new results that cannot be utilized fruitfully to develop a biomarker or a treatment for the disease.\\
    As for every biological study, a perfect validation would be based on reversing the "experiment" and, thus, using the transcriptomic signature learned from data to analyze ASD subjects at the phenotype level. Clearly, this would speed up the research of a biomarker for diagnosis by producing more reliable results.
\end{itemize}

\chapter{Imaging Genetics}
\section{Introduction}
Summarizing the concepts of the previous chapter, the gene expression is the process, mediated by environmental effects, with which the instructions encoded in the DNA are used to produce the proteins and the functional products necessary for the cellular functions. The macroscopic and joint effect of this phenomenon happening at a cellular level, is what is usually called phenotype. \\
The flux of information going from the genotype to the phenotype can be described at different levels, for instance, with DNA sequence for the genotype, transcriptome and proteome for the gene expression, blood test, histological exam, MRI and other medical exams for the characterization of some phenotypical features. 
Obviously, these descriptors can be compared to identifying causal associations between pairs of levels: one closer to the genotype and one to the phenotype.\\
For the research on mental disorders, comparing a neuroimaging-based phenotype with genetic information turned out to be a winning strategy, giving rise to the research field called \ac{IG}.

This brief chapter provides a description of the IG approach in general and on the advantages of using transcriptome as the genetic information to analyze. A review of the most relevant studies of this type of IG, called \ac{IT}, is also presented, being it employed in the present work.

\section{Imaging genetics and imaging transcriptomics}
Human imaging genetics is a rapidly growing field of study that has the potential to unveil the molecular basis of brain function and structure, analyzing the relationship between genetics and neuroimaging data \cite{meyer2006intermediate,munoz2009imaging,arslan2015genes,hashimoto2015imaging}. Its strength relies in the possibility to mutually validate the individual findings obtained either with neuroimaging or with genetic information.\\
IG studies can be set up in different ways:
\begin{itemize}
    \item searching neuroimaging differences between two groups of subjects defined based on their genetic information \cite{hashimoto2015imaging};
    \item using neuroimaging properties as the grouping factor and looking for correlated differences in genetic information \cite{hashimoto2015imaging};
    \item comparing the neuroimaging and genetic properties that, individually, allow to separate groups of data, defined by their clinical phenotype.
\end{itemize}
This versatility allows to select the most effective approach based on the specific experimental conditions and available domain knowledge.

In an IG study, the neuroimaging data can be structural or functional and the genetic ones are usually either DNA variations or transcriptome information. 
Pioneering IG studies \cite{heinz2000genotype, bookheimer2000patterns} were based on the analysis of neuroimaging properties of carriers and non-carriers of a genetic mutation supposedly associated to a certain condition.
However, this approach based on the causal association between DNA sequence variations and neuroimaging properties, does not take into account the environmental effects that may contribute to the resulting phenotype \cite{nestler2016epigenetic}, as it is supposed to happen for ASD etiology. On the contrary, the gene expression process is affected by regulatory mechanisms due to external conditions and in this sense transcriptome data would be more informative than DNA.
Furthermore, different DNA variations may result in similar phenotypes (as it happens for intellectual disability \cite{karam2015genetic}) or the same one may result in different outcomes (as it happens for Rett's syndrome) complicating the causal association analysis. Instead, it is commonly agreed that, being the transcriptome information closer to the phenotype with respect to the DNA, shared downstream mechanisms detectable at the gene expression level may separate more easily two phenotypic groups.  
Finally, given that many genes show strong variations in expression across the brain \cite{hawrylycz2015canonical}, observing a spatial correlation between a certain neuroimaging signature and the distribution of the gene expression across the brain strongly reinforces the conclusions of a causal association study. However, this spatial information cannot be inferred from DNA sequence. \\
For these reasons (summarized in Table \ref{tab:dna_transcriptome}), transcriptome provides a genetic measure more closely related to brain structure and function and easier to compare to neuroimaging data.
Its major drawback is that it can be measured only post-mortem, preventing the possibility of a human based study comparing functional brain images and gene expression data acquired from the same subjects. 
Despite this limitation, the release of the \ac{AHBA} \cite{ahba}, that collects the transcriptome profiles taken from 3702 spatially distinct samples along with their coordinates in the MRI usually adopted reference space\footnote{The \ac{MNI}}, allowed many researchers to investigate how the gene expression distribution across the brain is related to spatial variations in neuroimaging properties, giving rise to the nascent sub-field of imaging transcriptomics \cite{fornito2019bridging}.

\begin{table}
    \centering
    
    \begin{tabular}{|l|l|}
        \hline
         \cellcolor{lightgray}{\textbf{\;\;\;\;\;\;\;\;\;\;\;\;\;\;Neuroimaging $\rightleftarrows$ DNA}} & \cellcolor{lightgray}{\textbf{\;\;\;\;\;\;\;\;\;Neuroimaging $\rightleftarrows$ Transcriptome}}   \\[1mm]
         \hline
         \tabitem Can not detect environmental effects. & \tabitem Can detect environmental effects.\\[2mm]
         \tabitem \multirow{4}{0.45 \textwidth}{Difficult causal linkage: multiple DNA variations may cause the same phenotype and the same DNA variation may cause different phenotypes.} & \tabitem \multirow{3}{0.45\textwidth}{Easier causal linkage: being it closer to the phenotype, transcriptome is more likely to detect shared downstream mechanisms.}\\[2mm]
         & \\
         & \\
         & \\
         \tabitem \multirow{1}{0.45\textwidth}{DNA has no degree of spatial information.} & 
         \tabitem \multirow{3}{0.45\textwidth}{Transcriptome changes throughout the brain so can be easier to compare to neuroimaging properties.}\\
         
         & \\
         & \\[2mm]
         \tabitem \multirow{3}{0.45\textwidth}{Possibility to analyze functional neuroimaging and DNA information of the same patients.} 
         & \tabitem \multirow{3}{0.45\textwidth}{Impossibility to analyze functional neuroimaging and transcriptome information of the same patients.}\\
         & \\
         & \\
         \hline
    \end{tabular}

    \caption{Comparison between the use of DNA and transcriptome information in an imaging genetics study.}
    \label{tab:dna_transcriptome}
\end{table}

\section{Successful imaging transcriptomics studies}\label{successful_link}
\begin{figure}
    \centering
    \includegraphics[width=0.99 \textwidth]{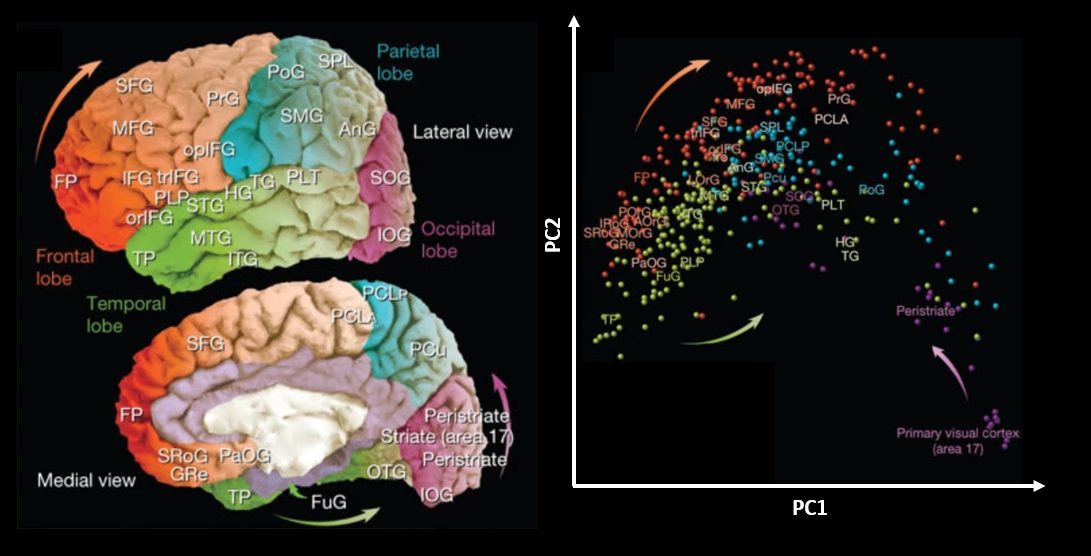}
    \caption{Multidimensional scaling applied to transcriptome data from different brain samples across the neocortex. The information on the brain structure is well preserved in the MDS plot. Reprinted by permission from Springer Nature Customer Service Centre GmbH: Springer Nature, Nature \cite{hawrylycz2012anatomically} (2012)}
    \label{fig:mds_transcriptome}
\end{figure}

The first study that proved the possibility to link brain-wide gene expression and structural human brain organization is presented in the AHBA announcing manuscript \cite{hawrylycz2012anatomically}. In this work, the authors applied a MDS analysis to the transcriptome data obtained from different samples across the neocortex, showing that their relative positions in the MDS plot mirror their actual positions on the cortical surface (estimated through neuroimaging) and that there is nearly a linear relationship between MDS-based distance among the samples and their physical distance (see Fig \ref{fig:mds_transcriptome}). This result is widely mentioned as a proof that brain structure organization is substantially determined at the gene expression level.

Another critical discovery is the experimental demonstration that human spatial organization of functional networks, defined with rsfMRI, is mirrored by correlation of gene expression profiles in the brain. 
In the pioneering work \cite{richiardi2015correlated} of this kind, the authors mapped the AHBA samples to the cortical functional networks defined using a distinct neuroimaging dataset of rsfMRI. They showed that the gene-expression based correlation between samples belonging to the same functional network is higher than the one calculated on samples belonging to different networks, irrespectively of their spatial distance. The study also identified a core set of genes highly involved in the functional network organization of the human brain and showed that polymorphisms of these genes modify brain connectivity, using an independent dataset of healthy adolescents. This work, despite being limited to the cortex, provides convergent and multimodal evidence that resting-state functional networks are supported by correlated gene expression.
Another subsequent study \cite{anderson2018gene} shows that the link between gene expression and functional networks can be found also in the striatum and the cortico-striatal system, thus suggesting that it is a feature of the whole brain, and not of the cortex alone.

All the aforementioned studies show a consensus in the existence of a correspondence between spatial gene expression distribution and brain morphological and functional organization in the human species. 
However, image transcriptomics has also been successfully used to understand the molecular mechanisms underpinning brain phenotypical differences between groups of subjects. For instance, a recent study \cite{liu2020integrative} has shown that brain sexual dimorphism is correlated with regional expression of sex-chromosome genes. The authors estimated the gray matter volume differences between males and females from two large structural neuroimaging datasets. Then, they coregistered this map of regional sex differences with the AHBA samples and computed, for each gene, the cross-region correlation between its expression profile and volume differences. This analysis showed that the gene-wide median correlation of the sex chromosome is higher than that of any other chromosome.

Finally, image transcriptomics has also proved to be a powerful approach to study complex neurological diseases. For instance, a recent study on ASD \cite{xie2020brain} isolated a set of genes whose expression is correlated with cortical volume variations between ASDs and healthy subjects. Notably, this set is enriched for genes known to be downregulated in ASD and involved in language-related neural processes, frequently impaired in the disease. This work is important because, as already said in chapter \ref{chap:bkg_neuro}, neuroimaging findings in ASD are usually inconsistent with each others \cite{haar2016anatomical} and thus lack an independent validation, but in this case the genetics and neuroimaging findings support each other. The analysis required a number of different data sources. Firstly, the authors conducted a meta-analysis across 30 neuroimaging datasets to find a set of brain regions with reliable cortical volume differences between ASDs and healthy subjects and then, using the AHBA dataset, they identified a list of genes whose spatial expression is significantly correlated to the regional volume differences previously found. Secondly, they replicated the analysis using a different neuroimaging dataset (ABIDE-I)\cite{di2014autism}, obtaining a second list of genes. Finally, the intersection of these two lists was filtered maintaining only the genes whose expression is also correlated to the volumes (not differences) of cortical regions of ASD or control subjects, using another independent neuroimaging dataset (ABIDE-II)\cite{di2017enhancing}. The robustness of this result is not only guaranteed by the numerous datasets involved, but also by the large overlap of the three gene sets, which cannot be randomly obtained, and, ultimately, by the convergence with the results of another IT study \cite{romero2019synaptic}.

Despite, as already said, the main limitation of imaging transcriptomics studies is the almost impossibility to compare neuroimaging and gene expression data of the same subjects, all the mentioned works (and many more \cite{forest2017gene,goel2014spatial,cioli2014differences,vertes2016gene,krienen2016transcriptional,parkes2017transcriptional,kirsch2016expression,whitaker2016adolescence,rittman2016regional,romme2017connectome,mccolgan2018brain,romero2019synaptic}) prove that these two data types are so intertwined that allow to obtain nonetheless meaningful results supported by either well-established biological evidences or more conventional studies. The advantages are straightforward: the reciprocal reinforcement of the discoveries obtained and a better understanding of the connection between genotype and phenotype.

\part{Contribution of this thesis}
\chapter{Contribution of this thesis}
\label{chap:contribution}

\newcommand{\RelatedText}[4][0.825\textwidth]{%
\begin{math}%
  \left.%
  \parbox{#1}{#3}\vphantom{\parbox{#1}{#4}}%
  \right#2%
  \parbox{#1}{#4}\vphantom{\parbox{#1}{#3}}%
\end{math}
}%
\subsection*{Motivations}
From the background chapters of this thesis, the following aspects of current ASD research emerge:
\begin{itemize}
    \RelatedText{\}}{
    \item ASD is a multifactorial neurodevelopmental disorder, still poorly understood and difficult to diagnose.
    \item Finding a biomarker to simplify and enable early diagnosis is of utmost importance.
    }{Ch. 1}
    \RelatedText{\}}{
    \item RsfMRI is a promising technique to identify an ASD biomarker, because it is safe and allows to analyze in-vivo brain activity. 
    \item The analysis of rsfMRI with AI has up to now delivered inconsistent results, often showing poor classification performance between ASDs and TDCs.
    \item Possible avenues of improvement for the research of a rsfMRI-based classifier are:
    \begin{itemize}
        \item[--] using an algorithm that is able to process the temporal dimension by extracting meaningful information on the blood flow;
        \item[--] using a classification algorithm resilient to confounders;
        \item[--] using explanation techniques to extract the biomarker learned by the classifier.
    \end{itemize} 
    }{Ch. 2}
    \vspace{0.05 cm}
    \RelatedText{\}}{
    \item Brain transcriptome can not provide a biomarker suitable for clinical practice, because it can not be obtained in-vivo.
    \item Despite the low number of samples analyzed, brain transcriptome studies on ASD found some reproducible results.
    \item ASD brain region-specific dysregulation of gene expression appears to be a common result across literature, despite not being thoroughly analyzed.
    \item Nonetheless, a wide number of findings from transcriptome studies have been reported only once, and their relevance to characterize ASD is still unclear.
    \item Brain transcriptome characterization of ASD could be improved by:
    \begin{itemize}
        \item[--] deepening on gene expression regional differences to understand whether this is a relevant characteristic of ASD.
        \item[--] finding an independent validation method to strengthen the results obtained and to reduce the uncertainties arising from the commonly small dataset size and the large number of arbitrary choices required in a typical transcriptomics study. 
    \end{itemize} 
    }{Ch. 3}
   \RelatedText{\}}{
    \item The imaging transcriptomics approach allows a mutual validation of the results.
    \item Imaging transcriptomics is a research field still in its infancy, but two studies on ASD analyzing transcriptome and sMRI data found consistent results \cite{xie2020brain,romero2019synaptic}.
    }{Ch. 4}
\end{itemize}

\subsection*{Objectives}
In light of what has been described in the previous paragraph, the main aim of this study is to find an rsfMRI-based ASD biomarker confirmed by transcriptome signatures, using an imaging transcriptomics approach. To accomplish this ambitious project, the following secondary objectives have been set:
\begin{enumerate}
    \item the development of a single AI-based classifier that is jointly resilient to confounders and able to process the temporal and spatial dimensions of the rsfMRI data in order to focus on variations of the blood flow in the brain.
    \item the characterization of the ASD brain at the transcriptome level by including a thorough analysis on regional gene expression differences to understand whether abnormal transcriptome patterning across the brain constitutes an important ASD signature.
    \item the extraction of the spatial characterization of the temporo-spatial pattern learned by the classifier described in the first aim and its comparison with the transcriptome-based characterization of the ASD brain obtained as per the second aim.
\end{enumerate}
The last objective constitutes the trait d'union between the imaging and transcriptomics parts of the study and allows the definition of the ASD biomarker as in the primary aim of this work.

\subsection*{Practical implementation}
\begin{figure}
    \centering
    \includegraphics[width=0.9\textwidth]{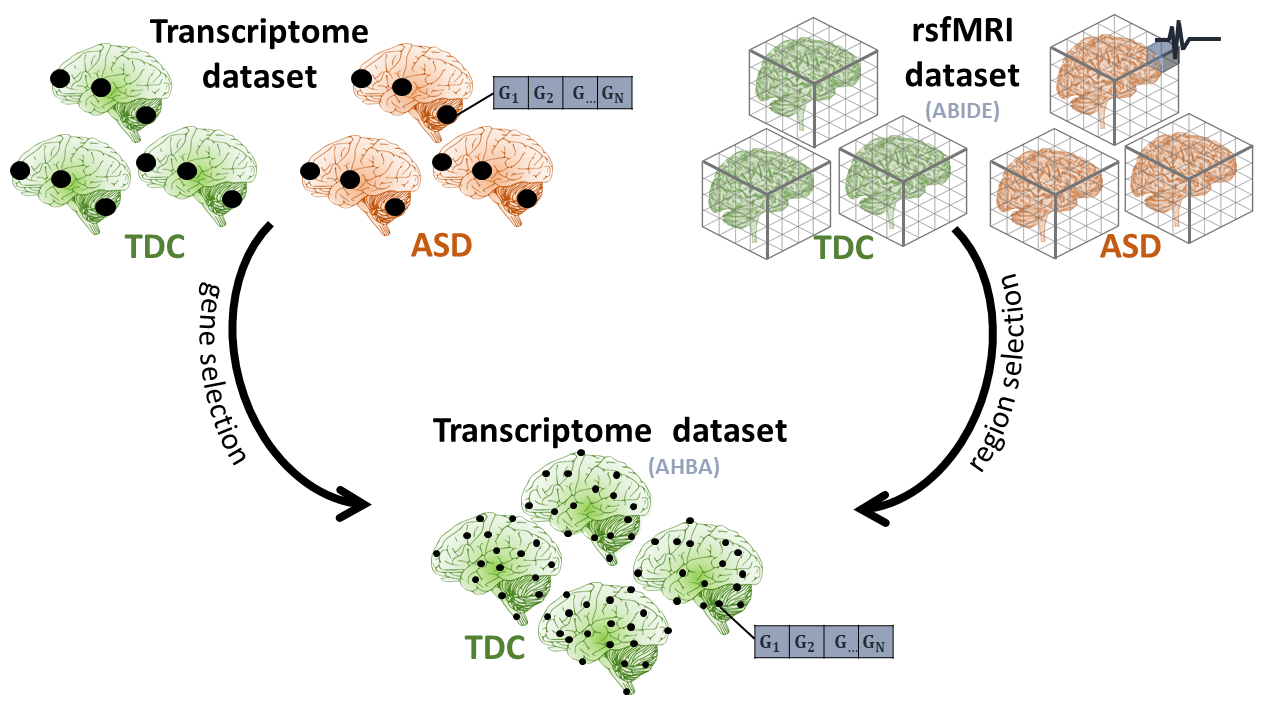}
    \caption{Schematic of the analysis workflow.}
    \label{fig:analysis_workflow}
\end{figure}

To pursue the objectives described in the previous paragraph, three data-sources have been used: 
\begin{itemize}
    \item all the rsfMRIs of the ABIDE collections \cite{di2014autism,di2017enhancing}, that together constitute the largest set of functional data freely available containing data from both ASDs and TDCs.
    \item The brain RNA-seq dataset already analyzed in \cite{parikshak2016genome}, which is one of the largest available, containing data from both ASDs and TDCs across three brain regions.
    \item The AHBA dataset \cite{ahba}, that contains brain microarray data of healthy subjects acquired across the whole brain with uniquely high spatial resolution.
\end{itemize}
The first dataset is used to train and test the rsfMRI-based classifier, which is developed by combining the 2S-CNN and the BR-NET (described in Sec.s \ref{subsec:bkg-2s-cnn} and \ref{subsec:bkg-br-nn}) to address the confounder and temporo-spatial issues, respectively. The spatial characterization of the pattern learned by the classifier is extracted using an explainability algorithm and on the basis of its output the set of brain regions critical for the prediction outcome is defined.\\
The second dataset is used to find a set of genes that characterizes the ASD brain. In this analysis, the genes that are differentially expressed between ASDs and TDCs in individual regions or in differences between pairs of regions are taken into consideration and compared. \\
Finally, the third dataset is used to verify if the genes, found analyzing the second dataset, are upregulated or downregulated in the regions found to be critical for the classification performed on the first dataset, with respect to the other ones. The intuition behind this approach is that if the genes involved in ASD take in these regions extreme GE values in healthy subjects, it is more likely that these brain regions would be found significantly differentially expressed in an ASD-vs-TDC study.
The regions relevant both for the rsfMRI and transcriptome analyses constitute then a valid ASD biomarker.\\
A schematic representation of this workflow is provided in Fig. \ref{fig:analysis_workflow}. 
Note that, despite this multi-dataset approach may appear dispersive, it is necessary. In fact, analyzing the brain transcriptome unavoidably requires to compare in-vivo imaging and ex-vivo transcriptomic data of different subjects. The decision to analyze two transcriptomic dataset instead is motivated by the necessity to have both ASDs and TDCs data for the selection of the genes of interest and a dataset with a fine spatial sampling, such as AHBA, to allow the evaluation of how these genes are expressed in the classifier's regions, with respect to other random genes. Obviously, a dataset with the AHBA spatial resolution containing data of both ASDs and TDCs would have been preferred, but it is not available yet.

\subsection*{Contributions}
Summarizing, the contributions of this thesis to the state-of-the-art are the following ones:
\begin{itemize}[topsep=0pt,itemsep=0pt,parsep=0pt,before=\vspace{1mm},after=\vspace{1mm}]
    \item this work represents the first imaging transcriptomics study on ASD using rsfMRI data.
    \item this work provides a high-performance neuroimaging biomarker that can be applied on minimally-preprocessed MRI data and that is confirmed by the gene expression characterization of ASD. The existence of this biomarker shows that the autism spectrum can be characterized as a whole by a set of neuroimaging features.
    \item the transcriptomic and imaging parts of this work, individually, represent an advancement of the state-of-the-art characterization of ASD and jointly shed light on the neuro-mechanisms that lead from the transcriptomic signature of the disorder to its phenotype.
    \item in the transcriptomic part of this thesis, a new area of investigation of ASD brain transcriptome based on individual regional differences is proposed, which can also be generalized to other mental disorders.
    \item in the neuroimaging part, a DL architecture that is jointly resilient to confounders and able to process temporal and spatial information using state-of-the-art techniques is developed. Again, this tool can be used for the study of other mental disorders.
\end{itemize}



\part{Materials and methods}
\chapter{Neuroimaging analysis}
\label{chap:mm_neuroimaging}
\section{Introduction}
\begin{figure}
    \centering
    \includegraphics[width=0.8\textwidth]{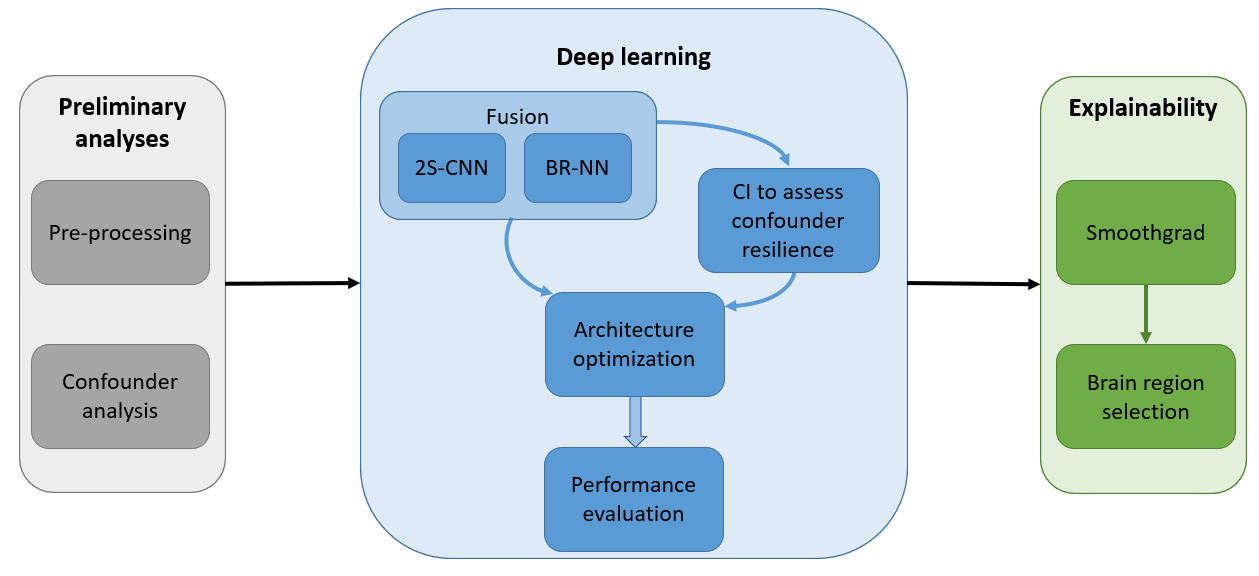}
    \caption{Schematic of the neuroimaging analysis workflow.}
    \label{fig:schema-mm-neuroimaging}
\end{figure}

This chapter contains the description of the neuroimaging dataset and the analysis pipeline. 
The analysis is based on rsfMRI data from the \ac{ABIDE} dataset and its main steps are outlined in Fig. \ref{fig:schema-mm-neuroimaging} and briefly explained below.

Firstly, the rsfMRIs are pre-processed with a series of operations such as skull stripping, motion correction and time and spatial re-sampling. Then, the confounder analysis is performed, using the CI method described in the background to quantify the confounding effect of variables such as age and sex.
Subsequently, a deep neural network is trained to distinguish ASDs from TDCs on the basis of the pre-processed rsfMRIs, while minimizing the effects of the confounder variables identified in the previous step. In order to do this, the BR-NET architecture described in Sec. \ref{subsec:bkg-br-nn} is used, combined with the 2S-CNN architecture described in Sec. \ref{subsec:bkg-2s-cnn} to best analyze the 4D information contained in rsfMRIs. The CI is used in this stage to optimize the architecture and the hyperparameters, measuring whether the potentially confounding variables are influencing the predictions.
Finally, the performance of the network is assessed, both in terms of its capability to identify ASDs, and in terms of its resilience to the identified confounding effects. An explainability analysis using the smoothgrad framework (desribed in Sec. \ref{subsec:bkg_smoothgrad}) is then performed to find the anatomical brain regions most relevant for the network predictions.

\section{The ABIDE database}
The ABIDE database is a multi-centric dataset containing phenotypical information, structural and functional MRIs of over 2,000 ASD and TDC subjects. It is the largest neuroimaging ASD database. 
It has been released in two parts, ABIDE-I in 2012 with about 1,000 subjects collected in 17 different sites and ABIDE-II in 2016 with a similar number of subjects collected in 19 sites. Note that many of the sites provided data for both ABIDE-I and ABIDE-II and some sites contributed by sharing multiple samples to the same dataset release. In another work \cite{ferrari2020dealing} made by the author of this thesis, it was observed that different samples acquired by the same site and, according to the documentation, with the same acquisition modalities, can be nonetheless distinguished with ML. This is probably due to overlooked hardware/software changes. In this work the word \textit{sample} is used to refer to the largest group of data supposedly acquired in the same conditions.
A summary of the phenotypical characteristics of the subjects for which rsfMRI data are available (the ones used in this work) is reported in Tab. \ref{tab:table_abide}, while Tab. \ref{tab:site-composition} describes the site/sample composition of the dataset.\\
There are several reasons for which the choice of this database is optimal for the kind of work described in this thesis. First, the relatively large amount of data enables the possibility of using deep-learning based approaches, while progress on ways to detect and eliminate confounding effects reduce the problems caused by a multi-centric database. Furthermore, the ASD subjects included in the ABIDE database have met stringent criteria for diagnosis \cite{di2017enhancing}, which enables the possibility to perform a dual-class study without worrying about the reliability of the diagnosis.
Finally, there is a large body of literature on machine and deep learning analyses conducted on the ABIDE database with which the results obtained and the methodologies followed can be compared.

\begin{table}\centering
\renewcommand{\arraystretch}{1}
{
\begin{tabular}{|ll|c|c|}

\hline                   
\multicolumn{2}{|c|}{}&  \textbf{\hspace{0.5cm} \textcolor{red}{ASDs} \hspace{0.5cm}} &  \textbf{\textcolor{red}{TDCs}} \\\hline

\multicolumn{2}{|l|}{\textbf{N subjects}} & 1010 & 1151 \\\hline

\textbf{Sex} & \textcolor{blue}{M/F} & 877/133 & 874/277 \\\hline

\textbf{Eye status at scan} & \textcolor{blue}{O/C/?} & 258/751/1 & 869/282/0 \\\hline

\textbf{Preferred hand} & \textcolor{blue}{L/R/A/?} & 76/716/46/172 & 55/883/38/175 \\\hline

\multicolumn{2}{|l|}{\textbf{Age}} & \multicolumn{2}{l|}{\centering \raisebox{-0.5\totalheight}{\includegraphics[width=0.4\linewidth]{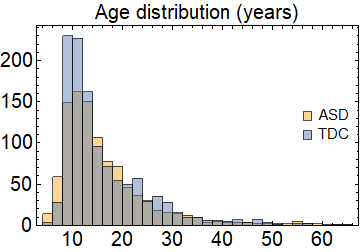}}} \\\hline

\multicolumn{2}{|l|}{\textbf{FIQ (missing in 169 subjects)}} & \multicolumn{2}{l|}{\centering \raisebox{-0.5\totalheight}{\includegraphics[width=0.4\linewidth]{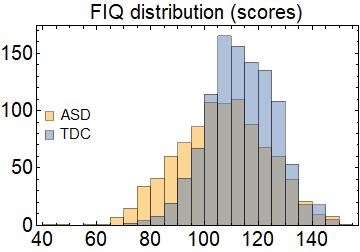}}} \\
\hline

\multicolumn{4}{|c|}{
\textbf{Largely available information}}
\\

\hline
\multicolumn{4}{|l|}{\hspace{3.5cm} \tabitem Age} \\
\multicolumn{4}{|l|}{\hspace{3.5cm} \tabitem Sex} \\
\multicolumn{4}{|l|}{\hspace{3.5cm} \tabitem Site} \\
\multicolumn{4}{|l|}{\hspace{3.5cm} \tabitem Sample} \\
\multicolumn{4}{|l|}{\hspace{3.5cm} \tabitem ADOS (missing in 339 ASDs)} \\
\multicolumn{4}{|l|}{\hspace{3.5cm} \tabitem Eye status (missing in 1 subject)} \\
\multicolumn{4}{|l|}{\hspace{3.5cm} \tabitem FIQ (missing in 169 subjects)} \\
\multicolumn{4}{|l|}{\hspace{3.5cm} \tabitem Handedness (missing in 347 subjects)}

\\

\hline

\end{tabular}}
\caption{Summary of the ABIDE dataset composition information.} 
\label{tab:table_abide}
\end{table}

\begin{table}
\centering
 \resizebox{\textwidth}{!}{\begin{tabular}{|c|c||c|c|}
  \hline
   \multicolumn{2}{|c||}{\cellcolor{gray!90} \textbf{\textcolor{white}{ABIDE I}}} &
    \multicolumn{2}{|c|}{\cellcolor{gray!90} \textbf{\textcolor{white}{ABIDE II}}} \\
     \hline
      \cellcolor{gray!10} \textbf{Site} & \cellcolor{gray!10} \textbf{Sample acronym} & \cellcolor{gray!10} \textbf{Site} & \cellcolor{gray!10} \textbf{Sample acronym} \\
      \hline
      \hline
      California Institute of Technology & Calthec-I & Barrow Neurological Institute & BNI-II\\
      \hline
      Carnegie Mellon University & CMU-I & Erasmus University Medical Center Rotterdam & EMC-II \\
      \hline
      \cellcolor{blue!50!green!30} Kennedy Krieger Institute & \cellcolor{blue!50!green!30} KKI-I & ETH Z\"urich & ETH-II\\
      \hline
      Ludwig Maximilians University Munich & MaxMun-I & Georgetown University & GU-II\\
      \hline
      \cellcolor{yellow!50!green!30} NYU Langone Medical Center & \cellcolor{yellow!50!green!30} NYU-I & Indiana University & IU-II\\
      \hline
      Olin, Institute of Living at Hartford Hospital & Olin-I & Institut Pasteur and Robert Debré Hospital & IP-II\\
      \hline
      \cellcolor{red!50!yellow!30} Oregon Health and Science University & \cellcolor{red!50!yellow!30} OHSU-I & Katholieke Universiteit Leuven & KUL-II\\
      \hline
      \cellcolor{red!30!blue!30} San Diego State University & \cellcolor{red!30!blue!30} SDSU-I &  \cellcolor{blue!50!green!30}Kennedy Krieger Institute & \cellcolor{blue!50!green!30} KKI\textsubscript{8ch}-II\\
      \hline
      Social Brain Lab BCN NIC UMC Groningen & \multirow{2}{*}{SBL-I} & Kennedy Krieger Institute & KKI\textsubscript{32ch}-II\\
      \cline{3-4}
      and Netherlands Institute for Neurosciences& & \cellcolor{yellow!50!green!30} NYU Langone Medical Center & \cellcolor{yellow!50!green!30} NYU\textsubscript{1}-II\\
      \hline
      Stanford University & Stanford-I & \cellcolor{yellow!50!green!30} NYU Langone Medical Center & \cellcolor{yellow!50!green!30} NYU\textsubscript{2}-II\\
      \hline
      Trinity Centre for Health Sciences & Trinity-I & Olin Neuropsychiatry Research Center, & \multirow{2}{*}{Olin-II}\\
      \cline{1-2}
      \cellcolor{yellow!40} University of California, Los Angeles & \cellcolor{yellow!40} UCLA\textsubscript{1}-I & Institute of Living at Hartford Hospital &\\
      \hline
      \cellcolor{yellow!40} University of California, Los Angeles & \cellcolor{yellow!40} UCLA\textsubscript{2}-I & \cellcolor{red!50!yellow!30} Oregon Health and Science University & \cellcolor{red!50!yellow!30} OHSU-II\\
      \hline
      \cellcolor{green!30} University of Leuven & \cellcolor{green!30} Leuven\textsubscript{1}-I & Trinity Centre for Health Sciences & Trinity-II\\
      \hline
      \cellcolor{green!30} University of Leuven & \cellcolor{green!30} Leuven\textsubscript{2}-I & \cellcolor{red!30!blue!30} San Diego State University & \cellcolor{red!30!blue!30} SDSU-II\\
      \hline
      \cellcolor{red!30}University of Michigan & \cellcolor{red!30} UM\textsubscript{1}-I& Stanford University & Stanford-II \\
      \hline
      \cellcolor{red!30}University of Michigan & \cellcolor{red!30} UM\textsubscript{2}-I & University of California Davis & UCD-II \\
      \hline
      University of Pittsburg School of Medicine & Pitt-I & \cellcolor{yellow!40} University of California, Los Angeles & \cellcolor{yellow!40} UCLA-II \\ 
      \hline
      \cellcolor{blue!20!white!70!}University of Utah School of Medicine &\cellcolor{blue!20!white!70!} USM-I & University of Miami & UMIA-II \\
      \hline
      Yale Child Study Center & Yale-I & \cellcolor{blue!20!white!70!} University of Utah School of Medicine & \cellcolor{blue!20!white!70!} USM-II\\
      \hline
  \end{tabular}}
 \caption{List of the samples of the ABIDE I and ABIDE II collections.
 The samples reported to be acquired with the same scanner type and acquisition parameters are highlighted with the same colour. The contents of this table have been extracted by various sources of documentation available at \cite{ABIDEIdoc,ABIDEIdocNITRC,ABIDEIIdoc}, which however present some missing data and discrepancies. Table previously published by the author of this thesis in \cite{ferrari2020dealing}.}
 \label{tab:site-composition}
\end{table}

\section{Preprocessing}
The pre-processing steps performed are minimal, coherently with the idea that deep learning studies should leave data as raw as possible and let the network learn how to best use the available information.
The steps applied are the following:
\begin{itemize}
    \item Temporal and spatial resampling: all images are resampled to have the same shape: 60 frames of $128 \times 128 \times 64$.
    \item Skull stripping.
    \item Spatial smoothing: a smoothing gaussian filter with a 5 mm kernel is applied to filter noise in the spatial dimension.
    \item Motion correction: the image sequences are corrected for motion and aligned to each other, to lock the postion of the brain in the video.
    \item Intensity normalization.
    \item Temporal filtering: only the frequencies between 0.001 Hz and 0.1 Hz are kept, to filter noise in the temporal dimension.
\end{itemize}
All the operations and the parameters used are standard in the pre-processing of rsfMRI and have been implemented with the Freesurfer package using the fs-FAST pipeline.
Note that, while alignment to a standard position is necessary to perform skull stripping, the images have been used in the neural network training in their original orientation, because detecting spatially invariant features is a key characteristic of DL algorithms and CNNs in particular \cite{cheng2016rifd}. \\
These pre-processing steps were run automatically on all subjects for which information on age, sex, FIQ, handedness, eye status, site and sample was available, because these are important confounding effects that must be investigated and potentially corrected for. Considering also that the pre-processing failed on some images because of bad image quality, the final processed dataset contains 1423 subjects. 

\section{The neural network architecture}
\begin{figure}
    \centering
    \includegraphics[width=0.8\textwidth]{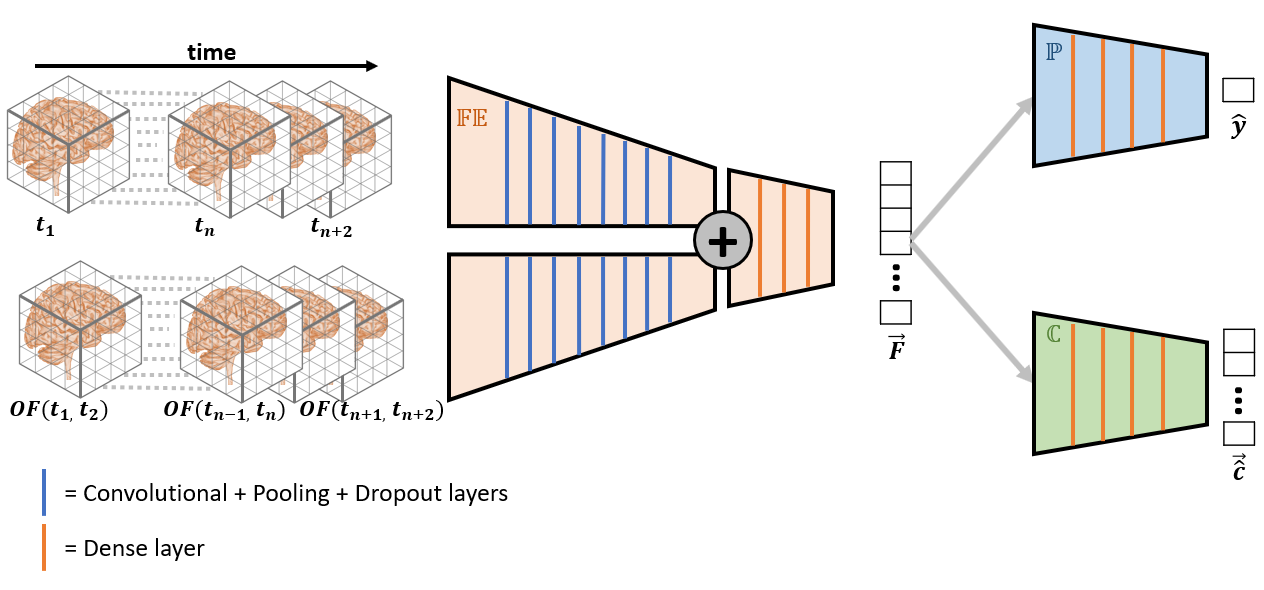}
    \caption{Schematic representation of the network structure.}
    \label{fig:network_structure}
\end{figure}

The architecture of this network reflects the two main challenges it has to face: dealing with confounders and appropriately exploiting the temporal information of rsfMRIs.
For this reason, as already mentioned, the network fuses the two ideas of BR-NN and 2S-CNN into a single architecture. The overall structure of the network is represented in Fig. \ref{fig:network_structure}. 
The two streams (standard video and optical flow) are processed with a series of alternating convolutional, pooling and dropout layers and then fused into a sequence of dense layers, the last of which coincides with the output of the feature extractor of the BR-NN architecture. From there, two series of dense layers are used for the task predictor and the confounder predictor respectively.

The convolutional layers perform 4D convolutions using a hyper-cubic kernel, i.e., with the same size in all dimensions. The size and shape of the kernel are difficult parameters to optimize, because they depend on the expected feature size in the 3D space and in the temporal dimension and are thus connected to the sampling. A cubic kernel is an educated guess based on the fact that spatial and temporal features are connected in the functional MRI case, because the features of interest are areas with increased blood flow. In fact, the time and spatial sampling are such that they approximately correspond to the smallest significant features that can be reliably detected by the instrument, and are well connected to each other by the dynamic of the BOLD signal: i.e., a BOLD signal change in one voxel takes about one time step to be detected \cite{simon2015understanding,buxton2013physics}. During the architecture optimization phase, the number of layers, the number of parameters and the kernel size of the convolutional layers have been varied, to improve performance.

Note that this architecture is slightly different from the one presented in \cite{carreira2017quo}, that was joining the optical flow stream and the standard stream at the last stage, averaging their predictions. Here, they are joined at a much earlier stage, at the end of the feature extractor. This choice is due to the particular structure of the BR-NN architecture, that separates the network into three chunks.\\
For the adversarial loss of the feature extractor, the Pearson's correlation coefficient was used for continuous confounders, while for binary categorical confounders, the point-biserial correlation was used. This figure of merit is a special case of the Pearson's correlation, designed to correlate a binary variable (the confounder class), with a continuous variable (the output of the confounder predictor, i.e., the probability of the input of belonging to a certain confounder class).\\
The sample variable, which is not binary but still categorical, was binarized by pooling all the wrong samples into a single class, and the correct one in the other class. To have a good balance of subjects belonging to both classes in each batch, the correct sample was assigned to class 0 for half the subjects and to class 1 for the other half.
Note that this binarization operation is performed only for the loss calculation, the confounder predictor actually works with all the samples as separate classes. 

\section{Confounder analysis}
\label{sec:mm_confounders}
\subsection{Confounding variables tested}
The first analysis performed is the confounder selection. Several variables have been investigated for their possible confounding effect using the CI method. Note that the method produces results that depend on the classifier, and thus the same network architecture described above and used throughout the rest of this work has been used. However, since the network is built for resilience to confounders, in order to assess confounding effects the adversarial part of the feature extractor was switched off, i.e., $\lambda$ in eq. \ref{eq:lc2} was set to 0.\\
The tested variables are: age, sex, sample, site, FIQ, handedness and eye status during scan.
Most of these variables (age, sex, sample, site and FIQ) have been selected because they have already proved to be confounding in sMRI data, as shown in a study published by the author of this thesis \cite{ferrari2020dealing}. Handedness was found to be non confounding for sMRI, but has been tested again for rsfMRI. Since this was a preliminary analysis, handedness was tested as a binary category, left/right, excluding ambidextrous subjects.
The last feature, eye status, was included because it was proved to have an important effect on rsfMRIs, as the visual cortex remains active even at rest if the eyes are open \cite{patriat2013effect}.

\subsection{Assessment of the confounding effect}
During the confounding effect evaluation, each confounder was tested individually, matching ASDs and TDCs for every other variable (except those found to have no effect). Categorical variables were matched perfectly, while continuous variables were considered matched within 3 years of age and 10 FIQ points. These values were measured to be sufficiently small to prevent confounding effects in sMRI analysis \cite{ferrari2020dealing}.


The calculation of the CI involves the assessment of the trend of the AUC with respect to increasing biases.
For all CI evaluations, the biases were increased in steps of 5\%, from 5\% to 95\%; small deviations from these exact values were caused by the finite dataset size. \\
The sample effect was tested considering subjects of the $NYU-I$ and $NYU_1-II$ samples, which are the largest samples of the same site in the database.
The confounding effect of the site variable, instead, was tested both with only two different sites appearing in the training dataset and with 4 total sites, binarizing them considering 1 site as one instance of the site variable, and the remaining 3 sites as the other instance (in this second configuration a smaller CI value is expected, but it is interesting to study how it diminishes when increasing the heterogeneity of the confounders).\\
The procedure for testing age and FIQ was more complex. In order to compute the CI for continuous variables, it is necessary to discretize their values and to compute the CI for the pairs of bins of interest. This involves selecting a discretization step $d$, representing the bin width within which it is expected to have a null value of CI. In this analysis $d$ was set to 15 scores for FIQ and 3 and 5 years for age. Two values are tested for age, because in the structural case age was shown to have a very strong confounding effect even for very small values of $d$, while only one value is used for FIQ, since it did not have a very strong effect in sMRIs \cite{ferrari2020dealing}.
To unequivocally identify a pair of bins to analyze, two quantities are used: the distance $l$ between the two bins and the starting point $a$ of the first bin. Both these quantities are expressed in the unit of measure of the continuous variable under examination. For the analysis of the FIQ and age variables $l \in \{10,20,30\}$ scores and $l \in \{3,5,10,20\}$ years were chosen, respectively. The value of $a$ was set to 90 scores and 4 years.\\
These decisions are mainly motivated by the necessity to have a sufficient quantity of subjects to calculate the CI.

Given that this is a preliminary analysis, the network architecture and the learning process were only lightly optimized, with a grid search exploring the parameters reported in Tab. \ref{tab:grid_search_conf}. The optimized configuration is described in Tab. \ref{tab:final_configuration_conf}.

\begin{table}
    \centering
    \begin{tabular}{|c|l|c|}
         \hline
         \textbf{Parameter group}         & \textbf{Parameter} & \textbf{Explored range} \\
         \hline
         \multirow{4}{*}{Convolutional block}     & Convolutional kernel size   & 3x3 - 5x5 \\
                             & Pooling kernel size   & 2x2 - 4x4 \\
                             & Dropout rate   & 0.05 - 0.20 \\
                             & Number of layers   & 6 - 12 \\
         \hline
         \multirow{3}{*}{Dense block}         & Number of dense layers in $\mathbb{FE}$ & 1 - 4 \\
                             & Size of the extracted features & 200 - 1,000 \\
                             & Number of dense layers in $\mathbb{P}$ & 3 - 6 \\
         \hline
         \multirow{2}{*}{Training process}    & Learning rates of the network parts & $10^{-6}$ - $10^{-2}$ \\
                             & Batch size & 5 - 20 \\
         \hline
    \end{tabular}
    \caption{Summary of the parameter ranges explored in the optimization of the neural network architecture and training process for the confounder selection. Note that the learning rate of $\mathbb{C}$ was set to 0 for the confounder identification. The activation function used for all layers was a rectified linear unit (ReLU).}
    \label{tab:grid_search_conf}
\end{table}
\begin{table}
    \centering
    \begin{tabular}{|c|l|c|}
         \hline
         \textbf{Parameter group}         & \textbf{Parameter} & \textbf{Value} \\
         \hline
         \multirow{4}{*}{Convolutional block}     & Convolutional kernel size   & 3x3 \\
                             & Pooling kernel size   & 3x3 \\
                             & Dropout rate   & 0.1 \\
                             & Number of layers   & 10 \\
         \hline
         \multirow{3}{*}{Dense block}         & Number of dense layers in $\mathbb{FE}$ & 1 \\
                             & Size of the extracted features & 400 \\
                             & Number of dense layers in $\mathbb{P}$ & 4 \\
         \hline
         \multirow{2}{*}{Training process}    & Learning rates of the three network parts & $10^{-5}$ \\
                             & Batch size & 10 \\
         \hline
    \end{tabular}
    \caption{Final configuration of the neural network used for the confounder analysis.}
    \label{tab:final_configuration_conf}
\end{table}

\subsection{Assessment of the network resilience to confounders}
After the confounder variables were identified (see Tab. \ref{tab:ci_results}), the capability of the BR-NN architecture to eliminate or reduce their confounding effects was tested. In order to do this, $\lambda$ in eq. \ref{eq:lc2} was no longer set to 0, but was instead optimized to reach the best compromise between eliminating confounding effects and reaching good AUC performance.
All the confounders were dealt with at the same time in the correction. Eq. \ref{eq:lc2} was calculated separately for each confounder and then all the individual contributions were summed up and multiplied by a single hyperparameter $\lambda$.
To evaluate the effectiveness of the network, the CI was measured for every tested variable and compared to the results previously obtained with $\lambda=0$. The network configuration and training process were optimized following Tab. \ref{tab:grid_search_conf}. The network architecture resulting from the optimization is described in Tab. \ref{tab:final_configuration_conf}.

\section{Optimization of the network architecture}
For the full training of the network for ASD/TDC classification, a much larger range of parameter was explored, outlined in Tab. \ref{tab:grid_search_last}.
The optimization of the learning rates is the most tricky one, because the net is composed by three different blocks ($\mathbb{FE}$, $\mathbb{P}$ and $\mathbb{C}$) trained at turn, and varying the learning rate of one of them may change the best setting for the others. 
Other critical parameters to choose are the batch size and the overall size of the network. While a larger batch size is desirable, both because it accelerates the training process and because it makes correlation-based losses more stable, it is often limited by the available memory in the GPU and, as a consequence, by the dimension of the network. 
The best performing network was selected both on the basis of its AUC performance in the task, and on its ability to ignore confounders, measured with the correlation between the predictions of $\mathbb{C}$ and the true confounder values.

\begin{table}
    \centering
    \begin{tabular}{|c|l|c|}
         \hline
         \textbf{Parameter group}         & \textbf{Parameter} & \textbf{Explored range} \\
         \hline
         \multirow{4}{*}{Convolutional block}     & Convolutional kernel size   & $3\times 3$ - $6\times 6$ \\
                             & Number of kernels in the first layer  & 10 - 40 \\
                             & Pooling kernel size & $2\times 2$ - $4 \times 4$ \\
                             & Dropout rate        & 0.05 - 0.35 \\
                             & Number of layers    & 5 - 20 \\
         \hline
         \multirow{3}{*}{Dense block}         & Number of dense layers in $\mathbb{FE}$ & 1 - 4 \\
                             & Size of the extracted features & 100 - 2,000 \\
                             & Number of dense layers in $\mathbb{P}$ and $\mathbb{C}$ & 2 - 8 \\
         \hline
         \multirow{1}{*}{Activation function}     & Leaky ReLU $\alpha$   & 0 - 0.2 \\
         \hline
         \multirow{2}{*}{Training process}    & Learning rates of the three network parts & $10^{-6}$ - $10^{-1}$ \\
                             & Batch size & 5 - 50 \\
         \hline
    \end{tabular}
    \caption{Summary of the parameter ranges explored in the optimization of the neural network architecture and training process for ASD/TDC classification.\\
    More information on the parameters, layers and activation functions of a DL algorithm can be found in Appendix \ref{app:neural_networks}.
    }
    \label{tab:grid_search_last}
\end{table}

\section{Network training and performance assessment}
The architectural optimization of the network took place on a 10-fold cross-validation.
The best performing model in terms of AUC was then trained on all the subjects that passed the pre-processing steps and tested on a hold-out dataset composed by the 169 subjects without FIQ information, that could not be used previously because they lacked information for $\mathbb{C}$. 
In addition to the performance on the test set, also the results of the correlation between the predictions of $\mathbb{C}$ and the true confounder values obtained during its cross-validation training process are reported, in order to evaluate whether the model has successfully avoided learning from confounders.\\
Note that the test set entirely contains the EMC-II sample, meaning that it contains subjects from a site that has never been used in training, thus representing another proof of the generalizability of the classifier.\\
During the optimization and cross-validation phase, the training process lasted for 100 epochs, while 500 epochs were used for training the final selected model.

Finally, the correlations between the errors of the network and the image quality was investigated. It is in fact possible that the wrongly classified subjects are those that have data of lower quality. 
In order to test this hypothesis, the quality assessment protocol designed by the preprocessed connectomes project was used \cite{shehzad2015preprocessed}. In particular, the figure of merit called mean distance index was taken into consideration, consisting in the average of $1 - \rho_s$ between each frame and the median frame, where $\rho_s$ is the Spearman's correlation coefficient. Lower values indicate better quality and thus, if the hypothesis above is true, the value of this index should be positively correlated with the errors of the network.

\section{Explainability and brain region selection}
The final aim of this analysis is to extract a set of regions that are relevant for the ASD/TDC classification. In order to do so, the \textit{Smoothgrad} framework was applied to identify the pixels that are most important for the network predictions.
\textit{Smoothgrad} was applied individually to every subject, but to understand which locations in the brain on average are most important, the brain of each subject must be warped and registered to a standard atlas, to prevent that one pixel in an image is located in a completely different brain region with respect to the same pixel in another image.
So, the transformation registering every subject to the MNI brain was saved, and applied to the corresponding Smoothgrad output.
Each pixel of these transformed saliency maps was assigned to a brain region depending on its MNI coordinates, according to the \ac{AAL} atlas \cite{rolls2020automated}. For every region of every subject, the average of its pixel activation values in the \textit{Smoothgrad} map was calculated.
The brain regions that had the highest average activation values throughout all the subjects were those deemed most relevant for the ASD/TDC classification task.

\chapter{Transcriptome analysis}
\label{chap:mm_transcriptome}

\section{Introduction}
\begin{figure}
    \centering
    \includegraphics[width=0.9\textwidth]{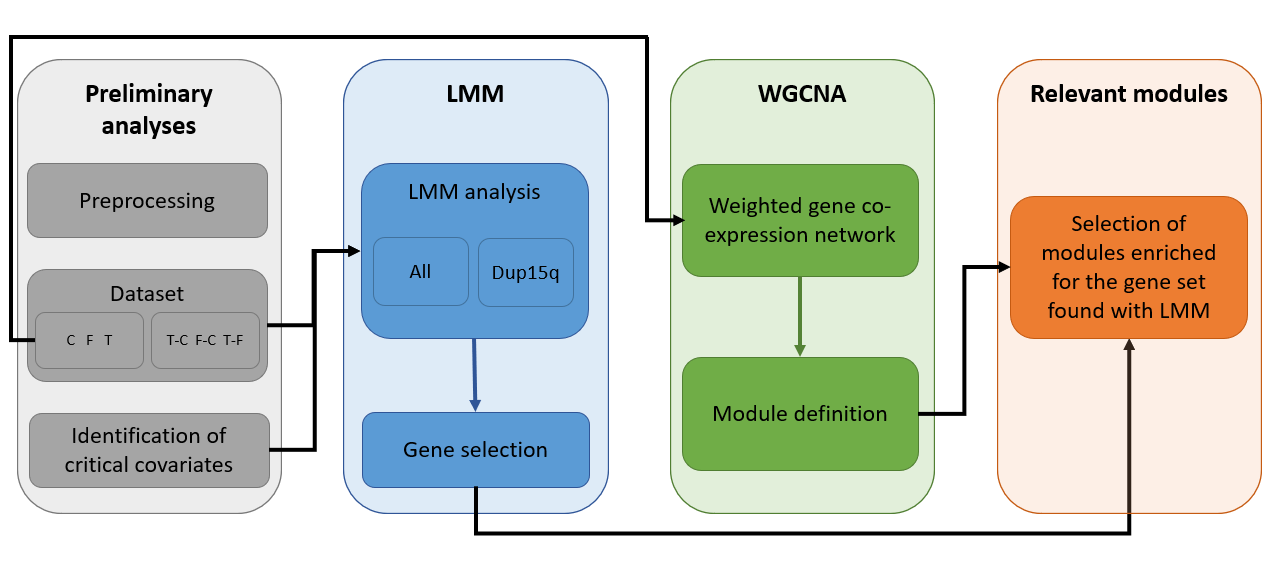}
    \caption{Schematic of the transcriptomic analysis workflow.}
    \label{fig:schema-mm-genetics}
\end{figure}

In this section, the genetic part of the study is described, which is aimed at identifying a set of genes that characterizes ASD at the transcriptome level. 
To achieve this goal, a large RNA-seq dataset \cite{gesch} is analyzed, containing the transcriptome of \ac{C}, \ac{F} and \ac{T} samples taken from 48 ASDs and 49 TDCs. This dataset has been already analyzed and subsequently made publicly available. To easily compare this work to the original one \cite{parikshak2016genome}, the latter is referred as the "Gheschwind's Lab Original" Work (GLO work).\\
The structure of this analysis is mainly composed by 4 steps, illustrated in Fig. \ref{fig:schema-mm-genetics}:
\begin{enumerate}
    \item \textit{Preliminary analyses.}\\
    These include standard pre-processing operations for RNA-seq data, the creation of the dataset that contains both data from single tissues and their paired differences and a study to identify which variables affect the GE values and thus should be included as covariates in the LMM described in the next step.
    \vspace{0.2cm}
    \item \textit{Linear mixed model.}\\ 
    A LMM was used to identify the genes that mostly differentiate the ASD subjects present in the dataset from the TDCs. In addition, the LMM analysis was applied to a subset of subjects with Dup15q and matched TDCs. This analysis represents a validity check and should highlight the genes in the Dup15q region. 
    \vspace{0.2cm}
    \item \textit{\ac{WGCNA}.}\\ 
    WGCNA was used to group all the genes into modules that share the same biological functions and regulatory mechanisms.
    \vspace{0.2cm}
    \item \textit{Selection of modules enriched for the gene set found in the second step.}\\
    The gene modules found in step 3 were then analyzed to identify those enriched for the gene set found in step 2.
    \end{enumerate}
A similar pipeline has been adopted also in the GLO work, which mainly focuses on the cortical data, considering frontal and temporal samples as belonging to the same tissue. In this study instead, the three tissues are analyzed separately, exploring also the inter-tissue gene expression differences, as already mentioned in Ch. \ref{chap:contribution}.\\
A more detailed description of the dataset, the sub-analyses and the methodological choices made in this work is provided in the following sections.
    
\section{Dataset description}
\label{sec:transcriptomic_dataset}
The dataset used in this analysis consists in the rRNA-depleted RNA sequencing (RNA-seq) of 252 post-mortem samples of C, F and T taken from 48 ASDs and 49 TDCs. Samples of the three tissues are not available for all individuals and sometimes a tissue has been sampled more than once in a subject. The brains have been stored in two different banks: the \ac{ATP} \cite{atp} and the "\ac{NICHD} brain and tissue bank" \cite{nichd}.
All the ASDs analyzed are considered idiopathic, except for nine subjects which were affected by Dup15q syndrome. A summary of the dataset composition is provided in Tab. \ref{tab:info_transcriptomic_dataset_1}, other available information is listed in Tab. \ref{tab:info_transcriptomic_dataset_2}. \\
Cause of death may influence post-mortem brain gene expression, and in particular there is a difference between sudden death and prolonged agony \cite{monoranu2009ph}. The causes of death that unambiguously belong to one of these two categories have been pooled as described in Tab. \ref{tab:cause_of_death}. These two categories were the only ones used to test whether cause of death had influenced gene expression levels.

The dataset contains gene expression levels estimated in three different ways: with Cufflinks\footnote{Cufflinks assembles the mapped reads into possible transcripts and then assigns them to their genes.} and with HTSeq using both the union-exon \footnote{The union-exon model assigns to each gene all the reads with a significant overlap with any of its exons.} and the whole-gene models\footnote{The whole-gene model assigns all the reads with a significant overlap with any portion of its DNA sequence, considering both its introns and exons.}. More details on the procedure to quantify GE can be found in the original paper \cite{parikshak2016genome}.

%
%
%
     
      

\begin{table}\centering
\renewcommand{\arraystretch}{1}
{
\begin{tabular}{|ll|c|c|}

\hline                   
\multicolumn{2}{|c|}{}&  \textbf{\hspace{0.5cm} \textcolor{red}{ASDs} \hspace{0.5cm}} &  \textbf{\textcolor{red}{TDCs}} \\\hline

\multicolumn{2}{|l|}{\textbf{N subjects}} & 48 & 49 \\\hline

\textbf{Sex} & \textcolor{blue}{M/F} & 39/9 & 41/8 \\\hline

\multicolumn{2}{|l|}{\textbf{Age}} & \multicolumn{2}{l|}{\centering \raisebox{-0.5\totalheight}{\includegraphics[width=0.45\linewidth]{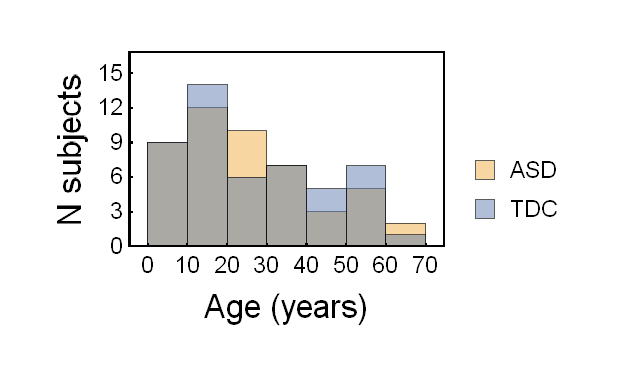}}} \\\hline

\multicolumn{2}{|l|}{\textbf{Post mortem interval (PMI)}} & \multicolumn{2}{l|}{\centering \raisebox{-0.5\totalheight}{\includegraphics[width=0.45\linewidth]{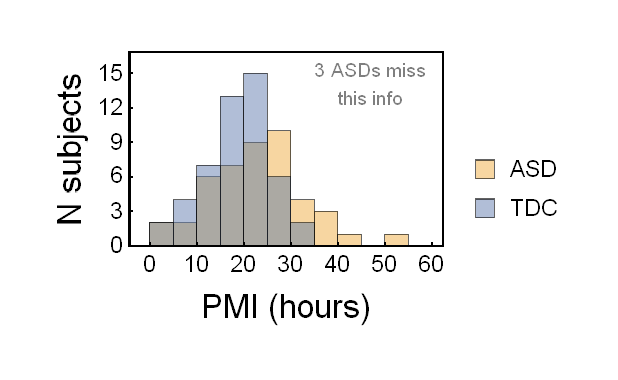}}} \\\hline

\textbf{Brain bank} & \textcolor{blue}{ATP/NICHD} & 34/14 & 24/25 \\\hline
\multicolumn{2}{|l|}{\textbf{Dup15q}} & 9 & - \\\hline

\textbf{Cerebellum} & \raisebox{-0.5\totalheight}{\includegraphics[width=0.1\textwidth]{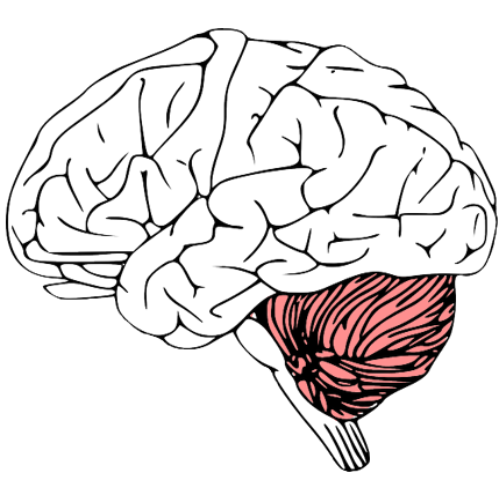}} & 38 (+1) & 41 (+4) \\\hline

\textbf{Frontal} & \raisebox{-0.5\totalheight}{\includegraphics[width=0.1\textwidth]{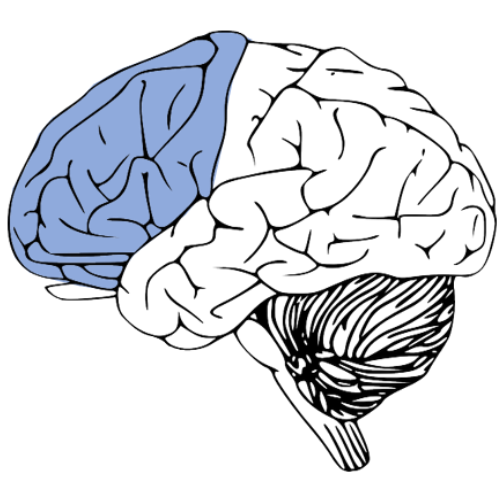}} & 42 (+1) & 42 (+3) \\\hline

\textbf{Temporal} & \raisebox{-0.5\totalheight}{\includegraphics[width=0.1\textwidth]{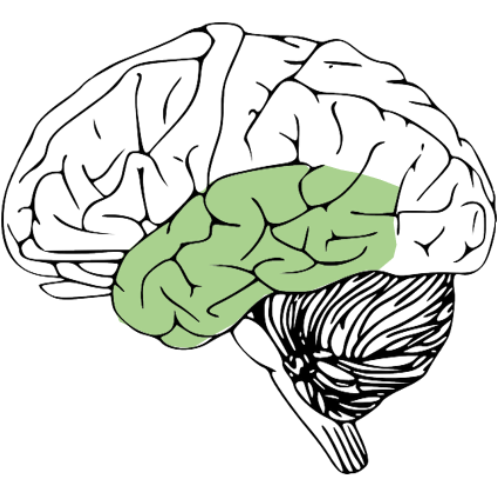}} & 40 (+2) & 32 (+5) \\\hline

\end{tabular}}
\caption{Summary of the first dataset composition information. In the last three rows the numbers out of parentheses refer to the number of subjects for which at least one sample is available, while the total of out and in parentheses numbers is the number of samples available (including multiple samples taken from the same subject and tissue).} 
\label{tab:info_transcriptomic_dataset_1}
\end{table}

\begin{table}\centering
\begin{tabular}{|l|}

\hline                   
\textbf{ \textcolor{red}{Subject information} (missing for some subjects) } \\
\hline
Cause of death\\
Seizure \\
Pyschiatric medications \\
Comorbidities\\
ADI-R score \cite{rutter2003adi} \\
Intelligence Quotient (IQ) score\\
Brain pH\\
Brain weight\\
\hline
\textbf{ \textcolor{red}{Sample information} (available for each sample) } \\
\hline
RNA integrity number (RIN)\\
Sequencing batch (Seq Batch)\\
RNA Sequencing indicators:\\

\hspace{1cm} \tabitem TotalReadsInBam\\
\hspace{1cm} \tabitem After.rmdup.samtools\\
\hspace{1cm} \tabitem Num.dup.samtools\\
\hspace{1cm} \tabitem Unique.Reads.samtools\\
\hspace{1cm} \tabitem PropExonicReads.HTSC\\
\hspace{1cm} \tabitem TotalReads.picard\\
\hspace{1cm} \tabitem Aligned.Reads.picard\\
\hspace{1cm} \tabitem HQ.Aligned.Reads.picard\\
\hspace{1cm} \tabitem PF.All.Bases.picard\\
\hspace{1cm} \tabitem Coding.Bases.picard\\
\hspace{1cm} \tabitem UTR.Bases.picard\\
\hspace{1cm} \tabitem Intronic.Bases.picard\\
\hspace{1cm} \tabitem Intergenic.bases.picard\\
\hspace{1cm} \tabitem Median.CV.Coverage.picard\\
\hspace{1cm} \tabitem Median.5prime.Bias.picard\\
\hspace{1cm} \tabitem Median.3prime.Bias.picard\\
\hspace{1cm} \tabitem Median.5to3prime.Bias.picard\\
\hspace{1cm} \tabitem AT.Dropout.picard\\
\hspace{1cm} \tabitem GC.Dropout.picard\\
\hline

\end{tabular}
\caption{Additional information available on the dataset, in addition to what is reported in Table \ref{tab:info_transcriptomic_dataset_1}} 
\label{tab:info_transcriptomic_dataset_2}
\end{table}

\begin{table}\centering
\begin{tabular}{|l|c|l|}

\hline                   
\textbf{Cause of death} & \textbf{N subjects} & \textbf{Specifics}  \\
\hline
\multirow{5}{*}{Sudden} & \multirow{5}{*}{24} & Accident\\
 & & Drowning\\
 & & Heart attack\\
 & & Motor vehicle accident\\
 & & Sudden unexpected death\\
\hline
\multirow{3}{*}{Prolonged (Cancer)} & \multirow{3}{*}{6} & Cancer\\
 & & Complications of pseuodmyxoma peritonei\\
 & & Pancreatic cancer\\
\hline
\end{tabular}
\caption{Summary of incidence and definition of the adopted death categories, that may affect differently the brain GE \cite{monoranu2009ph}.}
\label{tab:cause_of_death}
\end{table}

\section{Preliminary analysis}
\subsection{Preprocessing}
\label{subsec:transcriptomic_preproc}
Before running any of the preprocessing operations, given that the dataset presents some redundant data, only the samples with the highest RIN among those of the same subject and same brain region were considered in the subsequent analyses, discarding the others.\\
The preprocessing steps performed involve:
\begin{enumerate}
    \item Filtering out the genes that are too short to be correctly sequenced.
    \item Filtering out the genes that are not sufficiently expressed in the dataset, and thus whose counts may be heavily affected by statistical errors. The genes to exclude have been selected as those that satisfied at least one of the following conditions:
    \begin{itemize}
        \item a GE value of 0 in at least 80\% of the samples of one brain region, as measured by Cufflinks;
        \item less than 10 counts in at least 80\% of the samples of one brain region, as measured by HTSeq using the union-exon model;
        \item less than 10 counts in at least 80\% of the samples of one brain region, as measured by HTSeq using the whole-gene model.
    \end{itemize}
    \item Correcting for sequencing depth.
    \item Applying the GeTMM \cite{smid2018gene} algorithm, that combines the gene length normalization and the TMM correction for RNA composition; 
    \item Transforming the data with a log-2 normalization.
\end{enumerate} 
Then, two groups of datasets were created, each made of 3 different datasets. The first group (ST) contains the resulting normalized gene expressions levels of the 3 tissues without further modifications; the other one (DT) contains the differences between the gene expression levels of pairs of tissues.
This last group of datasets is therefore formed by subject-by-subject, gene-by-gene differences in the following tissue pairs: C-T, C-F, F-T.\\
All the following analyses were conducted on both groups, unless otherwise specified.

\subsection{Identification of the factors influencing gene expression levels}
\label{subsec:transcriptomic_ge_influence}
A machine learning based analysis was conducted to identify which variables in Tab. \ref{tab:info_transcriptomic_dataset_2} had significant influence on the gene expression levels, in order to include them as covariates in the ensuing LMM analysis.
This investigation was performed by training a machine learning model in 10-fold cross-validation on data from each tissue in ST or pair of tissues in DT (so C, F, T, C-T, C-F and F-T) to identify the specific variable under examination.
The model used was a logistic regression for categorical variables and a small, 3-layer NN for continuous ones.
The AUC (in case of categorical variables) or the mean squared error (MSE, in case of continuous ones) obtained by the classifier was compared with those obtained from a permutation test, in which the association between the input features (the gene expression levels of a certain sample) and the target label (the value of the variable under examination for that sample) was eliminated by random shuffling. The permutation test was repeated 100 times for each variable.
A variable was considered relevant if it achieved a BH corrected p-value $p_{BH}$ below 0.05 in at least one of the 6 tested datasets.

This analysis was conducted to identify covariates to regress out in the subsequent LMM analysis. In the particular case of age, which is known to have highly non-linear effects \cite{glass2013gene}, it was necessary not only to understand whether it was relevant or not, but also how to appropriately correct for it. For this reason, an analysis on the non-linear dependency of gene expression levels on age was conducted and is described in appendix \ref{app:genetics_preprocessing}.
The GLO work instead has resorted to completely eliminate the subjects below 10 years of age from the analysis \cite{parikshak2016genome}, however at the cost of a significantly reduction in sample size.

\section{LMM analysis}
\label{sec:mm_lmm_analysis}
\subsection{Derivation of the lists of relevant genes}
The LMM analysis was performed separately on all the 6 datasets from ST and DT, and on a subset of each of those, containing the ASDs with the dup15q syndrome with matched TDCs and only the genes known to be involved in the dup15q syndrome. In this analysis, the variables considered were the diagnosis and the covariates found as described in the previous section.
In a mixed model, it is necessary to identify both fixed and random effects. A variable was classified as a random effect if it satisfied the criteria described in section \ref{sec:bkg_lmm}. All other variables were considered fixed effects.
From the LMM analysis, a list of genes relevant for ASD/TDC classification was derived for each dataset. The genes were selected depending on their p-value (as reported by the LMM analysis and corrected for false-discovery-rate with the Benjamini-Hochberg method) and their effect size. In particular, genes were deemed relevant if they had an effect size above 0.8 and a FDR-corrected p-value below 0.05.

The results of the study on the ASDs with dup15q were used to validate the methodology of the analysis, both in the choice of the pre-processing operations and covariates, and in the unconventional use of the DT datasets. In fact, if a large fraction of dup15q genes is detected by the LMM as important to distinguish the ASDs with dup15q, it can be concluded that the study is well designed.
Further analyses were performed to show that ST and DT detect different sets of genes, to understand which biological pathways were represented the most in the genes found and to visualize their ASD/TDC discrimination capabilities. In particular, the visualization was made using multi-dimensional scaling with standardized euclidean distance as a dimension-reduction process to condense the information provided by the identified genes in only two coordinates per subject. 
Finally, a permutation test was performed to test whether the genes found were better than a random, equally sized set of genes at distinguishing ASDs from TDCs with a logistic regression model. 

\subsection{Validation and enrichment analyses}
The lists of genes detected as significantly dysregulated between all the ASD and TDC subjects with the LMM analysis have been tested for possible enrichements with the lists of:
\begin{enumerate}
    \item Genes possibly implicated in ASD according to current literature.
    \item Genes involved in the Angelman, Fragile X, Rett and Down syndromes, which all display important phenotypic commonalities with ASD. In fact, among the subjects with these syndromes the percentage of those that meet the ASD diagnostic criteria is of 61\% \cite{bonati2007evaluation}, 50\% \cite{abbeduto2014fragile}, 40\% \cite{mount2003features} and 18\% \cite{diguiseppi2010screening}, respectively.
    \item Genes known to be regulated by or to regulate MECP2, i.e. the gene causing Rett's syndrome. Given that several genes altered in MECP2 mouse models proved to cause ASD \cite{hudazoghbi_video}, genes belonging to this list may be involved in the disorder as well. 
\end{enumerate}
The objective of the first EA is to validate the methodological choices by verifying whether the genes found are related to ASD according to current knowledge of the disorder. The specificity of the genes found is instead investigated with the second EA. Furthermore, this analysis allows to understand which of the four syndromes mentioned is closer to ASD at the gene expression level. Finally, the last EA explores the hypothesis that genes in the MECP2 pathway may be involved in ASD.  

As explained in Sec. \ref{sec:bkg_transcriptome_soa} and in Appendix \ref{app:differences}, one important motivation for studying gene expression regional differences is that genes associated with psychiatric disorders (among which ASD) display a high \ac{DS} in healthy brain. The latter being computed as the mean Pearson correlation of pairs of gene expression profiles across different brain regions extracted from healthy subjects \cite{hawrylycz2015canonical}. Thus, the distribution of the DS values (computed in \cite{hawrylycz2015canonical}) of the genes found studying the ST and DT datasets has been compared, to check whether the proposed DT analysis and its practical implementation detect genes with high DS, as expected.

Finally a conventional GO enrichment analysis has been performed on all the lists of genes found analyzing the ST and DT datasets and their union, using the WebGestalt platform \cite{liao2019webgestalt}.

\section{Gene co-expression network analysis}
\label{sec:mm_network}

\subsection{Network construction}
The ST data were used for building a gene co-expression network using the WGCNA R package \cite{zhang2005general}. 
For every network, the adjacency matrix was created elevating the similarity matrix to a power chosen with the scale-free topology criterion (see Sec. \ref{sec:bkg_network} for more details). In particular, the $R^2$ of the scale-free topology fit was plotted against the mean connectivity of the network, and the best compromise between a scale-free topology and a dense enough connectivity was chosen, as shown in appendix \ref{app:network_construction}. 

Firstly, it was verified whether it was possible to combine the information from all 3 tissues. In fact, the networks built with gene expressions in different tissues may be very different to each other, and in this case it would not be sensible to build a single network.
Therefore, the DeltaCON measurement (\cite{koutra2013deltacon}, see also Sec. \ref{subsec:bkg-deltacon}) of network dissimilarity was used to evaluate whether two networks were comparable to each other. The index is unbound, therefore $\frac{1}{1+\delta_c}$ was used as a measure of similarity, bound between 0 and 1, as suggested by the authors.\\
Since this index does not have an easily interpretable meaning (i.e., there is not a widely recognized threshold value for declaring that two networks are similar) a permutation test was performed. The $\delta_c$ value of pairs of networks built with data from pairs of tissues (exploring all possible pairs) was compared to those of pairs coming from a set of 100 networks built with 80\% of the samples from a single tissue, each time randomly selected. All networks were built with the same soft power threshold.\\
If the values of $\delta_c$ for the networks built with different tissues is close to the values of the permutation test, it means that the information contained in the gene expression levels of the two tissues is sufficiently similar to justify their combination in a single network.\\
Sec. \ref{sec:res_genetics_network} of the results chapter shows that this is indeed the case, and thus one network was built with the data from all three tissues.

\subsection{Module selection}
From the network built as described in the previous section, modules of correlated genes were extracted using the blockwiseModules function of the WGCNA package. This function takes several parameters as input, to tweak the way in which genes are grouped into modules. There is not a commonly established way to choose these parameters, and thus this is a step in which somewhat arbitrary choices must unavoidably be made.
The parameters used were chosen by visual inspection of the dendrogram so that the resulting network contained a sensible number of sufficiently populated modules (10-30 modules, minimum 20 genes per module). In particular, the cut height at which modules are merged (\textit{$merge\_cut\_height$}) was set to 0.25, while the minimum module size was set to 20 genes (\textit{$minimum\_size$}).

These modules were then compared with the lists of genes found in the analysis described in Sec. \ref{sec:mm_lmm_analysis}, in order to select those enriched for such genes. The gene lists were both evaluated separately and merged together, for a total of 9 lists: C, F, T, F-C, T-C, T-F, all 3 ST, all 3 DT and all lists together. Module enrichment was evaluated with the FET and the significance threshold was set to a FDR-BH corrected p-value below 0.05.

Note that some of these lists were obtained using the DT datasets, while the network was built using GE values in single tissues. However, since the network is based on Pearson correlation $\rho$, analyzing the gene expression profile across single tissues or across differences among pairs of tissue is mathematically equivalent. In fact, given four vectors such as the following ones:
\begin{AutoMultiColItemize}
  \item $\vv{v_{S1}}=(a, b, c)$
  \item $\vv{v_{S2}}=(d, e, f)$
  \item $\vv{v_{D1}}=(a-b, b-c, c-a)$
  \item $\vv{v_{D2}}=(d-e, e-f, f-d)$
\end{AutoMultiColItemize}
It can be easily shown, using any symbolic solver, that $\rho(\vv{v_{S1}},\vv{v_{S2}})=\rho(\vv{v_{D1}},\vv{v_{D2}})$.

\subsection{GO enrichment analysis}
Finally, a conventional GO enrichment analysis has been performed using the WebGestalt platform \cite{liao2019webgestalt} on all the network modules found to be significantly enriched by the gene lists of interest defined with the LMM analysis.

\chapter{Imaging genetics}
\label{chap:mm_link}

\section{Introduction}
\label{sec:mm_link_intro}
In this section the imaging transcriptomics part of the study is described, which aims at investigating the possible connection between brain-wide expression distribution of the gene modules found in the analysis described in Ch. \ref{chap:mm_transcriptome} and the brain regions relevant for the ASD-vs-TDC classification as determined in Ch. \ref{chap:mm_neuroimaging}.\\
To achieve this goal, a dataset containing finely spatially sampled microarray data from 6 healthy subjects has been used to investigate whether these modules take more extreme expression values (upper or lower) in the brain regions identified with the neuroimaging analysis, with respect to the other ones.

As it will be better explained throughout this chapter, the peculiarities of this dataset and the complexities arising in a study linking neuroimaging and genetic findings make a large number of arbitrary choices necessary in the analysis. The lack of a standard processing pipeline for this kind of works has indeed been recognized as a problem \cite{arnatkeviciute2019practical}.

\begin{figure}
    \centering
    \includegraphics[trim={2.5cm 2.5cm 2.5cm 0cm},clip,width=0.8\textwidth]{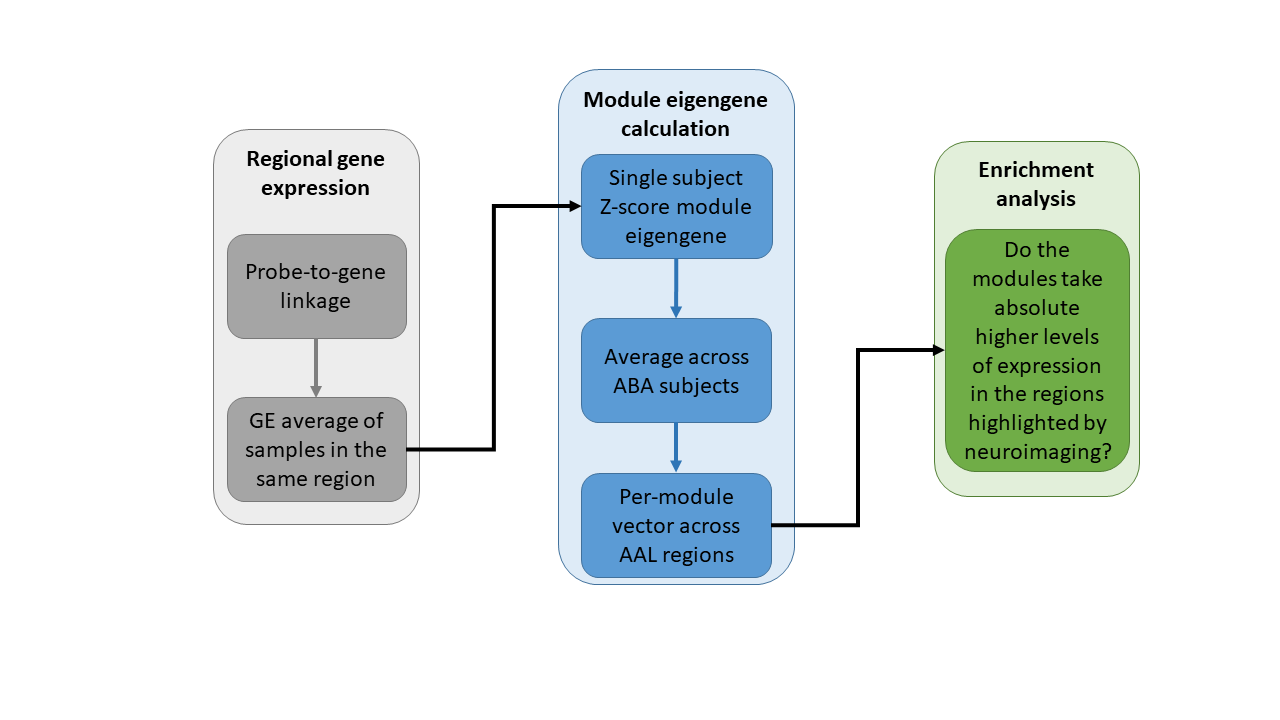}
    \caption{Schematic of the imaging genetics analysis workflow.}
    \label{fig:schema-mm-link}
\end{figure}

The structure of the analysis is illustrated in Fig. \ref{fig:schema-mm-link} and briefly summarized below.\\
Firstly, it must be noted that the microarray dataset contains information on probes, not on genes, and that multiple probes may detect the expression of the same gene. Therefore, for each gene a single value of expression was chosen, based on the probe with the highest DS value as measured by \cite{hawrylycz2015canonical}.
Then, based on its MNI coordinates (available in the dataset), each sample was associated to a brain region according to the AAL atlas and the transcriptomic profile of a region has been calculated by averaging the GE values of each gene across the samples belonging to the same region.
Subsequently, the first component of the eigengene was computed for each gene module and for each subject and Z-score transformed. In this way, six scaled vectors per module (one for each subject) have been obtained, containing the GE values of the module eigengene across all the brain regions of the AAL atlas (for which samples were available for the specific subject).
The inter-subject GE values of each brain region, made comparable with each other by the Z-score transformation, were then averaged together to obtain a single vector per module with one value per brain region. Note that not all subjects contributed to the average in every region, as some regions were missing in certain subjects.
These per-module vectors were then used to study whether these modules take significantly higher absolute expression values in the regions highlighted by the neuroimaging analysis with a permutation test.

\section{Dataset description}
\label{sec:mm_aba}
The AHBA dataset contains microarray data from roughly 45,000 probes (corresponding to approximately 20,000 genes) measured in 3,702 samples collected from 6 neurotypical adult subjects. Samples were taken from cortical, subcortical, brainstem and
cerebellar regions and gene expression data were extracted with a custom-developed Agilent chip consisting of a standard Whole Human Genome Microarray Kit and over 18,000 custom probes \cite{arnatkeviciute2019practical}. More probes can refer to the same gene, in fact, 93\% of all genes are annotated to more than one probe \cite{arnatkeviciute2019practical}.
Each sample is located in the brain of the donor subject with its native voxel coordinates in an MRI image acquired prior to dissection and its MNI coordinates.
The samples are not taken from the same brain locations across the 6 subjects. In particular, 4 of the subjects donated only the left brain hemisphere, thus right-hemisphere regions are missing in 2/3 of the samples.
The microarray expression values available have been already normalized by the curators of the AHBA dataset, with within-array, cross-batches and cross-brain corrections, as detailed in the documentation \cite{ahbadocs}.
In addition to the microarray GE values, a binary value for each probe in each sample indicates whether the GE value exceeded the background value during measurement, assessing the reliability of the reported value.
A summary of the dataset composition is reported in Tab. \ref{tab:aba_info}.

\begin{table}\centering
\resizebox{\textwidth}{!}{
\renewcommand{\arraystretch}{1.5}{
\begin{tabular}{|c|c|c|c|c|c|c|}

\hline 
\multicolumn{7}{|c|}{\textcolor{red}{\textbf{AHBA information}}}\\
\hline
\multirow{2}{*}{\textbf{Tot subjects}} & \multirow{2}{*}{\textbf{Tot samples}} & \textbf{Probes in} & \textbf{Genes in} & \multicolumn{3}{|c|}{\multirow{2}{*}{\textbf{Data for each sample}}} \\
 &  & \textbf{each sample} & \textbf{each sample} & \multicolumn{3}{|c|}{}\\\hline

\multirow{4}{*}{6} & 
\multirow{4}{*}{3,702} & 
\multirow{4}{*}{45,821} &
\multirow{4}{*}{20,232} & 
\multicolumn{3}{|l|}{ \hspace{0.25cm} \tabitem Microarray gene expression values} \\
&&&& \multicolumn{3}{|l|}{ \hspace{0.25cm} \tabitem Binary mask of expression exceeding the background} \\
&&&& \multicolumn{3}{|l|}{\hspace{0.25cm} \tabitem Native voxel coordinates of the sample} \\
&&&& \multicolumn{3}{|l|}{ \hspace{0.25cm} \tabitem MNI coordinates of the sample} \\

\hline
\multicolumn{7}{|c|}{\textcolor{red}{\textbf{Donors information}}}\\
\hline
\textbf{Donor ID} & \textbf{Age} & \textbf{Sex} &\textbf{Hemispheres} & \textbf{Cortex \& subcortex samples} & \textbf{Brainstem samples} & \textbf{Cerebellum samples}\\
\hline
 H0351\_1009 & 57 & M & L & 295 & 26 & 42\\
 H0351\_1012 & 31 & M & L & 401 & 80 & 68\\
 H0351\_1015 & 49 & F & L & 329 & 79 & 62\\
 H0351\_1016 & 55 & M & L & 362 & 59 & 80\\
 H0351\_2001 & 24 & M & L\&R & 739 & 154 & 53\\
 H0351\_2002 & 39 & M & L\&R & 622 & 188 & 83\\
\hline

\end{tabular}}}

\caption{Summary of the AHBA dataset composition information.}
\label{tab:aba_info}
\end{table}

\section{Determination of the gene expression levels across brain regions}
\label{sec:mm_regional_gene_expression}
As mentioned in Sec. \ref{sec:mm_link_intro}, this imaging genetics approach faces the problem of defining a meaningful quantity to represent the expression level of gene modules in the brain region of the AAL atlas, to enable an analysis to identify the regions mostly affected by the modules of interest.
This is not a straightforward task, because of the following reasons:
\begin{itemize}
    \item the AHBA dataset contains probe GE values, while the modules are lists of genes;
    \item there are multiple samples from multiple subjects corresponding to each AAL region;
    \item different regions are not covered by the same number of samples, nor by the same number of subjects;
    \item the same probe may be above background in a certain group of samples, and below background in other samples belonging to the same AAL region and subject.
\end{itemize}
These issues can all obviously be dealt with, however they force to introduce arbitrary choices in the analysis.

In this work, for each gene, only the probe with the highest DS was used. In fact, the end goal of the analysis is to detect the regions in which the genes belonging to a module take extreme expression values with respect to their brain-wide distribution. To this aim, using probes whose expression distribution is stable across subjects increases the chances of reliably identifying their extreme values, rather than simply picking up noise.
There is then the issue of how to combine the information coming from multiple measurements for the same probe, deriving from the many samples belonging to the same region. It was decided to group samples by subject, and to average above-background values across samples of the same subject, producing thus a vector of GE values per region, per subject.
Furthermore, these values were considered above background only when their corresponding probes were above background in at least 20\% of the samples.

As already mentioned, these choices are arbitrary. For example, one may wonder why the average was performed grouping by subject, since different subjects have widely different sampling in the various regions. It was considered the best course of action, since the subject variability is important and GE values in one subject may not be comparable to the same values in another subject. Similarly, many other alternative strategies could be suggested, but it is the opinion of the author that these arbitrary decisions do not weaken a strong, hypothesis-driven result, as long as this arbitrariness does not become a way to probe many combinations of such strategies until a positive result is found.

\section{Calculation of module eigengenes}
\label{sec:mm_eigengene_calculation}
After obtaining the regional gene expression vectors for each subject, a way must be found to compare modules between each other. In order to do this, for every gene module, the first eigengene vector across all brain regions was calculated for every subject. Note that it was necessary to calculate this eigengene for every subject, because each of them contributes with a different number of regions.\\
This operation generated one eigengene vector per subject, per module. All vectors have different lengths, depending on the number of regions sampled in each subject. These vectors were first Z-score transformed and then averaged together. In the average, each subject contributed only in those regions in which they had samples.\\
The reason for performing the Z-score transformation was because this allows to identify the regions of a certain subject where a particular module is over or under expressed, with respect to all other regions. This, rather than the overall GE value of the region, is in fact the metric of interest.

The output of this analysis is a vector of values representing how expressed a certain module is in a certain region, considering all available subjects. When this value is low (high) in a certain region, it means that the module was overall under-expressed (over-expressed) in that region with respect to all other regions. 

\section{Comparison between neuroimaging and transcriptomics results}
\label{sec:mm_enrichment_analysis}
The last step of the analysis consists in studying whether the modules of interest take higher absolute values of expression in the regions highlighted by neuroimaging, when compared to other regions and modules. 
This was verified with two analyses:
\begin{enumerate}
    \item \textit{Estimation of the joint effect of the modules of interest on the regions relevant for neuroimaging.}\\
    The distribution of the average absolute eigengene expression values of the modules of interest across all the brain regions was computed, comparing the values obtained in the regions relevant for neuroimaging and in those that are not. This comparison was performed by sorting the brain regions according to their absolute average eigengene value. If the regions highlighted by neuroimaging take higher values with respect to the other ones, it means that the modules of interest jointly affect these regions more than the other ones.
    \vspace{2mm}
    \item \textit{Comparison of the effect of each module on the brain regions relevant for classification.}\\
    For each module, the regions were sorted according to the module eigengene expression value and ranked. An enrichment of the regions of interest in the higher or lower ranks indicates that the expression values of that module in those regions are indeed extreme. Note that this analysis aims at proving a stronger association with respect to the first one. In fact, in this case modules are analyzed individually, rather than on average. If all (or many) relevant modules showed a significant higher absolute level of expression in the regions of interest for neuroimaging, while the others did not, this would constitute a stronger conclusion than what a similar analysis comparing the two averages could reach.
    This study is also complemented by a significance analysis, comparing for each module the average value of the eigengene in the regions of interest, with its average value on an equally sized group of regions randomly chosen 10,000 times from all regions.
    By comparing the p-values obtained it is also possible to understand if the modules identified with the transcriptomic analysis effectively affect the more the regions found with neuroimaging with respect to the other modules.  
\end{enumerate}

\part{Results}
\chapter{Neuroimaging results}
\label{chap:res_neuroimaging}

\section{Preprocessing}
\label{sec:res_neuroimaging_preproc}
The preprocessing operations succeeded in 1423 subjects. The resulting dataset is thus composed of 1423 rsfMRIs, each having $128 \times 128 \times 64$ voxels in the 3 spatial axes (x, y and z respectively) sampled every 6 seconds for 60 frames.

\FloatBarrier
\section{Confounder analysis}
\label{sec:res_neuroimaging_confounder}

\subsection{Assessment of the confounding effect}
Tab. \ref{tab:ci_results} contains the results of the confounder analysis performed with the CI. Results are reported both with the confounder correction disabled (column with $\lambda=0$, thus the proper confounder test) and enabled ($\lambda\neq0$). It can be noted that all the variables analyzed with the exception of handedness have a confounding effect. In particular, the confounding effect of age, site and eye status is very important, having a CI above 0.5, which has been shown to be a high value for experimental data \cite{ferrari2020measuring}.
The site effect has been measured both in a 1-vs-1 setup, with only two total sites in the dataset, and in a 1-vs-3 setup. Obviously, the confounding effect resulted to be stronger in the 1-vs-1 case, since the task of recognizing site is easier when the number of sites is small. The sample effect was also studied and still resulted to be confounding, albeit less than the site, confirming what was already observed in sMRI data \cite{ferrari2020dealing}.\\
The effect of continuous variables was studied for different values of $d$ and $l$ as described in Sec. \ref{sec:mm_confounders}. The table reports the minimum and maximum confounder effects observed and their $d$ and $l$ values. As expected, the minimum effect is obtained for high $d$ and small $l$, i.e., for example in case of age, groups made with a large age range and low age difference between the two tested groups.\\
From this analysis it can be concluded that sex, sample, site, eye status and age (for age differences above 3 years) are all strongly confounding, while FIQ, even for larger score differences is less confounding, but still relevant. On the contrary, handedness can be excluded as a confounder. \\
Appendix \ref{app:confounders} contains all the plots of the CI, for further inspection of these results.

\begin{table}
    \centering
    \resizebox{\textwidth}{!}{
    \begin{tabular}{|l|l|l|c|c|c|}
        \hline
        \textbf{Variable} & \textbf{Type} & \textbf{Notes} & \textbf{Confounder?} & \textbf{CI ($\lambda = 0$)} & \textbf{CI ($\lambda \neq 0$)} \\
        \hline
        Handedness & Binary & --- & No & $0.11 \pm 0.08$ & Not tested \\
        \hline
        
        \multirow{2}{*}{FIQ} & \multirow{2}{*}{Continuous} & Min for $d=15$ \& $l=10$ & Yes & $0.14 \pm 0.10$ & $0.01 \pm 0.01$ \\
         &  & Max for $d=15$ \& $l=30$ & Yes & $0.25 \pm 0.08$ & $0.09 \pm 0.03$ \\
        \hline        
        
        Sample & Categorical & 1 sample vs 1 sample& Yes & $0.34 \pm 0.06$ & $0.02 \pm 0.02$ \\
        \hline             
        Sex & Binary & --- & Yes & $0.45 \pm 0.05$ & $0.03 \pm 0.04$ \\
        \hline                
        Eye status & Binary & --- & Yes & $0.51 \pm 0.05$ & $0.03 \pm 0.02$ \\
        \hline
        Site (4 sites) & Categorical & 1 sites vs 3 sites & Yes & $0.60 \pm 0.04$ & $0.03 \pm 0.04$ \\
        \hline                
        Site (2 sites) & Categorical & 1 site vs 1 site & Yes & $0.64 \pm 0.02$ & $0.06 \pm 0.03$ \\
        \hline      
        \multirow{2}{*}{Age} & \multirow{2}{*}{Continuous} & Min for $d=5$ \& $l=3\hspace{0.2cm}$ & Yes & $0.17 \pm 0.10$ & $0.04 \pm 0.02$ \\
         &  & Max for $d=3$ \& $l=20$ & Yes & $0.74 \pm 0.08$ & $0.09 \pm 0.03$ \\
        \hline

    \end{tabular}
    }
    \caption{Results of the confounder analysis. Note that the table contains also the results with the confounder correction active ($\lambda\neq0$).}
    \label{tab:ci_results}
\end{table}

\subsection{Assessment of the network resilience to confounders}
After having determined which variables are confounders, the capability to deal with them of the bias-resilient approach taken in this work was evaluated. The adversarial part of the $\mathbb{FE}$ network component was activated (i.e., $\lambda$ was no longer set to 0) and the values of the CI obtained in this configuration (last column of Tab. \ref{tab:ci_results}) were compared to those described in the previous section. It can be observed that the CI value strongly diminishes for all the confounding variables, reaching magnitudes comparable with 0 for FIQ (at differences of 10 points), sample, sex, eye status and site (for multiple sites).
The remaining variables, FIQ for larger score differences, site when only two sites are included and age, see their CI value notably decreased. It must also be noted that the case of 1-vs-1 sites is not of real concern for the subsequent analyses, as the training set always contains more than two sites.\\
In summary, it can be concluded that the bias-resilient correction effectively reduces and in most cases completely destroys the relationship between the confounders and the task.

\FloatBarrier
\section{Optimization of the network architecture}
\label{sec:res_neuroimaging_optimization}
Table \ref{tab:final_architecture} contains the final architecture of the network, optimized during a 10-fold cross-validation varying the parameters as described in Tab. \ref{tab:grid_search_last}.
The resulting network has approximately 100 milion parameters. 

\begin{table}
    \centering
    \begin{tabular}{|l|l|c|}
         \hline
         \textbf{Parameter group}         & \textbf{Parameter} & \textbf{Value} \\
         \hline
         \multirow{4}{*}{Convolutional block}     & Convolutional kernel size   & $3\times 3$ \\
                             & Number of kernels in the first layer  & 25 \\
                             & Pooling kernel size   & $2\times 2$ \\
                             & Dropout rate   & 0.15 \\
                             & Number of layers   & 20 \\
         \hline
         \multirow{3}{*}{Dense block}         & Number of dense layers in $\mathbb{FE}$ & 3 \\
                             & Size of the extracted features & 400 \\
                             & Number of dense layers in $\mathbb{P}$ and $\mathbb{C}$ & 5 \\
         \hline
         Activation function & Leaky ReLU $\alpha$ & 0.05 \\
         \hline
         \multirow{4}{*}{Training process}    & Learning rate of $\mathbb{FE} + \mathbb{P}$ & $5 \times 10^{-3}$ \\
                             & Learning rate of $\mathbb{C}$ & $1 \times 10^{-4}$ \\
                             & Learning rate of $\mathbb{FE}$ (adversarial) & $7 \times 10^{-5}$ \\
                             & Batch size & 15 \\
         \hline
         \multirow{3}{*}{Global}    & Total parameters in $\mathbb{FE}$ & $\approx$ 90 millions \\
                                    & Total parameters in $\mathbb{P}$ & $\approx$ 2 millions \\
                                    & Total parameters in $\mathbb{C}$ & $\approx$ 2 millions \\
         \hline
    \end{tabular}
    \caption{Final configuration of the neural network used for the ASD/TDC classification task.}
    \label{tab:final_architecture}
\end{table}

\FloatBarrier
\section{Network training and performance assessment}
\label{sec:res_neuroimaging_performance}

The optimized network reported in Tab. \ref{tab:grid_search_last} was trained to distinguish ASDs from TDCs. Its performance was evaluated both in terms of confounder resilience and of the capability of accomplishing the task using the AUC metric.

Table \ref{tab:results_asd_tdc_tv} reports the performance obtained in the 10-fold cross-validation on 1253 subjects on both the training and validation datasets, in order to also evaluate over-fitting. The AUCs 0.87 $\pm$ 0.03 and 0.89 $\pm$ 0.04 obtained on the training and validation dataset respectively, are compatible with each other within the margin of error and are superior to most of the results obtained in literature and described in the background part (see Tab. \ref{tab:asd_studies} for a quick overview). Furthermore, the only studies that obtained better performance are affected by the issues explained in Sec. \ref{sec:bkg_studies_on_mri}.\\
The performance was then evaluated on the test set, containing 169 subjects not used in any optimization stage, obtaining an AUC of 0.89 (see Tab. \ref{tab:results_asd_tdc_test}). Fiftythree individuals of the test set constitute the entirety of the subjects acquired by one research center ($EMC-II$), that was thus never used by the network during training. Again, performance did not degrade with respect to the training and validation datasets, showing that the network did not overfit the data, nor used the site information to improve performance.

\begin{table}
    \centering
    \begin{tabular}{|l|c|c|c|c|}
    \hline
    \multicolumn{5}{|c|}{\textcolor{red}{\textbf{1254 subjects, 10-fold cross-validation}}}\\
    \hline
    \hline
    \cellcolor{gray!30}{\textbf{Dataset}} &  \cellcolor{gray!30}{\textbf{AUC}} &   \cellcolor{gray!30}{\textbf{Acc.}} &   \cellcolor{gray!30}{\textbf{Sens.}} &  \cellcolor{gray!30}{\textbf{Spec.}} \\
    \hline
    \textbf{Training}   & $ 0.87 \pm 0.03$ & $0.87 \pm 0.02$ &$0.88 \pm 0.03$ & $0.85 \pm 0.03$ \\
    \textbf{Validation} & $ 0.89 \pm 0.04$ & $0.87 \pm 0.04$&$0.87 \pm 0.04$ & $0.86 \pm 0.03$ \\
    \hline
    \end{tabular}
    \caption{Results of the ASD/TDC classification on the training and validation sets}
    \label{tab:results_asd_tdc_tv}
\end{table}
\begin{table}
    \centering
    \begin{tabular}{|l|c|c|c|c|}
    \hline
    \multicolumn{5}{|c|}{\textcolor{red}{\textbf{Test set}}}\\
    \hline
    \hline
    \cellcolor{gray!30}{\textbf{Dataset}} &  \cellcolor{gray!30}{\textbf{AUC}} &   \cellcolor{gray!30}{\textbf{Acc.}} &   \cellcolor{gray!30}{\textbf{Sens.}} &  \cellcolor{gray!30}{\textbf{Spec.}} \\
    \hline
    \textbf{Whole test set} (169 subjects) & $ 0.89$& $ 0.86$& $ 0.88$& $ 0.85$ \\
    \textbf{Single-site test set} (53 subjects) & $ 0.88$&$ 0.87$&$ 0.88$&$ 0.86$ \\
    \hline
    \end{tabular}
    \caption{Results of the ASD/TDC classification on the hold-out set. The single-site test set is a subset of the test set containing one site (EMC-II) that was not present in the training and validation sets.}
    \label{tab:results_asd_tdc_test}
\end{table}

Regarding confounder resilience instead, Tab. \ref{tab:final_confounder_correction_analysis} reports the correlations between the predictions of $\mathbb{C}$ and the true values of the confounders for the training, validation and test set together with both the confounder correction inactive and active. It can be seen that when $\lambda \neq 0$, the correlation is strongly reduced. The values are reported for the 3 datasets combined because, as shown also for the ASD/TDC classification performance, there is almost no difference between the three. 

Appendix \ref{app:resilience} contains further plots showing the correlations for each sample (Tab. \ref{tab:final_confounder_correction_analysis} reports only their average) and how the correlations varied during the training epochs.

\begin{table}
    \centering
    \begin{tabular}{|l|c|c|}
        \hline
        \textbf{Confounder} & \textbf{Correlation $\lambda=0$} & \textbf{Correlation $\lambda \neq 0$} \\
        \hline
        Sex                   & 0.28 & 0.05 \\
        Eye status            & 0.51 & 0.17 \\
        Sample/Site (average) & 0.48 & 0.13 \\
        FIQ                   & 0.09 & 0.06 \\
        Age                   & 0.47 & 0.12 \\
        \hline
    \end{tabular}
    \caption{Correlation between the predictions of $\mathbb{C}$ and the true value of the confounders for the final network architecture used for the ASD/TDC classification task. The sample/site correlation has been calculated as the average of the correlations for each sample (note that this include comparing different sites, as different sites are always different samples).}
    \label{tab:final_confounder_correction_analysis}
\end{table}

\begin{figure}
    \centering
    \includegraphics[width=0.65\textwidth]{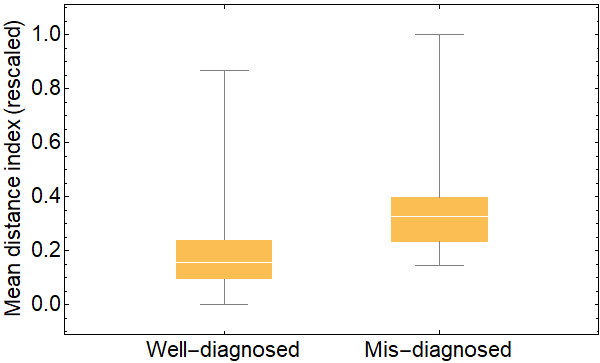}
    \caption{Whisker plot of the mean distance index. Values are rescaled between 0 and 1 to ease reading. Lower index values indicate higher image quality.}
    \label{fig:quality}
\end{figure}

Finally, Fig. \ref{fig:quality} contains the results of the investigation of the correlations between image quality and prediction errors. As the figure shows, images with a higher mean distance index (i.e., lower quality), are more often mis-classified, suggesting that the prediction is hampered by the low quality of the rsfMRIs. This result is encouraging, because it implies that it is possible to increase the diagnosis performance of this classifier by simply improving the quality of the images.

\FloatBarrier
\section{Explainability and brain region selection}
\label{sec:res_neuroimaging_explainability}
\begin{figure}
    \centering
    \includegraphics[width=0.6\textwidth]{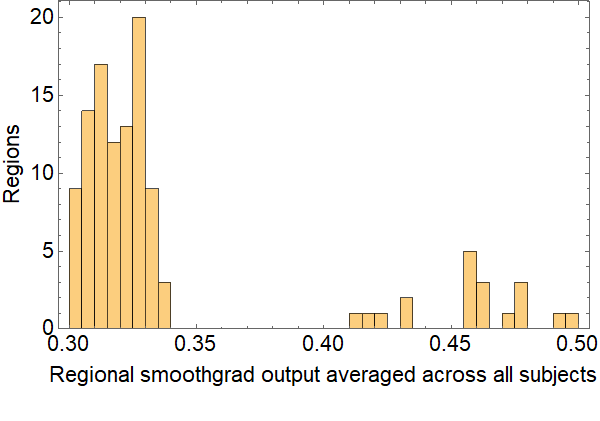}
    \caption{Histogram of the regional value of the smoothgrad output averaged across all subjects. 19 regions stand out on the right as having an abnormally high value.}
    \label{fig:smoothgrad_avg}
\end{figure}

This section reports the results of the explainability analysis performed with \textit{Smoothgrad} and the extraction of the most relevant regions.
Fig. \ref{fig:smoothgrad_avg} contains a histogram of the average regional activation value of the saliency map. The per-region value was calculated subject by subject averaging together the values of all voxels belonging to each region, then computing the average across all subjects.
It can be noted that 19 regions are completely separated from the bulk and have substantially larger average activation values.

\begin{figure}
    \centering
    \includegraphics[width=0.6\textwidth]{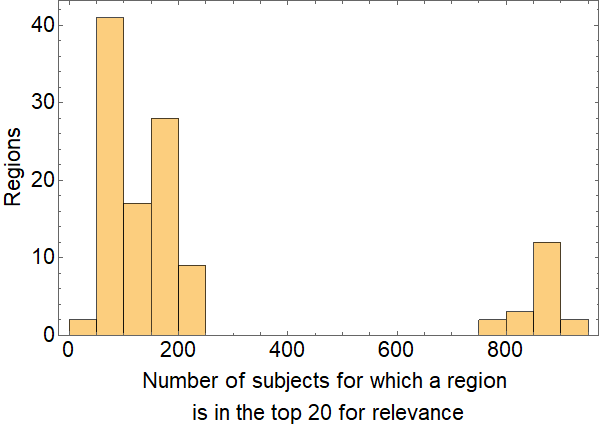}
    \caption{Histogram of the number of subjects in which each region is in the top 20 for relevance according to smoothgrad. Again, 19 regions stand out for being relevant for over 800 subjects. It cannot be seen from the plot, but these regions are the same ones that appear as outliers on the right in Fig. \ref{fig:smoothgrad_avg}.}
    \label{fig:smoothgrad_ranking}
\end{figure}

Similarly, Fig. \ref{fig:smoothgrad_ranking} shows a histogram binning regions by the number of subjects for which they are one of the top 20 regions for average activation. Again, 19 regions are relevant for a substantially larger number of subjects with respect to all the other ones.

\begin{figure}
    \centering
    \begin{subfigure}{0.75\textwidth}
        \includegraphics[width=1.0\textwidth]{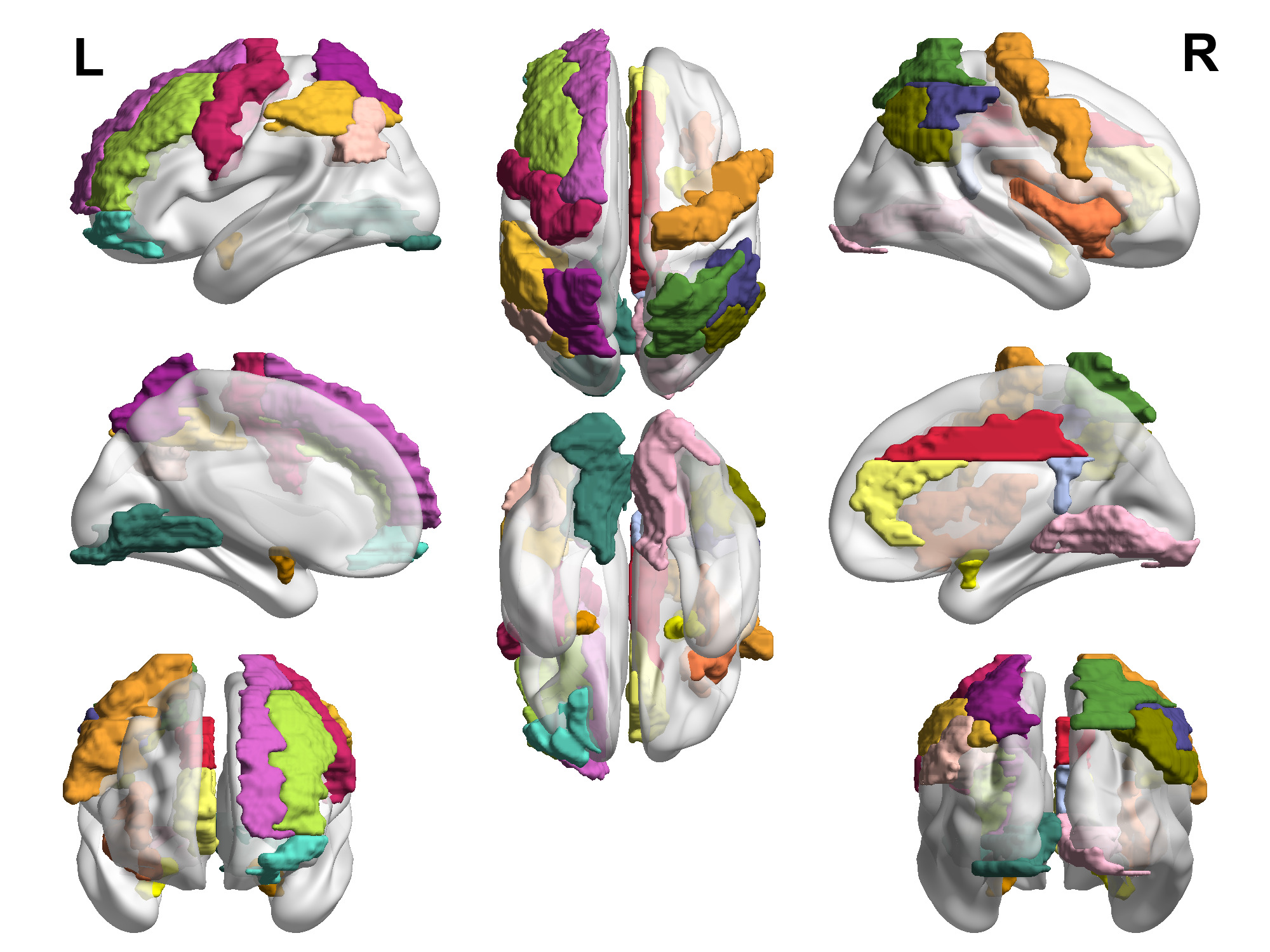}
    \end{subfigure}
    \begin{subfigure}{0.22\textwidth}
        \includegraphics[width=1.0\textwidth]{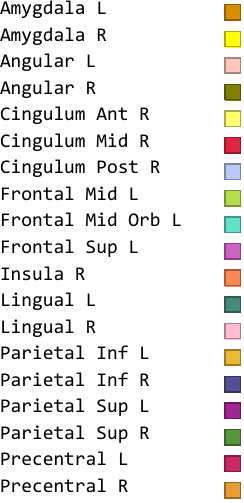}
    \end{subfigure}    
    \caption{View of the 19 regions found relevant.}
    \label{fig:immagine_19_regioni}
\end{figure}

It is interesting to note that the regions having higher values in the two plots are the same (they are reported in Tab. \ref{tab:relevant_regions} and shown highlighted with different colors in a reconstruction of the brain in Fig. \ref{fig:immagine_19_regioni}). These plots show therefore that there is a well-defined group of regions that have abnormally high importance in the ASD/TDC diagnosis according to \textit{Smoothgrad}. These regions are rather numerous, and there is not a single region relevant for all subjects (which are in fact 1423). This means that the pattern on which the network bases its predictions is quite complex and is not the same for all subjects.
It is possible that the pattern extracted analyzing the activation values could be better interpreted as the superposition of multiple patterns, each used for diagnosing different subtypes of ASD.

\begin{table}
    \centering
    \begin{tabular}{|l|l|}
        \hline
        \multicolumn{2}{|c|}{\textbf{19 relevant regions}}\\
        \hline
        Amygdala L & Amygdala R \\
        Angular L & Angular R \\
        Lingual L & Lingual R \\
        Parietal inf L & Parietal inf R \\
        Parietal sup L & Parietal sup R \\
        Precentral L & Precentral R \\
        \hline
        \multicolumn{2}{|l|}{\hspace{1.4cm} Cingulum ant R} \\
        \multicolumn{2}{|l|}{\hspace{1.4cm} Cingulum mid R} \\
        \multicolumn{2}{|l|}{\hspace{1.4cm} Cingulum post R} \\
        \multicolumn{2}{|l|}{\hspace{1.4cm} Frontal mid L} \\
        \multicolumn{2}{|l|}{\hspace{1.4cm} Frontal mid orb L} \\
        \multicolumn{2}{|l|}{\hspace{1.4cm} Frontal sup L} \\
        \multicolumn{2}{|l|}{\hspace{1.4cm} Insula R} \\
        \hline
    \end{tabular}
    \caption{Relevant regions selected with the explainability analysis. 6 pairs of regions are found on both hemispheres, while the remaining 7 ones are in only one hemisphere.}
    \label{tab:relevant_regions}
\end{table}

\begin{figure}
    \centering
    \begin{subfigure}{0.8\textwidth}
        \centering
        \includegraphics[width=0.9\textwidth]{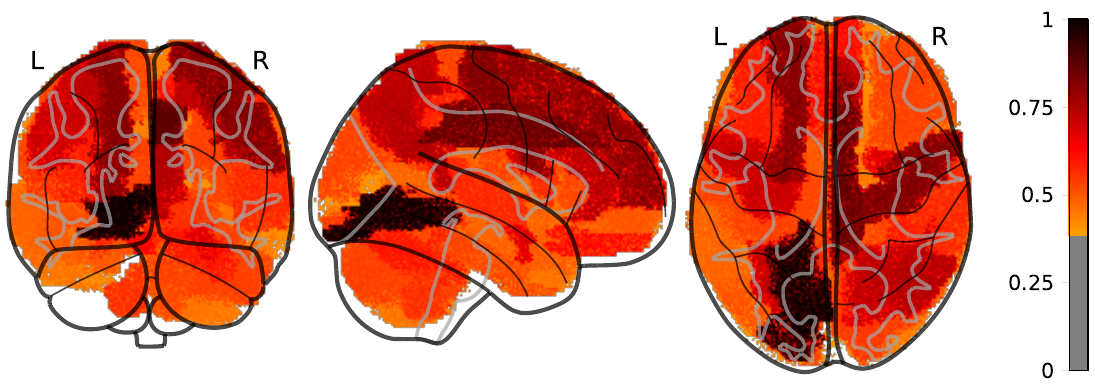}
        \caption{Threshold of 0.40}
        \label{fig:glass_brain_smooth_avg_40}
    \end{subfigure}\\
    \begin{subfigure}{0.8\textwidth}
        \centering    
        \includegraphics[width=0.9\textwidth]{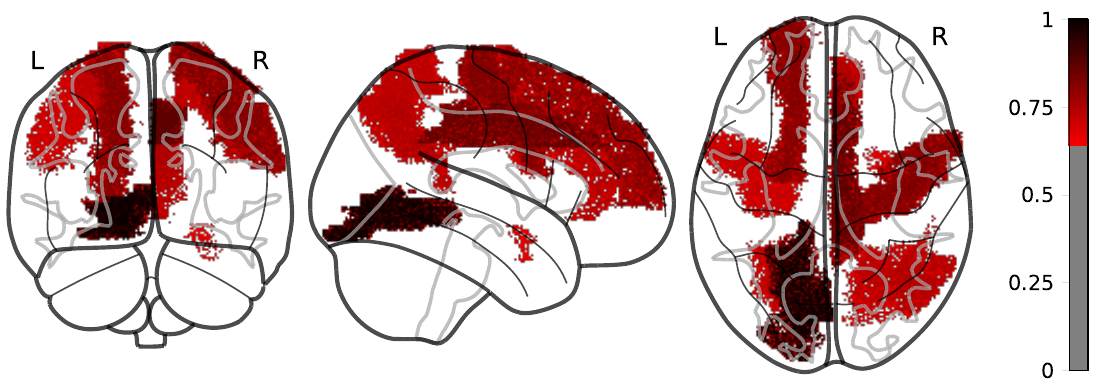}
        \caption{Threshold of 0.65}
        \label{fig:glass_brain_smooth_avg_65}
    \end{subfigure}\\    
    \caption{2-D views of the average brain as seen through \textit{Smoothgrad}. In Fig. (a), the pixels with a value below 0.40 have been set to white, while in Fig. (b) this has been done for the pixels below 0.65.}
    \label{fig:glass_brain_smooth_avg}
\end{figure}

To better understand the classification pattern, Fig. \ref{fig:glass_brain_smooth_avg} shows 2D views of the average \textit{Smoothgrad} pattern, with two different thresholds. From Fig. \ref{fig:glass_brain_smooth_avg_40}, the one with the lower threshold, it can be noticed that the whole brain is found more relevant than the image background, which is below threshold. This is an encouraging result, as it demonstrates that the network does not base its predictions on meaningless noise outside the brain.\\
The second picture, Fig. \ref{fig:glass_brain_smooth_avg_65}, shows that there are a few neatly separated regions that are more relevant than others. 

\begin{figure}
    \centering
    \begin{subfigure}{0.8\textwidth}
        \centering
        \includegraphics[width=0.9\textwidth]{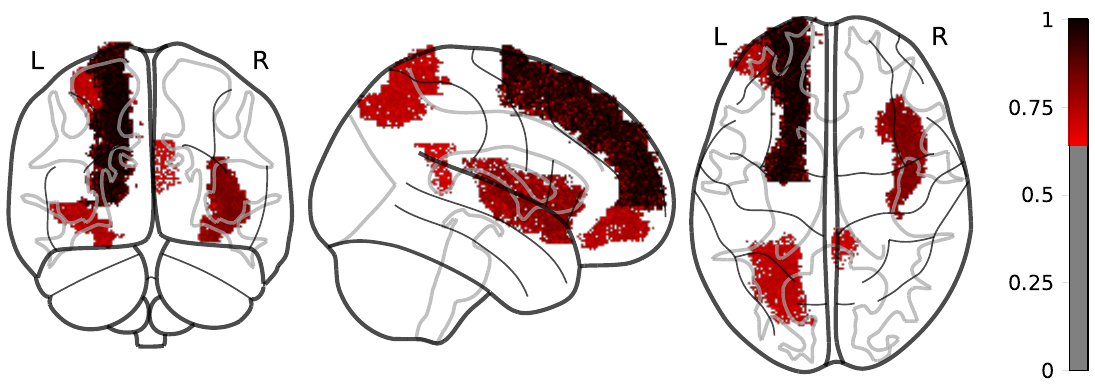}
        \caption{Subject ID: 30144}
        \label{fig:glass_brain_smooth_s1}
    \end{subfigure}\\
    \begin{subfigure}{0.8\textwidth}
        \centering    
        \includegraphics[width=0.9\textwidth]{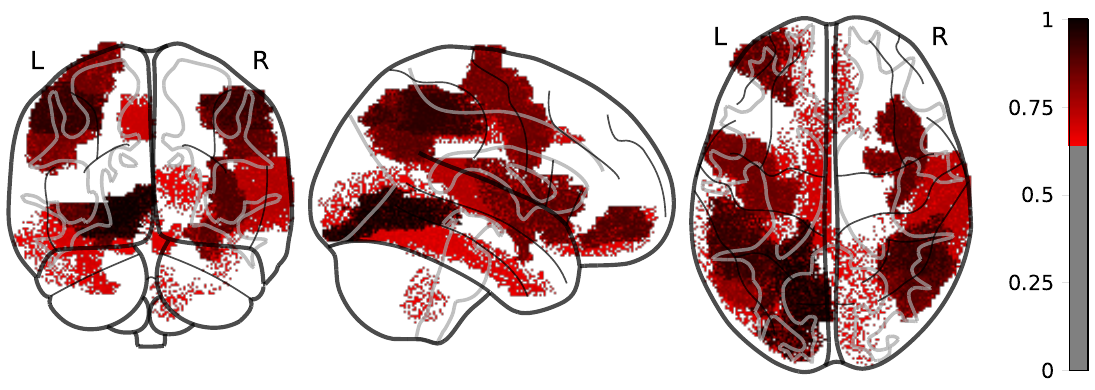}
        \caption{Subject ID: NYU\_0050962}
        \label{fig:glass_brain_smooth_s2}
    \end{subfigure}\\ 
     \begin{subfigure}{0.8\textwidth}
        \centering    
        \includegraphics[trim=3mm 0 -7mm 0, clip, width=0.9\textwidth]{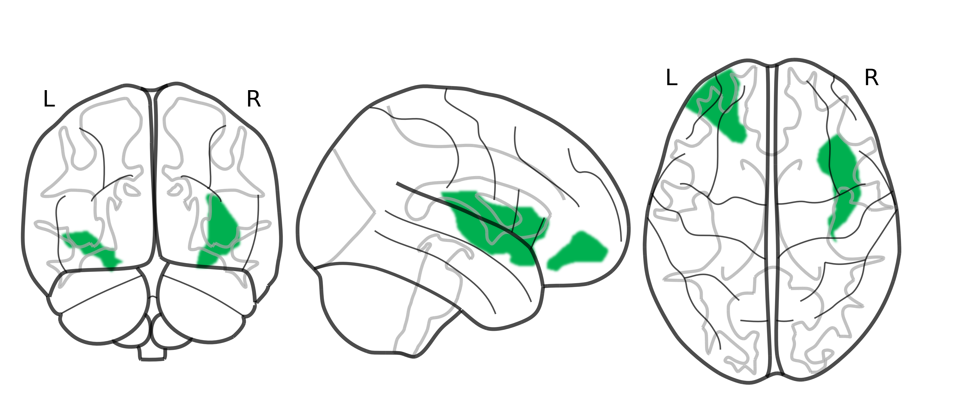}
        \caption{Common regions}
    \end{subfigure}\\ 
    \caption{The glass brains in (a) and (b) represents the saliency pattern extracted with \textit{Smoothgrad} from two ASD subjects. The classification patterns are very different, nonetheless, the two subjects have both been correctly classified. Pixels below 0.65 have been set to white. The glass brain in (c), instead, shows the common regions between (a) and (b).}
    \label{fig:glass_brain_smooth_subjects}
\end{figure}

Finally, Fig. \ref{fig:glass_brain_smooth_subjects} shows the application of \textit{Smoothgrad} to two different ASD subjects, both correctly classified. The classification pattern used by the network according to the explainability framework is very different in the two cases. Tab. \ref{tab:smoothgrad_2_subj}, reporting the main anagraphic and phenotypical information on the two subjects, shows that they do not differ significantly for their clinical profile and for age, but they are a male and a female.  This qualitative observation suggests that a multivariate regression analysis to explore the association between each brain region involved in the disorder and the subject characteristics available in the dataset is worth to be explored.

\begin{table}\centering
\setlength{\tabcolsep}{0.5pt}
\renewcommand{\arraystretch}{1}
\begin{tabular}{|lc|cc|}
\hline
Subject ID && 30144 & NYU\_0050962\\
\hline
Sex && Male & Female\\
Age at scan && 22 & 24.41\\
\multirow{2}{*}{Comorbidities} && \multirow{2}{*}{-} & Anxiety disorder NOS\\
&&& Mood disorder NOS\\
FIQ && 100 & 98\\
ADOS total &$[0,22]$&13 &15 \\
ADOS communication &$[0,8]$& 4&3 \\
ADOS social &$[0,14]$&9 &12 \\
ADOS SBRI &$[0,8]$& 2&0 \\
\hline
\end{tabular}
\caption{Information about the two subjects depicted in Fig. \ref{fig:glass_brain_smooth_subjects}. In this table the acronyms "NOS" and "SBRI" stand for "Not Otherwise Specified" and "Stereotyped Behaviors and Restricted Interests", respectively.  } 
\label{tab:smoothgrad_2_subj}
\end{table}

\FloatBarrier

\chapter{Transcriptomics results}
\label{chap:res_genetics}

\section{Preliminary analysis}
\label{sec:res_genetics_preliminary}

\subsection{Preprocessing}
\begin{figure} 
\centering
\begin{tabular}{cc}
\subfloat[]{\includegraphics[width = 1.5in]{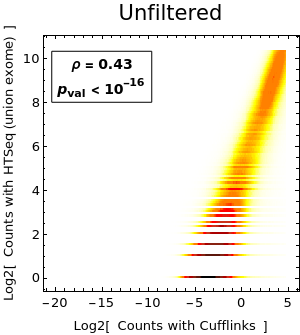}} &
\subfloat[]{\includegraphics[width = 1.5in]{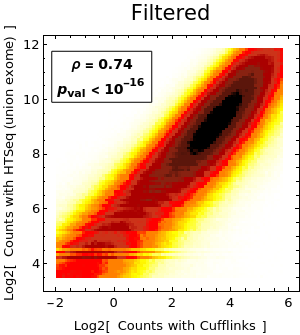}}\\
\subfloat[]{\includegraphics[width = 1.5in]{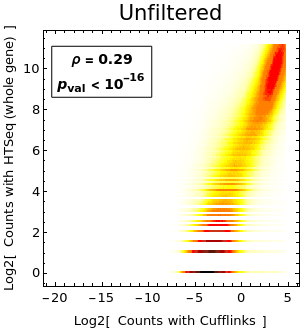}} &
\subfloat[]{\includegraphics[width = 1.5in]{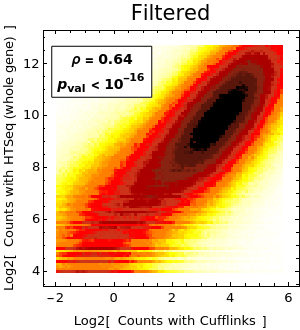}}\\
\subfloat[]{\includegraphics[width = 1.5in]{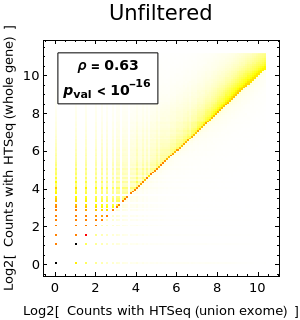}} &
\subfloat[]{\includegraphics[width = 1.5in]{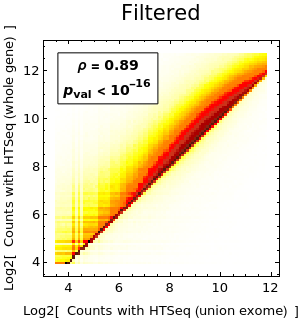}}\\
\end{tabular}
\caption{Density histograms showing the gene expression levels quantified with two different methods on the two axes. Each inset reports their Pearson's correlation $\rho$. Histograms on the left column include data of all the genes in all the samples, while on the right the genes with a low statistic have been excluded. Note that the striped aspect of histograms (a-d) is due to the fact that HTSeq provides a discrete measure (i.e. the number of counts) while Cufflinks provides a continuous estimation and this difference is exacerbated by $Log_2$-transformation, especially for low gene expression levels. In Fig. \ref{fig:pre-proc}e the point (0,0) is much more populated than any other one, explaining why this image shows a restricted palette of colours.}\label{fig:pre-proc}
\end{figure}

After the preprocessing steps, a total of 16,995 genes available in all the 235 selected samples was included in the final dataset. Fig. \ref{fig:pre-proc} underlines the importance of the filtering step, eliminating the less-expressed and shorter genes. The correlation between the gene expression as measured by the three different tools (Cufflinks, whole-gene HTSeq and union-exon HTSeq) markedly improves when those genes are eliminated, as it can be seen comparing the left column, containing the correlation values calculated on the unfiltered gene list (Fig.s \ref{fig:pre-proc}a, \ref{fig:pre-proc}c and \ref{fig:pre-proc}e) with the right one, calculated with the filtered list (Fig.s \ref{fig:pre-proc}b, \ref{fig:pre-proc}d and \ref{fig:pre-proc}f).

\subsection{Identification of the factors influencing gene expression levels}
Tab. \ref{tab:relevantQ_feat} contains the results of the investigation with LMM of the factors influencing gene expression in the dataset. Dark green rectangles indicate that a certain variable is found to influence gene expression in a certain tissue with a BH-corrected p-value below 0.05, while a light green rectangle represents an uncorrected p-value below 0.05 (but a corrected one above that threshold).
The variables considered to influence gene expression and thus included as covariates in the LMM analysis are those with at least one dark green rectangle: seq. batch, brain bank, sex, age and seizure.
Age is the only continuous variable and therefore the only one for which different functional forms could be used in the LMM. An in-depth analysis on the specific effect of age on gene expression are reported in appendix \ref{app:genetics_preprocessing}. The outcome of this analysis is that the best representation of age dependency for GE and $\Delta$GE values is with a linear+quadratic form.

\definecolor{myR}{rgb}{0.4, 0.8, 0.55}
\definecolor{myY}{rgb}{0.85, 0.95, 0}

\newcommand{\Rcell}{\fcolorbox{black}{myR}{\rule[0.3cm]{1.1cm}{0cm}}}

\newcommand{\Ycell}{\fcolorbox{black}{myY}{\rule[0.3cm]{1.1cm}{0cm}}}

\begin{table}\centering
\setlength{\tabcolsep}{0.5pt}
\renewcommand{\arraystretch}{1}
\begin{tabular}{|l|cccccc|}

\hline                   
 & \hspace*{1pt} \textbf{ \textcolor{black}{F}} \hspace*{1pt}
 & \hspace*{1pt} \textbf{ \textcolor{black}{T}} \hspace*{1pt}
 & \hspace*{1pt} \textbf{ \textcolor{black}{C}} \hspace*{1pt}
 & \hspace*{1pt} \textbf{ \textcolor{black}{F-C}} \hspace*{1pt}
 & \hspace*{1pt} \textbf{ \textcolor{black}{T-C}} \hspace*{1pt}
 & \hspace*{1pt} \textbf{ \textcolor{black}{T-F}} \hspace*{1pt}\\
\hline
\hspace*{1pt} Seq Batch \hspace*{1pt} &\Rcell&\Rcell&\Rcell&\Rcell&\Rcell&\Ycell\\
\hspace*{1pt} Brain Bank \hspace*{1pt} &\Rcell&\Rcell&\Rcell&\Rcell&\Ycell&\Rcell\\
\hspace*{1pt} Sex \hspace*{1pt} &\Rcell&\Rcell&\Rcell&\Ycell & &\\
\hspace*{1pt} Age \hspace*{1pt} &\Rcell&\Rcell&\Ycell&\Ycell &\Ycell&\Ycell\\
\hspace*{1pt} Seizure \hspace*{1pt} &\Rcell& &\Ycell&\Ycell & & \\
\hspace*{1pt} Cause of death \hspace*{1pt} & & & & & &  \\
\hspace*{1pt} Pyschiatric medications \hspace*{1pt} & & & & & & \\
\hspace*{1pt} Brain pH \hspace*{1pt} & & & & & &\\
\hspace*{1pt} Brain weight \hspace*{1pt} & & &\Ycell & & &\\
\hspace*{1pt} RIN \hspace*{1pt} & & &\Ycell & & &\\
\hspace*{1pt} TotalReadsInBam \hspace*{1pt} & & & & &\Ycell &\\
\hspace*{1pt} After.rmdup.samtools \hspace*{1pt} & & & & &\Ycell &\\
\hspace*{1pt} Num.dup.samtools \hspace*{1pt} &\Ycell & & & & &\Ycell\\
\hspace*{1pt} Unique.Reads.samtools \hspace*{1pt} & & & & & &\\
\hspace*{1pt} PropExonicReads.HTSC \hspace*{1pt} & & & & & &\\
\hspace*{1pt} TotalReads.picard \hspace*{1pt} & & & &\Ycell& &\\
\hspace*{1pt} Aligned.Reads.picard \hspace*{1pt} & &\Ycell & & & &\\
\hspace*{1pt} HQ.Aligned.Reads.picard \hspace*{1pt} & & & & & &\\
\hspace*{1pt} PF.All.Bases.picard \hspace*{1pt} & &\Ycell & &\Ycell & &\\
\hspace*{1pt} Coding.Bases.picard \hspace*{1pt} & & & & & &\\
\hspace*{1pt} UTR.Bases.picard \hspace*{1pt} & & & & & &\Ycell\\
\hspace*{1pt} Intronic.Bases.picard \hspace*{1pt} & & & & & &\\
\hspace*{1pt} Intergenic.bases.picard \hspace*{1pt} & &\Ycell & & & &\\
\hspace*{1pt} Median.CV.Coverage.picard \hspace*{1pt} & & & & & &\\
\hspace*{1pt} Median.5prime.Bias.picard \hspace*{1pt} & & & & & &\\
\hspace*{1pt} Median.3prime.Bias.picard \hspace*{1pt} & & &\Ycell & & &\\
\hspace*{1pt} Median.5to3prime.Bias.picard \hspace*{1pt} & & & & & &\\
\hspace*{1pt} AT.Dropout.picard \hspace*{1pt} &\Ycell &\Ycell & & & &\\
\hspace*{1pt} GC.Dropout.picard \hspace*{1pt} & & & & & &\\
\hline
\multicolumn{1}{|c|}{\multirow{2}{*}{\textbf{Legend}}} & \multicolumn{6}{|l|}{\Rcell \hspace*{1pt} : p-value $FDR_{BH}$ corrected $\leq$ 0.05}\\
& \multicolumn{6}{|l|}{\Ycell \hspace*{1pt} : p-value $\leq$ 0.05}\\
\hline

\end{tabular}
\caption{Results from the imputing analysis and the permutation test. Rectangular markers indicate whether the features, listed in the first column, can be imputed from the transcriptome alone taken from the tissues, listed in the first line, with performances statistically significant. The significance has been evaluated with a permutation test and p-values below 0.05 are depicted with yellow markers, while green markers indicate p-values that remains below 0.05 after $FDR_{BH}$ correction.}
\label{tab:relevantQ_feat}
\end{table}

\FloatBarrier
\section{LMM analysis}
\label{sec:res_genetics_lmm}

\subsection{Derivation of the lists of relevant genes}
This section contains the results of the LMM analysis, aimed at identifying a list of genes relevant for ASD/TDC classification. 
According to the results of the analysis described in the previous section, the formula used in the LMM fit is therefore:
\begin{equation}
    Y\;\;\sim\;\; 1 + DIAGNOSIS + AGE + AGE^2 + BRAIN\_BANK + SEIZURE + SEX + (1 | BATCH)
\end{equation}
Where $Y$ is the GE or $\Delta$GE value of a gene, and $(1 | BATCH)$ indicates that the batch was estimated as a random effect, as prescribed by the criteria reported in Sec. \ref{sec:mm_lmm_analysis}. In fact, the $BATCH$ variable is the only one for which all 4 criteria apply: it has a lot of different values and each value has relatively few samples, unevenly distributed.

\begin{figure} 
\centering
\begin{tabular}{ccc}
\subfloat[]{
\includegraphics[width =.31\textwidth]{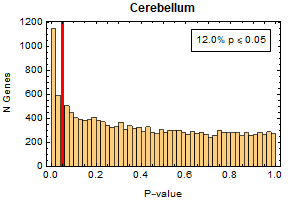}} 
&
\subfloat[]{\includegraphics[width =.31\textwidth]{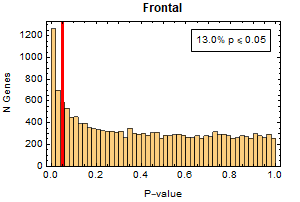}}
&
\subfloat[]{\includegraphics[width =.31\textwidth]{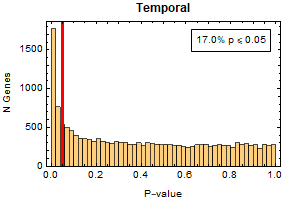}}
\\
\subfloat[]{\includegraphics[width =.31\textwidth]{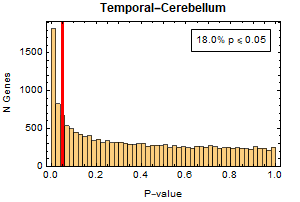}}
&
\subfloat[]{\includegraphics[width =.31\textwidth]{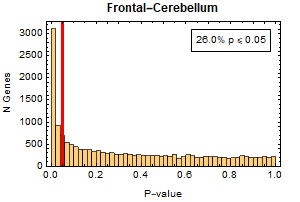}}
&
\subfloat[]{\includegraphics[width
=.31\textwidth]{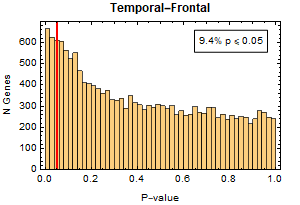}}

\end{tabular}
\caption{P-values of disease effect in the LMMs computed on all genes and all subjects under examination.}\label{fig:LMM_hist_pval_all}
\end{figure}

Fig. \ref{fig:LMM_hist_pval_all} shows the histogram of the p-values of the effect of diagnosis on the expression values of each gene reported by the LMM fit. The red line marks a 0.05 uncorrected threshold. The histograms, with the exception of those of temporal-frontal differences, are all peaked to the left, at small p-values, already providing a hint that, for a larger-than-random amount of genes, GE or $\Delta$GE significantly differ between ASDs and TDCs. It is not surprising that the temporal-frontal difference does not produce interesting results, as these two tissues are transcriptionally similar to each other, being both regions of the neocortex. In fact, the high transcriptomic homogeneity of the neocortex has been observed also in the first study on AHBA \cite{hawrylycz2012anatomically}.\\
Considering a threshold on the BH-corrected p-values of 0.05 and one on the effect size of 0.8 (i.e., the average entity of the difference between ASDs and TDCs found by the LMM), between 69 and 814 genes per dataset analyzed were differentially expressed (see Tab. \ref{tab:gene_numbers}). 

\begin{figure} 
\centering
\begin{tabular}{ccc}
\subfloat[]{
\includegraphics[width =.31\textwidth]{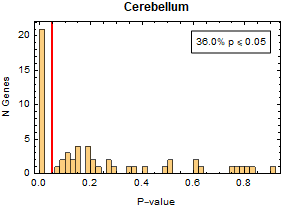}} 
&
\subfloat[]{\includegraphics[width =.31\textwidth]{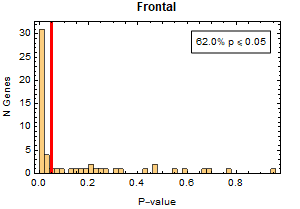}}
&
\subfloat[]{\includegraphics[width =.31\textwidth]{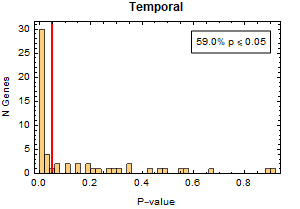}}
\\
\subfloat[]{\includegraphics[width =.31\textwidth]{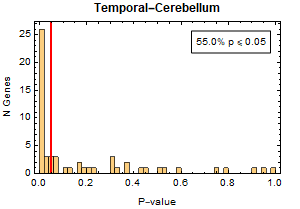}}
&
\subfloat[]{\includegraphics[width =.31\textwidth]{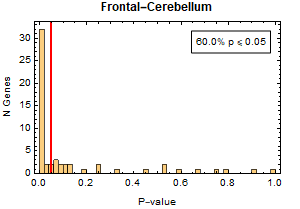}}
&
\subfloat[]{\includegraphics[width
=.31\textwidth]{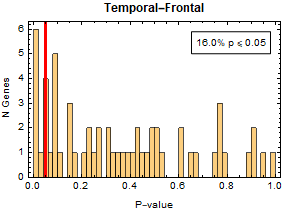}}

\end{tabular}
\caption{P-values of disease effect in the LMMs computed only on the genes of Dup15q and on the ASDs with that syndrome and an equal number of matched TDCs.}
\label{fig:LMM_hist_pval_15q}
\end{figure}

Fig. \ref{fig:LMM_hist_pval_15q} shows the same analysis, performed only on the genes involved in the Dup15q syndrome, considering only the ASDs with Dup15q and an equally-sized group of matched TDCs. In this case, the peak close to 0 in the histogram p-value is even sharper, with between 35\% and 60\% of the tested genes showing a p-value below 0.05. Note that not all genes are expected to be DE, because the duplicated portion of the 15th chromosome differs from subject to subject. Nonetheless, this shows that the LMM method is able to correctly identify differential expression with remarkable accuracy. As already described in Sec. \ref{sec:mm_lmm_analysis}, this finding validates the methodology of the study, both in the pre-processing steps and in the choice of the analysis of the DT datasets. In fact, all the analyses on the various datasets, with the exception of the temporal-frontal differences, find a similar number of DE genes (see Fig. \ref{fig:LMM_hist_pval_15q}).

\begin{table}[]
    \centering
    \begin{tabular}{|l|l|}
        \hline
        \textbf{Dataset} & \textbf{Number of DE genes} \\
        \hline
        C & 163 \\
        F & 85 \\
        T & 265 \\
        T-C & 203 \\
        F-C & 614 \\
        T-C & 69 \\
        All ST (union) & 432 \\
        All DT (union) & 751 \\
        All (union) & 1151 \\
        \hline
    \end{tabular}
    \caption{Number of DE genes found in each dataset.}
    \label{tab:gene_numbers}
\end{table}

\begin{figure} 
\centering
\begin{tabular}{cc}
\subfloat[]{
\includegraphics[width =.45\textwidth]{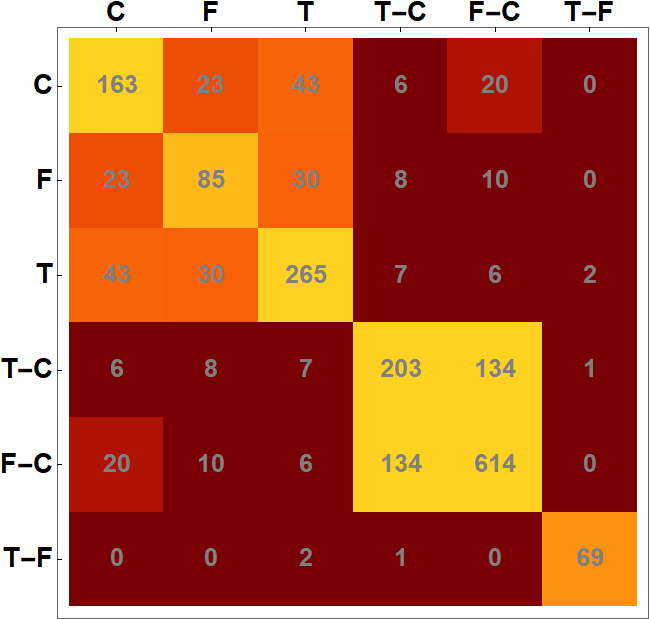}} 
&
\subfloat[]{\includegraphics[width =.45\textwidth]{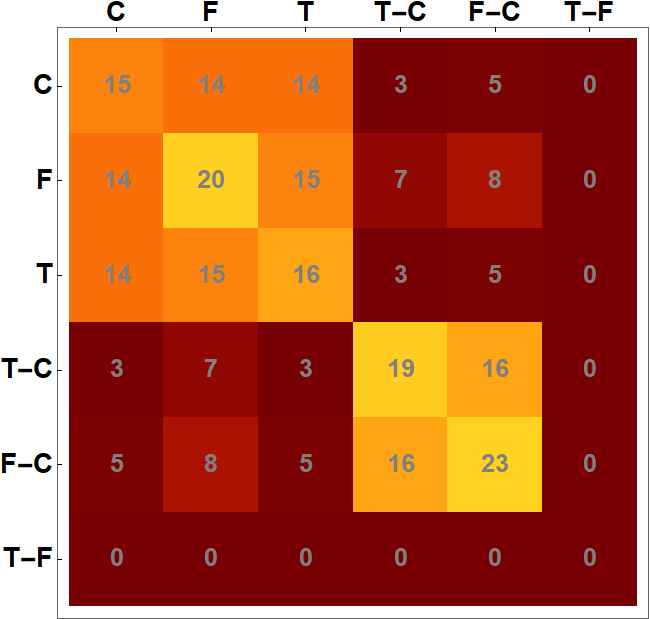}}

\end{tabular}
\caption{Intersections between the lists of relevant genes found in the whole dataset (Fig. a) and in the Dup15q matched cohort (Fig. b).}
\label{fig:intersections_relevant_lists}
\end{figure}

It is interesting to note that in both analyses, the genes found on the ST datasets are different from those found on the DT ones, highlighting the importance of considering both approaches. This is clearly shown in Fig. \ref{fig:intersections_relevant_lists}, that illustrates the size of the pair-wise intersection between the relevant gene lists obtained considering two different datasets. It can be observed that the lists found on C, F and T share many common genes, as the ones found on T-C and F-C do. However, lists found on ST datasets do not have large intersections with those found on DT ones.

\begin{figure} 
\centering
\begin{tabular}{ccc}
\subfloat[]{
\includegraphics[trim={0 0 2cm 0},clip,width =.3\textwidth]{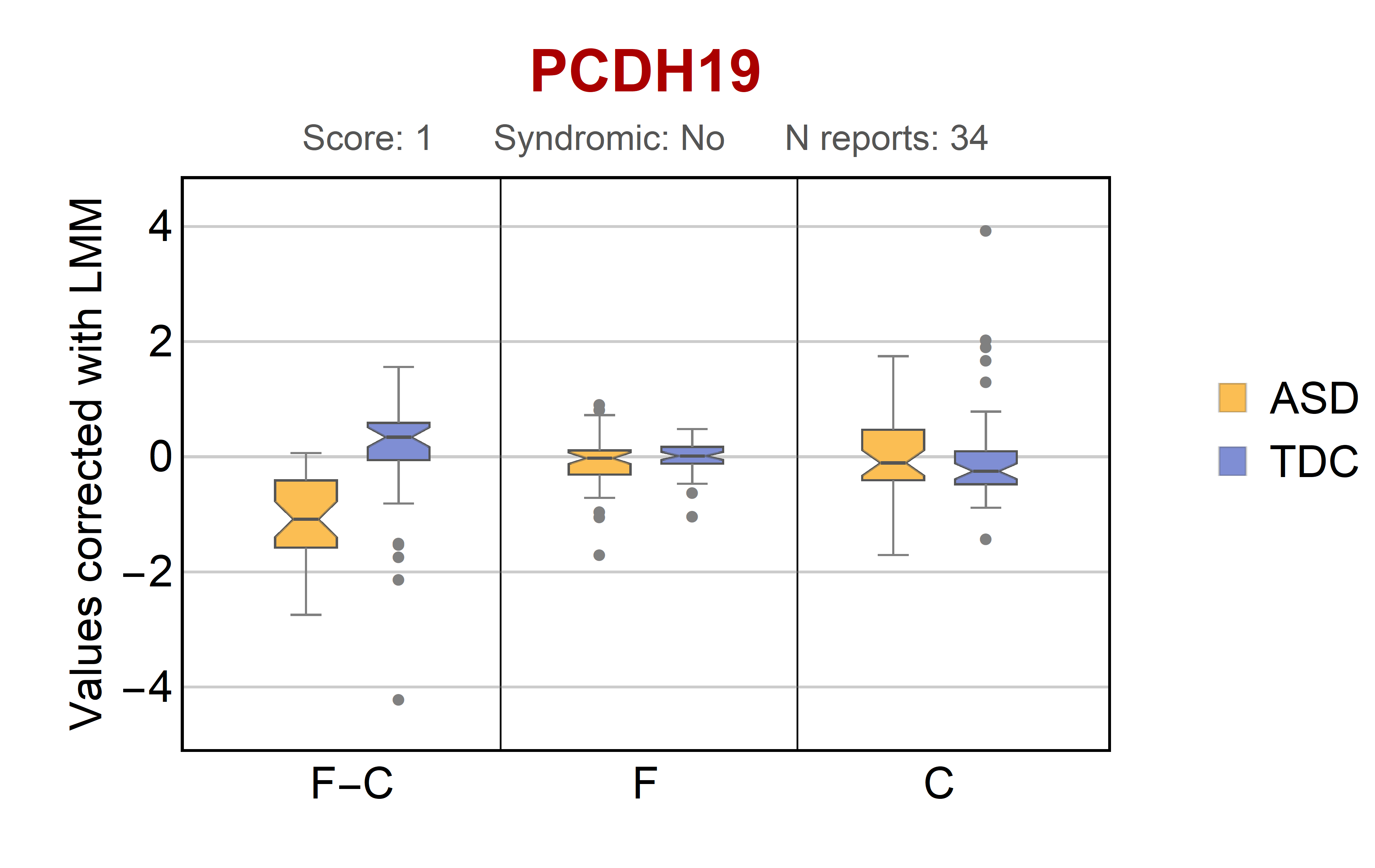}} 
&
\subfloat[]{\includegraphics[trim={0 0 2cm 0},clip, width =.3\textwidth]{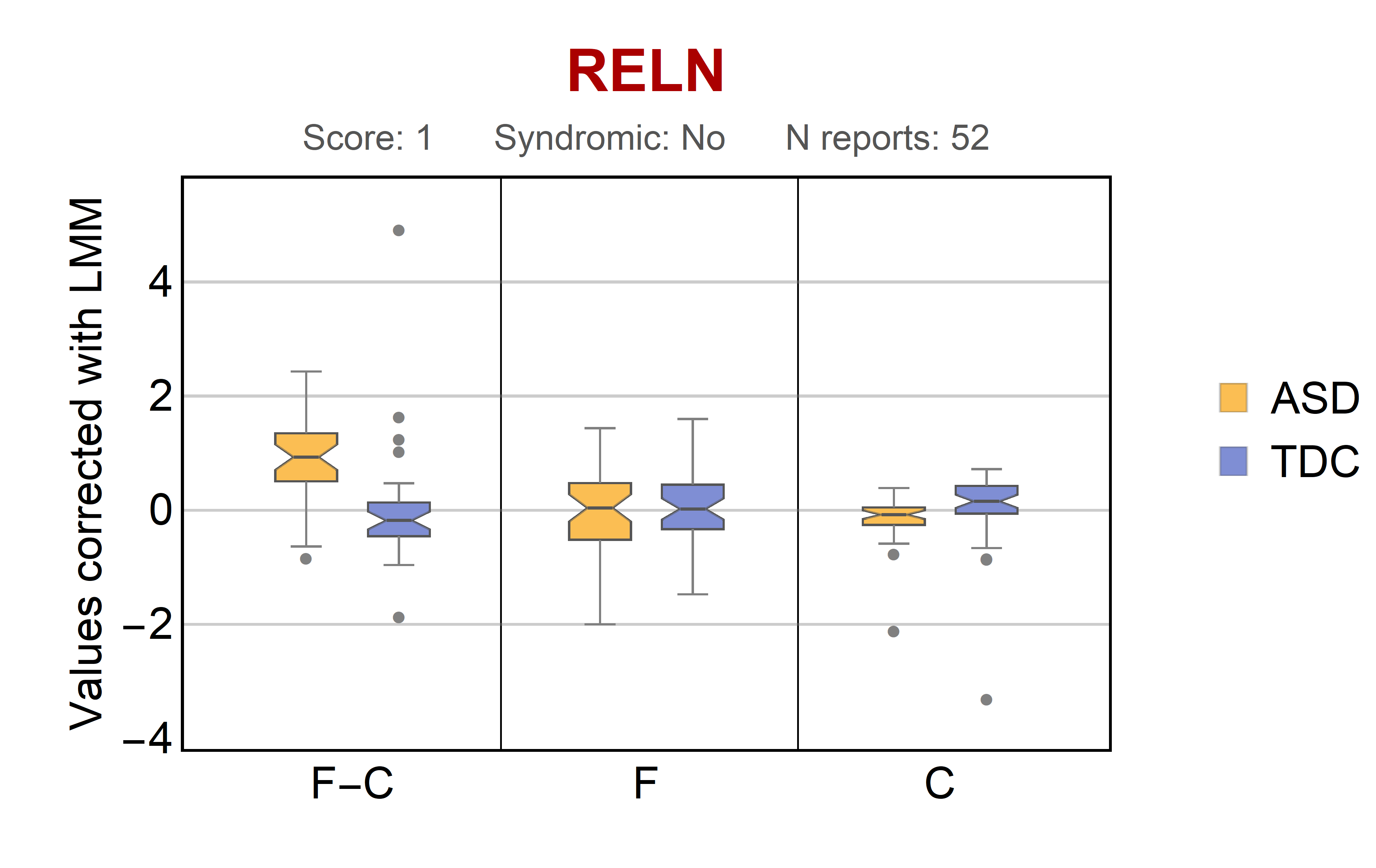}}
&
\subfloat[]{\includegraphics[width =.35\textwidth]{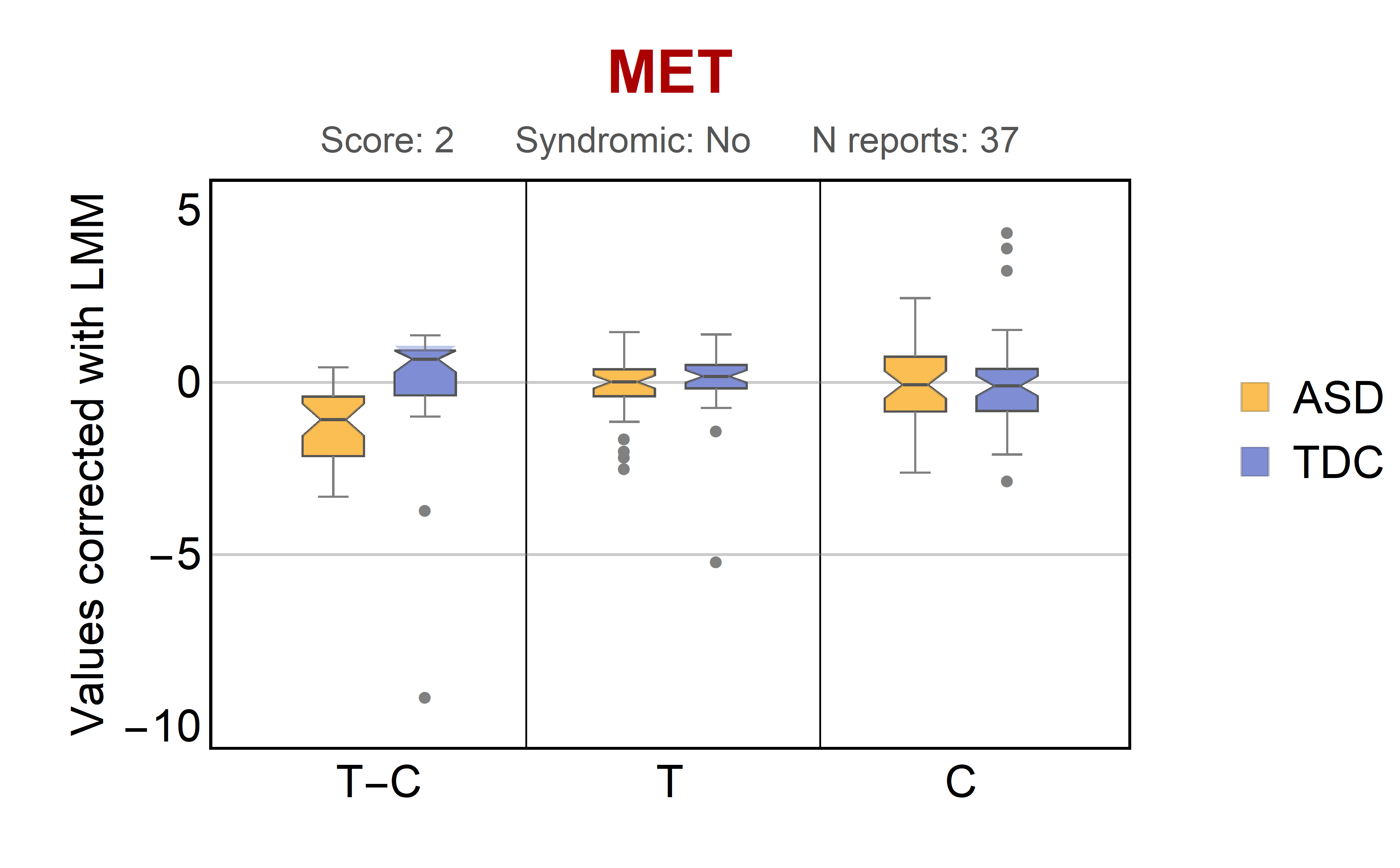}}
\\
\subfloat[]{\includegraphics[trim={0 0 2cm 0},clip,width =.3\textwidth]{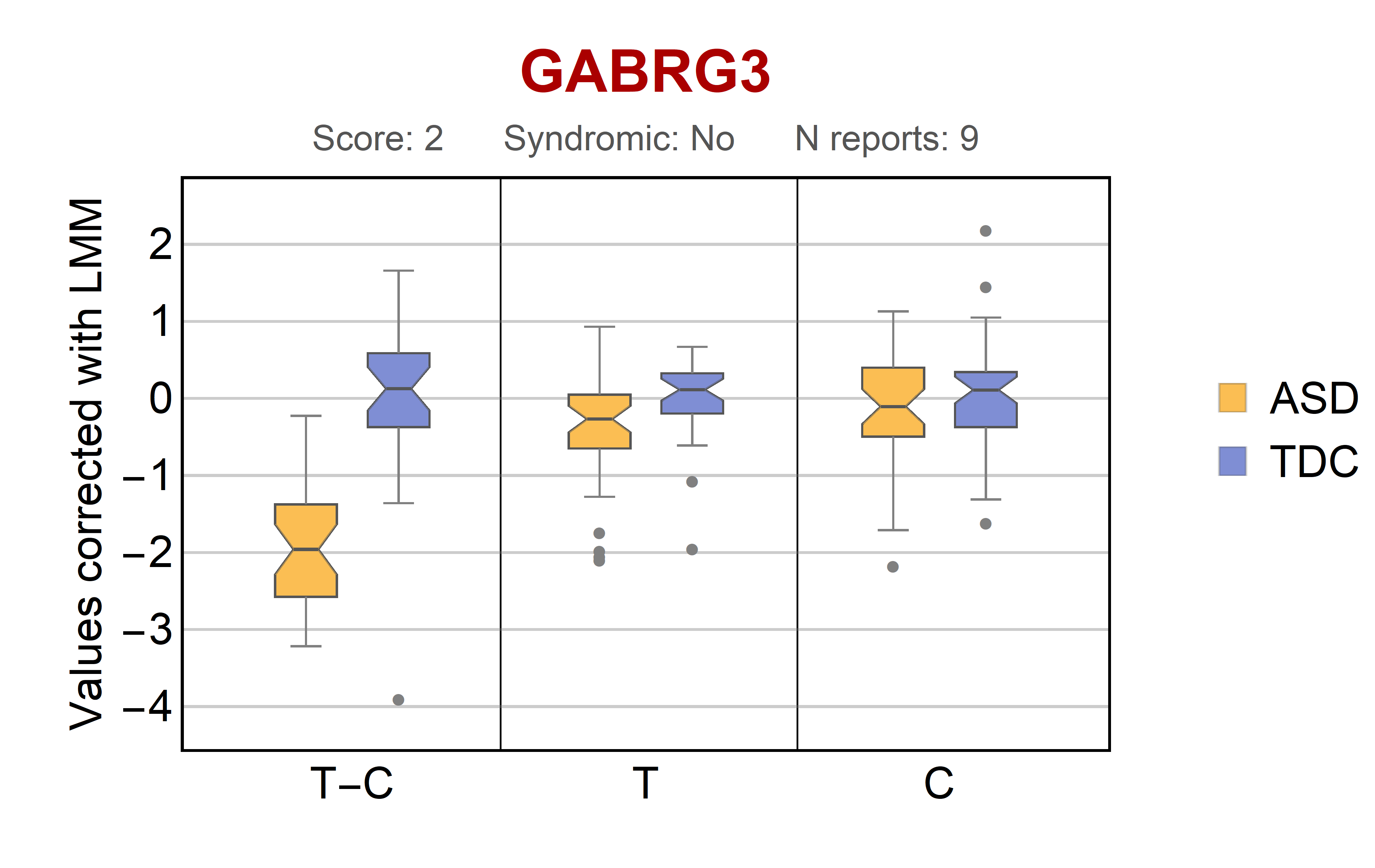}}
&
\subfloat[]{\includegraphics[trim={0 0 2cm 0},clip,width =.3\textwidth]{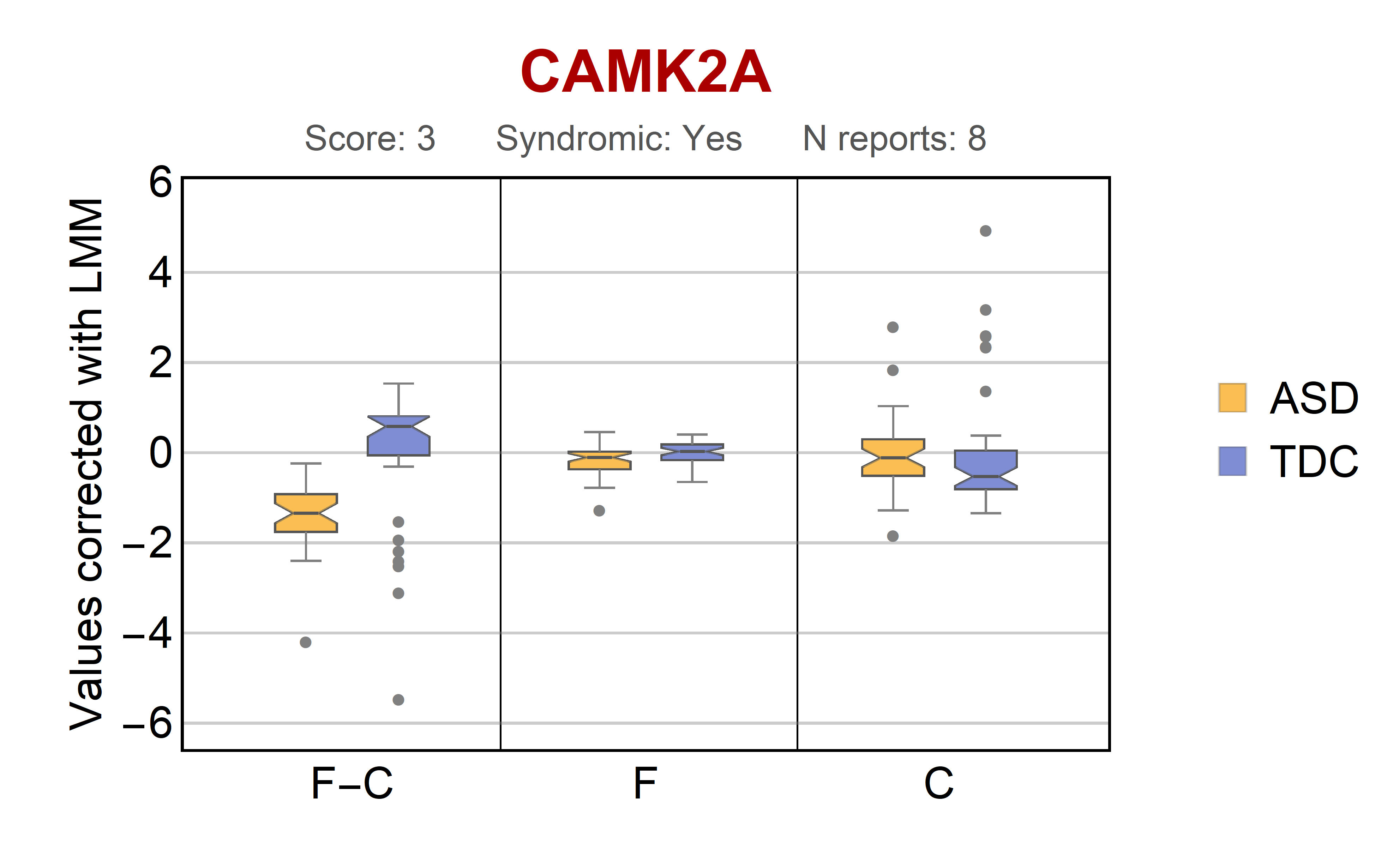}}
&
\subfloat[]{\includegraphics[width
=.35\textwidth]{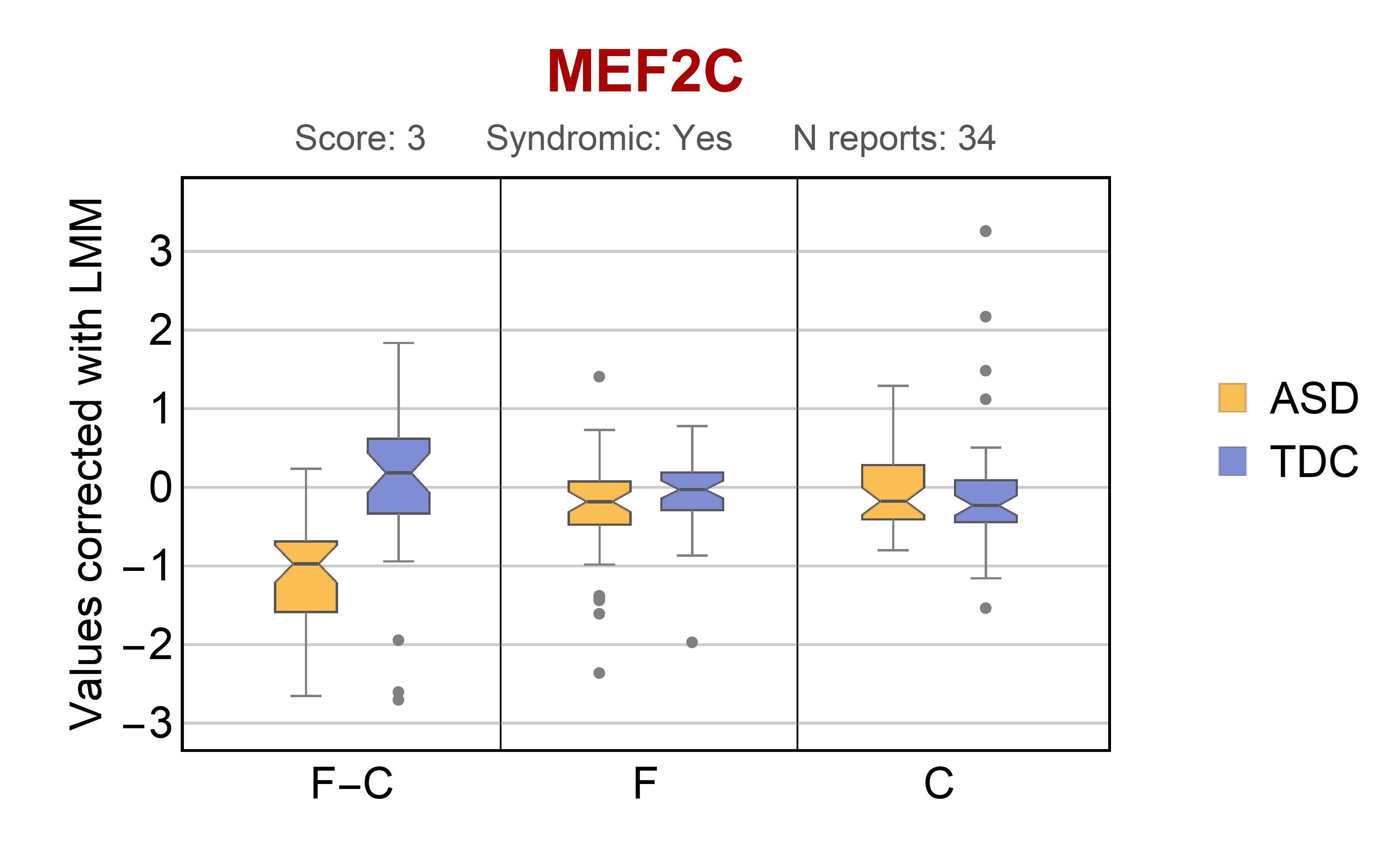}} 

\end{tabular}
\caption{Examples of genes found to be relevant for ASD analyzing $\Delta$GE and not analyzing GE in singular tissues. All the examples reported are included in the SFARI dataset and scored with 1 when the gene is considered linked to ASD with high confidence, with 2 when the gene is considered a strong candidate and 3 when there is a suggestive evidence of its role in the disease. Each box plots is tripartite, in the first part the $\Delta$GE distributions of the gene are reported, while in the remaining ones the GE distributions in the single tissues are illustrated. In every plot, the values are corrected for the coefficients found by the LMM and for this reason all the distributions reported are centered around 0.}
\label{fig:why_diff_impo}
\end{figure}

To further elaborate on the importance of also considering $\Delta$GE, Fig. \ref{fig:why_diff_impo} shows 6 examples of genes known to be relevant for ASD, that are detected by the LMM analysis on DT, but not on ST. In fact, in all 6 plots, the ASD and TDC groups are well-separated (on average) on the left part, that shows the $\Delta$GE, but very similar to each other on the remaining two parts, showing the two tissues taken individually.

\begin{figure} 
\centering
\begin{tabular}{cc}
\subfloat[]{
\includegraphics[trim={0 0 2.5cm 0},clip,width =.43\textwidth]{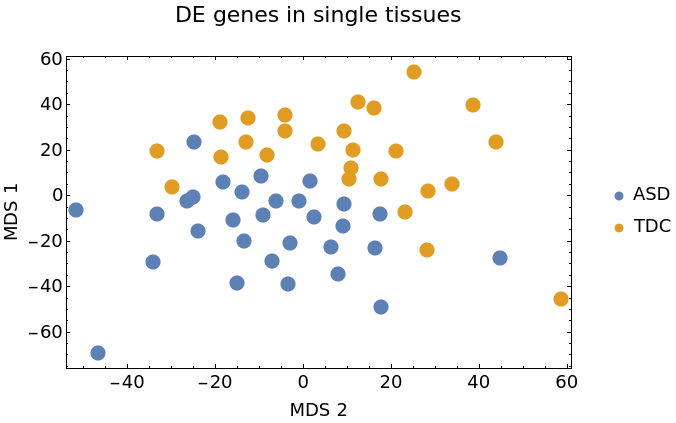}} 
&
\subfloat[]{\includegraphics[width =.48\textwidth]{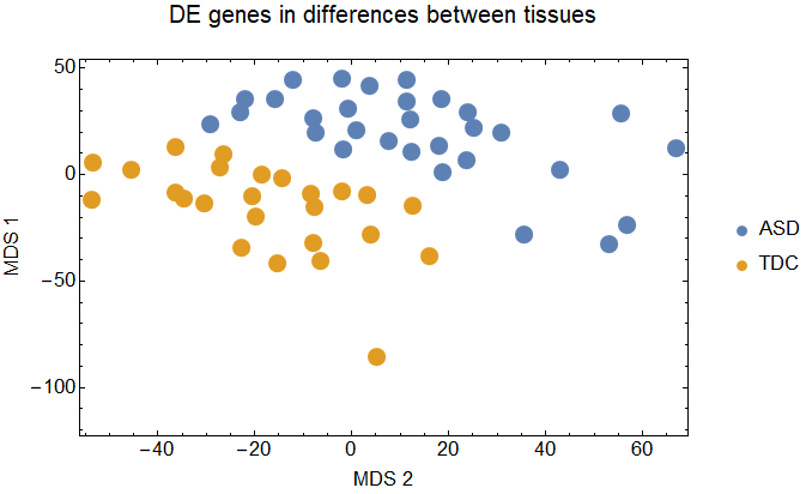}}

\end{tabular}
\caption{2D MDS plots obtained with the standardized euclidean distance of the GE values across C, F and T of all the genes found to be relevant in single tissues (Fig. a) and of the $\Delta$GE values computed between the pairs TC, FC, and TF of all the genes found to be relevant in these pais (Fig. b).}
\label{fig:MDS}
\end{figure}

\begin{figure}
    \centering
    \begin{subfigure}{0.47\textwidth}
        \includegraphics[trim={0 0 2.5cm 0},clip,width =.8\textwidth]{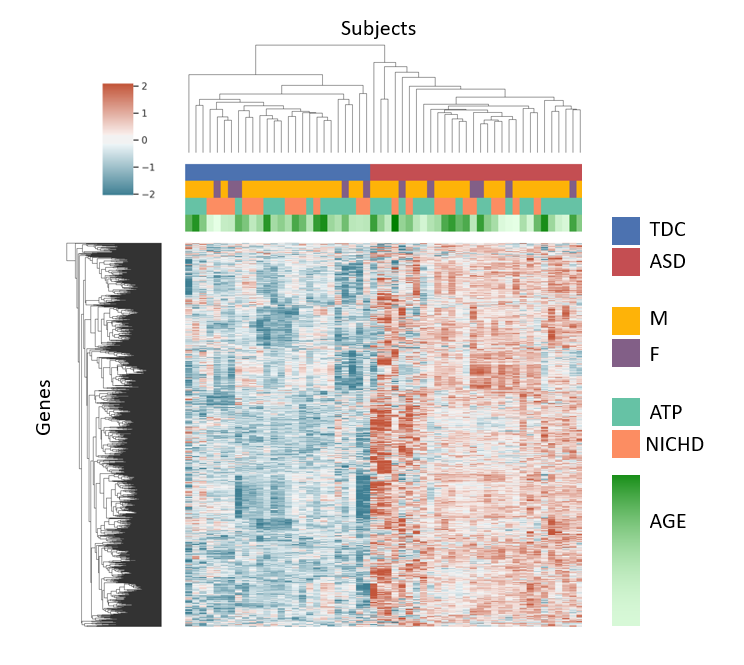}
        \caption{GE}
    \end{subfigure}
    \begin{subfigure}{0.47\textwidth}
       \includegraphics[width =.98\textwidth]{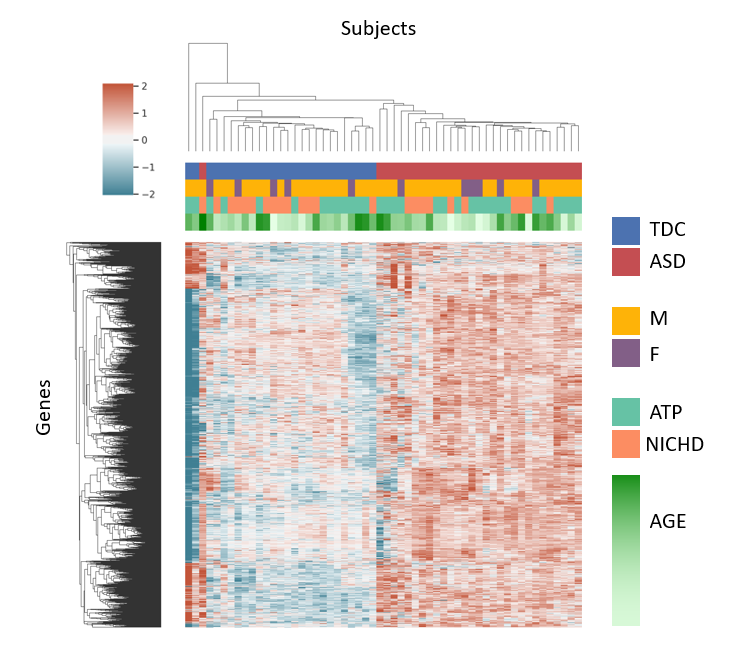}
       \caption{$\Delta$GE}
    \end{subfigure}    
\caption{Hierachically-clustered heatmaps of the GE values across C, F and T of all the genes found to be relevant in single tissues (Fig. a) and of the $\Delta$GE values of all the genes found to be relevant in the differences (Fig. b). }
    \label{fig:heatmaps}
\end{figure}

The gene lists found were then used for a MDS analysis to show their ASD/TDC diagnosis capabilities. Fig. \ref{fig:MDS}a and Fig. \ref{fig:MDS}b show the MDS calculated on all the genes found in at least one of the ST or DT datasets, respectively. In both cases it can be seen that ASDs and TDCs are clearly separated from each other. However, it must be noted that obviously this analysis is prone to over-fitting, having been made on the same dataset used to identify such genes. Also, it could not be used as a diagnosing tool, as these data can be gathered only post-mortem. \\
Similarly, Fig. \ref{fig:heatmaps} shows hierarchically clustered heatmaps of the same two lists of genes. These plots show that ASDs and TDCs (the blue/red band on top) are well separated, and that all the other variables (the other colored bands below the blue/red one) are instead scattered randomly between the various clusters, signaling that they did not act as confounders.

\begin{figure} 
\centering
\begin{tabular}{cc}
\subfloat[]{
\includegraphics[width =.45\textwidth]{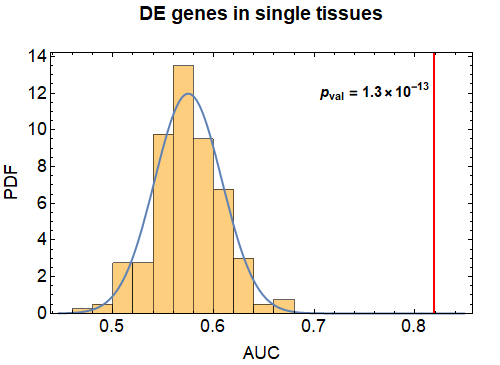}} 
&
\subfloat[]{\includegraphics[width =.45\textwidth]{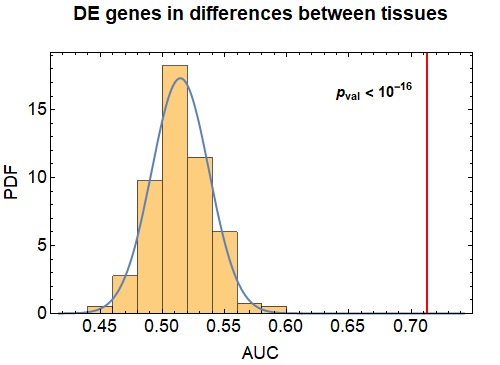}}

\end{tabular}
\caption{Results of the permutation analysis testing whether the sets of genes found with the LMM analysis allow a better discrimination between ASDs and TDCs than any other randomly taken set of genes of the same cardinality. The histograms show the AUCs obtained with the permutation analysis, while the red bars indicate the AUCs obtained with the list of genes obtained from the analysis of the single tissues (Fig. a) and of their differences (Fig. b)}
\label{fig:permutation_our_lists_relev}
\end{figure}

Finally, Fig. \ref{fig:permutation_our_lists_relev} shows a permutation test aimed at verifying whether the sets of genes found with the LMM analysis can be used by a logistic regression classifier to discriminate between ASDs and TDCs better than any other randomly taken set of genes of the same cardinality. The histograms show the permutation test results, while the AUC obtained with the gene set under test is marked with a red line. For both ST and DT gene lists, the results are remarkably better than random.

\FloatBarrier
\subsection{Validation and enrichment analyses}
Tab. \ref{tab:enrichment_with_disease} shows the results of the EA performed to check whether the lists found with the LMM are specific for ASD. As it emerges from the table, most of the lists are significantly enriched for genes reported to be implicated in ASD, according to the inclusion criteria of two different databases: SFARI \cite{abrahams2013sfari} and DISGENET \cite{pinero2020disgenet}. Among the four syndromes taken in consideration, the Angelman syndrome is the one that displays a more significant intersection with the lists of genes found. This result was expected and confirms the validity of the approach adopted. In fact, the percentage of subjects with this syndrome that meet the ASD criteria, is much larger with respect to the ones estimated for the other three syndromes. Furthermore, Angelman syndrome involves the same PWACR region that is duplicated in Dup15q, which is highly represented in this dataset (9 subjects out of 48 ASDs).  The intersections with genes involved in the other three syndromes are less important, although present as expected. Interestingly, the intersections with the genes in the MECP2 pathway seem to be more pronounced with respect to the one with the Rett's syndrome genes.\\
By taking an overall look to the table, the "T-F" list jumps out for its poor intersections with the other columns. As summarized in Tab. \ref{tab:summary_go}, this list is enriched by genes involved in immune response, which have been found consistently upregulated in ASD literature, as described in Sec. \ref{brainASD-s-o-a}, but also in many other mental disorders, as mentioned in Sec. \ref{causes_of_ASD}.
A possible explanation for this unexpected result may be that immune genes are not included in the list of genes associated to a disorder because inflammation is not considered a core feature of it, but just a secondary effect.

\begin{figure} 
    \centering
    \includegraphics[width =.95\textwidth]{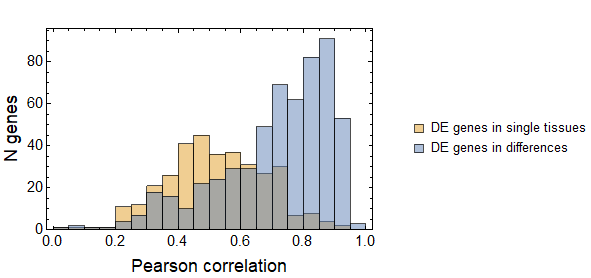}
    \caption{Distribution of the Pearson correlation computed by \cite{hawrylycz2015canonical} as a measure of gene DS, across the lists of genes found to be relevant for ASD analyzing the ST and DT datasets. }
    \label{image:DS}
\end{figure}

Fig. \ref{image:DS} shows the distributions of the DS scores, computed by \cite{hawrylycz2015canonical}, across the genes found to be relevant analyzing the ST datasets (orange distribution) and the DT ones (blue distribution). The two distributions appear clearly separated and the blue one is shifted and skewed towards high DS scores. This shows that the analysis of individual regional differences is more likely to detect of DS genes, that for their stability constitute a more reliable result, as they are less affected by individual differences. 

Finally, Tab. \ref{tab:summary_go} summarizes the results obtained with the GO enrichment analysis.
Interpreting the results of such type of analyses is always very complex and arbitrary, however, it can be noted that the lists obtained with the ST datasets are enriched by genes involved in cellular assembly, organization and gene expression. Instead, the genes found analyzing the differences between cortex and cerebellum are clearly enriched by synaptic functions. The difference between the two cortices also enhances a list of genes associated to immune response.\\
A more comprehensive picture of the GO terms relevant for these lists can be found in Appendix \ref{app:lmm_enrichment}, that includes the top ten most relevant GO terms and a visual summary of the most important biological processes, obtained with REVIGO.

\newcolumntype{Y}{>{\centering\arraybackslash}X}

\definecolor{myR}{rgb}{0.4, 0.8, 0.55}
\definecolor{myY}{rgb}{0.85, 0.95, 0}

\newcommand{\RcellB}{\fcolorbox{black}{myR}{\rule[0.5cm]{1.3cm}{0cm}}}

\newcommand{\YcellB}{\fcolorbox{black}{myY}{\rule[0.5cm]{1.3cm}{0cm}}}

\begin{table}\centering
\setlength{\tabcolsep}{0pt}
\renewcommand{\arraystretch}{1.5}
{
\begin{tabularx}{\textwidth}{|cc|YYYYYYY|}
\hline 

\multicolumn{2}{|c|}{}
&  
ASD
&
ASD
&
Angelman
&
Down
&
Fragile X
&
Rett
&
MECP2
\\
&  
&
\tiny{SFARI}
&
\tiny{DISGENET}
&
\tiny{DISGENET}
&
\tiny{DISGENET}
&
\tiny{DISGENET}
&
\tiny{DISGENET}
&
\tiny{WikiPathways}
\\\hline
\textbf{C} 
& 
\raisebox{-0.25\totalheight}{\includegraphics[width=0.12\textwidth]{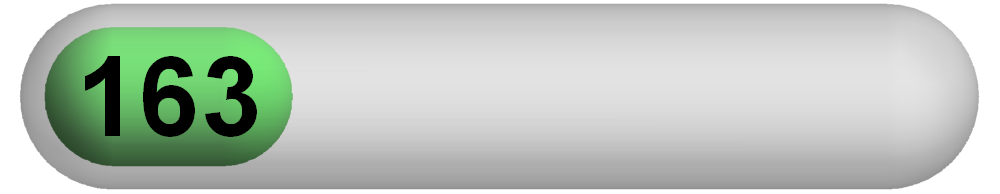}}  
&
\RcellB
&
\RcellB
&
\RcellB
&
&
&
&
\\
\textbf{F} 
&  
\raisebox{-0.25\totalheight}{\includegraphics[width=0.12\textwidth]{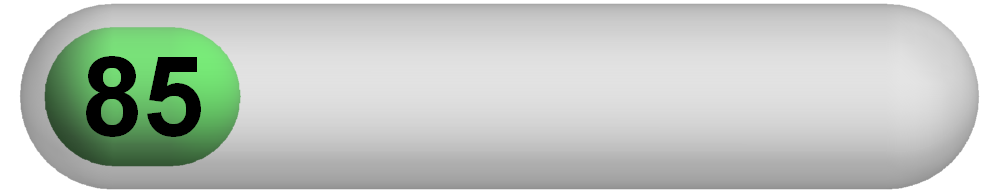}}  
&
\RcellB
&
\RcellB
&
\RcellB
&
&
&
&
\YcellB
\\
\textbf{T} 
& 
\raisebox{-0.25\totalheight}{\includegraphics[width=0.12\textwidth]{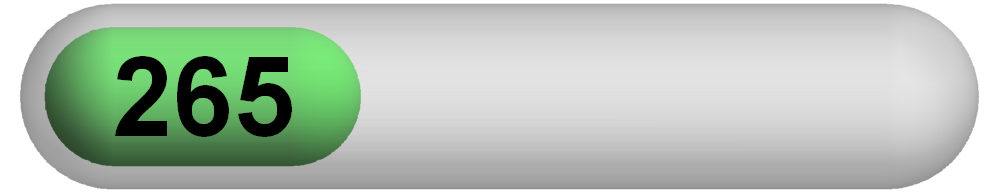}} 
&
\YcellB
&
\RcellB
&
\RcellB
&
\YcellB
&
&
&
\YcellB
\\
\textbf{T-C} 
& 
\raisebox{-0.25\totalheight}{\includegraphics[width=0.12\textwidth]{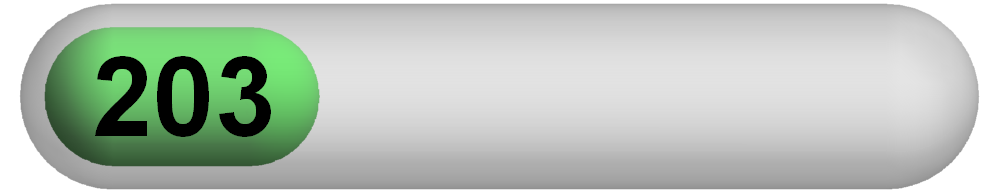}} 
&
\RcellB
&
\YcellB
&
\RcellB
&
&
&
\YcellB
&
\YcellB
\\
\textbf{F-C} 
& 
\raisebox{-0.25\totalheight}{\includegraphics[width=0.12\textwidth]{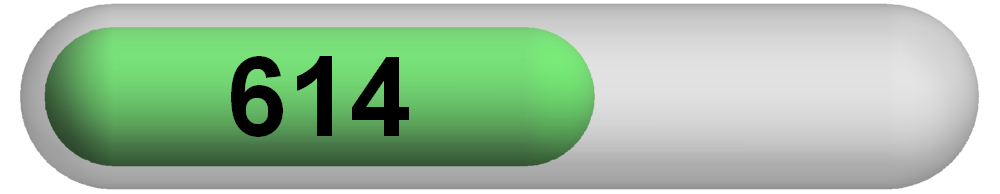}} 
&
\RcellB
&
\RcellB
&
\RcellB
&
&
\YcellB
&
\RcellB
&
\RcellB
\\
\textbf{T-F} 
& 
\raisebox{-0.25\totalheight}{\includegraphics[width=0.12\textwidth]{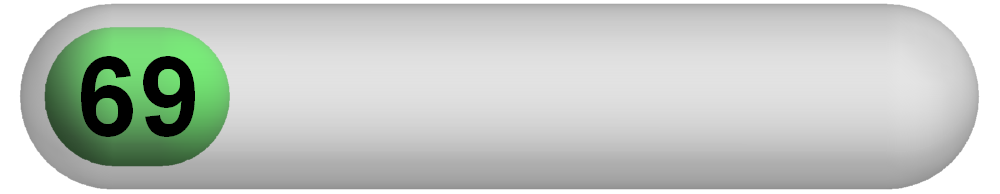}}  
&
&
&
&
\YcellB
&
&
&
\\
\textbf{All singles} 
& 
\raisebox{-0.25\totalheight}{\includegraphics[width=0.12\textwidth]{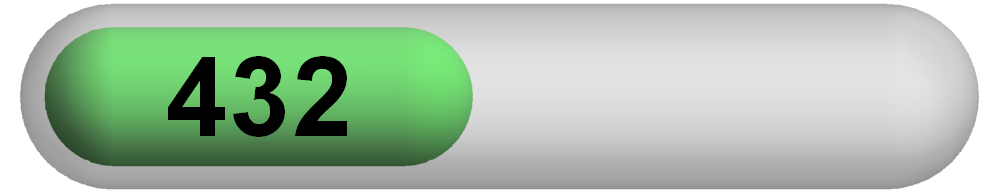}}  
&
\RcellB
&
\RcellB
&
\RcellB
&
&
&
&
\YcellB
\\
\textbf{All diffs.} 
& 
\raisebox{-0.25\totalheight}{\includegraphics[width=0.12\textwidth]{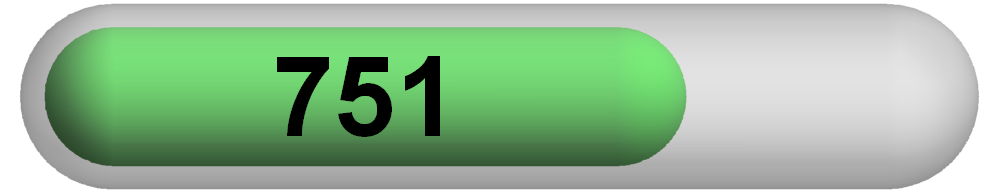}} 
&
\RcellB
&
\RcellB
&
\RcellB
&
&
&
\RcellB
&
\YcellB
\\
\textbf{All} 
& 
\raisebox{-0.25\totalheight}{\includegraphics[width=0.12\textwidth]{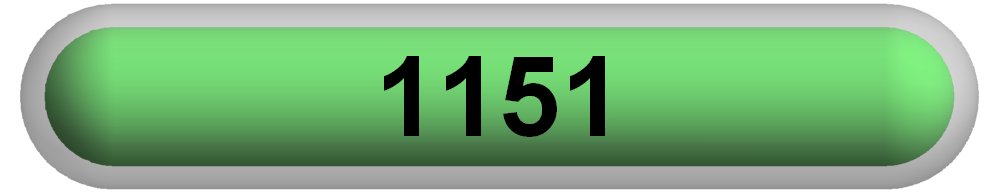}}  
&
\RcellB
&
\RcellB
&
\RcellB
&
&
&
\YcellB
&
\RcellB
\\\hline
\multicolumn{2}{|c|}{\multirow{2}{*}{\textbf{Legend}}} &
\RcellB
&
\multicolumn{6}{l|}{ \hspace*{1pt} : p-value $FDR_{BH}$ corrected $\leq$ 0.05}\\
&&\YcellB
&\multicolumn{6}{l|}{ \hspace*{1pt} : p-value $\leq$ 0.05}\\
\hline
\end{tabularx}}
\caption{Results of the enrichment analysis.
Rectangular markers indicate whether our lists of relevant genes (rows) are enriched by the list of genes known to be involved in the diseases (columns).
The significance has been evaluated with a permutation test and p-values below 0.05  are  depicted  with  yellow  markers, while green markers indicate p-values that remain below 0.05 after $FDR_{BH}$ correction.} 
\label{tab:enrichment_with_disease}
\end{table}
\begin{table}\centering
\setlength{\tabcolsep}{0pt}
\renewcommand{\arraystretch}{1.5}
{
\begin{tabularx}{1.1\textwidth}{|cc|Y|}
\hline 

\multicolumn{2}{|c|}{}
&  
\textbf{Main Functions}
\\\hline
\textbf{C} 
& 
\raisebox{-0.25\totalheight}{\includegraphics[width=0.12\textwidth]{Figures/Fig_Results/Fig_Report/Bar_C.png}}  
&
Chromatin organization\;\;\;\;Cell cycle\;\;\;\;Gene silencing
\\
\textbf{F} 
&  
\raisebox{-0.25\totalheight}{\includegraphics[width=0.12\textwidth]{Figures/Fig_Results/Fig_Report/Bar_F.png}}  
&
Chromatin organization\;\;\;\;Circadian rhythm\;\;\;\;DNA metabolic proc. 
\\
\textbf{T} 
& 
\raisebox{-0.25\totalheight}{\includegraphics[width=0.12\textwidth]{Figures/Fig_Results/Fig_Report/Bar_T.png}} 
&
RNA proc.\;\;\;\;Translation\;\;\;\;Ribonucleoprotein\;\;\;\;Protein targeting\;\;\;\;Viral proc.
\\
\textbf{T-C} 
& 
\raisebox{-0.25\totalheight}{\includegraphics[width=0.12\textwidth]{Figures/Fig_Results/Fig_Report/Bar_CT.png}} 
&
Cell-cell signaling\;\;\;\;Neurotransmitter\;\;\;\;Synaptic transmission
\\
\textbf{F-C} 
& 
\raisebox{-0.25\totalheight}{\includegraphics[width=0.12\textwidth]{Figures/Fig_Results/Fig_Report/Bar_CF.png}} 
&
Neuron\;\;\;\;Synapse\;\;\;Ion transport
\\
\textbf{T-F} 
& 
\raisebox{-0.25\totalheight}{\includegraphics[width=0.12\textwidth]{Figures/Fig_Results/Fig_Report/Bar_FT.png}}  
&
Immune system
\\
\textbf{All singles} 
& 
\raisebox{-0.25\totalheight}{\includegraphics[width=0.12\textwidth]{Figures/Fig_Results/Fig_Report/Bar_All_single.png}}  
&
RNA binding\;\;\;\;RNA processing\;\;\;\;RNA metabolic proc.\;\;\;\;Chromatin organization
\\
\textbf{All diffs.} 
& 
\raisebox{-0.25\totalheight}{\includegraphics[width=0.12\textwidth]{Figures/Fig_Results/Fig_Report/Bar_All_diff.png}} 
&
Neuron\;\;\;\;Synapse\;\;\;Ion transport
\\
\textbf{All} 
& 
\raisebox{-0.25\totalheight}{\includegraphics[width=0.12\textwidth]{Figures/Fig_Results/Fig_Report/Bar_All.png}}  
&
Synapse\;\;\;\;Chromatin organization\;\;\;\;Response to cytokines

\\\hline
\end{tabularx}}
\caption{Summary of the GO analysis applied to each relevant list. A more comprehensive description of the results is reported in Appendix \ref{app:lmm_enrichment}}
\label{tab:summary_go}
\end{table}

\FloatBarrier
\section{Gene co-expression network analysis}
\label{sec:res_genetics_network}

\subsection{Network construction}
\begin{figure} 
\centering
\includegraphics[width =0.7\textwidth]{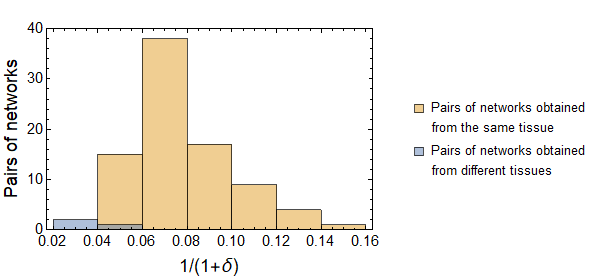}
\caption{Histogram of the distribution of the network similarities as measure with the deltacon index for pairs of networks obtained from the same tissue (yellow) and from different tissues (blue).}\label{fig:deltacon}
\end{figure}

Fig. \ref{fig:deltacon} shows the histogram of the similarity index derived from the delta-con values obtained comparing pairs of networks obtained from the same tissue (orange) and pairs of networks obtained from different tissues (blue). It can be seen that the blue histogram is on the very edge of the orange one, thus showing that networks built with GE values from different tissues are very similar to networks built with a slightly different dataset containing data from only one tissue. From this, it can be concluded that it is safe to merge the gene expression values obtained from different tissues.

\begin{figure} 
\centering
\includegraphics[width =0.7\textwidth]{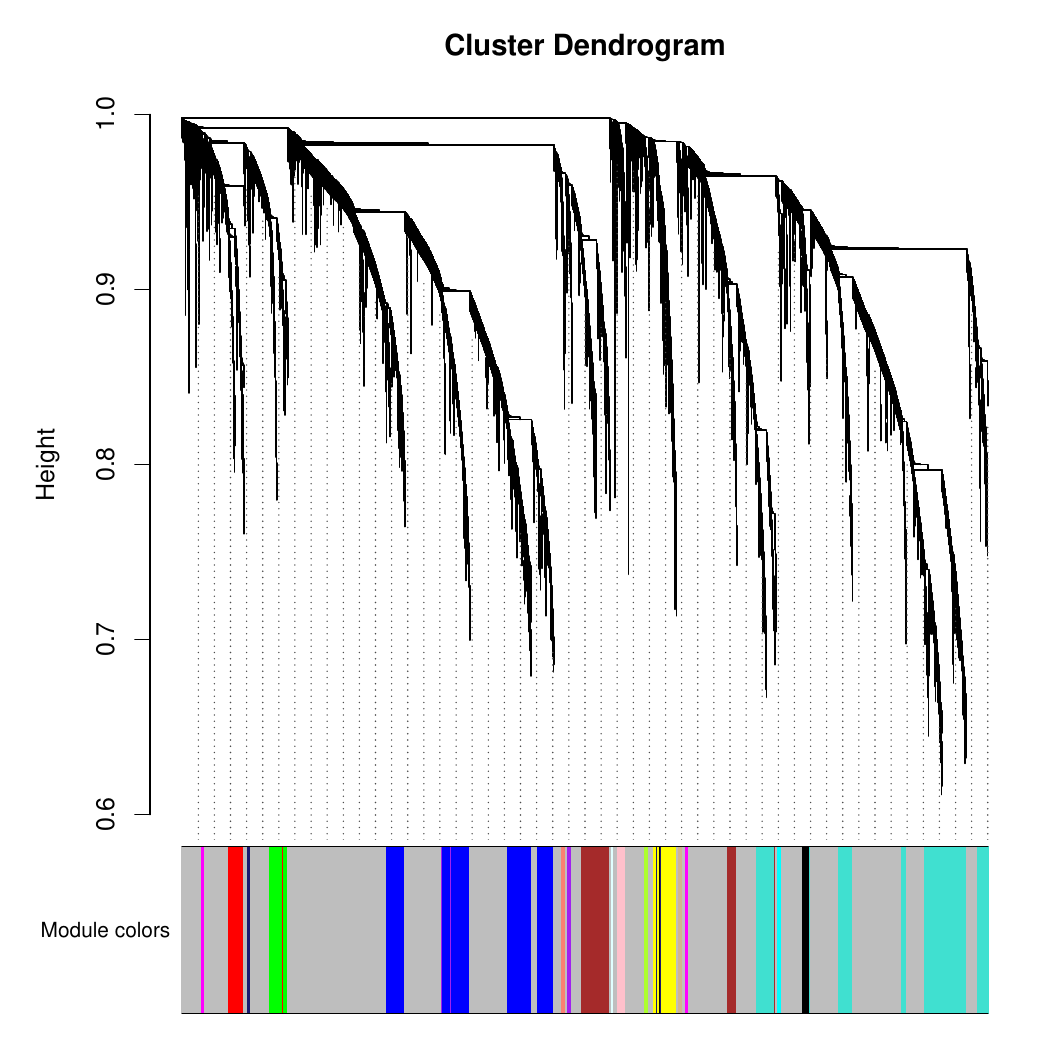}
\caption{Dendrogram representation of the computed network.}
\label{fig:dendrogram_network}
\end{figure}

Fig. \ref{fig:dendrogram_network} contains the dendrogram of the network computed on all GE values from ST datasets merged together. The dendrogram shows also the identified modules, highlighted in different colors at the bottom. The genes colored gray do not belong to any module. The modules and the number of genes they contain are also reported in Tab. \ref{tab:modules_found}.\\

\begin{table}
    \centering
    \begin{tabular}{|c|l|l|}
        \hline
        \textbf{Module name} & \textbf{Module color} & \textbf{Number of genes} \\
        \hline
        M1  & brown       & 797 \\ 
        M2  & blue        & 1786 \\ 
        M3  & tan         & 91 \\ 
        M4  & black       & 183 \\ 
        M5  & magenta     & 109 \\ 
        M6  & turquoise   & 1888 \\ 
        M7  & pink        & 164 \\ 
        M8  & green       & 351 \\ 
        M9  & yellow      & 433 \\ 
        M10 & red         & 328 \\ 
        M11 & greenyellow & 102 \\ 
        M12 & midnight blue & 77 \\
        M13 & cyan        & 83 \\ 
        M14 & light cyan  & 55 \\ 
        M15 & purple      & 108 \\ 
        M16 & salmon      & 89 \\          
        \hline
    \end{tabular}
    \caption{Gene modules found.}
    \label{tab:modules_found}
\end{table}

\subsection{Module selection}
Fig. \ref{fig:modules_enrichment} shows how each module is enriched by each list of relevant genes found in the LMM analysis. Most modules either contain very few genes of specific lists, or a large number of them. The modules whose enrichment is significant (BH-corrected p-value below 0.05, as estimated with a FET), are highlighted with a black border and their p-value is reported.
Five modules are found to be enriched by at least one list of genes: blue, black, turquoise, cyan and salmon.

\begin{figure} 
\centering
\includegraphics[trim={0 0 10cm 10cm}, width =1\textwidth]{Figures/Fig_Results/Fig_Report/allModules_enriched_test.pdf}
\caption{Image showing which modules of the network are enriched by each list of relevant genes.}\label{fig:modules_enrichment}
\end{figure}

\begin{figure} 
\centering
\begin{tabular}{c}
\subfloat[]{
\includegraphics[trim={0 0 7.2cm 0},clip,width =.75\textwidth]{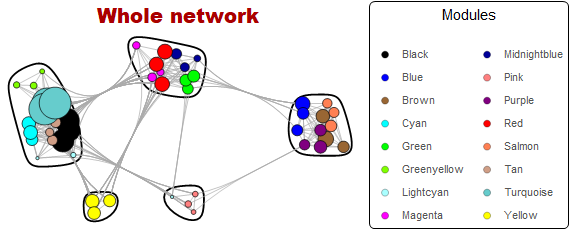}} 
\\
\subfloat{\includegraphics[trim={0 7.7cm 0 0},clip,width =.7\textwidth]{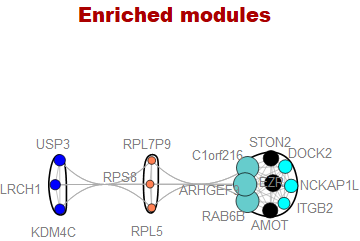}}
\\
\renewcommand{\thesubfigure}{b}
\subfloat[]{\includegraphics[trim={0 0 0 4.5cm},clip,width =.75\textwidth]{Figures/Fig_Results/Fig_Report/3hubs_relevModules.png}}
\\
\subfloat{\includegraphics[trim={13cm 0 0 -0.7cm},clip,width =.3\textwidth]{Figures/Fig_Results/Fig_Report/3hubs_allModules.png}} 
\end{tabular}

\caption{Representation of the whole network (Fig. a), and of the modules enriched for the relevant genes found in the LMM analysis (Fig. b). Each module is represented by its three most connected hubs. Dimensions of the hubs qualitatively represent their connectivity with respect to the whole network. Distances between groups of nodes do not have any particular meaning, while clustered groups indicate that their members are more interconnected.}
\label{fig:networks}
\end{figure}

Fig. \ref{fig:networks}a shows a graphical visualization of the whole network. Each module is represented by its three most connected hubs. Groups of hubs clustered together indicate that they are somewhat interconnected, and thus the represented modules are similar to each other. The dots size indicates their connectivity with respect to the whole network, while distances between groups have no particular meaning. It can be seen that if the network as a whole is observed, 5 clusters can be identified, two of which (the leftmost and the rightmost) contain the modules enriched by the genes of interest. If instead the enriched modules are analyzed individually (Fig. \ref{fig:networks}b), cyan, turquoise and black remain grouped together, while salmon and blue are split.

\subsection{GO enrichment analysis}
\label{sec:res_go_modules}
A summary of the results of the GO enrichment analysis on the 5 modules of interests is reported in Tab.\ref{tab:summary_go_modules}. \\
Most of the GO terms (or some other closely related) included in this table were already present in Tab. \ref{tab:summary_go}. However, the expansion of the lists performed with the network analysis allowed to better detect the importance of glial and translation-related genes. Among the 5 modules listed in the table, the last three are the ones that are more specific in their functions. This is particularly evident looking at Appendix \ref{app:network_enrichment} that for each module shows the top ten most relevant GO terms and a visual representation (obtained with Revigo) of the biological processes, molecular functions and cell components found to be enriched.

\begin{table}\centering
\setlength{\tabcolsep}{0pt}
\renewcommand{\arraystretch}{1.5}
{

\resizebox{\textwidth}{!}{
\begin{tabularx}{1.\textwidth}{|cc|Y|}
\hline 

\multicolumn{2}{|c|}{}
&  
\textbf{Main Functions}
\\\hline
\textbf{Blue} 
& 
\raisebox{-0.25\totalheight}{\includegraphics[width=0.12\textwidth]{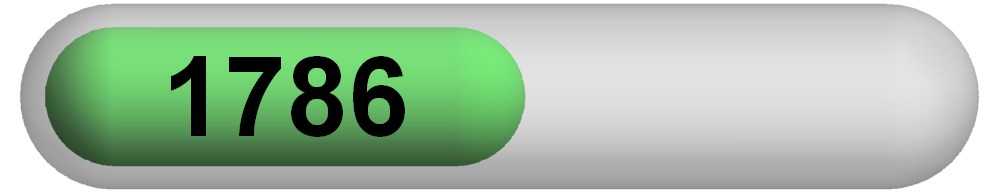}}  
&
Chromosome organization\;\;\;\;Methylation\;\;\;\;RNA processing\;\;\;\;Cell cycle
\\
\textbf{Black} 
&  
\raisebox{-0.25\totalheight}{\includegraphics[width=0.12\textwidth]{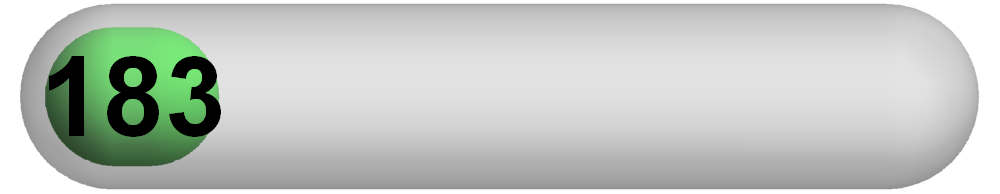}}  
&
Glia\;\; Regul. of apoptosis\;\; Regul. of cell communication\;\; Resp. to lipid
\\
\textbf{Turquoise} 
& 
\raisebox{-0.25\totalheight}{\includegraphics[width=0.12\textwidth]{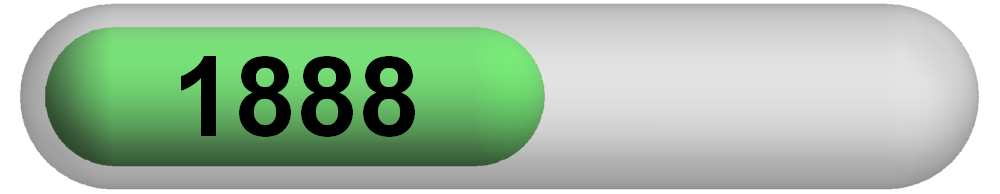}} 
&
Synapse\;\;\;\;Neuron\;\;\;\;Neurotransmitter
\\
\textbf{Cyan} 
& 
\raisebox{-0.25\totalheight}{\includegraphics[width=0.12\textwidth]{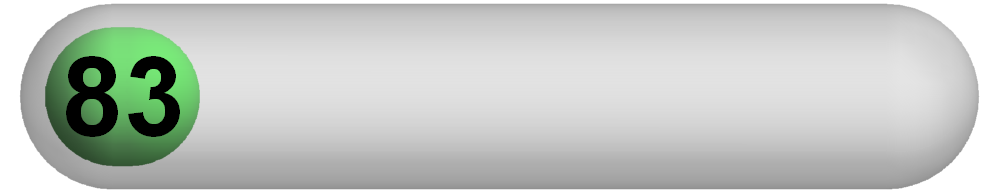}} 
&
Immune system\;\;\;\;Regul. of cell death
\\
\textbf{Salmon} 
& 
\raisebox{-0.25\totalheight}{\includegraphics[width=0.12\textwidth]{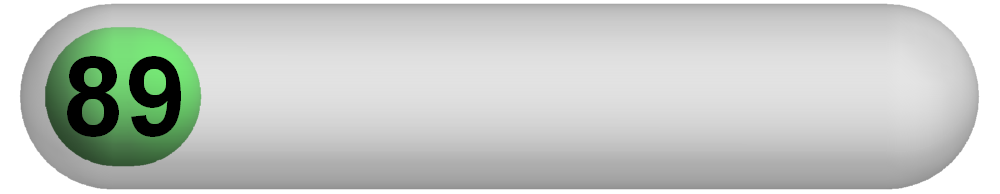}} 
&
Ribosome\;\;\;\;Translation
\\\hline
\end{tabularx}
}
}
\caption{Summary of the GO analysis applied to each relevant module of the network. A more comprehensive description of the results is reported in Appendix \ref{app:network_enrichment}.} 
\label{tab:summary_go_modules}
\end{table}

\chapter{Imaging genetics results}
\label{chap:res_link}

\section{Comparison between neuroimaging and transcriptomics results}
\label{sec:res_comparison_neuroimaging_transcriptomics}
This section contains the results of the imaging genetics analysis, comparing the findings of the neuroimaging and transcriptomics parts.
\subsubsection*{Estimation of the joint effect of the modules of interest on the regions relevant for neuroimaging.}
\begin{figure}[h!]
    \centering
    \includegraphics[width=0.65\textwidth]{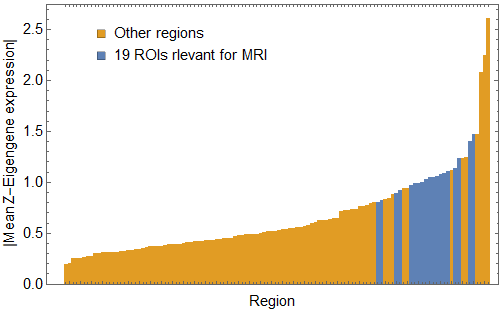}
    \caption{Absolute average Z-score-transformed expression value of the eigengenes of the 5 modules of interest in all regions.}
    \label{fig:mean_expr_our_5_modules}  
\end{figure}

\begin{figure}[h!]
    \centering
    \begin{subfigure}{0.65\textwidth}
        \centering
        \includegraphics[width=\textwidth]{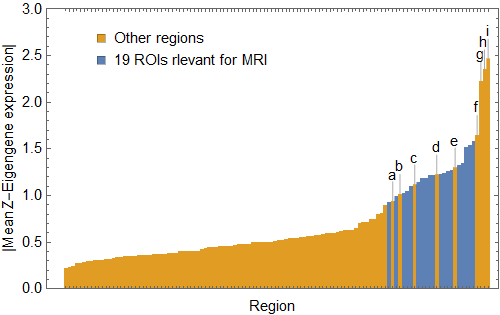}
    \end{subfigure}
    \begin{subfigure}{0.25\textwidth}
        \centering
        \includegraphics[width=\textwidth]{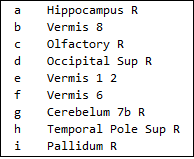}
    \end{subfigure}
    \caption{Absolute average Z-score-transformed expression value of the eigengenes of the modules of interest, except the cyan one, in all regions.}
    \label{fig:mean_expr_our_modules}  
\end{figure}

Fig. \ref{fig:mean_expr_our_5_modules} shows how the absolute Z-scored eigengene expression value averaged across the 5 modules of interest is distributed across the AAL brain regions, which are depicted along the x axis. The regions important for the neuroimaging-based classification are colored in blue, while the others are in orange. An enrichment of the 19 blue regions towards the right-most part of the plot signals that the 5 modules considered altogether affect more these regions than the other ones.\\
However, during the analysis it was observed that most of the genes belonging to the cyan module showed expression levels below background in most of the regions. This phenomenon may be easily explained by the fact that the cyan module has mainly a role in immune response (as shown in Sec. \ref{sec:res_go_modules}) and the AHBA donors are neurotypical and thus should not present any sign of neuroinflammation. Given the low reliability of the cyan module data, the analysis has been repeated considering only the other four modules of interest, whose results are shown in Fig. \ref{fig:mean_expr_our_modules}.
Fig.s \ref{fig:mean_expr_our_5_modules} and \ref{fig:mean_expr_our_modules} show very similar results. Nonetheless, excluding the data from the cyan module, 3 less regions with an average absolute Z-scored eigengene value above or compatible with the ones computed for the 19 regions of interest are observed.

\subsubsection*{Comparison of the effect of each module on the brain regions relevant for classification.}
\begin{figure}[h!]
    \centering
    \begin{subfigure}{0.45\textwidth}
        \centering
        \includegraphics[width=\textwidth]{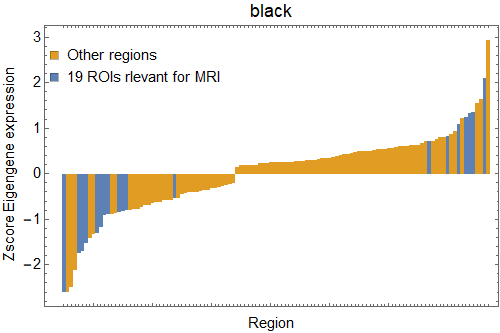}
        \caption{Black module, relevant for ASD according to the transcriptomic analysis.}
        \label{fig:link_1}
    \end{subfigure}
    \hfill
    \begin{subfigure}{0.45\textwidth}
        \centering
        \includegraphics[width=\textwidth]{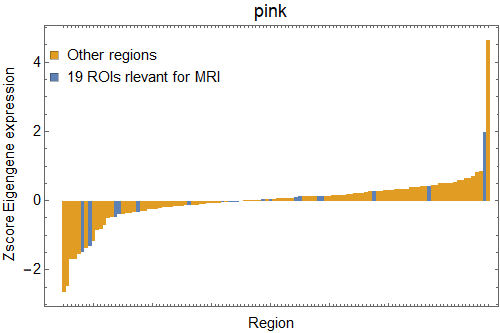}
        \caption{Pink module, not relevant for ASD according to the transcriptomic analysis.}
        \label{fig:link_2}
    \end{subfigure}
    \caption{Examples of the plots used to identify gene modules more expressed (in absolute terms) in the regions found relevant by neuroimaging.}
    \label{fig:link}    
\end{figure}

Fig. \ref{fig:link} shows an example of the plots made to assess whether the expression levels of a module are extreme in the regions found with neuroimaging. The sorted, Z-score-transformed value of the module eigengene expression in every region is reported. The regions of interest are colored in blue, while the others in orange. A clustering of the blue bars towards the leftmost or rightmost part of the plot indicates that the regions of interest are under- or over-expressed in the module, respectively.\\
Already at visual inspection, it is evident that Fig. \ref{fig:link_1}, representing a module enriched for the genes found in the LMM analysis, shows this clustering, while Fig. \ref{fig:link_2}, representing a non-enriched module, does not.

More rigorously, this clustering can be estimated with the permutation test described in Sec. \ref{sec:mm_enrichment_analysis}, the results of which are reported in Tab. \ref{tab:link_pval}. It can be seen from the table that 4 out of the 5 modules selected with the LMM analysis take a significantly higher value of expression (in absolute terms) in the regions highlighted by neuroimaging.\\
The only module identified with the transcriptomic analysis that does not result significantly expressed in the 19 regions relevant for classification is the cyan one. This fact confirms the previously made considerations on the reliability of the cyan module data.

Appendix \ref{app:imaging_genetics} contains plots like the ones shown in Fig. \ref{fig:link} for all the gene modules.

\begin{table}
    \centering
    \begin{tabular}{|c|c|c|}
    \hline
    \cellcolor{gray!30}{\textbf{Module}} & \cellcolor{gray!30}{\textbf{Mean Eigengene}} & \cellcolor{gray!30}{\textbf{BH-corrected p-value}}    \\               
    \hline
    {\color{red} \textbf{Blue}} & 1.27 &  \cellcolor{yellow!30}{$1 \times 10^{-4}$}              \\   
    {\color{red} \textbf{Black}} & 1.24 & \cellcolor{yellow!30}{$1 \times 10^{-4}$}           \\   
    {\color{red} \textbf{Turquoise}} & 1.09 & \cellcolor{yellow!30}{$1\times 10^{-4}$}          \\       
    {\color{red} \textbf{Salmon}} & 1.32 &  \cellcolor{yellow!30}{$6.2 \times 10^{-3}$ }          \\
    \hline
    Magenta & 0.50 & 0.27           \\       
    Pink & 0.40 & 0.55              \\   
    Lightcyan & 0.39 & 0.75         \\           
    {\color{red}  \textbf{Cyan}} & 0.38 & 0.98              \\   
    Purple & 0.37 & 0.84            \\       
    Green & 0.37 & 0.69             \\       
    Brown & 0.36 & 0.65             \\       
    Red & 0.35 & 0.82               \\   
    Greenyellow & 0.35 & 0.81       \\           
    Yellow & 0.33 & 0.70          \\       
    Grey & 0.29 & 0.90               \\   
    Midnightblue & 0.28 & 0.98      \\           
    Tan & 0.28 & 0.97\\
    \hline
    \end{tabular}
    \caption{Results of the permutation test analysis for identifying gene modules with an average absolute higher eigengene expression in the regions highlighted by neuroimaging. The line marks the BH-corrected p-value threshold of 0.05. Modules enriched for the gene lists found in the LMM analysis are highlighted in red. Note that $1\times 10 ^{-4}$ is the minimum measurable p-value, because of how the permutation test has been designed.}
    \label{tab:link_pval}
\end{table}

\part{Discussion and conclusions}
\chapter{Discussion}
\section{Introduction}
In this chapter, the previously illustrated results are discussed following the tri-partition adopted in this thesis: neuroimaging, genetics and imaging genetics. Finally, a section dedicated to the discussion of the overall most important results is provided.
\section{Discussion on neuroimaging results}
\subsection*{Confounder effects in neuroimaging and their corrections}
The confounder analysis, described in Sec. \ref{sec:res_neuroimaging_confounder}, shows that many confounders exist in MRI data analysis based on DL algorithms exists. Besides phenotypic characteristics and site, whose effects have been somewhat already highlighted by various works \cite{ferrari2020dealing,zhao2020training,dinsdale2021deep,patriat2013effect}, this analysis corroborate the strong effect of eye status, that is often overlooked although already observed \cite{spera2019evaluation}. This variable, in fact, shows a CI value even higher than sex (see Tab. \ref{tab:ci_results}). This implies that external stimuli can significantly affect the information in the rsfMRI and are strong confounders, whose traces may be lost during the acquisition process. For instance, whether the ASD subjects, which are notoriously scared by MRI machine and sounds, are reassured with speech, touch or by letting them hold toys may constitute a strong source of bias for these kind of analyses.\\
Thus, for future dataset collection initiatives, keeping trace of all the possible confounders is fundamental, because it is always possible to retrospectively correct their effect during data analysis using an adversarial learning framework, as this thesis shows. In fact, the proposed correction proved to strongly mitigate the confounding effects of both continuous and categorical variables, when considered individually (as shown in Tab \ref{tab:ci_results}), but also in their combination (as shown in Tab \ref{tab:final_confounder_correction_analysis}). \\
Concerning this last table, that reports the correlations between the confounders value and their predictions, note that the presence of correlations around 0.1-0.2 when $\lambda \neq 0$ should not be intended as the incapability of the approach to reach lower values. In fact, by increasing the value of $\lambda$ it is possible to completely eliminate any correlation. However, the objective of this work was not to completely remove any confounder dependency from the input of the ASD-vs-TDC predictor $\mathbb{P}$, but to ensure that it does not retain significant information about the confounders to mislead $\mathbb{P}$. In the opinion of the author, the correlations obtained were sufficiently low to satisfy this requirement.
\subsection*{Classifier and its performance}
As already stated in Sec. \ref{sec:bkg_ml_tools}, the high number of parameters of DNNs increases the probability to overfit the input data and this motivates many researchers to use more conservative approaches. Despite there is not a rule for the optimal ratio $r$ between the number of a DNN parameters and the dimensionality of its input data, Tab. \ref{tab:r_value} shows that the $r$ value of the DNN developed in this work is much smaller than the one of other popular and well regarded algorithms.
Again, this comparison is very naive and many other aspects should be taken into consideration. For instance, AlexNet has been developed for a multi-class problem with 1000 classes, but the classes are easily distinguished by humans, while ASDs and TDCs are not. Thus, the table simply shows that the dimension of the DNN implemented is reasonable. \\
A stronger proof that overfitting has not occurred comes from the comparison of the performance obtained on the training, validation, test and the site-out test sets (Tab.s \ref{tab:results_asd_tdc_tv} and \ref{tab:results_asd_tdc_test}), which all show compatible values.

\begin{table}
    \centering
    \begin{tabular}{|l|r|r|r|}
        \hline
        \textbf{Network} & \textbf{Parameters (millions)} & \textbf{Input data (millions)} & \textbf{Ratio} \\
        \hline
        ViT-G14 \cite{zhai2021scaling} & 1800 & 0.45 & 4000.00\\
        AlexNet \cite{krizhevsky2012imagenet} & 60 & 0.15 & 400.00\\
        This work & 100 & 30.00 & 3.33\\ 
        \hline
    \end{tabular}
    \caption{Summary of the number of parameters and the dimensionality of the input data of some well-known neural networks, as compared to those of the network presented in this thesis.}
    \label{tab:r_value}
\end{table}

Regarding the quality of the data, in this work, the typical filtering step used to discard the MRIs with a high degree of noise and with visible artifacts \cite{backhausen2016quality,yassin2020machine} was skipped for two reasons:
\begin{itemize}
    \item to maximize the number of data composing the training, validation and test sets. Note that the large size of the test and site-out test sets, which are often small or even missing in this kind of studies, is fundamental to prove the reliability of the performance obtained;
    \item because MRIs with artifacts and local noise in brain regions not relevant for classification may nonetheless contribute to the learning process and be well classified.
\end{itemize}
This choice was however risky, because it complicates the search for the classification pattern. As it can be observed in Fig. \ref{fig:quality}, that shows the different image quality of the mis- and well-classified data, the distribution of the latter ones also reaches very high values. This suggests that, as initially supposed, also data with low quality have positively contributed to the development of the classifier. However, the average higher degree of noise affecting mis-diagnosed data, indicates that both the learning process and the performance assessment could benefit from data cleaning.

In any case, the performance reached (AUC = 0.89) is above most of the results reported in literature, especially considering that the ASDs analyzed have not been selected as members of a well characterized subgroup of the disorder and should, thus, be representative of the full spectrum. It is not clear whether this good result owes more to the adoption of the 2S-NN, for improving the identification of spatio-temporal features, or to the BR-NET, for removing the effect of confounders. A future work comparing the results obtainable using these approaches separately may shed light on which one contributes the most to the outcome. However, it is also possible that only their joint implementation allows to reach similar performance. Surely, Tab. \ref{tab:final_confounder_correction_analysis} and the plots in Appendix \ref{app:resilience} show that without adversarial learning, the features $\vec{F}$, that represent an intermediate representation of the data, would clearly depend on many confounders. Thus, correcting for them seems crucial to identify a reliable pattern. Furthermore, considering that:
\begin{itemize}
    \item the correct identification of meaningful features in rsfMRI data is possible only by considering both  temporal and spatial characteristics of the data;
    \item the spatial pattern extracted with smoothgrad is very clear, meaning that the voxels of interest are only located inside the brain and distributed in known anatomical regions displaying a definite separation between relevant and non-relevant ones;
\end{itemize}
it is possible to speculate that the 2S-NN approach played an important role in the learning process, enabling the high performance registered, despite the noise in the data.

\subsection*{Explainability and pattern extracted}
Although Smoothgrad is not the most sophisticated and updated XAI algorithm, for this rsfMRI application, it provided an easily interpretable explanation of the spatial pattern that drives the classification for each image, as initially supposed in Sec. \ref{subsec:bkg_smoothgrad}.\\
Regarding the regions found to be relevant for classification, the following observation can be made:
\begin{itemize}
    \item \textit{No brain region has been found relevant for all the subjects.}\\
    As it can be seen for Fig. \ref{fig:smoothgrad_ranking}, the most recurrent brain region is relevant for only 920 subjects out of 1423 (around 60\% of the dataset).
    Although the brain regions that explain individual classification may not perfectly match the ones with a disrupted functioning in ASD (for instance, regions affected by artifacts or noise may not be discriminative for the algorithm even if they are in biological terms), this heterogeneity of results could reflect the heterogeneity of the disorder. 
    \vspace{2mm}
    \item \textit{The brain regions found to be relevant for classification are 19 out of 116 of the AAL (Tab \ref{tab:relevant_regions}).}\\
    This number is much higher than the number of relevant regions usually reported by other studies, that often cite 3-5 regions, and this discrepancy may have contributed to the difficulty in defining a common biomarker. In fact, if most of the studies report a list of a few regions, their intersection is likely to appear insignificant. It should be also noted that the total volume occupied by these regions constitute about 20\% of the total brain volume. As it can be noted from Fig. \ref{fig:immagine_19_regioni}, the ASD-vs-TDC discriminative pattern is widespread involving both the hemispheres, the cortex and and subcortical regions. The high number of brain regions found and their different roles may explain the variety of symptoms and comorbidities associated to the disorder.   
    \vspace{2mm}
    \item \textit{In most of the cases, both the left and the right side of a brain region are important for classification.}\\
    This observation constitute a sort of confirmation of the validity of the result obtained. In fact, many human functions are controlled by pairs of bilateral regions, thus it would be strange to find that only one is consistently disrupted in different subjects with the same disorder.  
    \vspace{2mm}
    \item \textit{Different patterns may be associated to different phenotypic characteristics of the subjects.}\\
    As already mentioned in Sec. \ref{sec:res_neuroimaging_explainability} about the Fig. \ref{fig:glass_brain_smooth_subjects}, a possible future work may focus on the identification of brain regions associated to different phenotypic and behavioural descriptors of the ASD subjects, by multivariate regression analysis. As already mentioned, the individual patterns extracted not necessarily match perfectly the brain regions effectively involved in the disorder, however, this analysis can potentially allow to explain the heterogeneity of the disorder in neurological terms, and is thus worth to be pursued.
\end{itemize}
In this work only the spatial dimension of the pattern have been extrapolated and analyzed. Even though the brain regions extracted may not be relevant per se, but being characterized by a temporal pattern that may occur also in other regions and that represents the core signature of ASD (for instance it could be the case of a slowdown or acceleration of oxygenated blood flux), this hypothesis seems to be very remote. In fact, the validation of the regions found obtained through the transcriptome part of the study, strongly suggests that the biomarker of the disorder can be effectively summarized by this list of brain regions. Nonetheless, the analysis of the temporal pattern represents another line of future work that is worth to be pursued, to better unveil the mechanisms of ASD.
\section{Discussion on transcriptomics results}
\subsection*{Validity of the lists of genes found with LMM}
The existence of a transcriptomic signature of ASD and the validity of the gene lists found with the LMM analysis are supported by various results of this work, listed below.
\begin{itemize}
    \item The LMM p-value distributions are not uniform and are peaked in proximity of zero (Fig. \ref{fig:LMM_hist_pval_all} and Fig. \ref{fig:LMM_hist_pval_15q}).
    \item The MDS and the hierarchical clustering confirm that the gene lists found allow to separate ASDs and TDCs in an unsupervised way (Fig. \ref{fig:MDS} and Fig. \ref{fig:heatmaps}).
    \item In addition to the previous point, the logistic regression analyses show, with a supervised approach, that the gene lists found are much more informative on the disorder than random lists of genes of the same length (Fig. \ref{fig:permutation_our_lists_relev}). 
    \item As Tab. \ref{tab:enrichment_with_disease} shows, the gene lists are more specific to ASD as compared to other disorders with similar symptomatology.
    \item The GO terms associated to the gene lists were associated to ASD by previous works.
    \item The gene lists found have a significant intersection with only 5 co-expression network modules out of 16 (Fig. \ref{fig:modules_enrichment}), which are related to biological and molecular functions already linked to ASD (Tab. \ref{tab:summary_go_modules}). 
    The fact that the genes found are involved in a few shared pathways and the low number intersections with other network modules strengthens the validity of the results. In fact, it is unlikely that such a coherent pattern would have emerged by randomly selecting genes because of noise or flawed experimental design.
\end{itemize}
\subsection*{Study of the regional GE differences}
One intermediate objective of this study was to understand if regional GE differences constitute an important transcriptomic feature of ASD. The previous paragraph already proved the relevance of the gene lists found in this work, both those obtained analyzing single regions and their differences. The following observations further  corroborate the idea that regional transcriptomic differences are worth to be considered in the characterization of ASD: 
    \begin{enumerate}
    \item The lists of genes obtained analyzing either single tissues or their differences are much more intersected with each other than between them (Fig. \ref{fig:intersections_relevant_lists}). This suggests that gene dysregulation in single tissues and dysregulation of regional differences are different phenomena that involve different genes.
    \item ASD-associated genes show a remarkable dysregulation of regional differences but are not dysregulated in single regions (Fig. \ref{fig:why_diff_impo}).
    \item The GO terms associated to the lists obtained with regional differences that are related to synapses and the immune system are the most commonly found in ASD literature. Our stringent selection criteria did not allow to detect these pathways analyzing separately C, F and T, even though they have been already identified in the same regions by other works \cite{parikshak2016genome,voineagu2011transcriptomic}. This suggests that the disruption of these pathways can be more easily identified by analyzing regional transcriptomic differences.
    \item Fig. \ref{image:DS} shows that the differential stability of the genes found analyzing DT datasets is significantly higher than the one of the genes detected with the ST. It is not clear whether the study of the DT dataset allows to spot genes with a high DS, i.e. whose regional expression differences are well preserved across subjects, because it analyzes regional differences or because with this approach subject-related noise is reduced. Anyway, finding a dysregulation of regional pattern of genes with a high differential stability is a meaningful result in itself and it may characterize the pathophysiology of ASD.
\end{enumerate}
All these analyses prove that the study of regional differences enriches the results obtainable with a standard DE analysis on single tissues. It should be also noted that under the same selection criteria the genes detected with the DT datasets are almost twice than those found with ST (i.e. 751 vs 432). This means that regional differences dysregulation is not a minor phenomenon. However, whether it plays an active role in causing ASD or it constitute just an observable effect is still unclear. Three scenarios are possible:
\begin{itemize}
    \item small GE variations that fall in the normal distribution for each single tissue create between-region dysregulations and this abnormal spatial pattern actively contributes to the onset of the disorder; 
    \item small GE dysregulations in single regions that contribute by themselves to the disorder are not detectable due to individual variability but can be appreciated by studying regional differences;
    \item GE regional differences not detectable in single tissues do not contribute to the disorder, but appear to be altered as an effect of the interaction with other dysregulated genes in single tissues. In fact, as it can be observed from Fig. \ref{fig:modules_enrichment} genes found with DT and ST datasets mainly intersect the same co-expression network modules.
\end{itemize}
Obviously, the regional differences dysregulations can also occur for a combination of these three causes, explaining also the high number of genes that display such behaviour. 

\subsection*{Co-expression network}
Finally, some interesting observations on the co-expression network developed are worth to be mentioned. From the representation of the network illustrated in Fig. \ref{fig:networks}, it can be noted that the cyan, black and turquoise modules are clustered together. While the correlation between the last two was somewhat expected given their associated GO terms (as Tab. \ref{tab:summary_go_modules} shows the black module is involved in glial functions and regulation of cell communication, and the turquoise one in synapse and neuron related functions), their proximity with the cyan module, that is exclusively involved in immune system response, is more interesting. In fact, there is a long-standing debate as to what comes first, whether neuroinflammation or synaptic functionality disruption, as discussed in the background part of this thesis. Although the connection between the cyan and turquoise modules does not answer this question, it confirms that the two pathways are highly correlated. 

Another interesting observation comes from the hubs of the 5 modules of interest illustrated in Fig. \ref{fig:modules_enriched}. Most of these were also included in the lists of genes found with the LMM, corroborating the idea that the pathways identified with the GO enrichment analysis applied to the network modules are effectively the ones involved in the original lists of genes. 

As a final remark, it should be noted that the pathways identified analyzing the network modules are in line with ASD literature and expand the ones found with the LMM. This confirms the utility of the network analysis as a tool to better explore group differences when the dataset has a limited size, as originally discussed in Sec. \ref{sec:bkg_network}. 

\section{Discussion on imaging genetics results}
Ch. \ref{chap:res_link} reports the most important results of this work, proving that the brain regions relevant for MRI classification are those in which four out of the five modules found with the transcriptome analysis of ASD assume higher absolute values. When the analyses described in this thesis were originally designed, this convergence of results was assumed to be visible only considering the average expression of the modules of interest, so their joint effect. Instead, this work revealed that each module (except the cyan one) alone takes a significantly absolute higher expression in the 19 regions found with neuroimaging, with respect to a random selection of 19 regions. Thanks to this finding, the results on the brain regions and those on the network modules strengthen themselves reciprocally.\\
As already mentioned in Sec. \ref{sec:res_comparison_neuroimaging_transcriptomics}, the cyan module, involved in immune system response was expected not to contribute to the neuroimaging-based biomarker of ASD, for various reasons:
\begin{itemize}
    \item due to the absence of a transcriptome dataset with ASD data and a fine brain-wide sampling, the regions of interest have been found using the data of healthy subjects, under the implied hypothesis that their brain-wide transcriptome profile grossly match that of the ASDs. However, neuroinflammation is presumably detectable only in subjects with mental disorders. In fact, during the data analysis, the GE values of most of the gene of the cyan module turned out to be below the background signal. So these data were not fully reliable.
    \item Even supposing that the signal present in the genes above background is sufficient to represent the transcriptome distribution of the entire module, the immunitary response in the AHBA individuals with supposedly no mental disorders may vary from subject to subject based on their cause of death.
    \item Neuroinflammation in mental disorders is usually defined as a low-level and chronic form of inflammation. Even though it is known that its long term effects include neurological problems \cite{gilhus2019neuroinflammation,schwartz2016neurological}, neuroinflammation per se, being low-level, is not expected to be detectable with MRI. 
\end{itemize}
Re-examining what brings the regions indicated with the letters f-i in Fig. \ref{fig:mean_expr_our_modules} to get such a high absolute average z-eigengene expression on the four network modules of interest, it turns out that in all of them, one of the four modules assumed a value almost always below the background level. So, their expression value has been set to zero (following the approach described in Sec.s \ref{sec:mm_regional_gene_expression} and \ref{sec:mm_eigengene_calculation}), which is a value significantly distant from the lowest reliable values of the module expression distributions. As a consequence their absolute z-scored value is very large.
Thus, the extraordinary mean absolute z-scored eigengene of these four regions is the results of the choice to include in the analysis also the regions in which the modules show an almost undetectable expression. This choice was known to have such a risk. However, the regions where a module is almost not expressed (or not expressed at all) in TDCs could have been highly relevant for classification if they had been instead expressed in ASDs. Since this was not known a priori, it was a sensible decision to adopt this data transformation strategy.\\
Regarding instead the regions indicated with the letters a-e in Fig. \ref{fig:mean_expr_our_modules}, not relevant for the neuroimaging pattern, but with a mean absolute z-score value that is within the lowest and the highest ones of the relevant regions, various hypotheses can be made:
\begin{itemize}
    \item they may be important for ASD, but their inclusion in the neuroimaging pattern could not add any classification improvement. As a consequence their salience value, which constituted the criterion for choosing the 19 relevant regions, is not high.
    \item They could be affected by noise or artifacts in the neuroimaging dataset that made impossible to use their signal for classification.
    \item They could be grossly different between ASDs and TDCs, but not specific of ASD. For example their signature could be present also in some TDCs with a low FIQ.
    \item These regions have a higher absolute average z-score value across the four relevant modules with respect to other regions. However, within each of them, the four relevant modules may not be the ones driving their functional behavior, because other modules may be more expressed.
    \item Their high absolute average z-score in the four modules of interest could arise from the fact that a few subjects may have below background GE values. When averaged across the 6 subjects, these values may generate a smaller value than what was observable for regions f-i, but still present.
\end{itemize}
\section{Overall discussion}
\subsection*{Neuroimaging-based classifier performance}
The performance of the classifier developed in this thesis (AUC = 89\%, Acc. = 86\%, Sens. = 88\%, Spec. = 85\%) is much higher than most reported in the literature review illustrated in Tab. \ref{tab:asd_studies}, whose average accuracy value is of 70\%. The two papers highlighted in yellow report better results, but as already discussed in Sec.\ref{soa_neuroimaging}, they overlooked (or did not mention in the paper) some good practices of ML applications necessary to consider their performance fully reliable. \\
A recent review on the use of ADOS-2, i.e. the second edition of the ADOS, which is currently considered the most reliable tool to evaluate ASD, revealed that its diagnosing sensitivity and specificity range from 89-92\% and 81-85\%, respectively \cite{lebersfeld2021systematic}. However, it should be considered that these values are unavoidably overestimated. In fact, the ground truth used in the experiments consists in the evaluation by trained psychiatrists of the ADOS-2 itself and other schedules, interviews and questionnaires that consider the same behavioural features. Thus, the diagnosing abilities of the classifier developed in this thesis can be considered competitive with those of human experts. Furthermore, the classifier can provide individual explanations of its predictions and the neuroimaging pattern learnt by the algorithm has been confirmed by transcriptomic analysis of ASD, so the validity of this tool is doubly grounded. As a consequence, after appropriate clinical trials, the developed classifier may become a valid computational tool to aid the physician in the diagnosis.\\
However, two important features of the classifier must be further tested to understand its clinical utility:
\begin{itemize}
    \item \textit{The ability to distinguish ASDs with respect to non-spectrum subjects with other neurodevelopmental disorders.}\\
    This is in fact the most common situation in medical practice. 
    Unfortunately, to the knowledge of the author of this thesis, databases with rsfMRIs of both ASDs and subjects with other neurodevelopmental disorders are not freely accessible.
    \vspace{2mm}
    \item \textit{The ability to classify infants for early diagnosis.}\\
    In fact, as already mentioned in Sec. \ref{sec:bkg_asd_notions}, early diagnosis is the main goal of this kind of classification studies. The high diagnosing performance across the wide age range of the ABIDE dataset (5-64 years), and the fact that only 1 out of the 13 subjects with age between 5 to 6 years has been misdiagnosed, may suggest that the developed classifier would hold its performance also on younger cohorts of subjects. However, \textit{early diagnosis} usually refers to the evaluation of ASD before the second year of age (possibly even before the first one), so in a pre-verbal age which can be substantially different from a functional point of view. As for the previous point, no databases with rsfMRIs from very young patients are freely accessible, to date, to test this classifier on an early diagnosis task.
\end{itemize}
Even if the developed classifier did not maintain its performance on this clinically more relevant tasks, it would be reasonable to suppose that re-training the adopted architecture with more appropriate data for these tasks would allow to obtain similarly high performance.

\subsection*{Convergence of the neuroimaging and transcriptomic signature of ASD}
From a biological point of view, the convergence of the neuroimaging and transcriptomic analyses of ASD is the most interesting result of this work.\\
Although convergence of results between imaging and genetics has already been reported in other works, they still constitute a niche type of research, mainly because they consist in multi-step analyses in which errors in a single step may undermine the detection of the convergence. 
Trivially, the convergence between two analyses can be observed only if both provide solid results minimally affected by noise.\\
During the development of this work, the author of this thesis identified some fundamental hypothesis and methodological points critical to make the observation of the convergence possible, which are:
\begin{itemize}
    \item \textit{The possibility to characterize the ASD spectrum as a unicum at the neuroimaging and transcriptomic levels.}\\
    Despite the phenotypical heterogeneity of the ASD spectrum, the disorder should have been characterized by a shared pattern both at the neuroimaging and transcriptome level. It is the opinion of the author that the success of the approach adopted in this thesis could not prescind from this hypothesis to be true.
    \vspace{2mm}
    \item \textit{The possibility to capture the transcriptomic signature of ASD from only 49 subjects.}\\
    While there were good possibilities to detect a single neuroimaging-based biomarker by analyzing around 665 ASD subjects, it was not straightforward that only 49 ASDs were sufficient to identify the genetic pathways characterizing an entire spectrum. 
    The convergence between the neuroimaging and the transcriptomic signatures, considering that the latter has been determined from a much smaller representation of ASD cases, suggests that ASD is effectively characterized by a few shared pathways (i.e. the 4 modules identified) that in each subject are disrupted due to the dysregulation of different genes. A similar conclusion has been drawn by a previous paper \cite{voineagu2011transcriptomic} when noticing that the pathways involved in subjects with Dup15q and in idiopathic cases are essentially the same, despite the disorder arises for different genetic causes.
    \vspace{2mm}
    \item \textit{The possibility for the classifier to learn most of and only the brain regions relevant for ASD.}\\
    The pattern learnt by the algorithm should have allowed to identify most of the brain regions whose rsfMRI signal is different between ASDs and TDCs and discard the others. This was not granted. In fact, the algorithm could have identified a subset of brain regions that was sufficient to minimize the prediction error or could have included some irrelevant regions, whose presence or absence in the pattern would not have had a significant impact on the performance of the classifier. Concerning this last possibility, the BR-NET is likely to have had a critical role in filtering out spurious regions. In fact, the adversarial approach that forces the net to discard any pattern correlated to the selected confounders, among which many biological features, makes it more rewarding to retain exclusively those features that are actually important for classification.
    \vspace{2mm}
     \item \textit{The comprehensibility of the neuroimaging pattern extracted from the classifier.}\\
     The pattern learnt by the classifier and the way it was represented by the explainability algorithm should have been intelligible and comparable with the information coming from the transcriptomic analysis. This has been an important assumption in the design of this thesis. The clear organization of the learnt pattern in brain regions was not granted, albeit supposed. This result maybe not guaranteed for other NNs or when applying this methodology to other disorders.
     \vspace{2mm}
     \item \textit{The almost equivalence of the module networks distribution across the brain in ASDs and TDCs.}\\
    The convergence of the transcriptomic and neuroimaging signatures has been evaluated by analyzing the brain-wide gene expression profile of the network modules of interest in TDC samples. This approach could have been successful only if the transcriptomic profiles of ASDs and TDCs had grossly matched in their regional distribution across brain. This has been the strongest assumption of this thesis, especially considering that the network modules of interest have been found by studying GE variations between ASDs and TDCs. To explain this apparent contradiction it must be specified that the underlying hypothesis of this work is that just a few of the genes in the network modules of interest are dysregulated in each subject with ASD, and they are not always the same. Furthermore, the degree of dysregulation due to the disorder was hypothesized to be much lower than the variability due to GE regional differences across the brain. Under these conditions, it was reasonable to expect that the network module eigengenes computed on the TDC samples would provide a correct information on the regional distribution of the GE values of these modules also in ASD subjects.\\
    In the opinion of the author, the possibility to observe the convergence on a TDC dataset indicates that the brain regions identified with neuroimaging do not show macroscopic GE differences among the two groups, but instead emerge as characteristic of the disorder, as a natural consequence of the fact that, in these regions, the transcriptomic pathways disrupted in ASD play an important role. This observation reminds a study \cite{werling2016gene}, mentioned in Sec. \ref{hp_driven_studies}, showing that sex-differential risk of ASD is not due to a higher expression of ASD risk genes in males with respect to females, but to their interaction with pathways that are differentially expressed by sex.
\end{itemize}
In conclusion, the observation of the convergence suggests that the listed underlying hypotheses or methodological choices adopted were correct, but does not constitute an absolute proof. Certainly, considering the high degree of noise, the multiple confounders and the artifacts affecting the MRI and gene expression data, the observed convergence of results indicates that the link between transcriptomic and neuroimaging information is striking.

\subsection*{Network modules involved in ASD}
The most interesting result on the transcriptomic part of the study is that, thanks to the double validation given by the imaging transcriptomic approach, the network modules and their associated pathways with a primary role in ASD have been delimited. Clearly, the exact list of genes belonging to these modules may slightly vary with the subjects and tissues analyzed, and strengthening the network analysis by expanding the dataset is an obvious future work. However, the results in Tab. \ref{tab:link_pval} do not leave any doubts as to which modules better explain the neuroimaging signature of ASD.\\
Two out of the four modules, that take a significant mean absolute expression across the 19 regions of interest, have well defined pathways (see Tab. \ref{tab:summary_go_modules}): the turquoise module in neuronal functions and the salmon module in ribosome and protein synthesis. Both these functions have already been reported in literature, and are discussed respectively in Sec. \ref{brainASD-s-o-a} and \ref{hp_driven_studies}. The functions associated to the black and blue modules are, instead, more heterogeneous, making it more difficult to understand whether these modules were already observed. Considering the GO terms of the black and blue modules in Tab. \ref{tab:summary_go_modules}, it can be noted that the functions related to cell cycle, glia and cell communication have been already highlighted by the literature review described in Sec. \ref{sec:bkg_summary_review_GE_ASD} and \ref{brainASD-s-o-a}. Furthermore, the RNA processing function, listed in the GO terms of the blue module, is strictly related to the alternative splicing term, which was also mentioned in literature (Sec. \ref{sec:bkg_summary_review_GE_ASD}).
Taken together, these elements indicate that all the modules associated to ASD in this work are somewhat already supported by other studies, despite comparing transcriptomic results is always a complex task, as already mentioned in Sec. \ref{premise}.\\
As a final remark, while the transcriptomic part of this study may appear simply a validation of the neuroimaging part, the neat definition of the network modules involved in the disorder allows to circumscribe the possible ASD susceptibility genes, opening a wide number of studies both in silico and in animal models.

\subsection*{Brain regions involved in ASD}
\label{subsec:discussion_brain_regions}
In the research on ASD, many brain regions have been imputed as possibly implicated in the disorder or some of its manifestations, but a consensual and reliable biomarker was not identified. Therefore, comparing the 19 brain regions identified with the ones that occurred the most across the vast and heterogeneous literature available would not strengthen their validity, especially because a second validation of the results of this work comes from the transcriptomic analysis. An in-depth discussion on the biological role that the brain regions identified could play in the pathogenesis of the disorder is out of the scope of this data science thesis. \\
Thus, this paragraph reports in the gray colorbox below a summary of the functions associated to these regions, according to literature, and a few points of reflection made by the author.  
\begin{tcolorbox}
\footnotesize
\textbf{Amygdala}\\
It has a primary role in memory and decision making in response to emotional situations, such as fear-related experiences \cite{gupta2011amygdala}, and in processing reactions to personal space violations \cite{kennedy2009personal}.
The amygdala has been also associated to the ability to socialize \cite{bickart2011amygdala} and to make judgements based on a person face \cite{bzdok2011ale}.
The left and right sides hold some peculiar functions \cite{baeken2014left} and seem to have independent memory systems \cite{blair2005unilateral}. The right amygdala is mainly involved in negative emotions \cite{barrett2007amygdala,berntson2007amygdala}, while the left one processes either pleasant or unpleasant emotions \cite{lanteaume2007emotion} and plays a role in the brain's reward system \cite{murray2009amygdala}.
The amygdala is also a well known sexually dimorphic region: larger in males than females \cite{goldstein2001normal} and rich in androgen receptors \cite{simerly1990distribution}.\\
\\
\textbf{Angular gyrus}\\
Its main role consists in transferring visual information to Wernicke's area to assign a meaning to visually perceived words or symbols \cite{hall2010guyton}. It is, thus, involved in complex language functions (i.e. reading, writing and interpreting). Injuries to this part of the brain include alexia, acalculia and agraphia \cite{henschen1919language,gerstmann1940syndrome}.
However, it seems to also play a role in 3D orienting \cite{chen2012neural} and it activates differently to intended and consequential movement, suggesting its involvement in self-awareness functions \cite{farrer2008angular}.\\
\\
\textbf{Cingulum bundle}\\
It is a critical white matter fiber tract in the brain, which forms connections between the frontal, parietal and temporal lobes \cite{wu2016segmentation}. Diverse functions have been attributed to this brain area: emotional response to sensory inputs \cite{cohen1999alteration}, regulation of aggressive behavior \cite{zhu2019brain}, communication \cite{devinsky1995contributions}, maternal bonding \cite{devinsky1995contributions} and decision making \cite{bubb2018cingulum}.
The functions of the anterior cingulum are mainly oriented towards emotions \cite{devinsky1995contributions} and, in fact, together with the anterior insula, the anterior cingulum is part of the salience network that contributes to a variety of complex emotional-related functions, including social behavior and self-awareness \cite{menon2015salience}.
The Posterior cingulum cortex, instead, has an important cognitive role, showing increased activity when individuals retrieve autobiographical memories or plan for the future \cite{leech2014role}.\\
\\
\textbf{Frontal Lobe}\\
It controls voluntary movements, expressive language and higher level executive functions \cite{stuss2013principles}, i.e. cognitive abilities such as the capacity to plan, initiate and monitor behaviours in order to achieve a goal. It is considered the control center of behaviour and emotions and the site that defines individual personality \cite{stuss2013principles}. For the variety of its high level functions, lesions to the frontal lobe can show several different symptoms \cite{sawyer2017diagnosing,chow2000personality}.\\
\\
\textbf{Insula}\\
It is involved in multimodal sensory processing \cite{bushara2003neural}, Motor control tasks (among which gastric motility \cite{penfield1955insula}, and speech articulation \cite{uddin2017structure}), homeostatic functions \cite{uddin2017structure} and immune system regulation \cite{ramirez1996insular,ramirez1999conditioned}. This region also constitutes the site where the pain sensation takes form \cite{uddin2017structure}. In addition, as part of the salience network, the insula displays various higher-level functions, such as interoceptive awareness, social emotions and empathy \cite{lamm2010role}, which probably arise as a consequence of the insula role in conveying homeostatic information to consciousness \cite{de2006frames,xue2010impact}.\\
\\
\textbf{Lingual gyrus}\\
It is involved in many vision related tasks, such as identification of words and letters \cite{mechelli2000differential} and face recognition and memorization \cite{mccarthy1999electrophysiological,kiselnikov2014activation}. Its cognitive higher functions include visual mental imagery \cite{de2015visual,zhang2016gray}, which plays a role also in visual-unrelated thoughts as well as in self-criticism \cite{kim2020attachment}.\\
\\
\textbf{Parietal lobe}\\
It integrates different sensory information and is critically involved in proprioception \cite{filimon2009multiple} and stereognosis \cite{bassetti1993sensory}. It receives somatosensory and visual inputs \cite{baldauf2008posterior}, and through motor signals, controls movement of the arms, hands, and eyes \cite{fogassi2005motor}.
Besides sensorymotor abilities, its right side seems to be involved also in conceptual processing of numbers \cite{cappelletti2010role} and the inferior parietal lobe in memory retrieval \cite{vilberg2008memory}.\\
\\
\textbf{Precentral gyrus}\\
It controls the voluntary movements of face and body parts.
Its damage can cause dysarthria and aphemia suggesting a contribution of this brain region also in movement articulation \cite{terao2007primary,itabashi2016damage}. 
Several evidences indicates that the precentral gyrus is involved in imitation and in the mirror neuron system \cite{kilner2013we,rizzolatti2010mirror,buccino2004neural}.
\end{tcolorbox}
About the composition of the 19 brain regions selected by the ASD classifier, it is interesting to note that they have very different roles such as sensorymotor functions, basic and instinctive reactions to fear, and higher cognitive abilities that defines our personality and consciousness. This clearly shows that ASD is a disorder affecting behaviour at multiple levels. Among the 19 regions two of them have been highly cited as possible implicated in ASD, which are:
\begin{itemize}
    \item The amygdala, that for being involved in fear response and showing a sexually dimorphic structure and function may explain some core ASD features. Among the various studies suggesting a link between this brain region and ASD, two are worth of mention. The first one \cite{herrero2020identification}, already described in Sec. \ref{hp_driven_studies}, shows that known ASD susceptibility genes are dynamically expressed in the amygdala during fetal and early postnatal stages of development, more than any other group of genes, which may elucidate the temporal window of ASD onset. The second one \cite{stilling2018social}, shows that host-microbes influence the gene expression amygdala with consequences to social behaviour, which may shed light on the complex relationship between gastrointestinal problems and a behavioural disorder such as ASD.
    \item The insula, that again is involved in several functions that are disrupted in ASD, such as empathy and regulation of gastric motility and immune response. Interestingly, only the right insula appears in the neuroimaging pattern. While it is known that anatomical differences exist between in the connectivity pattern of the right and left insula \cite{jakab2012connectivity}, less is known on their different functions. Based on the study of insular stroke effects, one work proposed the right insula to be the central site of human body scheme representation \cite{karnath2010right}.
    The right frontal cortex and the right insula have been repeatedly found co-activated in many neuroimaging studies on inhibition \cite{ghahremani2015role}. One study \cite{cai2014dissociable} in particular suggests that the latter is involved in detecting behaviorally salient events that are processed by the frontal cortex to implement an inhibitory response. Despite body perception and inhibition are not widely used labels to describe ASD, they can play an important role in the motor and social impairments associated to the disorder. 
\end{itemize}

\subsection*{A spectrum with a single complex biomarker}
In conclusion, the pattern learned by the classifier developed in this thesis may be considered a biomarker of ASD, although further investigations are still necessary to understand its temporal characterization. Two final considerations must be mentioned about this result.\\
Firstly, the neuroimaging biomarker found can identify ASDs in a large age range, and this was not at all obvious. In fact, some studies have been designed with the idea that an ASD biomarkers may be detectable only at certain developmental stages \cite{loth2016identification}. Supporting this idea, other features associated to the disorder or to some subtypes of it are detectable during the first two years of life, such as the known early brain overgrowth \cite{courchesne2004brain}, or at most within the age of four, such as the increase of extra-axial cerebrospinal fluid volume \cite{shen2018extra}.
Instead, contrary to these expectations, this work shows that shared rsfMRI and GE signatures are detectable at any age, despite they may vary during lifespan.\\
Secondly, these signatures are common to the heterogeneous ASD spectrum, which, again, was not granted at all. In fact, as already mentioned in the introduction, many other works explored the subgrouping avenue, supposing that the whole spectrum could not be distinguished from TDCs. Furthermore, even the ASD definition as a unitary spectrum has been subject of debate \cite{ritvo2012postponing,singer2012diagnosis}. In fact, before the release of the DSM-5 \cite{american2013diagnostic}, three distinct diagnostic categories existed (Asperger's syndrome, autism disorder, pervasive developmental disorder – not otherwise specified). These were merged into the single category of ASD, with the only motivation that they could not be reliably differentiated from each another \cite{tsai2014dsm}. 
The high classification performance obtained in this work and the convergent results between the neuroimaging and transcriptomic analyses suggest that indeed the ASDs share important biological features. While the classification part of this study can be considered a proof of the AI capability to discover complex biological patterns, it must be noted that this result was only possible thanks to years of observational research and to the human ability to analyze behaviour that allowed to define the current ASD diagnostic criteria.\\
\chapter{Conclusions and future works}
\section{Conclusions}
This work presented an imaging transcriptomic study of ASD articulated into three analyses:
\begin{enumerate}
    \item The development of a DL-based classifier, jointly resilient to confounders and able to process the temporal and spatial dimensions of rsfMRIs in order to focus on the brain blood flow, to diagnose ASD and the extraction of the brain regions most important for classification.
    \item The identification of the main gene co-expression network modules, and their associated pathways, whose expression is disrupted in ASD brain. 
    \item The comparison of the brain-wide spatial distribution of the expression of the modules identified at the second step with the brain regions relevant for diagnosis found at the first step.
\end{enumerate}
The classifier developed can predict ASD from rsfMRI with performance (AUC = 89\%, Sens. = 0.88, Spec. = 0.85) compatible with computer assisted diagnosis. The brain regions identified as most important for classification are 19 out of the 116 defined by the AAL atlas, located in both the hemispheres, in the cortex and subcortex. However, none of the subjects have been diagnosed taking into consideration all of them, corroborating the idea that the phenotypical heterogeneity of the disorder is explained by a heterogeneity of different impairments in brain functionalities. Furthermore, the 19 regions have been linked to a wide variety of functions in literature, from sensorymotor ones to higher cognitive abilities. \\
The transcriptomic part of the study, instead, allowed to identify five network modules of interest for ASD. Out of which three can be easily identified as individually associated to precise functions: ribosomal, neuronal and immunitary ones. The other two modules are involved in more heterogeneous pathways, with associated GO terms such as glial function, regulation of cell communication and cell cycle. All these functions have been already associated to ASD in literature, but along with many others, without a clear distinction between which are the core ones characterizing the disorder.\\
In this study, instead, the significant convergence of the results obtained between the neuroimaging and transcriptomic analyses provides a reciprocal validation of the two results. In fact, it has been demonstrated that four out of the five network modules identified (i.e. all except the one related to immune response) show an absolute gene expression value significantly higher in the 19 regions, that make up the neuroimaging signature of ASD, as compared to any other random combination of regions in equal number. This convergence makes it possible to disentangle the results obtained from unavoidably arbitrary choices (such as cut-off thresholds) and limitations (such as the dataset dimension) in the study, indicating that the 19 brain regions and the 4 gene co-expression network modules associated to them reliably constitute important biomarkers of the ASD brain functional and transcriptomic features, respectively. 

In conclusion, from a medical perspective, this study shows that, despite the heterogeneity of ASD, the disorder is characterizable by a shared complex biomarker, both at the neuroimaging and transcriptomic levels, even if single individuals display some singularities in the pattern. Furthermore, it demonstrated an important convergence between the two signatures.\\
On a practical level, this thesis also provides an interpretable tool for assisting the physicians in the diagnosis of ASD, based on a rsfMRI exam, which is a non-invasive procedure, and without requiring many data manipulation operations.

\section{Future works}
This work paves the way for a wide number of other analyses, some of which have been already mentioned throughout the discussion. \\
Future works that can be easily performed starting from the data and the code produced in this thesis include:
\begin{itemize}
    \item Comparing the performance of the classifier when using solely the BR-NET architecture or the 2S-NN one, to isolate the approach that has been essential in reaching the high performance registered. 
    \item Analyzing the temporal pattern found by the classifier developed by using an appropriate XAI algorithm \cite{rojat2021explainable}.
    \item Trying to understand which brain regions are associated to different phenotypic and behavioural descriptors of the ASD subjects, by multivariate regression analysis.  
    \item Testing the classifier's ability to distinguish ASDs with respect to non-spectrum subjects with other neurodevelopmental disorders and, if necessary, re-training the algorithm with more appropriate data for this task.
    \item Testing (or re-training) the algorithm on data belonging to infants within the second year of life, which is considered the optimal age range for early diagnosis.
\end{itemize}
Possible future studies, building on the results of this work, but that may require substantial additional effort include:
\begin{itemize}
    \item Extracting from the classifier a more precise information on the neuroimaging biomarker, by assigning a critical threshold for each rsfMRI feature (or combination of features) to distinguish ASDs and TDCs. This would allow to use the pattern learned even without applying the classifier to the data.
    \item Repeating the gene co-expression network analysis performed in this work including more data, possibly coming from a wider set of brain regions, in order to better define the genes belonging to the four network of interest and define, thus, a list of candidate ASD susceptibility genes.
    \item Studying these candidate genes with reverse genetics animal models, to better elucidate their role in the ASD pathogenesis.
    \item Studying how the neuroimaging signature changes in response to variations in the gene expression network modules associated to the 19 regions by using animal models.
\end{itemize}

In conclusion, the results obtained in this thesis provide countless ideas to reach a better comprehension of the neuroimaging biomarker of ASD and to the genetic causes leading to brain functional alterations.

{
\footnotesize
\begin{spacing}{0.5}
\bibliographystyle{Style/IEEEtran}
\bibliography{references}

\begin{thebibliography}{100}
\providecommand{\url}[1]{#1}
\csname url@samestyle\endcsname
\providecommand{\newblock}{\relax}
\providecommand{\bibinfo}[2]{#2}
\providecommand{\BIBentrySTDinterwordspacing}{\spaceskip=0pt\relax}
\providecommand{\BIBentryALTinterwordstretchfactor}{4}
\providecommand{\BIBentryALTinterwordspacing}{\spaceskip=\fontdimen2\font plus
\BIBentryALTinterwordstretchfactor\fontdimen3\font minus
  \fontdimen4\font\relax}
\providecommand{\BIBforeignlanguage}[2]{{%
\expandafter\ifx\csname l@#1\endcsname\relax
\typeout{** WARNING: IEEEtran.bst: No hyphenation pattern has been}%
\typeout{** loaded for the language `#1'. Using the pattern for}%
\typeout{** the default language instead.}%
\else
\language=\csname l@#1\endcsname
\fi
#2}}
\providecommand{\BIBdecl}{\relax}
\BIBdecl

\bibitem{thurm2012importance}
A.~Thurm and S.~E. Swedo, ``The importance of autism research,''
  \emph{Dialogues in clinical neuroscience}, vol.~14, no.~3, p. 219, 2012.

\bibitem{harris2018leo}
J.~Harris, ``Leo kanner and autism: a 75-year perspective,''
  \emph{International Review of Psychiatry}, vol.~30, no.~1, pp. 3--17, 2018.

\bibitem{american2013diagnostic}
A.~P. Association \emph{et~al.}, \emph{Diagnostic and statistical manual of
  mental disorders (DSM-5{\textregistered})}.\hskip 1em plus 0.5em minus
  0.4em\relax American Psychiatric Pub, 2013.

\bibitem{cohmer2014early}
S.~Cohmer, ``Early infantile autism and the refrigerator mother theory
  (1943-1970),'' \emph{Embryo project encyclopedia}, 2014.

\bibitem{folstein1977infantile}
S.~Folstein and M.~Rutter, ``Infantile autism: a genetic study of 21 twin
  pairs,'' \emph{Journal of Child psychology and Psychiatry}, vol.~18, no.~4,
  pp. 297--321, 1977.

\bibitem{rutter2000genetic}
M.~Rutter, ``Genetic studies of autism: from the 1970s into the millennium,''
  \emph{Journal of Abnormal Child Psychology}, vol.~28, no.~1, pp. 3--14, 2000.

\bibitem{bailey1995autism}
A.~Bailey, A.~Le~Couteur, I.~Gottesman, P.~Bolton, E.~Simonoff, E.~Yuzda, and
  M.~Rutter, ``Autism as a strongly genetic disorder: evidence from a british
  twin study,'' \emph{Psychological medicine}, vol.~25, no.~1, pp. 63--77,
  1995.

\bibitem{sandin2014familial}
S.~Sandin, P.~Lichtenstein, R.~Kuja-Halkola, H.~Larsson, C.~M. Hultman, and
  A.~Reichenberg, ``The familial risk of autism,'' \emph{Jama}, vol. 311,
  no.~17, pp. 1770--1777, 2014.

\bibitem{spiker2002behavioral}
D.~Spiker, L.~J. Lotspeich, S.~Dimiceli, R.~M. Myers, and N.~Risch,
  ``Behavioral phenotypic variation in autism multiplex families: evidence for
  a continuous severity gradient,'' \emph{American Journal of Medical
  Genetics}, vol. 114, no.~2, pp. 129--136, 2002.

\bibitem{ivanov2015autism}
H.~Y. Ivanov, V.~K. Stoyanova, N.~T. Popov, and T.~I. Vachev, ``Autism spectrum
  disorder-a complex genetic disorder,'' \emph{Folia medica}, vol.~57, no.~1,
  pp. 19--28, 2015.

\bibitem{hornig2002infectious}
M.~Hornig, R.~Mervis, K.~Hoffman, and W.~Lipkin, ``Infectious and immune
  factors in neurodevelopmental damage,'' \emph{Molecular psychiatry}, vol.~7,
  no.~2, pp. S34--S35, 2002.

\bibitem{rodier2011environmental}
P.~Rodier, ``Environmental exposures that increase the risk of autism spectrum
  disorders,'' \emph{Autism spectrum disorders}, pp. 863--874, 2011.

\bibitem{eaves2006screening}
L.~C. Eaves, H.~D. Wingert, H.~H. Ho, and E.~C. Mickelson, ``Screening for
  autism spectrum disorders with the social communication questionnaire,''
  \emph{Journal of Developmental \& Behavioral Pediatrics}, vol.~27, no.~2, pp.
  S95--S103, 2006.

\bibitem{zander2016objectivity}
E.~Zander, C.~Willfors, S.~Berggren, N.~Choque-Olsson, C.~Coco, A.~Elmund,
  {\AA}.~H. Moretti, A.~Holm, I.~Jif{\"a}lt, R.~Kosieradzki \emph{et~al.},
  ``The objectivity of the autism diagnostic observation schedule (ados) in
  naturalistic clinical settings,'' \emph{European child \& adolescent
  psychiatry}, vol.~25, no.~7, pp. 769--780, 2016.

\bibitem{apaaverage}
\BIBentryALTinterwordspacing
A.~P. Association. Average age at asd diagnosis. [Online]. Available:
  \url{https://www.psychiatry.org/newsroom/news-releases/children-diagnosed-with-autism-at-earlier-age-more-likely-to-receive-evidence-based-treatments}
\BIBentrySTDinterwordspacing

\bibitem{kentrou2019delayed}
V.~Kentrou, D.~M. de~Veld, K.~J. Mataw, and S.~Begeer, ``Delayed autism
  spectrum disorder recognition in children and adolescents previously
  diagnosed with attention-deficit/hyperactivity disorder,'' \emph{Autism},
  vol.~23, no.~4, pp. 1065--1072, 2019.

\bibitem{estes2015long}
A.~Estes, J.~Munson, S.~J. Rogers, J.~Greenson, J.~Winter, and G.~Dawson,
  ``Long-term outcomes of early intervention in 6-year-old children with autism
  spectrum disorder,'' \emph{Journal of the American Academy of Child \&
  Adolescent Psychiatry}, vol.~54, no.~7, pp. 580--587, 2015.

\bibitem{reichow2012early}
B.~Reichow, E.~E. Barton, B.~A. Boyd, and K.~Hume, ``Early intensive behavioral
  intervention (eibi) for young children with autism spectrum disorders
  (asd),'' \emph{Cochrane Database of Systematic Reviews}, no.~10, 2012.

\bibitem{fernandez2017syndromic}
B.~A. Fernandez and S.~W. Scherer, ``Syndromic autism spectrum disorders:
  moving from a clinically defined to a molecularly defined approach,''
  \emph{Dialogues in clinical neuroscience}, vol.~19, no.~4, p. 353, 2017.

\bibitem{rylaarsdam2019genetic}
L.~E. Rylaarsdam and A.~Guemez~Gamboa, ``Genetic causes and modifiers in autism
  spectrum disorder,'' \emph{Frontiers in cellular neuroscience}, vol.~13, p.
  385, 2019.

\bibitem{giovedi2014involvement}
S.~Gioved{\'\i}, A.~Corradi, A.~Fassio, and F.~Benfenati, ``Involvement of
  synaptic genes in the pathogenesis of autism spectrum disorders: the case of
  synapsins,'' \emph{Frontiers in pediatrics}, vol.~2, p.~94, 2014.

\bibitem{gupta2014transcriptome}
S.~Gupta, S.~E. Ellis, F.~N. Ashar, A.~Moes, J.~S. Bader, J.~Zhan, A.~B. West,
  and D.~E. Arking, ``Transcriptome analysis reveals dysregulation of innate
  immune response genes and neuronal activity-dependent genes in autism,''
  \emph{Nature communications}, vol.~5, no.~1, pp. 1--8, 2014.

\bibitem{pua2017autism}
E.~P.~K. Pua, S.~C. Bowden, and M.~L. Seal, ``Autism spectrum disorders:
  Neuroimaging findings from systematic reviews,'' \emph{Research in Autism
  Spectrum Disorders}, vol.~34, pp. 28--33, 2017.

\bibitem{goodfellow2016deep}
I.~Goodfellow, Y.~Bengio, and A.~Courville, \emph{Deep learning}.\hskip 1em
  plus 0.5em minus 0.4em\relax MIT press, 2016.

\bibitem{donovan2017neuroanatomy}
A.~P. Donovan and M.~A. Basson, ``The neuroanatomy of autism--a developmental
  perspective,'' \emph{Journal of anatomy}, vol. 230, no.~1, pp. 4--15, 2017.

\bibitem{richiardi2015correlated}
J.~Richiardi, A.~Altmann, A.-C. Milazzo, C.~Chang, M.~M. Chakravarty,
  T.~Banaschewski, G.~J. Barker, A.~L. Bokde, U.~Bromberg, C.~B{\"u}chel
  \emph{et~al.}, ``Correlated gene expression supports synchronous activity in
  brain networks,'' \emph{Science}, vol. 348, no. 6240, pp. 1241--1244, 2015.

\bibitem{hawrylycz2012anatomically}
M.~J. Hawrylycz, E.~S. Lein, A.~L. Guillozet-Bongaarts, E.~H. Shen, L.~Ng,
  J.~A. Miller, L.~N. Van De~Lagemaat, K.~A. Smith, A.~Ebbert, Z.~L. Riley
  \emph{et~al.}, ``An anatomically comprehensive atlas of the adult human brain
  transcriptome,'' \emph{Nature}, vol. 489, no. 7416, p. 391, 2012.

\bibitem{wang2015correspondence}
G.-Z. Wang, T.~G. Belgard, D.~Mao, L.~Chen, S.~Berto, T.~M. Preuss, H.~Lu,
  D.~H. Geschwind, and G.~Konopka, ``Correspondence between resting-state
  activity and brain gene expression,'' \emph{Neuron}, vol.~88, no.~4, pp.
  659--666, 2015.

\bibitem{lowd2005adversarial}
D.~Lowd and C.~Meek, ``Adversarial learning,'' in \emph{Proceedings of the
  eleventh ACM SIGKDD international conference on Knowledge discovery in data
  mining}, 2005, pp. 641--647.

\bibitem{burkart2021survey}
N.~Burkart and M.~F. Huber, ``A survey on the explainability of supervised
  machine learning,'' \emph{Journal of Artificial Intelligence Research},
  vol.~70, pp. 245--317, 2021.

\bibitem{Ferrari2020autismChapter}
\BIBentryALTinterwordspacing
E.~Ferrari, \emph{Artificial Intelligence for Autism Spectrum Disorders}.\hskip
  1em plus 0.5em minus 0.4em\relax Cham: Springer International Publishing,
  2020, pp. 1--15. [Online]. Available:
  \url{https://doi.org/10.1007/978-3-030-58080-3_249-1}
\BIBentrySTDinterwordspacing

\bibitem{georgiades2017editorial}
S.~Georgiades, S.~L. Bishop, and T.~Frazier, ``Editorial perspective:
  Longitudinal research in autism--introducing the concept of
  ‘chronogeneity’,'' \emph{Journal of Child Psychology and Psychiatry},
  vol.~58, no.~5, pp. 634--636, 2017.

\bibitem{baio2018prevalence}
J.~Baio, L.~Wiggins, D.~L. Christensen, M.~J. Maenner, J.~Daniels, Z.~Warren,
  M.~Kurzius-Spencer, W.~Zahorodny, C.~R. Rosenberg, T.~White \emph{et~al.},
  ``Prevalence of autism spectrum disorder among children aged 8 years—autism
  and developmental disabilities monitoring network, 11 sites, united states,
  2014,'' \emph{MMWR Surveillance Summaries}, vol.~67, no.~6, p.~1, 2018.

\bibitem{loomes2017male}
R.~Loomes, L.~Hull, and W.~P.~L. Mandy, ``What is the male-to-female ratio in
  autism spectrum disorder? a systematic review and meta-analysis,''
  \emph{Journal of the American Academy of Child \& Adolescent Psychiatry},
  vol.~56, no.~6, pp. 466--474, 2017.

\bibitem{maestro2002attentional}
S.~Maestro, F.~Muratori, M.~C. Cavallaro, F.~Pei, D.~Stern, B.~Golse, and
  F.~Palacio-Espasa, ``Attentional skills during the first 6 months of age in
  autism spectrum disorder,'' \emph{Journal of the American Academy of Child \&
  Adolescent Psychiatry}, vol.~41, no.~10, pp. 1239--1245, 2002.

\bibitem{rogers2009infant}
S.~J. Rogers, ``What are infant siblings teaching us about autism in infancy?''
  \emph{Autism Research}, vol.~2, no.~3, pp. 125--137, 2009.

\bibitem{myers2007management}
S.~M. Myers, C.~P. Johnson \emph{et~al.}, ``Management of children with autism
  spectrum disorders,'' \emph{Pediatrics}, vol. 120, no.~5, pp. 1162--1182,
  2007.

\bibitem{mandell2009nih}
D.~S. Mandell, ``Nih-pa author manuscript,'' \emph{Lancet}, vol. 374, no. 9701,
  pp. 1627--1638, 2009.

\bibitem{fernell2013early}
E.~Fernell, M.~A. Eriksson, and C.~Gillberg, ``Early diagnosis of autism and
  impact on prognosis: a narrative review,'' \emph{Clinical epidemiology},
  vol.~5, p.~33, 2013.

\bibitem{bryson2004early}
S.~E. Bryson, L.~Zwaigenbaum, and W.~Roberts, ``The early detection of autism
  in clinical practice,'' \emph{Paediatrics \& child health}, vol.~9, no.~4,
  pp. 219--221, 2004.

\bibitem{koegel2014importance}
L.~K. Koegel, R.~L. Koegel, K.~Ashbaugh, and J.~Bradshaw, ``The importance of
  early identification and intervention for children with or at risk for autism
  spectrum disorders,'' \emph{International journal of speech-language
  pathology}, vol.~16, no.~1, pp. 50--56, 2014.

\bibitem{orinstein2014intervention}
A.~J. Orinstein, M.~Helt, E.~Troyb, K.~E. Tyson, M.~L. Barton, I.-M. Eigsti,
  L.~Naigles, and D.~A. Fein, ``Intervention for optimal outcome in children
  and adolescents with a history of autism,'' \emph{Journal of Developmental \&
  Behavioral Pediatrics}, vol.~35, no.~4, pp. 247--256, 2014.

\bibitem{frye2019emerging}
R.~E. Frye, S.~Vassall, G.~Kaur, C.~Lewis, M.~Karim, and D.~Rossignol,
  ``Emerging biomarkers in autism spectrum disorder: a systematic review,''
  \emph{Annals of translational medicine}, vol.~7, no.~23, 2019.

\bibitem{london2007role}
E.~London, ``The role of the neurobiologist in redefining the diagnosis of
  autism,'' \emph{Brain pathology}, vol.~17, no.~4, pp. 408--411, 2007.

\bibitem{werling2013sex}
D.~M. Werling and D.~H. Geschwind, ``Sex differences in autism spectrum
  disorders,'' \emph{Current opinion in neurology}, vol.~26, no.~2, p. 146,
  2013.

\bibitem{geschwind2008autism}
D.~H. Geschwind, ``Autism: many genes, common pathways?'' \emph{Cell}, vol.
  135, no.~3, pp. 391--395, 2008.

\bibitem{wei2014apoptotic}
H.~Wei, I.~Alberts, and X.~Li, ``The apoptotic perspective of autism,''
  \emph{International journal of developmental neuroscience}, vol.~36, pp.
  13--18, 2014.

\bibitem{estes2015immune}
M.~L. Estes and A.~K. McAllister, ``Immune mediators in the brain and
  peripheral tissues in autism spectrum disorder,'' \emph{Nature Reviews
  Neuroscience}, vol.~16, no.~8, pp. 469--486, 2015.

\bibitem{piven2017toward}
J.~Piven, J.~T. Elison, and M.~J. Zylka, ``Toward a conceptual framework for
  early brain and behavior development in autism,'' \emph{Molecular
  psychiatry}, vol.~22, no.~10, pp. 1385--1394, 2017.

\bibitem{amaral2008neuroanatomy}
D.~G. Amaral, C.~M. Schumann, and C.~W. Nordahl, ``Neuroanatomy of autism,''
  \emph{Trends in neurosciences}, vol.~31, no.~3, pp. 137--145, 2008.

\bibitem{bonnet2018autism}
F.~Bonnet-Brilhault, T.~A. Rajerison, C.~Paillet, M.~Guimard-Brunault, A.~Saby,
  L.~Ponson, G.~Tripi, J.~Malvy, and S.~Roux, ``Autism is a prenatal disorder:
  Evidence from late gestation brain overgrowth,'' \emph{Autism Research},
  vol.~11, no.~12, pp. 1635--1642, 2018.

\bibitem{vasa2016disrupted}
R.~A. Vasa, S.~H. Mostofsky, and J.~B. Ewen, ``The disrupted connectivity
  hypothesis of autism spectrum disorders: time for the next phase in
  research,'' \emph{Biological Psychiatry: Cognitive Neuroscience and
  Neuroimaging}, vol.~1, no.~3, pp. 245--252, 2016.

\bibitem{raznahan2013compared}
A.~Raznahan, G.~L. Wallace, L.~Antezana, D.~Greenstein, R.~Lenroot, A.~Thurm,
  M.~Gozzi, S.~Spence, A.~Martin, S.~E. Swedo \emph{et~al.}, ``Compared to
  what? early brain overgrowth in autism and the perils of population norms,''
  \emph{Biological Psychiatry}, vol.~74, no.~8, pp. 563--575, 2013.

\bibitem{penzol2019functional}
M.~J. Penzol, G.~Salazar~de Pablo, C.~Llorente, C.~Moreno, P.~Hern{\'a}ndez,
  M.~L. Dorado, and M.~Parellada, ``Functional gastrointestinal disease in
  autism spectrum disorder: a retrospective descriptive study in a clinical
  sample,'' \emph{Frontiers in psychiatry}, vol.~10, p. 179, 2019.

\bibitem{lotrich2009risk}
F.~E. Lotrich, R.~E. Ferrell, M.~Rabinovitz, and B.~G. Pollock, ``Risk for
  depression during interferon-alpha treatment is affected by the serotonin
  transporter polymorphism,'' \emph{Biological psychiatry}, vol.~65, no.~4, pp.
  344--348, 2009.

\bibitem{dowlati2010meta}
Y.~Dowlati, N.~Herrmann, W.~Swardfager, H.~Liu, L.~Sham, E.~K. Reim, and K.~L.
  Lanct{\^o}t, ``A meta-analysis of cytokines in major depression,''
  \emph{Biological psychiatry}, vol.~67, no.~5, pp. 446--457, 2010.

\bibitem{goldstein2009inflammation}
B.~I. Goldstein, D.~E. Kemp, J.~K. Soczynska, and R.~S. McIntyre,
  ``Inflammation and the phenomenology, pathophysiology, comorbidity, and
  treatment of bipolar disorder: a systematic review of the literature,''
  \emph{The Journal of clinical psychiatry}, vol.~70, no.~8, pp. 1078--1090,
  2009.

\bibitem{nuzzo2014inflammatory}
D.~Nuzzo, P.~Picone, L.~Caruana, S.~Vasto, A.~Barera, C.~Caruso, and
  M.~Di~Carlo, ``Inflammatory mediators as biomarkers in brain disorders,''
  \emph{Inflammation}, vol.~37, no.~3, pp. 639--648, 2014.

\bibitem{potvin2008inflammatory}
S.~Potvin, E.~Stip, A.~A. Sepehry, A.~Gendron, R.~Bah, and E.~Kouassi,
  ``Inflammatory cytokine alterations in schizophrenia: a systematic
  quantitative review,'' \emph{Biological psychiatry}, vol.~63, no.~8, pp.
  801--808, 2008.

\bibitem{whyatt2013sensory}
C.~Whyatt and C.~Craig, ``Sensory-motor problems in autism,'' \emph{Frontiers
  in integrative neuroscience}, vol.~7, p.~51, 2013.

\bibitem{gowen2013motor}
E.~Gowen and A.~Hamilton, ``Motor abilities in autism: a review using a
  computational context,'' \emph{Journal of autism and developmental
  disorders}, vol.~43, no.~2, pp. 323--344, 2013.

\bibitem{mosconi2015sensorimotor}
M.~W. Mosconi and J.~A. Sweeney, ``Sensorimotor dysfunctions as primary
  features of autism spectrum disorders,'' \emph{Science China Life Sciences},
  vol.~58, no.~10, pp. 1016--1023, 2015.

\bibitem{bhat2020motor}
A.~N. Bhat, ``Is motor impairment in autism spectrum disorder distinct from
  developmental coordination disorder? a report from the spark study,''
  \emph{Physical therapy}, vol. 100, no.~4, pp. 633--644, 2020.

\bibitem{robertson2017sensory}
C.~E. Robertson and S.~Baron-Cohen, ``Sensory perception in autism,''
  \emph{Nature Reviews Neuroscience}, vol.~18, no.~11, pp. 671--684, 2017.

\bibitem{balasco2020sensory}
L.~Balasco, G.~Provenzano, and Y.~Bozzi, ``Sensory abnormalities in autism
  spectrum disorders: a focus on the tactile domain, from genetic mouse models
  to the clinic,'' \emph{Frontiers in psychiatry}, vol.~10, p. 1016, 2020.

\bibitem{marco2011sensory}
E.~J. Marco, L.~B. Hinkley, S.~S. Hill, and S.~S. Nagarajan, ``Sensory
  processing in autism: a review of neurophysiologic findings,''
  \emph{Pediatric research}, vol.~69, no.~8, pp. 48--54, 2011.

\bibitem{tavassoli2014sensory}
T.~Tavassoli, R.~A. Hoekstra, and S.~Baron-Cohen, ``The sensory perception
  quotient (spq): development and validation of a new sensory questionnaire for
  adults with and without autism,'' \emph{Molecular autism}, vol.~5, no.~1, pp.
  1--10, 2014.

\bibitem{williams2008self}
J.~H. Williams, ``Self--other relations in social development and autism:
  multiple roles for mirror neurons and other brain bases,'' \emph{Autism
  Research}, vol.~1, no.~2, pp. 73--90, 2008.

\bibitem{iacoboni2006mirror}
M.~Iacoboni and M.~Dapretto, ``The mirror neuron system and the consequences of
  its dysfunction,'' \emph{Nature Reviews Neuroscience}, vol.~7, no.~12, pp.
  942--951, 2006.

\bibitem{hamilton2008emulation}
A.~F. d.~C. Hamilton, ``Emulation and mimicry for social interaction: a
  theoretical approach to imitation in autism,'' \emph{The Quarterly Journal of
  Experimental Psychology}, vol.~61, no.~1, pp. 101--115, 2008.

\bibitem{american2000diagnostic}
A.~American Psychiatric~Association \emph{et~al.}, \emph{Diagnostic and
  statistical manual of mental disorders, text revision (DSM-IV-TR)}.\hskip 1em
  plus 0.5em minus 0.4em\relax Washington, DC: American psychiatric association
  Washington, 2000.

\bibitem{hagberg2002clinical}
B.~Hagberg, ``Clinical manifestations and stages of rett syndrome,''
  \emph{Mental retardation and developmental disabilities research reviews},
  vol.~8, no.~2, pp. 61--65, 2002.

\bibitem{huppke2006very}
P.~Huppke, E.~M. Maier, A.~Warnke, C.~Brendel, F.~Laccone, and J.~G{\"a}rtner,
  ``Very mild cases of rett syndrome with skewed x inactivation,''
  \emph{Journal of medical genetics}, vol.~43, no.~10, pp. 814--816, 2006.

\bibitem{bienvenu2006molecular}
T.~Bienvenu and J.~Chelly, ``Molecular genetics of rett syndrome: when dna
  methylation goes unrecognized,'' \emph{Nature Reviews Genetics}, vol.~7,
  no.~6, pp. 415--426, 2006.

\bibitem{della2016mecp2}
F.~Della~Ragione, M.~Vacca, S.~Fioriniello, G.~Pepe, and M.~D'Esposito,
  ``Mecp2, a multi-talented modulator of chromatin architecture,''
  \emph{Briefings in functional genomics}, vol.~15, no.~6, pp. 420--431, 2016.

\bibitem{lasalle2009evolving}
J.~M. LaSalle and D.~H. Yasui, ``Evolving role of mecp2 in rett syndrome and
  autism,'' \emph{Epigenomics}, vol.~1, no.~1, pp. 119--130, 2009.

\bibitem{rarediseases}
\BIBentryALTinterwordspacing
NORD. Rare diseases. [Online]. Available:
  \url{https://rarediseases.org/rare-diseases/dup15q-syndrome/}
\BIBentrySTDinterwordspacing

\bibitem{finucane201615q}
B.~M. Finucane, L.~Lusk, D.~Arkilo, S.~Chamberlain, O.~Devinsky, S.~Dindot,
  S.~S. Jeste, J.~M. LaSalle, L.~T. Reiter, N.~C. Schanen \emph{et~al.}, ``15q
  duplication syndrome and related disorders,''
  \emph{GeneReviews{\textregistered}[Internet]}, 2016.

\bibitem{ortiz202115q}
E.~Ortiz-Prado, A.~L. Iturralde, K.~Simba{\~n}a-Rivera, L.~G{\'o}mez-Barreno,
  I.~Hidalgo, M.~Rubio-Neira, N.~Espinosa, J.~Izquierdo-Condoy, M.~E.
  Arteaga-Espinosa, A.~Lister \emph{et~al.}, ``15q duplication syndrome: Report
  on the first patient from ecuador with an unusual clinical presentation,''
  \emph{Case Reports in Medicine}, vol. 2021, 2021.

\bibitem{mao2000characteristics}
R.~Mao, S.~M. Jalal, K.~Snow, V.~V. Michels, S.~M. Szabo, and
  D.~Babovic-Vuksanovic, ``Characteristics of two cases with dup (15)(q 11.2-q
  12): one of maternal and one of paternal origin,'' \emph{Genetics in
  Medicine}, vol.~2, no.~2, pp. 131--135, 2000.

\bibitem{frohlich2016quantitative}
J.~Frohlich, D.~Senturk, V.~Saravanapandian, P.~Golshani, L.~T. Reiter,
  R.~Sankar, R.~L. Thibert, C.~DiStefano, S.~Huberty, E.~H. Cook \emph{et~al.},
  ``A quantitative electrophysiological biomarker of duplication 15q11. 2-q13.
  1 syndrome,'' \emph{PloS one}, vol.~11, no.~12, p. e0167179, 2016.

\bibitem{pizzo2019rare}
L.~Pizzo, M.~Jensen, A.~Polyak, J.~A. Rosenfeld, K.~Mannik, A.~Krishnan,
  E.~McCready, O.~Pichon, C.~Le~Caignec, A.~Van~Dijck \emph{et~al.}, ``Rare
  variants in the genetic background modulate cognitive and developmental
  phenotypes in individuals carrying disease-associated variants,''
  \emph{Genetics in Medicine}, vol.~21, no.~4, pp. 816--825, 2019.

\bibitem{na2013impact}
E.~S. Na, E.~D. Nelson, E.~T. Kavalali, and L.~M. Monteggia, ``The impact of
  mecp2 loss-or gain-of-function on synaptic plasticity,''
  \emph{Neuropsychopharmacology}, vol.~38, no.~1, pp. 212--219, 2013.

\bibitem{cassidy2000prader}
S.~B. Cassidy, E.~Dykens, and C.~A. Williams, ``Prader-willi and angelman
  syndromes: Sister imprinted disorders,'' \emph{American journal of medical
  genetics}, vol.~97, no.~2, pp. 136--146, 2000.

\bibitem{zaky2017autism}
E.~Zaky, ``Autism spectrum disorder (asd),'' \emph{The Past, The Present, and
  The Future. J Child Adolesc Behav}, vol.~5, p. e116, 2017.

\bibitem{lombardo2019big}
M.~V. Lombardo, M.-C. Lai, and S.~Baron-Cohen, ``Big data approaches to
  decomposing heterogeneity across the autism spectrum,'' \emph{Molecular
  psychiatry}, vol.~24, no.~10, pp. 1435--1450, 2019.

\bibitem{faro2006functional}
S.~H. Faro and F.~B. Mohamed, \emph{Functional MRI: basic principles and
  clinical applications}.\hskip 1em plus 0.5em minus 0.4em\relax Springer
  Science \& Business Media, 2006.

\bibitem{jezzard1999sources}
P.~Jezzard and S.~Clare, ``Sources of distortion in functional mri data,''
  \emph{Human brain mapping}, vol.~8, no. 2-3, pp. 80--85, 1999.

\bibitem{deoni2010quantitative}
S.~C. Deoni, ``Quantitative relaxometry of the brain,'' \emph{Topics in
  magnetic resonance imaging: TMRI}, vol.~21, no.~2, p. 101, 2010.

\bibitem{basser2011microstructural}
P.~J. Basser and C.~Pierpaoli, ``Microstructural and physiological features of
  tissues elucidated by quantitative-diffusion-tensor mri,'' \emph{Journal of
  magnetic resonance}, vol. 213, no.~2, pp. 560--570, 2011.

\bibitem{mosconi2006structural}
M.~Mosconi, L.~Zwaigenbaum, and J.~Piven, ``Structural mri in autism: findings
  and future directions,'' \emph{Clinical Neuroscience Research}, vol.~6, no.
  3-4, pp. 135--144, 2006.

\bibitem{goodfellow2016machine}
I.~Goodfellow, Y.~Bengio, and A.~Courville, ``Machine learning basics,''
  \emph{Deep learning}, vol.~1, pp. 98--164, 2016.

\bibitem{deng2014deep}
L.~Deng and D.~Yu, ``Deep learning: methods and applications,''
  \emph{Foundations and trends in signal processing}, vol.~7, no. 3--4, pp.
  197--387, 2014.

\bibitem{scheinost2019ten}
D.~Scheinost, S.~Noble, C.~Horien, A.~S. Greene, E.~M. Lake, M.~Salehi, S.~Gao,
  X.~Shen, D.~O'Connor, D.~S. Barron \emph{et~al.}, ``Ten simple rules for
  predictive modeling of individual differences in neuroimaging,''
  \emph{NeuroImage}, vol. 193, pp. 35--45, 2019.

\bibitem{quaak2021deep}
M.~Quaak, L.~van~de Mortel, R.~M. Thomas, and G.~van Wingen, ``Deep learning
  applications for the classification of psychiatric disorders using
  neuroimaging data: systematic review and meta-analysis,'' \emph{NeuroImage:
  Clinical}, vol.~30, 2021.

\bibitem{biswas2020prediction}
S.~D. Biswas, R.~Chakraborty, and A.~Pramanik, ``On prediction models for the
  detection of autism spectrum disorder,'' in \emph{Computational Intelligence
  in Pattern Recognition}.\hskip 1em plus 0.5em minus 0.4em\relax Springer,
  2020, pp. 359--371.

\bibitem{khodatars2021deep}
M.~Khodatars, A.~Shoeibi, D.~Sadeghi, N.~Ghaasemi, M.~Jafari, P.~Moridian,
  A.~Khadem, R.~Alizadehsani, A.~Zare, Y.~Kong \emph{et~al.}, ``Deep learning
  for neuroimaging-based diagnosis and rehabilitation of autism spectrum
  disorder: A review,'' \emph{Computers in Biology and Medicine}, vol. 139, p.
  104949, 2021.

\bibitem{fischl2012freesurfer}
B.~Fischl, ``Freesurfer,'' \emph{Neuroimage}, vol.~62, no.~2, pp. 774--781,
  2012.

\bibitem{penny2011statistical}
W.~D. Penny, K.~J. Friston, J.~T. Ashburner, S.~J. Kiebel, and T.~E. Nichols,
  \emph{Statistical parametric mapping: the analysis of functional brain
  images}.\hskip 1em plus 0.5em minus 0.4em\relax Elsevier, 2011.

\bibitem{jenkinson2012fsl}
M.~Jenkinson, C.~F. Beckmann, T.~E. Behrens, M.~W. Woolrich, and S.~M. Smith,
  ``Fsl,'' \emph{Neuroimage}, vol.~62, no.~2, pp. 782--790, 2012.

\bibitem{ferrari2020dealing}
E.~Ferrari, P.~Bosco, S.~Calderoni, P.~Oliva, L.~Palumbo, G.~Spera, M.~E.
  Fantacci, and A.~Retico, ``Dealing with confounders and outliers in
  classification medical studies: The autism spectrum disorders case study,''
  \emph{Artificial Intelligence in Medicine}, vol. 108, p. 101926, 2020.

\bibitem{hrdlicka2005subtypes}
M.~Hrdlicka, I.~Dudova, I.~Beranova, J.~Lisy, T.~Belsan, J.~Neuwirth,
  V.~Komarek, L.~Faladova, M.~Havlovicova, Z.~Sedlacek \emph{et~al.},
  ``Subtypes of autism by cluster analysis based on structural mri data,''
  \emph{European child \& adolescent psychiatry}, vol.~14, no.~3, pp. 138--144,
  2005.

\bibitem{van2018cortical}
D.~Van~Rooij, E.~Anagnostou, C.~Arango, G.~Auzias, M.~Behrmann, G.~F. Busatto,
  S.~Calderoni, E.~Daly, C.~Deruelle, A.~Di~Martino \emph{et~al.}, ``Cortical
  and subcortical brain morphometry differences between patients with autism
  spectrum disorder and healthy individuals across the lifespan: results from
  the enigma asd working group,'' \emph{American Journal of Psychiatry}, vol.
  175, no.~4, pp. 359--369, 2018.

\bibitem{monk2009abnormalities}
C.~S. Monk, S.~J. Peltier, J.~L. Wiggins, S.-J. Weng, M.~Carrasco, S.~Risi, and
  C.~Lord, ``Abnormalities of intrinsic functional connectivity in autism
  spectrum disorders,'' \emph{Neuroimage}, vol.~47, no.~2, pp. 764--772, 2009.

\bibitem{rausch2016altered}
A.~Rausch, W.~Zhang, K.~V. Haak, M.~Mennes, E.~J. Hermans, E.~van Oort, G.~van
  Wingen, C.~F. Beckmann, J.~K. Buitelaar, and W.~B. Groen, ``Altered
  functional connectivity of the amygdaloid input nuclei in adolescents and
  young adults with autism spectrum disorder: a resting state fmri study,''
  \emph{Molecular autism}, vol.~7, no.~1, pp. 1--13, 2016.

\bibitem{alaerts2016sex}
K.~Alaerts, S.~P. Swinnen, and N.~Wenderoth, ``Sex differences in autism: a
  resting-state fmri investigation of functional brain connectivity in males
  and females,'' \emph{Social cognitive and affective neuroscience}, vol.~11,
  no.~6, pp. 1002--1016, 2016.

\bibitem{goodfellow2013challenges}
I.~J. Goodfellow, D.~Erhan, P.~L. Carrier, A.~Courville, M.~Mirza, B.~Hamner,
  W.~Cukierski, Y.~Tang, D.~Thaler, D.-H. Lee \emph{et~al.}, ``Challenges in
  representation learning: A report on three machine learning contests,'' in
  \emph{International conference on neural information processing}.\hskip 1em
  plus 0.5em minus 0.4em\relax Springer, 2013, pp. 117--124.

\bibitem{mundhenk2016large}
T.~N. Mundhenk, G.~Konjevod, W.~A. Sakla, and K.~Boakye, ``A large contextual
  dataset for classification, detection and counting of cars with deep
  learning,'' in \emph{European Conference on Computer Vision}.\hskip 1em plus
  0.5em minus 0.4em\relax Springer, 2016, pp. 785--800.

\bibitem{hyde2019applications}
K.~K. Hyde, M.~N. Novack, N.~LaHaye, C.~Parlett-Pelleriti, R.~Anden, D.~R.
  Dixon, and E.~Linstead, ``Applications of supervised machine learning in
  autism spectrum disorder research: a review,'' \emph{Review Journal of Autism
  and Developmental Disorders}, vol.~6, no.~2, pp. 128--146, 2019.

\bibitem{rashid2020use}
M.~Rashid, H.~Singh, and V.~Goyal, ``The use of machine learning and deep
  learning algorithms in functional magnetic resonance imaging—a systematic
  review,'' \emph{Expert Systems}, vol.~37, no.~6, p. e12644, 2020.

\bibitem{suthaharan2016support}
S.~Suthaharan, ``Support vector machine,'' in \emph{Machine learning models and
  algorithms for big data classification}.\hskip 1em plus 0.5em minus
  0.4em\relax Springer, 2016, pp. 207--235.

\bibitem{ingalhalikar2021functional}
M.~Ingalhalikar, S.~Shinde, A.~Karmarkar, A.~Rajan, D.~Rangaprakash, and
  G.~Deshpande, ``Functional connectivity-based prediction of autism on site
  harmonized abide dataset,'' \emph{IEEE Transactions on Biomedical
  Engineering}, 2021.

\bibitem{krizhevsky2012imagenet}
A.~Krizhevsky, I.~Sutskever, and G.~E. Hinton, ``Imagenet classification with
  deep convolutional neural networks,'' \emph{Advances in neural information
  processing systems}, vol.~25, pp. 1097--1105, 2012.

\bibitem{ciregan2012multi}
D.~Ciregan, U.~Meier, and J.~Schmidhuber, ``Multi-column deep neural networks
  for image classification,'' in \emph{2012 IEEE conference on computer vision
  and pattern recognition}.\hskip 1em plus 0.5em minus 0.4em\relax IEEE, 2012,
  pp. 3642--3649.

\bibitem{esteva2017dermatologist}
A.~Esteva, B.~Kuprel, R.~A. Novoa, J.~Ko, S.~M. Swetter, H.~M. Blau, and
  S.~Thrun, ``Dermatologist-level classification of skin cancer with deep
  neural networks,'' \emph{nature}, vol. 542, no. 7639, pp. 115--118, 2017.

\bibitem{taigman2014deepface}
Y.~Taigman, M.~Yang, M.~Ranzato, and L.~Wolf, ``Deepface: Closing the gap to
  human-level performance in face verification,'' in \emph{Proceedings of the
  IEEE conference on computer vision and pattern recognition}, 2014, pp.
  1701--1708.

\bibitem{he2015delving}
K.~He, X.~Zhang, S.~Ren, and J.~Sun, ``Delving deep into rectifiers: Surpassing
  human-level performance on imagenet classification,'' in \emph{Proceedings of
  the IEEE international conference on computer vision}, 2015, pp. 1026--1034.

\bibitem{lecun1998gradient}
Y.~LeCun, L.~Bottou, Y.~Bengio, and P.~Haffner, ``Gradient-based learning
  applied to document recognition,'' \emph{Proceedings of the IEEE}, vol.~86,
  no.~11, pp. 2278--2324, 1998.

\bibitem{sherstinsky2020fundamentals}
A.~Sherstinsky, ``Fundamentals of recurrent neural network (rnn) and long
  short-term memory (lstm) network,'' \emph{Physica D: Nonlinear Phenomena},
  vol. 404, p. 132306, 2020.

\bibitem{meng2017relational}
Q.~Meng, D.~Catchpoole, D.~Skillicom, and P.~J. Kennedy, ``Relational
  autoencoder for feature extraction,'' in \emph{2017 International Joint
  Conference on Neural Networks (IJCNN)}.\hskip 1em plus 0.5em minus
  0.4em\relax IEEE, 2017, pp. 364--371.

\bibitem{wang2016auto}
Y.~Wang, H.~Yao, and S.~Zhao, ``Auto-encoder based dimensionality reduction,''
  \emph{Neurocomputing}, vol. 184, pp. 232--242, 2016.

\bibitem{khosla2019ensemble}
M.~Khosla, K.~Jamison, A.~Kuceyeski, and M.~R. Sabuncu, ``Ensemble learning
  with 3d convolutional neural networks for functional connectome-based
  prediction,'' \emph{NeuroImage}, vol. 199, pp. 651--662, 2019.

\bibitem{heinsfeld2018identification}
A.~S. Heinsfeld, A.~R. Franco, R.~C. Craddock, A.~Buchweitz, and F.~Meneguzzi,
  ``Identification of autism spectrum disorder using deep learning and the
  abide dataset,'' \emph{NeuroImage: Clinical}, vol.~17, pp. 16--23, 2018.

\bibitem{ghiassian2016using}
S.~Ghiassian, R.~Greiner, P.~Jin, and M.~R. Brown, ``Using functional or
  structural magnetic resonance images and personal characteristic data to
  identify adhd and autism,'' \emph{PloS one}, vol.~11, no.~12, p. e0166934,
  2016.

\bibitem{arbabshirani2017single}
M.~R. Arbabshirani, S.~Plis, J.~Sui, and V.~D. Calhoun, ``Single subject
  prediction of brain disorders in neuroimaging: Promises and pitfalls,''
  \emph{Neuroimage}, vol. 145, pp. 137--165, 2017.

\bibitem{auzias2016influence}
G.~Auzias, S.~Takerkart, and C.~Deruelle, ``On the influence of confounding
  factors in multisite brain morphometry studies of developmental pathologies:
  Application to autism spectrum disorder,'' \emph{IEEE journal of biomedical
  and health informatics}, vol.~20, no.~3, pp. 810--817, 2016.

\bibitem{xiaoxiao2018channel}
X.~Li, N.~C. Dvornek, X.~Papademetris, J.~Zhuang, L.~H. Staib, P.~Ventola, and
  J.~S. Duncan, ``2-channel convolutional 3d deep neural network (2cc3d) for
  fmri analysis: Asd classification and feature learning,'' in \emph{2018 IEEE
  15th International Symposium on Biomedical Imaging (ISBI 2018)}, 2018, pp.
  1252--1255.

\bibitem{ferrari2020measuring}
E.~Ferrari, A.~Retico, and D.~Bacciu, ``Measuring the effects of confounders in
  medical supervised classification problems: the confounding index (ci),''
  \emph{Artificial Intelligence in Medicine}, vol. 103, p. 101804, 2020.

\bibitem{stanfield2008towards}
A.~C. Stanfield, A.~M. McIntosh, M.~D. Spencer, R.~Philip, S.~Gaur, and S.~M.
  Lawrie, ``Towards a neuroanatomy of autism: a systematic review and
  meta-analysis of structural magnetic resonance imaging studies,''
  \emph{European Psychiatry}, vol.~23, no.~4, pp. 289--299, 2008.

\bibitem{cody2002structural}
H.~Cody, K.~Pelphrey, and J.~Piven, ``Structural and functional magnetic
  resonance imaging of autism,'' \emph{International Journal of Developmental
  Neuroscience}, vol.~20, no. 3-5, pp. 421--438, 2002.

\bibitem{courchesne2007mapping}
E.~Courchesne, K.~Pierce, C.~M. Schumann, E.~Redcay, J.~A. Buckwalter, D.~P.
  Kennedy, and J.~Morgan, ``Mapping early brain development in autism,''
  \emph{Neuron}, vol.~56, no.~2, pp. 399--413, 2007.

\bibitem{haar2014anatomical}
S.~Haar, S.~Berman, M.~Behrmann, and I.~Dinstein, ``Anatomical abnormalities in
  autism?'' \emph{Cerebral cortex}, vol.~26, no.~4, pp. 1440--1452, 2014.

\bibitem{anagnostou2011review}
E.~Anagnostou and M.~J. Taylor, ``Review of neuroimaging in autism spectrum
  disorders: what have we learned and where we go from here,'' \emph{Molecular
  autism}, vol.~2, no.~1, p.~4, 2011.

\bibitem{abraham2015learning}
A.~Abraham, ``Learning functional brain atlases modeling inter-subject
  variability,'' Ph.D. dissertation, Universit{\'e} Paris-Saclay, 2015.

\bibitem{courchesne2011brain}
E.~Courchesne, K.~Campbell, and S.~Solso, ``Brain growth across the life span
  in autism: age-specific changes in anatomical pathology,'' \emph{Brain
  research}, vol. 1380, pp. 138--145, 2011.

\bibitem{shen2018extra}
M.~D. Shen, C.~W. Nordahl, D.~D. Li, A.~Lee, K.~Angkustsiri, R.~W. Emerson,
  S.~J. Rogers, S.~Ozonoff, and D.~G. Amaral, ``Extra-axial cerebrospinal fluid
  in high-risk and normal-risk children with autism aged 2--4 years: a
  case-control study,'' \emph{The Lancet Psychiatry}, vol.~5, no.~11, pp.
  895--904, 2018.

\bibitem{wang2019identification}
C.~Wang, Z.~Xiao, B.~Wang, and J.~Wu, ``Identification of autism based on
  svm-rfe and stacked sparse auto-encoder,'' \emph{IEEE Access}, vol.~7, pp.
  118\,030--118\,036, 2019.

\bibitem{ahmed2020single}
M.~R. Ahmed, Y.~Zhang, Y.~Liu, and H.~Liao, ``Single volume image generator and
  deep learning-based asd classification,'' \emph{IEEE Journal of Biomedical
  and Health Informatics}, vol.~24, no.~11, pp. 3044--3054, 2020.

\bibitem{sen2018general}
B.~Sen, N.~C. Borle, R.~Greiner, and M.~R. Brown, ``A general prediction model
  for the detection of adhd and autism using structural and functional mri,''
  \emph{PloS one}, vol.~13, no.~4, p. e0194856, 2018.

\bibitem{eslami2019asd}
T.~Eslami, V.~Mirjalili, A.~Fong, A.~R. Laird, and F.~Saeed, ``Asd-diagnet: a
  hybrid learning approach for detection of autism spectrum disorder using fmri
  data,'' \emph{Frontiers in neuroinformatics}, vol.~13, p.~70, 2019.

\bibitem{sharif2019novel}
H.~Sharif and R.~A. Khan, ``A novel machine learning based framework for
  detection of autism spectrum disorder (asd),'' \emph{arXiv preprint
  arXiv:1903.11323}, 2019.

\bibitem{xing2018convolutional}
X.~Xing, J.~Ji, and Y.~Yao, ``Convolutional neural network with element-wise
  filters to extract hierarchical topological features for brain networks,'' in
  \emph{2018 IEEE International Conference on Bioinformatics and Biomedicine
  (BIBM)}.\hskip 1em plus 0.5em minus 0.4em\relax IEEE, 2018, pp. 780--783.

\bibitem{rane2017developing}
S.~Rane, E.~Jolly, A.~Park, H.~Jang, and C.~Craddock, ``Developing predictive
  imaging biomarkers using whole-brain classifiers: Application to the abide i
  dataset,'' \emph{Research Ideas and Outcomes}, vol.~3, p. e12733, 2017.

\bibitem{leming2020ensemble}
M.~Leming, J.~M. G{\'o}rriz, and J.~Suckling, ``Ensemble deep learning on
  large, mixed-site fmri datasets in autism and other tasks,'' \emph{arXiv
  preprint arXiv:2002.07874}, 2020.

\bibitem{dvornek2017identifying}
N.~C. Dvornek, P.~Ventola, K.~A. Pelphrey, and J.~S. Duncan, ``Identifying
  autism from resting-state fmri using long short-term memory networks,'' in
  \emph{International Workshop on Machine Learning in Medical Imaging}.\hskip
  1em plus 0.5em minus 0.4em\relax Springer, 2017, pp. 362--370.

\bibitem{el2019simple}
A.~El~Gazzar, L.~Cerliani, G.~van Wingen, and R.~M. Thomas, ``Simple 1-d
  convolutional networks for resting-state fmri based classification in
  autism,'' in \emph{2019 International Joint Conference on Neural Networks
  (IJCNN)}.\hskip 1em plus 0.5em minus 0.4em\relax IEEE, 2019, pp. 1--6.

\bibitem{sherkatghanad2020automated}
Z.~Sherkatghanad, M.~Akhondzadeh, S.~Salari, M.~Zomorodi-Moghadam, M.~Abdar,
  U.~R. Acharya, R.~Khosrowabadi, and V.~Salari, ``Automated detection of
  autism spectrum disorder using a convolutional neural network,''
  \emph{Frontiers in neuroscience}, vol.~13, p. 1325, 2020.

\bibitem{sairam2019computer}
K.~Sairam, J.~Naren, G.~Vithya, and S.~Srivathsan, ``Computer aided system for
  autism spectrum disorder using deep learning methods,'' \emph{International
  Journal of Psychosocial Rehabilitation}, vol.~23, no.~01, 2019.

\bibitem{shahamat2020brain}
H.~Shahamat and M.~S. Abadeh, ``Brain mri analysis using a deep learning based
  evolutionary approach,'' \emph{Neural Networks}, vol. 126, pp. 218--234,
  2020.

\bibitem{thomas2020classifying}
R.~M. Thomas, S.~Gallo, L.~Cerliani, P.~Zhutovsky, A.~El-Gazzar, and G.~van
  Wingen, ``Classifying autism spectrum disorder using the temporal statistics
  of resting-state functional mri data with 3d convolutional neural networks,''
  \emph{Frontiers in Psychiatry}, vol.~11, p. 440, 2020.

\bibitem{jiao2020improving}
Z.~Jiao, H.~Li, and Y.~Fan, ``Improving diagnosis of autism spectrum disorder
  and disentangling its heterogeneous functional connectivity patterns using
  capsule networks,'' in \emph{2020 IEEE 17th International Symposium on
  Biomedical Imaging (ISBI)}.\hskip 1em plus 0.5em minus 0.4em\relax IEEE,
  2020, pp. 1331--1334.

\bibitem{eslami2021explainable}
T.~Eslami, J.~S. Raiker, and F.~Saeed, ``Explainable and scalable machine
  learning algorithms for detection of autism spectrum disorder using fmri
  data,'' in \emph{Neural Engineering Techniques for Autism Spectrum
  Disorder}.\hskip 1em plus 0.5em minus 0.4em\relax Elsevier, 2021, pp. 39--54.

\bibitem{bengs20204d}
M.~Bengs, N.~Gessert, and A.~Schlaefer, ``4d spatio-temporal deep learning with
  4d fmri data for autism spectrum disorder classification,'' \emph{arXiv
  preprint arXiv:2004.10165}, 2020.

\bibitem{abbas2018machine}
H.~Abbas, F.~Garberson, E.~Glover, and D.~P. Wall, ``Machine learning approach
  for early detection of autism by combining questionnaire and home video
  screening,'' \emph{Journal of the American Medical Informatics Association},
  vol.~25, no.~8, pp. 1000--1007, 2018.

\bibitem{carreira2017quo}
J.~Carreira and A.~Zisserman, ``Quo vadis, action recognition? a new model and
  the kinetics dataset,'' in \emph{proceedings of the IEEE Conference on
  Computer Vision and Pattern Recognition}, 2017, pp. 6299--6308.

\bibitem{adeli2021representation}
E.~Adeli, Q.~Zhao, A.~Pfefferbaum, E.~V. Sullivan, L.~Fei-Fei, J.~C. Niebles,
  and K.~M. Pohl, ``Representation learning with statistical independence to
  mitigate bias,'' in \emph{Proceedings of the IEEE/CVF Winter Conference on
  Applications of Computer Vision}, 2021, pp. 2513--2523.

\bibitem{smilkov2017smoothgrad}
D.~Smilkov, N.~Thorat, B.~Kim, F.~Vi{\'e}gas, and M.~Wattenberg, ``Smoothgrad:
  removing noise by adding noise,'' \emph{arXiv preprint arXiv:1706.03825},
  2017.

\bibitem{tanberk2020hybrid}
S.~Tanberk, Z.~H. Kilimci, D.~B. T{\"u}kel, M.~Uysal, and S.~Akyoku{\c{s}}, ``A
  hybrid deep model using deep learning and dense optical flow approaches for
  human activity recognition,'' \emph{IEEE Access}, vol.~8, pp.
  19\,799--19\,809, 2020.

\bibitem{zach2007duality}
C.~Zach, T.~Pock, and H.~Bischof, ``A duality based approach for realtime tv-l
  1 optical flow,'' in \emph{Joint pattern recognition symposium}.\hskip 1em
  plus 0.5em minus 0.4em\relax Springer, 2007, pp. 214--223.

\bibitem{geiger2012we}
A.~Geiger, P.~Lenz, and R.~Urtasun, ``Are we ready for autonomous driving? the
  kitti vision benchmark suite,'' in \emph{2012 IEEE conference on computer
  vision and pattern recognition}.\hskip 1em plus 0.5em minus 0.4em\relax IEEE,
  2012, pp. 3354--3361.

\bibitem{szeliski2010computer}
R.~Szeliski, \emph{Computer vision: algorithms and applications}.\hskip 1em
  plus 0.5em minus 0.4em\relax Springer Science \& Business Media, 2010.

\bibitem{yue2015beyond}
J.~Yue-Hei~Ng, M.~Hausknecht, S.~Vijayanarasimhan, O.~Vinyals, R.~Monga, and
  G.~Toderici, ``Beyond short snippets: Deep networks for video
  classification,'' in \emph{Proceedings of the IEEE conference on computer
  vision and pattern recognition}, 2015, pp. 4694--4702.

\bibitem{goodale1992separate}
M.~A. Goodale and A.~D. Milner, ``Separate visual pathways for perception and
  action,'' \emph{Trends in neurosciences}, vol.~15, no.~1, pp. 20--25, 1992.

\bibitem{simonyan2014two}
K.~Simonyan and A.~Zisserman, ``Two-stream convolutional networks for action
  recognition in videos,'' \emph{arXiv preprint arXiv:1406.2199}, 2014.

\bibitem{neto2019detecting}
E.~C. Neto, A.~Pratap, T.~M. Perumal, M.~Tummalacherla, P.~Snyder, B.~M. Bot,
  A.~D. Trister, S.~H. Friend, L.~Mangravite, and L.~Omberg, ``Detecting the
  impact of subject characteristics on machine learning-based diagnostic
  applications,'' \emph{NPJ digital medicine}, vol.~2, no.~1, pp. 1--6, 2019.

\bibitem{mehrabi2021survey}
N.~Mehrabi, F.~Morstatter, N.~Saxena, K.~Lerman, and A.~Galstyan, ``A survey on
  bias and fairness in machine learning,'' \emph{ACM Computing Surveys (CSUR)},
  vol.~54, no.~6, pp. 1--35, 2021.

\bibitem{ghassami2018fairness}
A.~Ghassami, S.~Khodadadian, and N.~Kiyavash, ``Fairness in supervised
  learning: An information theoretic approach,'' in \emph{2018 IEEE
  International Symposium on Information Theory (ISIT)}.\hskip 1em plus 0.5em
  minus 0.4em\relax IEEE, 2018, pp. 176--180.

\bibitem{singh2018fairness}
A.~Singh and T.~Joachims, ``Fairness of exposure in rankings,'' in
  \emph{Proceedings of the 24th ACM SIGKDD International Conference on
  Knowledge Discovery \& Data Mining}, 2018, pp. 2219--2228.

\bibitem{blum2019recovering}
A.~Blum and K.~Stangl, ``Recovering from biased data: Can fairness constraints
  improve accuracy?'' \emph{arXiv preprint arXiv:1912.01094}, 2019.

\bibitem{kukar1998cost}
M.~Kukar, I.~Kononenko \emph{et~al.}, ``Cost-sensitive learning with neural
  networks.'' in \emph{ECAI}, vol.~15, no.~27.\hskip 1em plus 0.5em minus
  0.4em\relax Citeseer, 1998, pp. 88--94.

\bibitem{mandal2020ensuring}
D.~Mandal, S.~Deng, S.~Jana, J.~M. Wing, and D.~Hsu, ``Ensuring fairness beyond
  the training data,'' \emph{arXiv preprint arXiv:2007.06029}, 2020.

\bibitem{ferrari2021addressing}
E.~Ferrari and D.~Bacciu, ``Addressing fairness, bias and class imbalance in
  machine learning: the fbi-loss,'' \emph{arXiv preprint arXiv:2105.06345},
  2021.

\bibitem{zhang2018mitigating}
B.~H. Zhang, B.~Lemoine, and M.~Mitchell, ``Mitigating unwanted biases with
  adversarial learning,'' in \emph{Proceedings of the 2018 AAAI/ACM Conference
  on AI, Ethics, and Society}, 2018, pp. 335--340.

\bibitem{elazar2018adversarial}
Y.~Elazar and Y.~Goldberg, ``Adversarial removal of demographic attributes from
  text data,'' \emph{arXiv preprint arXiv:1808.06640}, 2018.

\bibitem{wang2019balanced}
T.~Wang, J.~Zhao, M.~Yatskar, K.-W. Chang, and V.~Ordonez, ``Balanced datasets
  are not enough: Estimating and mitigating gender bias in deep image
  representations,'' in \emph{2019 IEEE/CVF International Conference on
  Computer Vision (ICCV)}.\hskip 1em plus 0.5em minus 0.4em\relax IEEE, pp.
  5309--5318.

\bibitem{sadeghi2019global}
B.~Sadeghi, R.~Yu, and V.~Boddeti, ``On the global optima of kernelized
  adversarial representation learning,'' in \emph{Proceedings of the IEEE/CVF
  International Conference on Computer Vision}, 2019, pp. 7971--7979.

\bibitem{adeli_git}
\BIBentryALTinterwordspacing
Q.~Zhao. Git repository. GitHub. [Online]. Available:
  \url{https://github.com/QingyuZhao/BR-Net}
\BIBentrySTDinterwordspacing

\bibitem{linardatos2021explainable}
P.~Linardatos, V.~Papastefanopoulos, and S.~Kotsiantis, ``Explainable ai: A
  review of machine learning interpretability methods,'' \emph{Entropy},
  vol.~23, no.~1, p.~18, 2021.

\bibitem{snustad2015principles}
D.~P. Snustad and M.~J. Simmons, \emph{Principles of genetics}.\hskip 1em plus
  0.5em minus 0.4em\relax John Wiley \& Sons, 2015.

\bibitem{lander1994genetic}
E.~S. Lander and N.~J. Schork, ``Genetic dissection of complex traits,''
  \emph{Science}, vol. 265, no. 5181, pp. 2037--2048, 1994.

\bibitem{josephs2021gene}
E.~B. Josephs, ``Gene expression links genotype and phenotype during rapid
  adaptation,'' 2021.

\bibitem{crick1970central}
F.~Crick, ``Central dogma of molecular biology,'' \emph{Nature}, vol. 227, no.
  5258, pp. 561--563, 1970.

\bibitem{collins2004finishing}
F.~Collins, E.~Lander, J.~Rogers, R.~Waterston, and I.~Conso, ``Finishing the
  euchromatic sequence of the human genome,'' \emph{Nature}, vol. 431, no.
  7011, pp. 931--945, 2004.

\bibitem{wang2015mechanism}
Y.~Wang, J.~Liu, B.~Huang, Y.-M. Xu, J.~Li, L.-F. Huang, J.~Lin, J.~Zhang,
  Q.-H. Min, W.-M. Yang \emph{et~al.}, ``Mechanism of alternative splicing and
  its regulation,'' \emph{Biomedical reports}, vol.~3, no.~2, pp. 152--158,
  2015.

\bibitem{berg2002biochemistry}
J.~M. Berg, J.~L. Tymoczko, L.~Stryer \emph{et~al.}, ``Biochemistry,'' 2002.

\bibitem{mattick2006non}
J.~S. Mattick and I.~V. Makunin, ``Non-coding rna,'' \emph{Human molecular
  genetics}, vol.~15, no. suppl\_1, pp. R17--R29, 2006.

\bibitem{statello2021gene}
L.~Statello, C.-J. Guo, L.-L. Chen, and M.~Huarte, ``Gene regulation by long
  non-coding rnas and its biological functions,'' \emph{Nature Reviews
  Molecular Cell Biology}, vol.~22, no.~2, pp. 96--118, 2021.

\bibitem{saxena2011long}
A.~Saxena and P.~Carninci, ``Long non-coding rna modifies chromatin: epigenetic
  silencing by long non-coding rnas,'' \emph{Bioessays}, vol.~33, no.~11, pp.
  830--839, 2011.

\bibitem{bartel2018metazoan}
D.~P. Bartel, ``Metazoan micrornas,'' \emph{Cell}, vol. 173, no.~1, pp. 20--51,
  2018.

\bibitem{scott2000transcripts}
H.~Scott, ``What transcripts are found in a human cell?'' \emph{Genome
  biology}, vol.~1, no.~1, pp. 1--4, 2000.

\bibitem{schroeder2006rin}
A.~Schroeder, O.~Mueller, S.~Stocker, R.~Salowsky, M.~Leiber, M.~Gassmann,
  S.~Lightfoot, W.~Menzel, M.~Granzow, and T.~Ragg, ``The rin: an rna integrity
  number for assigning integrity values to rna measurements,'' \emph{BMC
  molecular biology}, vol.~7, no.~1, pp. 1--14, 2006.

\bibitem{slomovic2006polyadenylation}
S.~Slomovic, D.~Laufer, D.~Geiger, and G.~Schuster, ``Polyadenylation of
  ribosomal rna in human cells,'' \emph{Nucleic acids research}, vol.~34,
  no.~10, pp. 2966--2975, 2006.

\bibitem{yang2011genomewide}
L.~Yang, M.~O. Duff, B.~R. Graveley, G.~G. Carmichael, and L.-L. Chen,
  ``Genomewide characterization of non-polyadenylated rnas,'' \emph{Genome
  biology}, vol.~12, no.~2, pp. 1--14, 2011.

\bibitem{arezi2003amplification}
B.~Arezi, W.~Xing, J.~A. Sorge, and H.~H. Hogrefe, ``Amplification efficiency
  of thermostable dna polymerases,'' \emph{Analytical biochemistry}, vol. 321,
  no.~2, pp. 226--235, 2003.

\bibitem{degrelle2008amplification}
S.~A. Degrelle, C.~Hennequet-Antier, H.~Chiapello, K.~Piot-Kaminski, F.~Piumi,
  S.~Robin, J.-P. Renard, and I.~Hue, ``Amplification biases: possible
  differences among deviating gene expressions,'' \emph{BMC genomics}, vol.~9,
  no.~1, pp. 1--15, 2008.

\bibitem{sudo2012use}
H.~Sudo, A.~Mizoguchi, J.~Kawauchi, H.~Akiyama, and S.~Takizawa, ``Use of
  non-amplified rna samples for microarray analysis of gene expression,''
  \emph{PLoS One}, vol.~7, no.~2, p. e31397, 2012.

\bibitem{sykacek2011impact}
P.~Sykacek, D.~P. Kreil, L.~A. Meadows, R.~P. Auburn, B.~Fischer, S.~Russell,
  and G.~Micklem, ``The impact of quantitative optimization of hybridization
  conditions on gene expression analysis,'' \emph{BMC bioinformatics}, vol.~12,
  no.~1, pp. 1--16, 2011.

\bibitem{koltai2008specificity}
H.~Koltai and C.~Weingarten-Baror, ``Specificity of dna microarray
  hybridization: characterization, effectors and approaches for data
  correction,'' \emph{Nucleic acids research}, vol.~36, no.~7, pp. 2395--2405,
  2008.

\bibitem{munier2014physicochemical}
M.~Munier, S.~Jubeau, A.~Wijaya, M.~Morancais, J.~Dumay, L.~Marchal, P.~Jaouen,
  and J.~Fleurence, ``Physicochemical factors affecting the stability of two
  pigments: R-phycoerythrin of grateloupia turuturu and b-phycoerythrin of
  porphyridium cruentum,'' \emph{Food chemistry}, vol. 150, pp. 400--407, 2014.

\bibitem{johnson2007adjusting}
W.~E. Johnson, C.~Li, and A.~Rabinovic, ``Adjusting batch effects in microarray
  expression data using empirical bayes methods,'' \emph{Biostatistics},
  vol.~8, no.~1, pp. 118--127, 2007.

\bibitem{kozarewa2009amplification}
I.~Kozarewa, Z.~Ning, M.~A. Quail, M.~J. Sanders, M.~Berriman, and D.~J.
  Turner, ``Amplification-free illumina sequencing-library preparation
  facilitates improved mapping and assembly of (g+ c)-biased genomes,''
  \emph{Nature methods}, vol.~6, no.~4, pp. 291--295, 2009.

\bibitem{zhao2014comparison}
S.~Zhao, W.-P. Fung-Leung, A.~Bittner, K.~Ngo, and X.~Liu, ``Comparison of
  rna-seq and microarray in transcriptome profiling of activated t cells,''
  \emph{PloS one}, vol.~9, no.~1, p. e78644, 2014.

\bibitem{wang2009rna}
Z.~Wang, M.~Gerstein, and M.~Snyder, ``Rna-seq: a revolutionary tool for
  transcriptomics,'' \emph{Nature reviews genetics}, vol.~10, no.~1, pp.
  57--63, 2009.

\bibitem{dillies2013comprehensive}
M.-A. Dillies, A.~Rau, J.~Aubert, C.~Hennequet-Antier, M.~Jeanmougin,
  N.~Servant, C.~Keime, G.~Marot, D.~Castel, J.~Estelle \emph{et~al.}, ``A
  comprehensive evaluation of normalization methods for illumina
  high-throughput rna sequencing data analysis,'' \emph{Briefings in
  bioinformatics}, vol.~14, no.~6, pp. 671--683, 2013.

\bibitem{robinson2010scaling}
M.~D. Robinson and A.~Oshlack, ``A scaling normalization method for
  differential expression analysis of rna-seq data,'' \emph{Genome biology},
  vol.~11, no.~3, pp. 1--9, 2010.

\bibitem{anders2010differential}
S.~Anders and W.~Huber, ``Differential expression analysis for sequence count
  data,'' \emph{Nature Precedings}, pp. 1--1, 2010.

\bibitem{cox2005delta}
C.~Cox, ``Delta method,'' \emph{Encyclopedia of biostatistics}, vol.~2, 2005.

\bibitem{jaskowiak2018clustering}
P.~A. Jaskowiak, I.~G. Costa, and R.~J. Campello, ``Clustering of rna-seq
  samples: Comparison study on cancer data,'' \emph{Methods}, vol. 132, pp.
  42--49, 2018.

\bibitem{bioconductor}
\BIBentryALTinterwordspacing
Bioconductor. Bioconductor guide. [Online]. Available:
  \url{https://www.bioconductor.org/packages/release/BiocViews.html#___DifferentialExpression}
\BIBentrySTDinterwordspacing

\bibitem{langfelder2013hub}
P.~Langfelder, P.~S. Mischel, and S.~Horvath, ``When is hub gene selection
  better than standard meta-analysis?'' \emph{PloS one}, vol.~8, no.~4, p.
  e61505, 2013.

\bibitem{horvath2006analysis}
S.~Horvath, B.~Zhang, M.~Carlson, K.~Lu, S.~Zhu, R.~Felciano, M.~Laurance,
  W.~Zhao, S.~Qi, Z.~Chen \emph{et~al.}, ``Analysis of oncogenic signaling
  networks in glioblastoma identifies aspm as a molecular target,''
  \emph{Proceedings of the National Academy of Sciences}, vol. 103, no.~46, pp.
  17\,402--17\,407, 2006.

\bibitem{carlson2006gene}
M.~R. Carlson, B.~Zhang, Z.~Fang, P.~S. Mischel, S.~Horvath, and S.~F. Nelson,
  ``Gene connectivity, function, and sequence conservation: predictions from
  modular yeast co-expression networks,'' \emph{BMC genomics}, vol.~7, no.~1,
  p.~40, 2006.

\bibitem{oldham2006conservation}
M.~C. Oldham, S.~Horvath, and D.~H. Geschwind, ``Conservation and evolution of
  gene coexpression networks in human and chimpanzee brains,''
  \emph{Proceedings of the National Academy of Sciences}, vol. 103, no.~47, pp.
  17\,973--17\,978, 2006.

\bibitem{keller2008gene}
M.~P. Keller, Y.~Choi, P.~Wang, D.~B. Davis, M.~E. Rabaglia, A.~T. Oler, D.~S.
  Stapleton, C.~Argmann, K.~L. Schueler, S.~Edwards \emph{et~al.}, ``A gene
  expression network model of type 2 diabetes links cell cycle regulation in
  islets with diabetes susceptibility,'' \emph{Genome research}, vol.~18,
  no.~5, pp. 706--716, 2008.

\bibitem{ivliev2010coexpression}
A.~E. Ivliev, P.~AC't~Hoen, and M.~G. Sergeeva, ``Coexpression network analysis
  identifies transcriptional modules related to proastrocytic differentiation
  and sprouty signaling in glioma,'' \emph{Cancer Research}, vol.~70, no.~24,
  pp. 10\,060--10\,070, 2010.

\bibitem{gargalovic2006identification}
P.~S. Gargalovic, M.~Imura, B.~Zhang, N.~M. Gharavi, M.~J. Clark, J.~Pagnon,
  W.-P. Yang, A.~He, A.~Truong, S.~Patel \emph{et~al.}, ``Identification of
  inflammatory gene modules based on variations of human endothelial cell
  responses to oxidized lipids,'' \emph{Proceedings of the National Academy of
  Sciences}, vol. 103, no.~34, pp. 12\,741--12\,746, 2006.

\bibitem{ghazalpour2006integrating}
A.~Ghazalpour, S.~Doss, B.~Zhang, S.~Wang, C.~Plaisier, R.~Castellanos,
  A.~Brozell, E.~E. Schadt, T.~A. Drake, A.~J. Lusis \emph{et~al.},
  ``Integrating genetic and network analysis to characterize genes related to
  mouse weight,'' \emph{PLoS Genet}, vol.~2, no.~8, p. e130, 2006.

\bibitem{fuller2007weighted}
T.~F. Fuller, A.~Ghazalpour, J.~E. Aten, T.~A. Drake, A.~J. Lusis, and
  S.~Horvath, ``Weighted gene coexpression network analysis strategies applied
  to mouse weight,'' \emph{Mammalian Genome}, vol.~18, no. 6-7, pp. 463--472,
  2007.

\bibitem{presson2008integrated}
A.~P. Presson, E.~M. Sobel, J.~C. Papp, C.~J. Suarez, T.~Whistler, M.~S.
  Rajeevan, S.~D. Vernon, and S.~Horvath, ``Integrated weighted gene
  co-expression network analysis with an application to chronic fatigue
  syndrome,'' \emph{BMC systems biology}, vol.~2, no.~1, p.~95, 2008.

\bibitem{langfelder2012systems}
P.~Langfelder, L.~W. Castellani, Z.~Zhou, E.~Paul, R.~Davis, E.~E. Schadt,
  A.~J. Lusis, S.~Horvath, and M.~Mehrabian, ``A systems genetic analysis of
  high density lipoprotein metabolism and network preservation across mouse
  models,'' \emph{Biochimica et Biophysica Acta (BBA)-Molecular and Cell
  Biology of Lipids}, vol. 1821, no.~3, pp. 435--447, 2012.

\bibitem{miller2010divergence}
J.~A. Miller, S.~Horvath, and D.~H. Geschwind, ``Divergence of human and mouse
  brain transcriptome highlights alzheimer disease pathways,''
  \emph{Proceedings of the National Academy of Sciences}, vol. 107, no.~28, pp.
  12\,698--12\,703, 2010.

\bibitem{dawson2012r}
J.~A. Dawson, S.~Ye, and C.~Kendziorski, ``R/ebcoexpress: an empirical bayesian
  framework for discovering differential co-expression,''
  \emph{Bioinformatics}, vol.~28, no.~14, pp. 1939--1940, 2012.

\bibitem{cellerino2018transcriptome}
A.~Cellerino and M.~Sanguanini, \emph{Transcriptome Analysis: Introduction and
  Examples from the Neurosciences}.\hskip 1em plus 0.5em minus 0.4em\relax
  Springer, 2018, vol.~17.

\bibitem{love2014moderated}
M.~I. Love, W.~Huber, and S.~Anders, ``Moderated estimation of fold change and
  dispersion for rna-seq data with deseq2,'' \emph{Genome biology}, vol.~15,
  no.~12, pp. 1--21, 2014.

\bibitem{hawinkel2020sequence}
S.~Hawinkel, J.~Rayner, L.~Bijnens, and O.~Thas, ``Sequence count data are
  poorly fit by the negative binomial distribution,'' \emph{PloS one}, vol.~15,
  no.~4, p. e0224909, 2020.

\bibitem{ritchie2015limma}
M.~E. Ritchie, B.~Phipson, D.~Wu, Y.~Hu, C.~W. Law, W.~Shi, and G.~K. Smyth,
  ``limma powers differential expression analyses for rna-sequencing and
  microarray studies,'' \emph{Nucleic acids research}, vol.~43, no.~7, pp.
  e47--e47, 2015.

\bibitem{pimentel2017differential}
H.~Pimentel, N.~L. Bray, S.~Puente, P.~Melsted, and L.~Pachter, ``Differential
  analysis of rna-seq incorporating quantification uncertainty,'' \emph{Nature
  methods}, vol.~14, no.~7, pp. 687--690, 2017.

\bibitem{jiang2007linear}
J.~Jiang and T.~Nguyen, \emph{Linear and generalized linear mixed models and
  their applications}.\hskip 1em plus 0.5em minus 0.4em\relax Springer, 2007,
  vol.~1.

\bibitem{gelman2005analysis}
A.~Gelman, ``Analysis of variance—why it is more important than ever,''
  \emph{The annals of statistics}, vol.~33, no.~1, pp. 1--53, 2005.

\bibitem{fox2015ecological}
G.~A. Fox, S.~Negrete-Yankelevich, and V.~J. Sosa, \emph{Ecological statistics:
  contemporary theory and application}.\hskip 1em plus 0.5em minus 0.4em\relax
  Oxford University Press, USA, 2015.

\bibitem{espin2018comparison}
A.~Esp{\'\i}n-P{\'e}rez, C.~Portier, M.~Chadeau-Hyam, K.~van Veldhoven, J.~C.
  Kleinjans, and T.~M. de~Kok, ``Comparison of statistical methods and the use
  of quality control samples for batch effect correction in human transcriptome
  data,'' \emph{PloS one}, vol.~13, no.~8, p. e0202947, 2018.

\bibitem{zhou2014efficient}
X.~Zhou and M.~Stephens, ``Efficient multivariate linear mixed model algorithms
  for genome-wide association studies,'' \emph{Nature methods}, vol.~11, no.~4,
  pp. 407--409, 2014.

\bibitem{jo2013statistical}
J.~Jo and U.~J. Chang, ``A statistical analysis of the fat mass repeated
  measures data using mixed model,'' \emph{Journal of the Korean Data and
  Information Science Society}, vol.~24, no.~2, pp. 303--310, 2013.

\bibitem{sullivan2012using}
G.~M. Sullivan and R.~Feinn, ``Using effect size—or why the p value is not
  enough,'' \emph{Journal of graduate medical education}, vol.~4, no.~3, pp.
  279--282, 2012.

\bibitem{lee2016alternatives}
D.~K. Lee, ``Alternatives to p value: confidence interval and effect size,''
  \emph{Korean journal of anesthesiology}, vol.~69, no.~6, p. 555, 2016.

\bibitem{marot2009moderated}
G.~Marot, J.-L. Foulley, C.-D. Mayer, and F.~Jaffr{\'e}zic, ``Moderated effect
  size and p-value combinations for microarray meta-analyses,''
  \emph{Bioinformatics}, vol.~25, no.~20, pp. 2692--2699, 2009.

\bibitem{goodman2019proposed}
W.~M. Goodman, S.~E. Spruill, and E.~Komaroff, ``A proposed hybrid effect size
  plus p-value criterion: empirical evidence supporting its use,'' \emph{The
  American Statistician}, vol.~73, no. sup1, pp. 168--185, 2019.

\bibitem{zhang2005general}
B.~Zhang and S.~Horvath, ``A general framework for weighted gene co-expression
  network analysis,'' \emph{Statistical applications in genetics and molecular
  biology}, vol.~4, no.~1, 2005.

\bibitem{barabasi1999emergence}
A.-L. Barab{\'a}si and R.~Albert, ``Emergence of scaling in random networks,''
  \emph{science}, vol. 286, no. 5439, pp. 509--512, 1999.

\bibitem{albert2000topology}
R.~Albert and A.-L. Barab{\'a}si, ``Topology of evolving networks: local events
  and universality,'' \emph{Physical review letters}, vol.~85, no.~24, p. 5234,
  2000.

\bibitem{ravasz2002hierarchical}
E.~Ravasz, A.~L. Somera, D.~A. Mongru, Z.~N. Oltvai, and A.-L. Barab{\'a}si,
  ``Hierarchical organization of modularity in metabolic networks,''
  \emph{science}, vol. 297, no. 5586, pp. 1551--1555, 2002.

\bibitem{d2005does}
P.~D'haeseleer, ``How does gene expression clustering work?'' \emph{Nature
  biotechnology}, vol.~23, no.~12, pp. 1499--1501, 2005.

\bibitem{yip2007gene}
A.~M. Yip and S.~Horvath, ``Gene network interconnectedness and the generalized
  topological overlap measure,'' \emph{BMC bioinformatics}, vol.~8, no.~1, pp.
  1--14, 2007.

\bibitem{kerr2008techniques}
G.~Kerr, H.~J. Ruskin, M.~Crane, and P.~Doolan, ``Techniques for clustering
  gene expression data,'' \emph{Computers in biology and medicine}, vol.~38,
  no.~3, pp. 283--293, 2008.

\bibitem{murtagh2012algorithms}
F.~Murtagh and P.~Contreras, ``Algorithms for hierarchical clustering: an
  overview,'' \emph{Wiley Interdisciplinary Reviews: Data Mining and Knowledge
  Discovery}, vol.~2, no.~1, pp. 86--97, 2012.

\bibitem{langfelder2007eigengene}
P.~Langfelder and S.~Horvath, ``Eigengene networks for studying the
  relationships between co-expression modules,'' \emph{BMC systems biology},
  vol.~1, no.~1, pp. 1--17, 2007.

\bibitem{liu2016weighted}
J.~Liu, L.~Jing, and X.~Tu, ``Weighted gene co-expression network analysis
  identifies specific modules and hub genes related to coronary artery
  disease,'' \emph{BMC cardiovascular disorders}, vol.~16, no.~1, pp. 1--8,
  2016.

\bibitem{ringner2008principal}
M.~Ringn{\'e}r, ``What is principal component analysis?'' \emph{Nature
  biotechnology}, vol.~26, no.~3, pp. 303--304, 2008.

\bibitem{tantardini2019comparing}
M.~Tantardini, F.~Ieva, L.~Tajoli, and C.~Piccardi, ``Comparing methods for
  comparing networks,'' \emph{Scientific reports}, vol.~9, no.~1, pp. 1--19,
  2019.

\bibitem{koutra2013deltacon}
D.~Koutra, J.~T. Vogelstein, and C.~Faloutsos, ``Deltacon: A principled
  massive-graph similarity function,'' in \emph{Proceedings of the 2013 SIAM
  International Conference on Data Mining}.\hskip 1em plus 0.5em minus
  0.4em\relax SIAM, 2013, pp. 162--170.

\bibitem{matusita1953estimation}
K.~Matusita, ``On the estimation by the minimum distance method,'' \emph{Annals
  of the Institute of Statistical Mathematics}, vol.~5, no.~2, pp. 59--65,
  1953.

\bibitem{bland1995multiple}
J.~M. Bland and D.~G. Altman, ``Multiple significance tests: the bonferroni
  method,'' \emph{Bmj}, vol. 310, no. 6973, p. 170, 1995.

\bibitem{narum2006beyond}
S.~R. Narum, ``Beyond bonferroni: less conservative analyses for conservation
  genetics,'' \emph{Conservation genetics}, vol.~7, no.~5, pp. 783--787, 2006.

\bibitem{benjamini1995controlling}
Y.~Benjamini and Y.~Hochberg, ``Controlling the false discovery rate: a
  practical and powerful approach to multiple testing,'' \emph{Journal of the
  Royal statistical society: series B (Methodological)}, vol.~57, no.~1, pp.
  289--300, 1995.

\bibitem{white2019beyond}
T.~White, J.~van~der Ende, and T.~E. Nichols, ``Beyond bonferroni revisited:
  concerns over inflated false positive research findings in the fields of
  conservation genetics, biology, and medicine,'' \emph{Conservation Genetics},
  vol.~20, no.~4, pp. 927--937, 2019.

\bibitem{supek2011revigo}
F.~Supek, M.~Bo{\v{s}}njak, N.~{\v{S}}kunca, and T.~{\v{S}}muc, ``Revigo
  summarizes and visualizes long lists of gene ontology terms,'' \emph{PloS
  one}, vol.~6, no.~7, p. e21800, 2011.

\bibitem{gene2019gene}
G.~O. Consortium, ``The gene ontology resource: 20 years and still going
  strong,'' \emph{Nucleic acids research}, vol.~47, no.~D1, pp. D330--D338,
  2019.

\bibitem{ansel2017variation}
A.~Ansel, J.~P. Rosenzweig, P.~D. Zisman, M.~Melamed, and B.~Gesundheit,
  ``Variation in gene expression in autism spectrum disorders: an extensive
  review of transcriptomic studies,'' \emph{Frontiers in neuroscience},
  vol.~10, p. 601, 2017.

\bibitem{voineagu2011transcriptomic}
I.~Voineagu, X.~Wang, P.~Johnston, J.~K. Lowe, Y.~Tian, S.~Horvath, J.~Mill,
  R.~M. Cantor, B.~J. Blencowe, and D.~H. Geschwind, ``Transcriptomic analysis
  of autistic brain reveals convergent molecular pathology,'' \emph{Nature},
  vol. 474, no. 7351, pp. 380--384, 2011.

\bibitem{parikshak2016genome}
N.~N. Parikshak, V.~Swarup, T.~G. Belgard, M.~Irimia, G.~Ramaswami, M.~J.
  Gandal, C.~Hartl, V.~Leppa, L.~de~la Torre~Ubieta, J.~Huang \emph{et~al.},
  ``Genome-wide changes in lncrna, splicing, and regional gene expression
  patterns in autism,'' \emph{Nature}, vol. 540, no. 7633, pp. 423--427, 2016.

\bibitem{anitha2012brain}
A.~Anitha, K.~Nakamura, I.~Thanseem, K.~Yamada, Y.~Iwayama, T.~Toyota,
  H.~Matsuzaki, T.~Miyachi, S.~Yamada, M.~Tsujii \emph{et~al.}, ``Brain
  region-specific altered expression and association of mitochondria-related
  genes in autism,'' \emph{Molecular autism}, vol.~3, no.~1, pp. 1--12, 2012.

\bibitem{khan2014disrupted}
A.~Khan, J.~Harney, A.~Zavacki, and E.~Sajdel-Sulkowska, ``Disrupted brain
  thyroid hormone homeostasis and altered thyroid hormone-dependent brain gene
  expression in autism spectrum disorders,'' \emph{J Physiol Pharmacol},
  vol.~65, no.~2, pp. 257--272, 2014.

\bibitem{garbett2008immune}
K.~Garbett, P.~J. Ebert, A.~Mitchell, C.~Lintas, B.~Manzi, K.~Mirnics, and
  A.~M. Persico, ``Immune transcriptome alterations in the temporal cortex of
  subjects with autism,'' \emph{Neurobiology of disease}, vol.~30, no.~3, pp.
  303--311, 2008.

\bibitem{provenzano2016comparative}
G.~Provenzano, Z.~Corradi, K.~Monsorno, T.~Fedrizzi, L.~Ricceri, M.~L.
  Scattoni, and Y.~Bozzi, ``Comparative gene expression analysis of two mouse
  models of autism: transcriptome profiling of the btbr and en2-/-
  hippocampus,'' \emph{Frontiers in neuroscience}, vol.~10, p. 396, 2016.

\bibitem{vargas2005neuroglial}
D.~L. Vargas, C.~Nascimbene, C.~Krishnan, A.~W. Zimmerman, and C.~A. Pardo,
  ``Neuroglial activation and neuroinflammation in the brain of patients with
  autism,'' \emph{Annals of Neurology: Official Journal of the American
  Neurological Association and the Child Neurology Society}, vol.~57, no.~1,
  pp. 67--81, 2005.

\bibitem{voineagu2013converging}
I.~Voineagu and V.~Eapen, ``Converging pathways in autism spectrum disorders:
  interplay between synaptic dysfunction and immune responses,''
  \emph{Frontiers in human neuroscience}, vol.~7, p. 738, 2013.

\bibitem{werling2016gene}
D.~M. Werling, N.~N. Parikshak, and D.~H. Geschwind, ``Gene expression in human
  brain implicates sexually dimorphic pathways in autism spectrum disorders,''
  \emph{Nature communications}, vol.~7, no.~1, pp. 1--11, 2016.

\bibitem{herrero2020identification}
M.~J. Herrero, D.~Velmeshev, D.~Hernandez-Pineda, S.~Sethi, S.~Sorrells,
  P.~Banerjee, C.~Sullivan, A.~R. Gupta, A.~R. Kriegstein, and J.~G. Corbin,
  ``Identification of amygdala-expressed genes associated with autism spectrum
  disorder,'' \emph{Molecular autism}, vol.~11, pp. 1--14, 2020.

\bibitem{young2020video}
G.~S. Young, J.~N. Constantino, S.~Dvorak, A.~Belding, D.~Gangi, A.~Hill,
  M.~Hill, M.~Miller, C.~Parikh, A.~Schwichtenberg \emph{et~al.}, ``A
  video-based measure to identify autism risk in infancy,'' \emph{Journal of
  Child Psychology and Psychiatry}, vol.~61, no.~1, pp. 88--94, 2020.

\bibitem{mckinnon2019restricted}
C.~J. McKinnon, A.~T. Eggebrecht, A.~Todorov, J.~J. Wolff, J.~T. Elison, C.~M.
  Adams, A.~Z. Snyder, A.~M. Estes, L.~Zwaigenbaum, K.~N. Botteron
  \emph{et~al.}, ``Restricted and repetitive behavior and brain functional
  connectivity in infants at risk for developing autism spectrum disorder,''
  \emph{Biological Psychiatry: Cognitive Neuroscience and Neuroimaging},
  vol.~4, no.~1, pp. 50--61, 2019.

\bibitem{griesi2021transcriptome}
K.~Griesi-Oliveira, M.~Fogo, B.~Pinto, A.~Alves, A.~Suzuki, A.~Morales,
  S.~Ezquina, O.~Sosa, G.~Sutton, D.~Sunaga-Franze \emph{et~al.},
  ``Transcriptome of ipsc-derived neuronal cells reveals a module of
  co-expressed genes consistently associated with autism spectrum disorder,''
  \emph{Molecular psychiatry}, vol.~26, no.~5, pp. 1589--1605, 2021.

\bibitem{willsey2013coexpression}
A.~J. Willsey, S.~J. Sanders, M.~Li, S.~Dong, A.~T. Tebbenkamp, R.~A. Muhle,
  S.~K. Reilly, L.~Lin, S.~Fertuzinhos, J.~A. Miller \emph{et~al.},
  ``Coexpression networks implicate human midfetal deep cortical projection
  neurons in the pathogenesis of autism,'' \emph{Cell}, vol. 155, no.~5, pp.
  997--1007, 2013.

\bibitem{lombardo2017hierarchical}
M.~V. Lombardo, E.~Courchesne, N.~E. Lewis, and T.~Pramparo, ``Hierarchical
  cortical transcriptome disorganization in autism,'' \emph{Molecular autism},
  vol.~8, no.~1, pp. 1--17, 2017.

\bibitem{lombardo2021ribosomal}
M.~V. Lombardo, ``Ribosomal protein genes in post-mortem cortical tissue and
  ipsc-derived neural progenitor cells are commonly upregulated in expression
  in autism,'' \emph{Molecular psychiatry}, vol.~26, no.~5, pp. 1432--1435,
  2021.

\bibitem{kelleher2008autistic}
R.~J. Kelleher~III and M.~F. Bear, ``The autistic neuron: troubled
  translation?'' \emph{Cell}, vol. 135, no.~3, pp. 401--406, 2008.

\bibitem{meyer2006intermediate}
A.~Meyer-Lindenberg and D.~R. Weinberger, ``Intermediate phenotypes and genetic
  mechanisms of psychiatric disorders,'' \emph{Nature Reviews Neuroscience},
  vol.~7, no.~10, p. 818, 2006.

\bibitem{munoz2009imaging}
M.~K.~E. MU{\~N}OZ, M.~L.~W. HYDE, and A.~R. Hariri, ``Imaging genetics,''
  \emph{Journal of the American Academy of Child and Adolescent Psychiatry},
  vol.~48, no.~4, p. 356, 2009.

\bibitem{arslan2015genes}
A.~Arslan, ``Genes, brains, and behavior: imaging genetics for neuropsychiatric
  disorders,'' \emph{The Journal of Neuropsychiatry and Clinical
  Neurosciences}, vol.~27, no.~2, pp. 81--92, 2015.

\bibitem{hashimoto2015imaging}
R.~Hashimoto, K.~Ohi, H.~Yamamori, Y.~Yasuda, M.~Fujimoto, S.~Umeda-Yano,
  Y.~Watanabe, M.~Fukunaga, and M.~Takeda, ``Imaging genetics and psychiatric
  disorders,'' \emph{Current molecular medicine}, vol.~15, no.~2, pp. 168--175,
  2015.

\bibitem{heinz2000genotype}
A.~Heinz, D.~Goldman, D.~W. Jones, R.~Palmour, D.~Hommer, J.~G. Gorey, K.~S.
  Lee, M.~Linnoila, and D.~R. Weinberger, ``Genotype influences in vivo
  dopamine transporter availability in human striatum,''
  \emph{Neuropsychopharmacology}, vol.~22, no.~2, p. 133, 2000.

\bibitem{bookheimer2000patterns}
S.~Y. Bookheimer, M.~H. Strojwas, M.~S. Cohen, A.~M. Saunders, M.~A.
  Pericak-Vance, J.~C. Mazziotta, and G.~W. Small, ``Patterns of brain
  activation in people at risk for alzheimer's disease,'' \emph{New England
  journal of medicine}, vol. 343, no.~7, pp. 450--456, 2000.

\bibitem{nestler2016epigenetic}
E.~J. Nestler, C.~J. Pe{\~n}a, M.~Kundakovic, A.~Mitchell, and S.~Akbarian,
  ``Epigenetic basis of mental illness,'' \emph{The Neuroscientist}, vol.~22,
  no.~5, pp. 447--463, 2016.

\bibitem{karam2015genetic}
S.~M. Karam, M.~Riegel, S.~L. Segal, T.~M. F{\'e}lix, A.~J. Barros, I.~S.
  Santos, A.~Matijasevich, R.~Giugliani, and M.~Black, ``Genetic causes of
  intellectual disability in a birth cohort: A population-based study,''
  \emph{American Journal of Medical Genetics Part A}, vol. 167, no.~6, pp.
  1204--1214, 2015.

\bibitem{hawrylycz2015canonical}
M.~Hawrylycz, J.~A. Miller, V.~Menon, D.~Feng, T.~Dolbeare, A.~L.
  Guillozet-Bongaarts, A.~G. Jegga, B.~J. Aronow, C.-K. Lee, A.~Bernard
  \emph{et~al.}, ``Canonical genetic signatures of the adult human brain,''
  \emph{Nature neuroscience}, vol.~18, no.~12, pp. 1832--1844, 2015.

\bibitem{ahba}
``Allen human brain atlas,'' \url{https://human.brain-map.org/}, Nov. 2020.

\bibitem{fornito2019bridging}
A.~Fornito, A.~Arnatkeviciute, and B.~Fulcher, ``Bridging the gap between
  connectome and transcriptome,'' \emph{Trends in Cognitive Sciences}, vol.~23,
  no.~1, pp. 34--50, 2019.

\bibitem{anderson2018gene}
K.~M. Anderson, F.~M. Krienen, E.~Y. Choi, J.~M. Reinen, B.~T. Yeo, and A.~J.
  Holmes, ``Gene expression links functional networks across cortex and
  striatum,'' \emph{Nature communications}, vol.~9, no.~1, pp. 1--14, 2018.

\bibitem{liu2020integrative}
S.~Liu, J.~Seidlitz, J.~D. Blumenthal, L.~S. Clasen, and A.~Raznahan,
  ``Integrative structural, functional, and transcriptomic analyses of
  sex-biased brain organization in humans,'' \emph{Proceedings of the National
  Academy of Sciences}, vol. 117, no.~31, pp. 18\,788--18\,798, 2020.

\bibitem{xie2020brain}
Y.~Xie, X.~Zhang, F.~Liu, W.~Qin, J.~Fu, K.~Xue, and C.~Yu, ``Brain mrna
  expression associated with cortical volume alterations in autism spectrum
  disorder,'' \emph{Cell Reports}, vol.~32, no.~11, p. 108137, 2020.

\bibitem{haar2016anatomical}
S.~Haar, S.~Berman, M.~Behrmann, and I.~Dinstein, ``Anatomical abnormalities in
  autism?'' \emph{Cerebral cortex}, vol.~26, no.~4, pp. 1440--1452, 2016.

\bibitem{di2014autism}
A.~Di~Martino, C.-G. Yan, Q.~Li, E.~Denio, F.~X. Castellanos, K.~Alaerts, J.~S.
  Anderson, M.~Assaf, S.~Y. Bookheimer, M.~Dapretto \emph{et~al.}, ``The autism
  brain imaging data exchange: towards large-scale evaluation of the intrinsic
  brain architecture in autism,'' \emph{Molecular psychiatry}, vol.~19, no.~6,
  p. 659, 2014.

\bibitem{di2017enhancing}
A.~Di~Martino, D.~O’connor, B.~Chen, K.~Alaerts, J.~S. Anderson, M.~Assaf,
  J.~H. Balsters, L.~Baxter, A.~Beggiato, S.~Bernaerts \emph{et~al.},
  ``Enhancing studies of the connectome in autism using the autism brain
  imaging data exchange ii,'' \emph{Scientific data}, vol.~4, no.~1, pp. 1--15,
  2017.

\bibitem{romero2019synaptic}
R.~Romero-Garcia, V.~Warrier, E.~T. Bullmore, S.~Baron-Cohen, and R.~A.
  Bethlehem, ``Synaptic and transcriptionally downregulated genes are
  associated with cortical thickness differences in autism,'' \emph{Molecular
  psychiatry}, vol.~24, no.~7, pp. 1053--1064, 2019.

\bibitem{forest2017gene}
M.~Forest, Y.~Iturria-Medina, J.~S. Goldman, C.~L. Kleinman, A.~Lovato,
  K.~Oros~Klein, A.~Evans, A.~Ciampi, A.~Labbe, and C.~M. Greenwood, ``Gene
  networks show associations with seed region connectivity,'' \emph{Human Brain
  Mapping}, vol.~38, no.~6, pp. 3126--3140, 2017.

\bibitem{goel2014spatial}
P.~Goel, A.~Kuceyeski, E.~LoCastro, and A.~Raj, ``Spatial patterns of
  genome-wide expression profiles reflect anatomic and fiber connectivity
  architecture of healthy human brain,'' \emph{Human brain mapping}, vol.~35,
  no.~8, pp. 4204--4218, 2014.

\bibitem{cioli2014differences}
C.~Cioli, H.~Abdi, D.~Beaton, Y.~Burnod, and S.~Mesmoudi, ``Differences in
  human cortical gene expression match the temporal properties of large-scale
  functional networks,'' \emph{PloS one}, vol.~9, no.~12, p. e115913, 2014.

\bibitem{vertes2016gene}
P.~E. V{\'e}rtes, T.~Rittman, K.~J. Whitaker, R.~Romero-Garcia,
  F.~V{\'a}{\v{s}}a, M.~G. Kitzbichler, K.~Wagstyl, P.~Fonagy, R.~J. Dolan,
  P.~B. Jones \emph{et~al.}, ``Gene transcription profiles associated with
  inter-modular hubs and connection distance in human functional magnetic
  resonance imaging networks,'' \emph{Philosophical Transactions of the Royal
  Society B: Biological Sciences}, vol. 371, no. 1705, p. 20150362, 2016.

\bibitem{krienen2016transcriptional}
F.~M. Krienen, B.~T. Yeo, T.~Ge, R.~L. Buckner, and C.~C. Sherwood,
  ``Transcriptional profiles of supragranular-enriched genes associate with
  corticocortical network architecture in the human brain,'' \emph{Proceedings
  of the National Academy of Sciences}, vol. 113, no.~4, pp. E469--E478, 2016.

\bibitem{parkes2017transcriptional}
L.~Parkes, B.~Fulcher, M.~Y{\"u}cel, and A.~Fornito, ``Transcriptional
  signatures of connectomic subregions of the human striatum,'' \emph{Genes,
  Brain and Behavior}, vol.~16, no.~7, pp. 647--663, 2017.

\bibitem{kirsch2016expression}
L.~Kirsch and G.~Chechik, ``On expression patterns and developmental origin of
  human brain regions,'' \emph{PLoS computational biology}, vol.~12, no.~8, p.
  e1005064, 2016.

\bibitem{whitaker2016adolescence}
K.~J. Whitaker, P.~E. V{\'e}rtes, R.~Romero-Garcia, F.~V{\'a}{\v{s}}a,
  M.~Moutoussis, G.~Prabhu, N.~Weiskopf, M.~F. Callaghan, K.~Wagstyl,
  T.~Rittman \emph{et~al.}, ``Adolescence is associated with genomically
  patterned consolidation of the hubs of the human brain connectome,''
  \emph{Proceedings of the National Academy of Sciences}, vol. 113, no.~32, pp.
  9105--9110, 2016.

\bibitem{rittman2016regional}
T.~Rittman, M.~Rubinov, P.~E. V{\'e}rtes, A.~X. Patel, C.~E. Ginestet, B.~C.
  Ghosh, R.~A. Barker, M.~G. Spillantini, E.~T. Bullmore, and J.~B. Rowe,
  ``Regional expression of the mapt gene is associated with loss of hubs in
  brain networks and cognitive impairment in parkinson disease and progressive
  supranuclear palsy,'' \emph{Neurobiology of aging}, vol.~48, pp. 153--160,
  2016.

\bibitem{romme2017connectome}
I.~A. Romme, M.~A. de~Reus, R.~A. Ophoff, R.~S. Kahn, and M.~P. van~den Heuvel,
  ``Connectome disconnectivity and cortical gene expression in patients with
  schizophrenia,'' \emph{Biological psychiatry}, vol.~81, no.~6, pp. 495--502,
  2017.

\bibitem{mccolgan2018brain}
P.~McColgan, S.~Gregory, K.~K. Seunarine, A.~Razi, M.~Papoutsi, E.~Johnson,
  A.~Durr, R.~A. Roos, B.~R. Leavitt, P.~Holmans \emph{et~al.}, ``Brain regions
  showing white matter loss in huntington’s disease are enriched for synaptic
  and metabolic genes,'' \emph{Biological psychiatry}, vol.~83, no.~5, pp.
  456--465, 2018.

\bibitem{ABIDEIdoc}
``{ABIDE I} documentation,''
  \url{http://fcon_1000.projects.nitrc.org/indi/abide/abide_I.html}, Accessed:
  12.04.2019.

\bibitem{ABIDEIdocNITRC}
``{ABIDE I} documentation, available for registered users only, on the {NITRC}
  platform,''
  \url{https://www.nitrc.org/ir/app/template/XDATScreen_report_xnat_projectData.vm/search_element/xnat:projectData/search_field/xnat:projectData.ID/search_value/ABIDE},
  Accessed: 12.04.2019.

\bibitem{ABIDEIIdoc}
``{ABIDE II} documentation,''
  \url{http://fcon_1000.projects.nitrc.org/indi/abide/abide_II.html}, Accessed:
  12.04.2019.

\bibitem{cheng2016rifd}
G.~Cheng, P.~Zhou, and J.~Han, ``Rifd-cnn: Rotation-invariant and fisher
  discriminative convolutional neural networks for object detection,'' in
  \emph{Proceedings of the IEEE conference on computer vision and pattern
  recognition}, 2016, pp. 2884--2893.

\bibitem{simon2015understanding}
A.~B. Simon and R.~B. Buxton, ``Understanding the dynamic relationship between
  cerebral blood flow and the bold signal: Implications for quantitative
  functional mri,'' \emph{Neuroimage}, vol. 116, pp. 158--167, 2015.

\bibitem{buxton2013physics}
R.~B. Buxton, ``The physics of functional magnetic resonance imaging (fmri),''
  \emph{Reports on Progress in Physics}, vol.~76, no.~9, p. 096601, 2013.

\bibitem{patriat2013effect}
R.~Patriat, E.~K. Molloy, T.~B. Meier, G.~R. Kirk, V.~A. Nair, M.~E. Meyerand,
  V.~Prabhakaran, and R.~M. Birn, ``The effect of resting condition on
  resting-state fmri reliability and consistency: a comparison between resting
  with eyes open, closed, and fixated,'' \emph{Neuroimage}, vol.~78, pp.
  463--473, 2013.

\bibitem{shehzad2015preprocessed}
Z.~Shehzad, S.~Giavasis, Q.~Li, Y.~Benhajali, C.~Yan, Z.~Yang, M.~Milham,
  P.~Bellec, and C.~Craddock, ``The preprocessed connectomes project quality
  assessment protocol-a resource for measuring the quality of mri data,''
  \emph{Frontiers in neuroscience}, vol.~47, 2015.

\bibitem{rolls2020automated}
E.~T. Rolls, C.-C. Huang, C.-P. Lin, J.~Feng, and M.~Joliot, ``Automated
  anatomical labelling atlas 3,'' \emph{Neuroimage}, vol. 206, p. 116189, 2020.

\bibitem{gesch}
G.~Lab, ``Git repository containing data from study "genome-wide changes in
  lncrna, splicing, and regional gene expression patterns in autism",''
  \url{https://github.com/dhglab/Genome-wide-changes-in-lncRNA-alternative-splicing-and-cortical-patterning-in-autism},
  Nov. 2020.

\bibitem{atp}
S.~Foundation and A.~Speaks, ``Autism brainnet, that since 2016 incorporates
  the tissue and the data previously collected by the autism tissue program
  (atp),'' \url{https://www.autismbrainnet.org/}, Feb. 2021.

\bibitem{nichd}
U.~of~Maryland, ``Nih neurobiobank,'' \url{https://neurobiobank.nih.gov/}, Feb.
  2021.

\bibitem{monoranu2009ph}
C.~M. Monoranu, M.~Apfelbacher, E.~Gr{\"u}nblatt, B.~Puppe, I.~Alafuzoff,
  I.~Ferrer, S.~Al-Saraj, K.~Keyvani, A.~Schmitt, P.~Falkai \emph{et~al.}, ``ph
  measurement as quality control on human post mortem brain tissue: a study of
  the brainnet europe consortium,'' \emph{Neuropathology and applied
  neurobiology}, vol.~35, no.~3, pp. 329--337, 2009.

\bibitem{rutter2003adi}
M.~Rutter, A.~Le~Couteur, and C.~Lord, ``Adi-r,'' \emph{Autism diagnostic
  interview revised. Manual. Los Angeles: Western Psychological Services},
  2003.

\bibitem{smid2018gene}
M.~Smid, R.~R.~C. van~den Braak, H.~J. van~de Werken, J.~van Riet, A.~van
  Galen, V.~de~Weerd, M.~van~der Vlugt-Daane, S.~I. Bril, Z.~S. Lalmahomed,
  W.~P. Kloosterman \emph{et~al.}, ``Gene length corrected trimmed mean of
  m-values (getmm) processing of rna-seq data performs similarly in intersample
  analyses while improving intrasample comparisons,'' \emph{BMC
  bioinformatics}, vol.~19, no.~1, pp. 1--13, 2018.

\bibitem{glass2013gene}
D.~Glass, A.~Vi{\~n}uela, M.~N. Davies, A.~Ramasamy, L.~Parts, D.~Knowles,
  A.~A. Brown, {\AA}.~K. Hedman, K.~S. Small, A.~Buil \emph{et~al.}, ``Gene
  expression changes with age in skin, adipose tissue, blood and brain,''
  \emph{Genome biology}, vol.~14, no.~7, pp. 1--12, 2013.

\bibitem{bonati2007evaluation}
M.~T. Bonati, S.~Russo, P.~Finelli, M.~R. Valsecchi, F.~Cogliati, F.~Cavalleri,
  W.~Roberts, M.~Elia, and L.~Larizza, ``Evaluation of autism traits in
  angelman syndrome: a resource to unfold autism genes,'' \emph{Neurogenetics},
  vol.~8, no.~3, pp. 169--178, 2007.

\bibitem{abbeduto2014fragile}
L.~Abbeduto, A.~McDuffie, and A.~J. Thurman, ``The fragile x syndrome--autism
  comorbidity: what do we really know?'' \emph{Frontiers in genetics}, vol.~5,
  p. 355, 2014.

\bibitem{mount2003features}
R.~H. Mount, T.~Charman, R.~P. Hastings, S.~Reilly, and H.~Cass, ``Features of
  autism in rett syndrome and severe mental retardation,'' \emph{Journal of
  autism and developmental disorders}, vol.~33, no.~4, pp. 435--442, 2003.

\bibitem{diguiseppi2010screening}
C.~DiGuiseppi, S.~Hepburn, J.~M. Davis, D.~J. Fidler, S.~Hartway, N.~R. Lee,
  L.~Miller, M.~Ruttenber, and C.~Robinson, ``Screening for autism spectrum
  disorders in children with down syndrome: population prevalence and screening
  test characteristics,'' \emph{Journal of developmental and behavioral
  pediatrics: JDBP}, vol.~31, no.~3, p. 181, 2010.

\bibitem{hudazoghbi_video}
\BIBentryALTinterwordspacing
H.~Zoghbi. Huda zoghbi (bcm/tch) 1: Rett syndrome: Genomes, epigenomes and
  neuropsychiatric conditions. iBiology. [Online]. Available:
  \url{https://www.youtube.com/watch?v=-mKHhPb1q9Q&ab_channel=iBiology}
\BIBentrySTDinterwordspacing

\bibitem{liao2019webgestalt}
Y.~Liao, J.~Wang, E.~J. Jaehnig, Z.~Shi, and B.~Zhang, ``Webgestalt 2019: gene
  set analysis toolkit with revamped uis and apis,'' \emph{Nucleic acids
  research}, vol.~47, no.~W1, pp. W199--W205, 2019.

\bibitem{arnatkeviciute2019practical}
A.~Arnatkeviciute, B.~D. Fulcher, and A.~Fornito, ``A practical guide to
  linking brain-wide gene expression and neuroimaging data,''
  \emph{Neuroimage}, vol. 189, pp. 353--367, 2019.

\bibitem{ahbadocs}
A.~H.~B. Atlas, ``Ahba documentation,''
  \url{http://help.brain-map.org/display/humanbrain/Documentation}, accessed:
  2021-08-17.

\bibitem{abrahams2013sfari}
B.~S. Abrahams, D.~E. Arking, D.~B. Campbell, H.~C. Mefford, E.~M. Morrow,
  L.~A. Weiss, I.~Menashe, T.~Wadkins, S.~Banerjee-Basu, and A.~Packer, ``Sfari
  gene 2.0: a community-driven knowledgebase for the autism spectrum disorders
  (asds),'' \emph{Molecular autism}, vol.~4, no.~1, pp. 1--3, 2013.

\bibitem{pinero2020disgenet}
J.~Pi{\~n}ero, J.~M. Ram{\'\i}rez-Anguita, J.~Sa{\"u}ch-Pitarch, F.~Ronzano,
  E.~Centeno, F.~Sanz, and L.~I. Furlong, ``The disgenet knowledge platform for
  disease genomics: 2019 update,'' \emph{Nucleic acids research}, vol.~48,
  no.~D1, pp. D845--D855, 2020.

\bibitem{zhao2020training}
Q.~Zhao, E.~Adeli, and K.~M. Pohl, ``Training confounder-free deep learning
  models for medical applications,'' \emph{Nature communications}, vol.~11,
  no.~1, pp. 1--9, 2020.

\bibitem{dinsdale2021deep}
N.~K. Dinsdale, M.~Jenkinson, and A.~I. Namburete, ``Deep learning-based
  unlearning of dataset bias for mri harmonisation and confound removal,''
  \emph{NeuroImage}, vol. 228, p. 117689, 2021.

\bibitem{spera2019evaluation}
G.~Spera, A.~Retico, P.~Bosco, E.~Ferrari, L.~Palumbo, P.~Oliva, F.~Muratori,
  and S.~Calderoni, ``Evaluation of altered functional connections in male
  children with autism spectrum disorders on multiple-site data optimized with
  machine learning,'' \emph{Frontiers in psychiatry}, vol.~10, p. 620, 2019.

\bibitem{zhai2021scaling}
X.~Zhai, A.~Kolesnikov, N.~Houlsby, and L.~Beyer, ``Scaling vision
  transformers,'' \emph{arXiv preprint arXiv:2106.04560}, 2021.

\bibitem{backhausen2016quality}
L.~L. Backhausen, M.~M. Herting, J.~Buse, V.~Roessner, M.~N. Smolka, and N.~C.
  Vetter, ``Quality control of structural mri images applied using
  freesurfer—a hands-on workflow to rate motion artifacts,'' \emph{Frontiers
  in neuroscience}, vol.~10, p. 558, 2016.

\bibitem{yassin2020machine}
W.~Yassin, H.~Nakatani, Y.~Zhu, M.~Kojima, K.~Owada, H.~Kuwabara, W.~Gonoi,
  Y.~Aoki, H.~Takao, T.~Natsubori \emph{et~al.}, ``Machine-learning
  classification using neuroimaging data in schizophrenia, autism, ultra-high
  risk and first-episode psychosis,'' \emph{Translational psychiatry}, vol.~10,
  no.~1, pp. 1--11, 2020.

\bibitem{gilhus2019neuroinflammation}
N.~E. Gilhus and G.~Deuschl, ``Neuroinflammation—a common thread in
  neurological disorders,'' \emph{Nature Reviews Neurology}, vol.~15, no.~8,
  pp. 429--430, 2019.

\bibitem{schwartz2016neurological}
M.~Schwartz and A.~Deczkowska, ``Neurological disease as a failure of
  brain--immune crosstalk: the multiple faces of neuroinflammation,''
  \emph{Trends in immunology}, vol.~37, no.~10, pp. 668--679, 2016.

\bibitem{lebersfeld2021systematic}
J.~B. Lebersfeld, M.~Swanson, C.~D. Clesi, and S.~E. O’Kelley, ``Systematic
  review and meta-analysis of the clinical utility of the ados-2 and the adi-r
  in diagnosing autism spectrum disorders in children,'' \emph{Journal of
  Autism and Developmental Disorders}, pp. 1--14, 2021.

\bibitem{gupta2011amygdala}
R.~Gupta, T.~R. Koscik, A.~Bechara, and D.~Tranel, ``The amygdala and
  decision-making,'' \emph{Neuropsychologia}, vol.~49, no.~4, pp. 760--766,
  2011.

\bibitem{kennedy2009personal}
D.~P. Kennedy, J.~Gl{\"a}scher, J.~M. Tyszka, and R.~Adolphs, ``Personal space
  regulation by the human amygdala,'' \emph{Nature neuroscience}, vol.~12,
  no.~10, pp. 1226--1227, 2009.

\bibitem{bickart2011amygdala}
K.~C. Bickart, C.~I. Wright, R.~J. Dautoff, B.~C. Dickerson, and L.~F. Barrett,
  ``Amygdala volume and social network size in humans,'' \emph{Nature
  neuroscience}, vol.~14, no.~2, pp. 163--164, 2011.

\bibitem{bzdok2011ale}
D.~Bzdok, R.~Langner, S.~Caspers, F.~Kurth, U.~Habel, K.~Zilles, A.~Laird, and
  S.~B. Eickhoff, ``Ale meta-analysis on facial judgments of trustworthiness
  and attractiveness,'' \emph{Brain Structure and Function}, vol. 215, no.~3,
  pp. 209--223, 2011.

\bibitem{baeken2014left}
C.~Baeken, D.~Marinazzo, P.~Van~Schuerbeek, G.-R. Wu, J.~De~Mey, R.~Luypaert,
  and R.~De~Raedt, ``Left and right amygdala-mediofrontal cortical functional
  connectivity is differentially modulated by harm avoidance,'' \emph{PLoS
  One}, vol.~9, no.~4, p. e95740, 2014.

\bibitem{blair2005unilateral}
H.~T. Blair, V.~K. Huynh, V.~T. Vaz, J.~Van, R.~R. Patel, A.~K. Hiteshi, J.~E.
  Lee, and J.~W. Tarpley, ``Unilateral storage of fear memories by the
  amygdala,'' \emph{Journal of Neuroscience}, vol.~25, no.~16, pp. 4198--4205,
  2005.

\bibitem{barrett2007amygdala}
L.~F. Barrett, E.~Bliss-Moreau, S.~L. Duncan, S.~L. Rauch, and C.~I. Wright,
  ``The amygdala and the experience of affect,'' \emph{Social cognitive and
  affective neuroscience}, vol.~2, no.~2, pp. 73--83, 2007.

\bibitem{berntson2007amygdala}
G.~G. Berntson, A.~Bechara, H.~Damasio, D.~Tranel, and J.~T. Cacioppo,
  ``Amygdala contribution to selective dimensions of emotion,'' \emph{Social
  cognitive and affective neuroscience}, vol.~2, no.~2, pp. 123--129, 2007.

\bibitem{lanteaume2007emotion}
L.~Lanteaume, S.~Khalfa, J.~R{\'e}gis, P.~Marquis, P.~Chauvel, and
  F.~Bartolomei, ``Emotion induction after direct intracerebral stimulations of
  human amygdala,'' \emph{Cerebral cortex}, vol.~17, no.~6, pp. 1307--1313,
  2007.

\bibitem{murray2009amygdala}
E.~A. Murray, A.~Izquierdo, and L.~Malkova, \emph{Amygdala function in positive
  reinforcement: Contributions from studies of nonhuman primates.}\hskip 1em
  plus 0.5em minus 0.4em\relax The Guilford Press, 2009.

\bibitem{goldstein2001normal}
J.~M. Goldstein, L.~J. Seidman, N.~J. Horton, N.~Makris, D.~N. Kennedy, V.~S.
  Caviness~Jr, S.~V. Faraone, and M.~T. Tsuang, ``Normal sexual dimorphism of
  the adult human brain assessed by in vivo magnetic resonance imaging,''
  \emph{Cerebral cortex}, vol.~11, no.~6, pp. 490--497, 2001.

\bibitem{simerly1990distribution}
R.~Simerly, L.~Swanson, C.~Chang, and M.~Muramatsu, ``Distribution of androgen
  and estrogen receptor mrna-containing cells in the rat brain: an in situ
  hybridization study,'' \emph{Journal of Comparative Neurology}, vol. 294,
  no.~1, pp. 76--95, 1990.

\bibitem{hall2010guyton}
J.~Hall, \emph{Guyton and Hall textbook of medical physiology}.\hskip 1em plus
  0.5em minus 0.4em\relax Elsevier Health Sciences, 2010.

\bibitem{henschen1919language}
S.~Henschen, ``On language, music and calculation mechanisms and their
  localisation in the cerebrum,'' \emph{Zeitschrift fur die gesamte Neurologie
  und Psychiatrie}, vol.~52, pp. 273--298, 1919.

\bibitem{gerstmann1940syndrome}
J.~Gerstmann, ``Syndrome of finger agnosia, disorientation for right and left,
  agraphia and acalculia: local diagnostic value,'' \emph{Archives of Neurology
  \& Psychiatry}, vol.~44, no.~2, pp. 398--408, 1940.

\bibitem{chen2012neural}
Q.~Chen, R.~Weidner, S.~Vossel, P.~H. Weiss, and G.~R. Fink, ``Neural
  mechanisms of attentional reorienting in three-dimensional space,''
  \emph{Journal of Neuroscience}, vol.~32, no.~39, pp. 13\,352--13\,362, 2012.

\bibitem{farrer2008angular}
C.~Farrer, S.~H. Frey, J.~D. Van~Horn, E.~Tunik, D.~Turk, S.~Inati, and S.~T.
  Grafton, ``The angular gyrus computes action awareness representations,''
  \emph{Cerebral cortex}, vol.~18, no.~2, pp. 254--261, 2008.

\bibitem{wu2016segmentation}
Y.~Wu, D.~Sun, Y.~Wang, Y.~Wang, and S.~Ou, ``Segmentation of the cingulum
  bundle in the human brain: a new perspective based on dsi tractography and
  fiber dissection study,'' \emph{Frontiers in neuroanatomy}, vol.~10, p.~84,
  2016.

\bibitem{cohen1999alteration}
R.~A. Cohen, R.~F. Kaplan, P.~Zuffante, D.~J. Moser, M.~A. Jenkins,
  S.~Salloway, and H.~Wilkinson, ``Alteration of intention and self-initiated
  action associated with bilateral anterior cingulotomy,'' \emph{The Journal of
  neuropsychiatry and clinical neurosciences}, vol.~11, no.~4, pp. 444--453,
  1999.

\bibitem{zhu2019brain}
W.~Zhu, X.~Zhou, and L.-X. Xia, ``Brain structures and functional connectivity
  associated with individual differences in trait proactive aggression,''
  \emph{Scientific reports}, vol.~9, no.~1, pp. 1--12, 2019.

\bibitem{devinsky1995contributions}
O.~Devinsky, M.~J. Morrell, and B.~A. Vogt, ``Contributions of anterior
  cingulate cortex to behaviour,'' \emph{Brain}, vol. 118, no.~1, pp. 279--306,
  1995.

\bibitem{bubb2018cingulum}
E.~J. Bubb, C.~Metzler-Baddeley, and J.~P. Aggleton, ``The cingulum bundle:
  anatomy, function, and dysfunction,'' \emph{Neuroscience \& Biobehavioral
  Reviews}, vol.~92, pp. 104--127, 2018.

\bibitem{menon2015salience}
V.~Menon, ``Salience network,'' \emph{Brain mapping}, 2015.

\bibitem{leech2014role}
R.~Leech and D.~J. Sharp, ``The role of the posterior cingulate cortex in
  cognition and disease,'' \emph{Brain}, vol. 137, no.~1, pp. 12--32, 2014.

\bibitem{stuss2013principles}
D.~T. Stuss and R.~T. Knight, \emph{Principles of frontal lobe function}.\hskip
  1em plus 0.5em minus 0.4em\relax Oxford University Press, 2013.

\bibitem{sawyer2017diagnosing}
R.~P. Sawyer, F.~Rodriguez-Porcel, M.~Hagen, R.~Shatz, and A.~J. Espay,
  ``Diagnosing the frontal variant of alzheimer’s disease: a clinician’s
  yellow brick road,'' \emph{Journal of clinical movement disorders}, vol.~4,
  no.~1, pp. 1--9, 2017.

\bibitem{chow2000personality}
T.~W. Chow, ``Personality in frontal lobe disorders,'' \emph{Current psychiatry
  reports}, vol.~2, no.~5, pp. 446--451, 2000.

\bibitem{bushara2003neural}
K.~O. Bushara, T.~Hanakawa, I.~Immisch, K.~Toma, K.~Kansaku, and M.~Hallett,
  ``Neural correlates of cross-modal binding,'' \emph{Nature neuroscience},
  vol.~6, no.~2, pp. 190--195, 2003.

\bibitem{penfield1955insula}
W.~Penfield and M.~Faulk~Jr, ``The insula: further observations on its
  function,'' \emph{Brain}, vol.~78, no.~4, pp. 445--470, 1955.

\bibitem{uddin2017structure}
L.~Q. Uddin, J.~S. Nomi, B.~H{\'e}bert-Seropian, J.~Ghaziri, and O.~Boucher,
  ``Structure and function of the human insula,'' \emph{Journal of clinical
  neurophysiology: official publication of the American Electroencephalographic
  Society}, vol.~34, no.~4, p. 300, 2017.

\bibitem{ramirez1996insular}
V.~Ram{\i}rez-Amaya, B.~Alvarez-Borda, C.~E. Ormsby, R.~D. Mart{\i}nez,
  R.~P{\'e}rez-Montfort, and F.~Berm{\'u}dez-Rattoni, ``Insular cortex lesions
  impair the acquisition of conditioned immunosuppression,'' \emph{Brain,
  behavior, and immunity}, vol.~10, no.~2, pp. 103--114, 1996.

\bibitem{ramirez1999conditioned}
V.~Ram{\i}rez-Amaya and F.~Bermudez-Rattoni, ``Conditioned enhancement of
  antibody production is disrupted by insular cortex and amygdala but not
  hippocampal lesions,'' \emph{Brain, behavior, and immunity}, vol.~13, no.~1,
  pp. 46--60, 1999.

\bibitem{lamm2010role}
C.~Lamm and T.~Singer, ``The role of anterior insular cortex in social
  emotions,'' \emph{Brain Structure and Function}, vol. 214, no. 5-6, pp.
  579--591, 2010.

\bibitem{de2006frames}
B.~De~Martino, D.~Kumaran, B.~Seymour, and R.~J. Dolan, ``Frames, biases, and
  rational decision-making in the human brain,'' \emph{Science}, vol. 313, no.
  5787, pp. 684--687, 2006.

\bibitem{xue2010impact}
G.~Xue, Z.~Lu, I.~P. Levin, and A.~Bechara, ``The impact of prior risk
  experiences on subsequent risky decision-making: the role of the insula,''
  \emph{Neuroimage}, vol.~50, no.~2, pp. 709--716, 2010.

\bibitem{mechelli2000differential}
A.~Mechelli, G.~W. Humphreys, K.~Mayall, A.~Olson, and C.~J. Price,
  ``Differential effects of word length and visual contrast in the fusiform and
  lingual gyri during,'' \emph{Proceedings of the Royal Society of London.
  Series B: Biological Sciences}, vol. 267, no. 1455, pp. 1909--1913, 2000.

\bibitem{mccarthy1999electrophysiological}
G.~McCarthy, A.~Puce, A.~Belger, and T.~Allison, ``Electrophysiological studies
  of human face perception. ii: Response properties of face-specific potentials
  generated in occipitotemporal cortex,'' \emph{Cerebral cortex}, vol.~9,
  no.~5, pp. 431--444, 1999.

\bibitem{kiselnikov2014activation}
A.~A. Kiselnikov, A.~A. Sergeev, A.~P. Dolgorukova, A.~V. Vartanov, J.~M.
  Glozman, S.~A. Kozlovskiy, and M.~M. Pyasik, ``Activation of left lingual
  gyrus related to working memory for schematic faces,'' \emph{International
  Journal of Psychophysiology}, vol.~94, pp. 120--261, 2014.

\bibitem{de2015visual}
B.~de~Gelder, M.~Tamietto, A.~J. Pegna, and J.~Van~den Stock, ``Visual imagery
  influences brain responses to visual stimulation in bilateral cortical
  blindness,'' \emph{Cortex}, vol.~72, pp. 15--26, 2015.

\bibitem{zhang2016gray}
L.~Zhang, L.~Qiao, Q.~Chen, W.~Yang, M.~Xu, X.~Yao, J.~Qiu, and D.~Yang, ``Gray
  matter volume of the lingual gyrus mediates the relationship between
  inhibition function and divergent thinking,'' \emph{Frontiers in psychology},
  vol.~7, p. 1532, 2016.

\bibitem{kim2020attachment}
J.~J. Kim, K.~M. Kent, R.~Cunnington, P.~Gilbert, and J.~N. Kirby, ``Attachment
  styles modulate neural markers of threat and imagery when engaging in
  self-criticism,'' \emph{Scientific reports}, vol.~10, no.~1, pp. 1--10, 2020.

\bibitem{filimon2009multiple}
F.~Filimon, J.~D. Nelson, R.-S. Huang, and M.~I. Sereno, ``Multiple parietal
  reach regions in humans: cortical representations for visual and
  proprioceptive feedback during on-line reaching,'' \emph{Journal of
  Neuroscience}, vol.~29, no.~9, pp. 2961--2971, 2009.

\bibitem{bassetti1993sensory}
C.~Bassetti, J.~Bogousslavsky, and F.~Regli, ``Sensory syndromes in parietal
  stroke,'' \emph{Neurology}, vol.~43, no.~10, pp. 1942--1942, 1993.

\bibitem{baldauf2008posterior}
D.~Baldauf, H.~Cui, and R.~A. Andersen, ``The posterior parietal cortex encodes
  in parallel both goals for double-reach sequences,'' \emph{Journal of
  Neuroscience}, vol.~28, no.~40, pp. 10\,081--10\,089, 2008.

\bibitem{fogassi2005motor}
L.~Fogassi and G.~Luppino, ``Motor functions of the parietal lobe,''
  \emph{Current opinion in neurobiology}, vol.~15, no.~6, pp. 626--631, 2005.

\bibitem{cappelletti2010role}
M.~Cappelletti, H.~L. Lee, E.~D. Freeman, and C.~J. Price, ``The role of right
  and left parietal lobes in the conceptual processing of numbers,''
  \emph{Journal of Cognitive Neuroscience}, vol.~22, no.~2, pp. 331--346, 2010.

\bibitem{vilberg2008memory}
K.~L. Vilberg and M.~D. Rugg, ``Memory retrieval and the parietal cortex: a
  review of evidence from a dual-process perspective,''
  \emph{Neuropsychologia}, vol.~46, no.~7, pp. 1787--1799, 2008.

\bibitem{terao2007primary}
Y.~Terao, Y.~Ugawa, T.~Yamamoto, Y.~Sakurai, T.~Masumoto, O.~Abe, Y.~Masutani,
  S.~Aoki, and S.~Tsuji, ``Primary face motor area as the motor representation
  of articulation,'' \emph{Journal of neurology}, vol. 254, no.~4, pp.
  442--447, 2007.

\bibitem{itabashi2016damage}
R.~Itabashi, Y.~Nishio, Y.~Kataoka, Y.~Yazawa, E.~Furui, M.~Matsuda, and
  E.~Mori, ``Damage to the left precentral gyrus is associated with apraxia of
  speech in acute stroke,'' \emph{Stroke}, vol.~47, no.~1, pp. 31--36, 2016.

\bibitem{kilner2013we}
J.~M. Kilner and R.~N. Lemon, ``What we know currently about mirror neurons,''
  \emph{Current biology}, vol.~23, no.~23, pp. R1057--R1062, 2013.

\bibitem{rizzolatti2010mirror}
G.~Rizzolatti and M.~Fabbri-Destro, ``Mirror neurons: from discovery to
  autism,'' \emph{Experimental brain research}, vol. 200, no.~3, pp. 223--237,
  2010.

\bibitem{buccino2004neural}
G.~Buccino, S.~Vogt, A.~Ritzl, G.~R. Fink, K.~Zilles, H.-J. Freund, and
  G.~Rizzolatti, ``Neural circuits underlying imitation learning of hand
  actions: an event-related fmri study,'' \emph{Neuron}, vol.~42, no.~2, pp.
  323--334, 2004.

\bibitem{stilling2018social}
R.~M. Stilling, G.~M. Moloney, F.~J. Ryan, A.~E. Hoban, T.~F. Bastiaanssen,
  F.~Shanahan, G.~Clarke, M.~J. Claesson, T.~G. Dinan, and J.~F. Cryan,
  ``Social interaction-induced activation of rna splicing in the amygdala of
  microbiome-deficient mice,'' \emph{Elife}, vol.~7, p. e33070, 2018.

\bibitem{jakab2012connectivity}
A.~Jakab, P.~P. Moln{\'a}r, P.~Bogner, M.~B{\'e}res, and E.~L. Ber{\'e}nyi,
  ``Connectivity-based parcellation reveals interhemispheric differences in the
  insula,'' \emph{Brain topography}, vol.~25, no.~3, pp. 264--271, 2012.

\bibitem{karnath2010right}
H.-O. Karnath and B.~Baier, ``Right insula for our sense of limb ownership and
  self-awareness of actions,'' \emph{Brain Structure and Function}, vol. 214,
  no. 5-6, pp. 411--417, 2010.

\bibitem{ghahremani2015role}
A.~Ghahremani, A.~Rastogi, and S.~Lam, ``The role of right anterior insula and
  salience processing in inhibitory control,'' \emph{Journal of Neuroscience},
  vol.~35, no.~8, pp. 3291--3292, 2015.

\bibitem{cai2014dissociable}
W.~Cai, S.~Ryali, T.~Chen, C.-S.~R. Li, and V.~Menon, ``Dissociable roles of
  right inferior frontal cortex and anterior insula in inhibitory control:
  evidence from intrinsic and task-related functional parcellation,
  connectivity, and response profile analyses across multiple datasets,''
  \emph{Journal of Neuroscience}, vol.~34, no.~44, pp. 14\,652--14\,667, 2014.

\bibitem{loth2016identification}
E.~Loth, W.~Spooren, L.~M. Ham, M.~B. Isaac, C.~Auriche-Benichou,
  T.~Banaschewski, S.~Baron-Cohen, K.~Broich, S.~Boelte, T.~Bourgeron
  \emph{et~al.}, ``Identification and validation of biomarkers for autism
  spectrum disorders,'' \emph{Nature reviews Drug discovery}, vol.~15, no.~1,
  pp. 70--70, 2016.

\bibitem{courchesne2004brain}
E.~Courchesne, ``Brain development in autism: early overgrowth followed by
  premature arrest of growth,'' \emph{Mental retardation and developmental
  disabilities research reviews}, vol.~10, no.~2, pp. 106--111, 2004.

\bibitem{ritvo2012postponing}
E.~R. Ritvo, ``Postponing the proposed changes in dsm 5 for autistic spectrum
  disorder until new scientific evidence adequately supports them,''
  \emph{Journal of autism and developmental disorders}, vol.~42, no.~9, pp.
  2021--2022, 2012.

\bibitem{singer2012diagnosis}
E.~Singer, ``Diagnosis: redefining autism,'' \emph{Nature}, vol. 491, no. 7422,
  pp. S12--S13, 2012.

\bibitem{tsai2014dsm}
L.~Y. Tsai and M.~Ghaziuddin, ``Dsm-5 asd moves forward into the past,''
  \emph{Journal of autism and developmental disorders}, vol.~44, no.~2, pp.
  321--330, 2014.

\bibitem{rojat2021explainable}
T.~Rojat, R.~Puget, D.~Filliat, J.~Del~Ser, R.~Gelin, and
  N.~D{\'\i}az-Rodr{\'\i}guez, ``Explainable artificial intelligence (xai) on
  timeseries data: A survey,'' \emph{arXiv preprint arXiv:2104.00950}, 2021.

\bibitem{schnack2010mapping}
H.~G. Schnack, N.~E. van Haren, R.~M. Brouwer, G.~C.~M. van Baal, M.~Picchioni,
  M.~Weisbrod, H.~Sauer, T.~D. Cannon, M.~Huttunen, C.~Lepage \emph{et~al.},
  ``Mapping reliability in multicenter mri: Voxel-based morphometry and
  cortical thickness,'' \emph{Human brain mapping}, vol.~31, no.~12, pp.
  1967--1982, 2010.

\bibitem{heinen2016robustness}
R.~Heinen, W.~H. Bouvy, A.~M. Mendrik, M.~A. Viergever, G.~J. Biessels, and
  J.~De~Bresser, ``Robustness of automated methods for brain volume
  measurements across different mri field strengths,'' \emph{PloS one},
  vol.~11, no.~10, p. e0165719, 2016.

\bibitem{guo2018repeatability}
C.~Guo, D.~Ferreira, K.~Fink, E.~Westman, and T.~Granberg, ``Repeatability and
  reproducibility of freesurfer, fsl-sienax and spm brain volumetric
  measurements and the effect of lesion filling in multiple sclerosis,''
  \emph{European radiology}, pp. 1--10, 2018.

\bibitem{fischl2004sequence}
B.~Fischl, D.~H. Salat, A.~J. Van Der~Kouwe, N.~Makris, F.~S{\'e}gonne, B.~T.
  Quinn, and A.~M. Dale, ``Sequence-independent segmentation of magnetic
  resonance images,'' \emph{Neuroimage}, vol.~23, pp. S69--S84, 2004.

\bibitem{ferrari2018common}
E.~Ferrari, P.~Bosco, G.~Spera, M.~E. Fantacci, and A.~Retico, ``Common
  pitfalls in machine learning applications to multi-center data: tests on the
  abide i and abide ii collections,'' in \emph{Joint Annual Meeting
  ISMRM-ESMRMB}, 2018.

\bibitem{american1994diagnostic}
A.~American Psychiatric~Association \emph{et~al.}, \emph{Diagnostic and
  statistical manual of mental disorders (DSM-IV)}.\hskip 1em plus 0.5em minus
  0.4em\relax Washington, DC: American psychiatric association Washington,
  1994.

\bibitem{marshall2008structural}
C.~R. Marshall, A.~Noor, J.~B. Vincent, A.~C. Lionel, L.~Feuk, J.~Skaug,
  M.~Shago, R.~Moessner, D.~Pinto, Y.~Ren \emph{et~al.}, ``Structural variation
  of chromosomes in autism spectrum disorder,'' \emph{The American Journal of
  Human Genetics}, vol.~82, no.~2, pp. 477--488, 2008.

\bibitem{weiss2009genome}
L.~A. Weiss and D.~E. Arking, ``A genome-wide linkage and association scan
  reveals novel loci for autism,'' \emph{Nature}, vol. 461, no. 7265, pp.
  802--808, 2009.

\bibitem{sebat2007strong}
J.~Sebat, B.~Lakshmi, D.~Malhotra, J.~Troge, C.~Lese-Martin, T.~Walsh,
  B.~Yamrom, S.~Yoon, A.~Krasnitz, J.~Kendall \emph{et~al.}, ``Strong
  association of de novo copy number mutations with autism,'' \emph{Science},
  vol. 316, no. 5823, pp. 445--449, 2007.

\bibitem{sanfeliu2019transcriptomic}
A.~Sanfeliu, K.~Hokamp, M.~Gill, and D.~Tropea, ``Transcriptomic analysis of
  mecp2 mutant mice reveals differentially expressed genes and altered
  mechanisms in both blood and brain,'' \emph{Frontiers in psychiatry},
  vol.~10, p. 278, 2019.

\bibitem{ross2016exclusive}
P.~D. Ross, J.~Guy, J.~Selfridge, B.~Kamal, N.~Bahey, K.~E. Tanner, T.~H.
  Gillingwater, R.~A. Jones, C.~M. Loughrey, C.~S. McCarroll \emph{et~al.},
  ``Exclusive expression of mecp2 in the nervous system distinguishes between
  brain and peripheral rett syndrome-like phenotypes,'' \emph{Human molecular
  genetics}, vol.~25, no.~20, pp. 4389--4404, 2016.

\bibitem{ben2009mouse}
S.~Ben-Shachar, M.~Chahrour, C.~Thaller, C.~A. Shaw, and H.~Y. Zoghbi, ``Mouse
  models of mecp2 disorders share gene expression changes in the cerebellum and
  hypothalamus,'' \emph{Human molecular genetics}, vol.~18, no.~13, pp.
  2431--2442, 2009.

\end{thebibliography}
\end{spacing}
}

\begin{appendices}

\chapter{Neural networks}
\label{app:neural_networks}

\section{Introduction}
\begin{figure}
    \centering
    \includegraphics[width=0.6\textwidth]{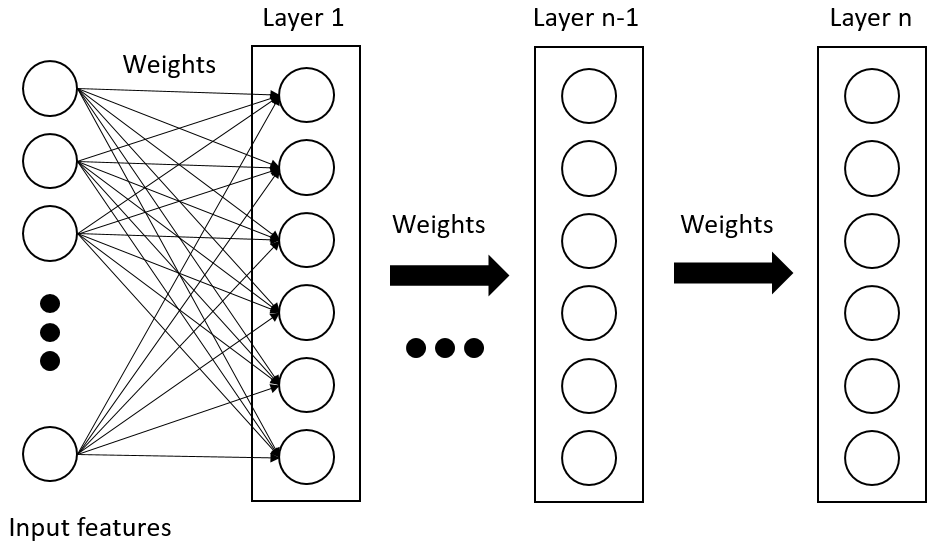}
    \caption{Illustration of the functioning of a neural network.}
    \label{fig:neural_network}
\end{figure}

A neural network is a kind of supervised machine learning algorithm. It functions by transforming a set of input features into a specific output. It is a computational model organized in layers of elements, called neurons, connected to each other by a series of weights (see Fig. \ref{fig:neural_network}). Each layer performs a mathematical operation on the output of the previous layer, with the first layer operating on the input features and the last layer representing the network prediction. The network is trained on a set of data in which the correct output is known, by adjusting the weights to minimize a cost function, which represents the error made by the network on that particular prediction.
In the context of research on ASD with rsfMRIs, the input features could be the rsfMRIs, while the output would be the class of the subject: ASD or TDC.
Given the large number of different available layers and the theoretically infinite ways in which these layers can be stacked and combined with each other, NNs represent an incredibly versatile class of models. The following section presents the key components used in this thesis.

\section{Basic notions on neural networks}
\label{subsec:networks_explainability}
Many layers and other operations performing different kinds of computations have been designed and their review is outside the scope of this work. Here, only the components used in this thesis are described, with a focus on the different layer types and activation functions.

\subsection{Layers}
\paragraph{\textbf{Dense layer.}} In a dense layer, every neuron is connected to every neuron of the previous layer by a different weight. The output of neuron $x_{i,j}$ (i-th neuron of the j-th layer) is:
\begin{equation}
    x_{i,j} = \vec{w}_{(i,j-1)}\cdot \vec{x}_{j-1}
\end{equation}
Where $\vec{x}_{j-1}$ is the vector containing the output of the neurons of layer $j-1$ and $\vec{w}_{(i,j-1)}$ is the vector representing the weights connecting the neurons of layer $j-1$ to the neuron $x_{(i,j)}$. This layer is one of the most general, since it connects every input neuron with every output neuron with a different weight. However, being it general, it is also inefficient, as no prior knowledge about the data structure is incorporated into the layer architecture and thus it is harder to train.

\paragraph{\textbf{Convolutional layer.}} A convolutional layer mathematically works by performing a convolution between a small subset of its input and a set of weights of equal size, called kernel. The weights are the same for every subset, thus significantly reducing the number of parameters required by the layer and therefore its memory footprint. More precisely, the layer operates as described by the following equation:
\begin{equation}
    x_{i,j,k} = \vec{w_k} \circledast \vec{x}_{(p_i,j-1)}
\end{equation}
Where $\circledast$ represents the convolution operation, $\vec{w_k}$ is the layer k-th kernel and $p_i$ indicates the partition of the input centered on the i-th neuron. The number of different kernels, their size (and thus the size of the partition to which they are applied) and the way the input is partitioned, such as whether the partitions overlap or not, are the layer hyperparameters. Note that this layer creates $k$ new versions of the input, each transformed by a different kernel.
Given that the same set of weights is scanned through the whole input, this layer is particularly apt at recognizing features independently of their position in the input (e.g, spatially invariant features if the input is an image).

\paragraph{\textbf{Pooling layer.}} This layer combines a set of input features into a single value, according to a certain function. The most commonly used ones are \textit{max}, \textit{mean} and \textit{median}. This layer is used to reduce the dimension of the input, preserving its most characteristic features. The formula to compute its output is:
\begin{equation}
     x_{i,j} = f(\vec{x}_{(p_i,j-1)})
\end{equation}
Where $p_i$ indicates the partition of the input centered on the i-th neuron and $f$ is one of the aforementioned functions, that transform a set of values into a single one.

\paragraph{\textbf{Dropout layer.}} The dropout layer is different from the other ones. It only takes effect during the training process and produces a copy of its input, with some random neurons set to 0, instead of their original value. Each neuron has the same probability to be set to 0, which is called the dropout rate. More precisely, given a random number $n\in[0,1]$, the output of this layer is: 

\begin{equation}
x_{i,j} = \left\{
        \begin{array}{ll}
            x_{(i,j-1)} & \quad n > r \\
            0 & \quad n \leq r
        \end{array}
    \right.
\end{equation}
Where $r$ is the dropout rate.\\
This layer is useful to counter the overfitting problem, i.e., the tendency of the network to \textit{memorize} the training dataset, by forcing the network to build a structure that is robust to the elimination of individual elements.

\subsection{Activation functions}
If one where to build a network composed of dense layers, without any other modification, the whole network could be described by a sequence of matrix multiplications, and thus would become a simple linear fit of the inputs into the output, unable to model non-linear behavior. In order to avoid this, \textit{activation functions} are used to introduce non-linearities into the model. 
Activation functions work by operating on individual neurons, transforming their output into a different one, according to the function definition. They operate on individual neurons, according to the following formula:
\begin{equation}
    x'_{(i,j)} = f(x_{(i,j)})
\end{equation}
Where $x_{(i,j)}$ is the i-th neuron of the j-th layer and $f$ is the activation function. Commonly used functional forms for $f$ are \textit{hyperbolic tangent}, \textit{sigmoid} and \textit{rectified linear units} (ReLU). This last form is calculated as in the following equation:
\begin{equation}
f(x) = \left\{
        \begin{array}{ll}
            x & \quad x > 0 \\
            0 & \quad x \leq 0
        \end{array}
    \right.
\end{equation}
All these functions have one thing in common: they are strongly non linear. However, all of them suffer from the so-called vanishing gradient effect. For certain values of x in fact, typically very large or very small (all negative values for the ReLU activation), the gradient of f tends to 0 or vanishes completely. This causes issues in the gradient-based training of the network. One possible solution, which was adopted in this thesis, is to slightly modify the ReLU function so that it has a non-zero gradient also for negative inputs:
\begin{equation}
    f(x) = \left\{
        \begin{array}{ll}
            x & \quad x > 0 \\
            \alpha \cdot x & \quad x \leq 0
        \end{array}
    \right.
\end{equation}
This function is called leaky ReLU, and has a hyperparameter $\alpha$ that is used to introduce a slope also in the negative part of the function domain. Typical values of $\alpha$ are very small, in the range of $10^{-3} - 10^{-2}$, to preserve non-linearity.
\chapter[Site dependency in neuroimaging data]{Site dependency in neuroimaging data\footnote{A large portion of this analysis has already been published by the author of this thesis in \cite{ferrari2020dealing}.}}
\label{site_dependency}

\section{Preamble}
In Section \ref{MRI_classification_ASD} it has been mentioned that the expected removal of acquisition dependency from raw MRI images using heavy pre-processing pipelines for feature extraction is not assured. 
Site dependency is in fact still largely misunderstood and poorly investigated \cite{auzias2016influence}.
Some studies show that the abstraction performed by feature extraction algorithms produces accurate data, insensible to scanner parameters \cite{schnack2010mapping,heinen2016robustness,guo2018repeatability}, as declared also in the Freesurfer release paper \cite{fischl2004sequence}. Few of them pinpoint scanner-related variations even on highly pre-processed data \cite{auzias2016influence}, that could be dangerous in a multivariate analysis.\\
In this appendix the site dependency of morphological features obtained from sMRIs using FreeSurfer\footnote{Freesufer is the most widely used software for MRI segmentation and extraction of brain features.} is investigated by presenting the main results published by the author of this thesis in \cite{ferrari2020dealing, ferrari2020measuring}.
Despite these works were based on sMRI, while this thesis relies on rsfMRI, their results show how complex it is to regress out of data site dependency  eliminating completely the possibility to stumble into CEs. Thus, the following analyses motivates the necessity to develop an algorithm resilient to CEs of site and other sources of variation, which is the choice made in this thesis.

\section{Site dependency evaluation in features extracted from sMRIs}
To estimate site dependency of morphological brain features extracted with Freesufer, four analyses have been conducted:
\begin{enumerate}
    \item \textit{Univariate assessment of site effect}\\
    It verifies whether the difference in individual feature values of subjects from different sites is significantly greater with respect to that of subjects belonging to the same site.
    \item \textit{Multivariate assessment of site effect}\\
    It evaluates the capability to distinguish subjects acquired by different sites with a multivariate approach.
    \item \textit{Number of features necessary to detect site effect}\\
    It investigates how many features are necessary to distinguish subjects belonging to different sites.
    \item \textit{Confounding effect of site}\\
    It studies the magnitude of the site confounding effect with respect to other confounders in the ASD/TDC classification.
\end{enumerate}
For all these analyses data have been extracted from sMRI images taken from ABIDE using Freesurfer v6.0. Each subject was represented by a vector of 296 features comprising: whole brain (called "global") features, volume of sub-cortical regions and 5 features of the cortical ones: volume, surface area, mean thickness and mean curvature.

\subsection{Univariate assessment of site effect}
\begin{figure}
    \centering
    \includegraphics[scale=0.28]{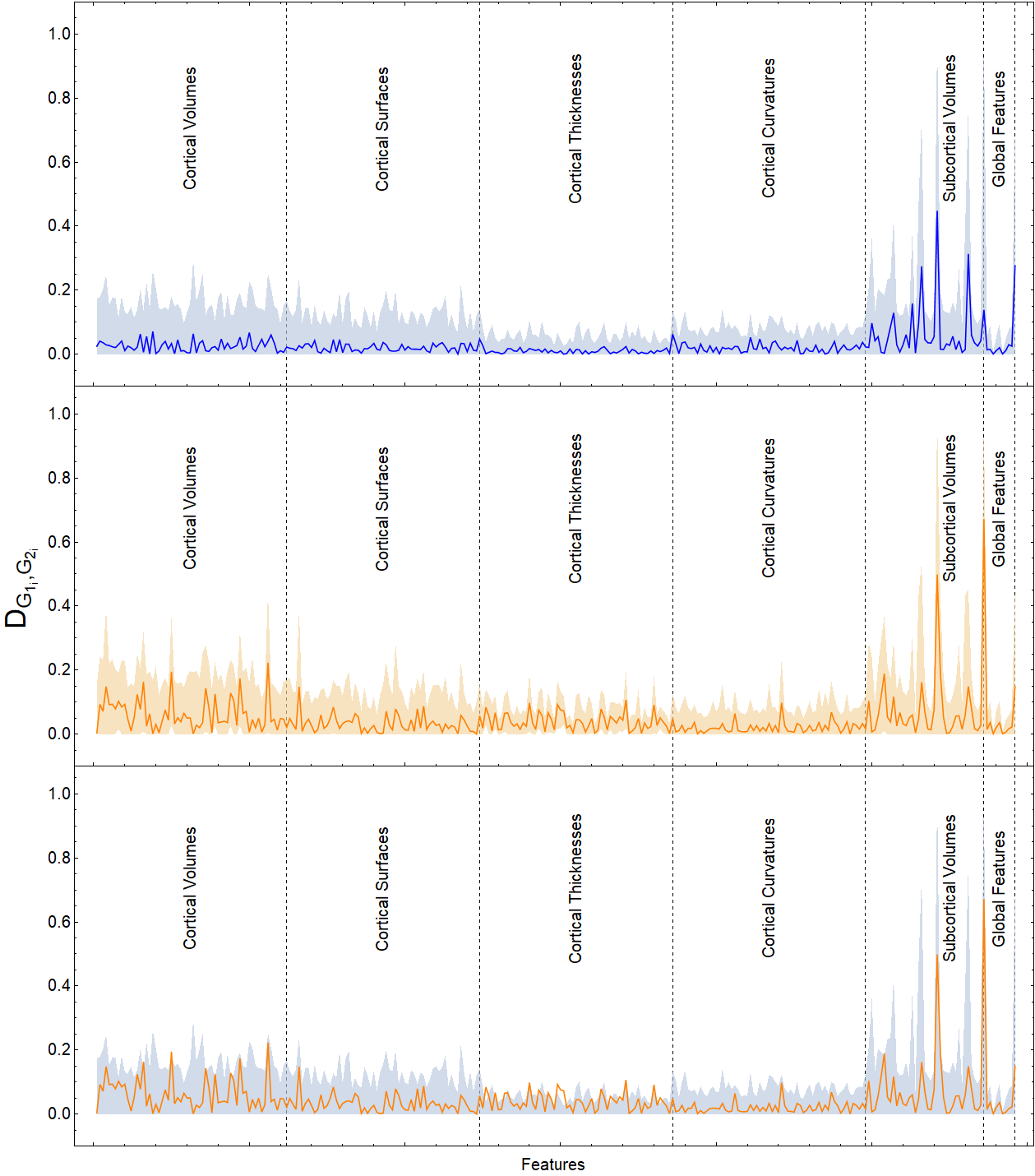}
    \caption{Differences across the feature vectors of subjects belonging to the same site NYU-I (blue line) and of subjects belonging to two different sites (orange line): NYU-I and $KKI_{8ch}-II$.\\ \mypictures{ferrari2020dealing}}
    \label{fig:graficoBLUeARANCIO}
\end{figure}

The objective of this analysis is to compare the differences between the individual features of two subgroups of subjects, $X_1$ and $X_2$, coming from the same site $X$, with the differences between two subgroups of subjects belonging to two different sites, $X$ and $Y$. A rather simple method to evaluate the differences between two groups of subjects $G_1$ and $G_2$ consists in computing the following quantity:
\begin{equation}
\centering
\vv{D_{G_1,G_2}}= {\left\{  D_{G_1,G_{2_i}}   \right\}}_{i=0}^{n} = {\left \{ \dfrac{|\mu_{G_{1_i}} - \mu_{G_{2_i}}|}{\mu_{G_{1}\cup G_{2} }}\right \}}_{i=0}^{n}
\end{equation}
Where $\mu_{G_{j_i}}$ is the average value of the feature $i$ between all the subjects in group $j$, while $\mu_{G_{1}\cup G_{2} }$ is the average feature obtained from both $G_1$ and $G_2$ taken as a single set.
In this way, $\vv{D_{G_1,G_2}}$ is a vector with the same length of the feature vectors that represent each subject; its components show the normalized difference between the two groups for each feature. \\
In order to understand whether the variability between groups is due to subject characteristics or site effect, three subgroups of subjects ($X_1$, $X_2$ and $Y$) have been analyzed; each composed by an equal amount of matched TDC subjects. In this way, by computing the two quantities $\vv{D_{X_1,X_2}}$ and $\vv{D_{X_1,Y}}$, it is possible to compare the physiological difference between two random groups acquired in the same way (the first quantity) and the difference between two groups differently acquired (the second quantity).\\
The vectors $\vv{D_{X_1,X_2}}$ and $\vv{D_{X_1,Y}}$ are depicted with a blue and an orange line in Fig. \ref{fig:graficoBLUeARANCIO}. The light blue band and the orange band represent respectively the standard deviations of the features computed on the sets $X_{1}\cup X_{2}$ and $X_{1}\cup Y$.
The first two plots show significant differences, however looking at the third one, in which $\vv{D_{X_1,Y}}$ is superposed to the standard deviation band of the set $X_{1}\cup X_{2}$, it can be noted that the differences between the two sites are almost never greater than the natural variability of the features within a single site.\\ 
The few orange peaks above the light blue band are in the cortical features. Nonetheless, this study shows that a univariate analysis is not able to identify significant differences imputable to site effect.
This result confirms previous studies \cite{schnack2010mapping,heinen2016robustness,guo2018repeatability,fischl2004sequence} that analyzed the site dependency with univariate methods.

\subsection{Multivariate assessment of site effect}
To verify if it is possible to detect site effect in a multivariate way, a \ac{LR} classifier has been trained to distinguish the TDC subjects of two different sites matched by age and gender. 
In Tab. \ref{tab:AUCsitidiversi} the mean AUCs obtained training 10 classifiers in cross-validation are reported.
The white cells of the table are those in which the two sites used a different scanner model and/or different acquisition protocols. It can be observed that in these cases, the classifier is able to predict the subject site very accurately. 
The yellow cells report the results obtained on samples coming from the same site, thus using the same scanner model and acquisition protocol, but that have been acquired  either in different data collection rounds or with a different physical scanner (but of the same model). The AUCs obtained on these data are lower, but still significantly higher than the null value of 0.5.\\
Finally, the results obtained comparing matched subjects coming from the same site yielded the AUCs highlighted in grey, with random guess performances.

\begin{table}[b]
   \centering
\resizebox{0.8\textwidth}{!}{
\begin{tabular}{|
>{\columncolor{white}}c|
>{\columncolor{white}}c|
>{\columncolor{white}}c|
>{\columncolor{white}}c|
>{\columncolor{white}}c|
>{\columncolor{white}}c|
>{\columncolor{white}}c|
>{\columncolor{white}}c|
>{\columncolor{white}}c|
>{\columncolor{white}}c|}
\hline
Site ID & $NYU$-I & $NYU_1$-II & $NYU_2$-II & $OHSU$-I & $OHSU$-II & $USM$-I & $USM$-II & $UM_1$-I & $UM_2$-II\\
\hline
\hline
$NYU$-I & 
\multicolumn{1}{>{\columncolor{gray}}c|}{0.51\textpm 0.12} &
\multicolumn{1}{>{\columncolor{yellow}}c|}{0.78\textpm 0.08} &  
\multicolumn{1}{>{\columncolor{yellow}}c|}{0.89\textpm 0.15} &
0.99\textpm 0.04 & 1.00\textpm 0.01 & 0.99\textpm 0.02 & 1.00\textpm 0.02 & 0.99\textpm 0.02 & 0.98\textpm 0.06\\
\hline
$NYU_1$-II &  & 
\multicolumn{1}{>{\columncolor{gray}}c|}{0.48\textpm 0.21} &
\multicolumn{1}{>{\columncolor{yellow}}c|}{0.70\textpm 0.23} &  
0.99\textpm 0.05 & 1.00\textpm 0.01 & 1.00\textpm 0.02 & 1.00\textpm 0.02 & 0.99\textpm 0.02 & 0.98\textpm 0.05\\
\hline
$NYU_2$-II &  &  & 
\multicolumn{1}{>{\columncolor{gray}}c|}{0.57\textpm 0.09} &
1.00\textpm 0.03 & 0.98\textpm 0.10 & 0.99\textpm 0.04 & 0.99\textpm 0.05 & 1.00\textpm 0.00 & 1.00\textpm 0.00 \\
\hline
$OHSU$-I &  &  &  &
\multicolumn{1}{>{\columncolor{gray}}c|}{0.55\textpm 0.19} &
\multicolumn{1}{>{\columncolor{yellow}}c|}{0.63\textpm 0.24} &  
0.97\textpm 0.08 & 0.96\textpm 0.08 & 1.00\textpm 0.00 & 1.00\textpm 0.00\\
\hline
$OHSU$-II &  &  &  &  &
\multicolumn{1}{>{\columncolor{gray}}c|}{0.43\textpm 0.21} &
0.99\textpm 0.03 & 0.96\textpm 0.10 & 0.98\textpm 0.05 & 0.98\textpm 0.06\\
\hline
$USM$-I &  &  &  &  &  &
\multicolumn{1}{>{\columncolor{gray}}c|}{0.50\textpm 0.15} &
\multicolumn{1}{>{\columncolor{yellow}}c|}{0.75\textpm 0.21} &  
0.99\textpm 0.03 & 0.99\textpm 0.04\\
\hline
$USM$-II &  &  &  &  &  &  &
\multicolumn{1}{>{\columncolor{gray}}c|}{0.55\textpm 0.20} &
0.97\textpm 0.06 & 0.98\textpm 0.06\\
\hline
$UM_1$-I &  &  &  &  &  &  &  &
\multicolumn{1}{>{\columncolor{gray}}c|}{0.48\textpm 0.17} &
\multicolumn{1}{>{\columncolor{yellow}}c|}{0.96\textpm 0.09}\\
\hline
$UM_2$-I &  &  &  &  &  &  &  &  &
\multicolumn{1}{>{\columncolor{gray}}c|}{0.57\textpm 0.15}\\
\hline
\end{tabular}}
    \caption{Performance (mean AUC \textpm standard deviation) of classification obtained training a Logistic Regression to distinguish between data acquired by different sites. \\
    \mypictures{ferrari2018common}}
    \label{tab:AUCsitidiversi}
\end{table}

\subsection{Number of features necessary to detect site effect}
\begin{figure}
    \centering
    \includegraphics[scale=0.45]{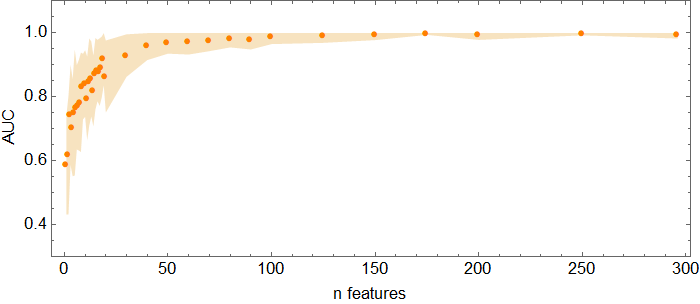}
    \caption{AUC obtained in the classification between subjects coming from NYU-I and $KKI_{8ch}-II$ samples, with a $n$ number of randomly chosen features under consideration, where $n$ vary along the x axis. \mypictures{ferrari2020dealing}}
    \label{AUC_along_vector}
\end{figure}

With the previous analyses, it has been shown that multivariate approaches are able to detect site effect with very high performance, while univariate ones fail completely.
So, the analysis described in this section aims at understanding the minimum number of features necessary to spot site dependency.
To this end, several classifiers were trained to distinguish matched TDCs belonging to the two largest samples (NYU-I and $KKI_{8ch}-II$). In each training set, every subject was represented with a fixed number $n$ of features.
Then, the performances of the classifiers obtained using different combinations of $n$ features, randomly chosen from the original 296 features were estimated.
The plot in Fig. \ref{AUC_along_vector} is obtained repeating this operation for $n$ decreasing from 296 to 1. The orange dots represent the mean AUC of these classifiers and the orange band shows the standard deviation of these AUC values.
As it can be observed, the classifier performances start to decrease substantially only below 10-20 features, reaching 0.5 only when the subjects are represented by a single feature.
So, to avoid detecting site effect, the analysis should be limited to a number of features so low as to completely eliminate the advantage of a multivariate approach. It is also interesting to note that the standard deviation is very low for $n>10$, which means that irrespectively of the chosen features, the classifier is able to distinguish the site with extremely good precision.

\subsection{Confounding effect of site}
\begin{figure}
    \centering
    \includegraphics[width=0.6\textwidth]{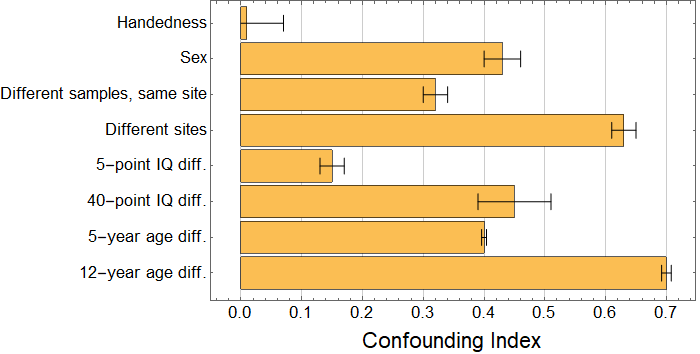}
    \caption{CI value of several possible confounders for ASD. The CI is an index that ranks from 0 to 1 the effect of a confounder variable on a certain binary classification task. For ASD, sex, acquisition modalities, acquisition site, large IQ differences and even small age differences are all strongly confounding. Data from \cite{ferrari2020dealing}.
    \mypictures{Ferrari2020autismChapter}}
    \label{fig:ci}
\end{figure}

Until now, it has been shown that a multivariate approach can easily recognize acquisition site even when analyzing few features, but this doesn't necessarily imply that site dependency has a strong confounding effect. For instance, if distinguishing ASD from TDC was a trivial task, the data dependency from other variables would not mislead the training process. In this analysis thus, the CI \cite{ferrari2020measuring}, already described in Sec. \ref{sec:confounders_ci}, has been applied to quantify and compare the effects of commonly cited sources of heterogeneity in sMRI, including site dependency, when trying to classify ASDs and TDCs subjects with a Logistic Regression algorithm. The results (Fig. \ref{fig:ci}) show that site dependency has a very strong CE, which is significantly higher than that of sex and a 5-year age difference and only slightly lower to the one of a 12-year age difference. Similarly to what has been observed in the second analysis, also the acquisition sample shows a CI significantly different from zero, although more modest with respect to the one of the acquisition site.

\section{Conclusions}
Summarizing, intra-site and inter-site variability of highly pre-processed MRI features are not significantly different if analyzed with a univariate analysis.
On the other hand, a multivariate analysis with more than 10 randomly selected features easily captures site data dependency. Thus, finding a criterion to correct each feature singularly can be very difficult and modeling multivariate site dependency is even more complicated because acquisition modalities can be detected by multiple patterns involving different combinations of features. 
Furthermore, the last analysis proved that site dependency is not only easily detectable from data but can really mislead the learning process when training to distinguish ASDs and TDCs, which is a hard task. Thus interventions to avoid learning from this strong confounder are necessary. Given that removing site dependency from data can be very challenging and matching them for all the sites and other possible confounders may severely limit the number of suitable training data; the only remaining option consists in avoiding learning from confounders by operating at the model level.
In this thesis this has been achieved through an adversarial learning framework that searches a saddle point between maximizing the classification performance and minimizing the ability to recognize site provenience (see Sec. \ref{sec:mm_network} for the details of the model adopted).

\chapter{Why studying gene expression regional variations between ASDs and TDCs?}
\label{app:differences}
\begin{figure} 
\centering
\begin{tabular}{ccc}
\subfloat[]{\includegraphics[width = 0.26\textwidth]{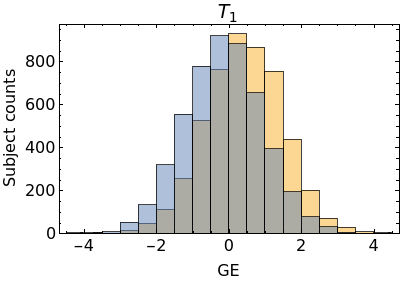}} &
\subfloat[]{\includegraphics[width = 0.26\textwidth]{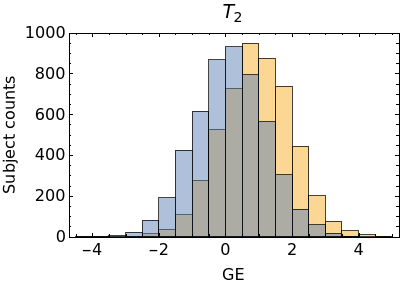}} &
\subfloat[]{\includegraphics[width = 0.3\textwidth]{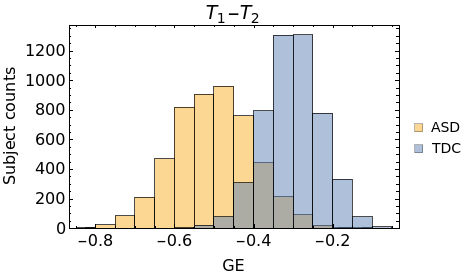}}
\end{tabular}
\caption{Example of a simulation in which the conditions in Eq. \ref{cond} hold. Fig (a) and (b) show the distribution of the GE expression simulated in two separate tissues, while Fig. (c) shows the distributions of their differences. }
\label{fig:sim_3}
\end{figure}

\begin{figure} 
\centering
\begin{tabular}{cc}
\subfloat[]{\includegraphics[width = 0.45\textwidth]{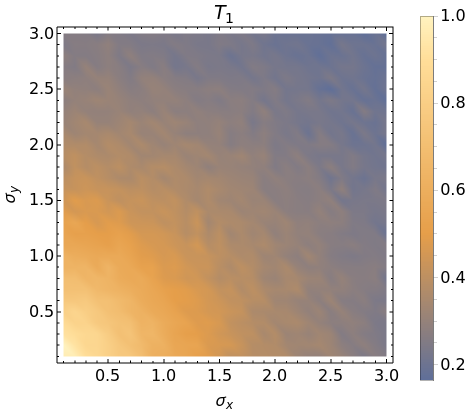}} &
\subfloat[]{\includegraphics[width = 0.45\textwidth]{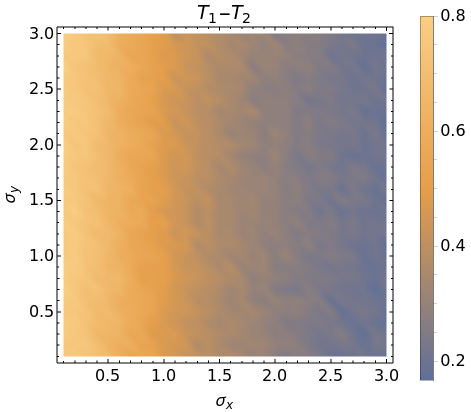}} 
\end{tabular}
\caption{Heatmaps showing the separation between transcriptomics data of ASDs and TDCs, measured with effect size, varying the effect of subject identity $\sigma_{y}$ and the tissue inhomogeneity $\sigma_{x}$. The results obtained analyzing GE data from one sample are illustrated in Fig. (a), while the ones from regional GE difference are reported in Fig. (b).}\label{fig:heat_1}
\end{figure}

\todo[inline]{mettere le correzioni di cellerino}
Most of the gene expression studies on ASD focus on finding a list of genes DE between ASDs and TDCs by comparing their transcriptome profile acquired in the same tissue or brain region. Instead, as declared in chapter \ref{chap:contribution}, one of the objectives of this work is to analyze whether inter-tissue gene expression differences characterizing the disorder exist.\\
Besides the fact that as it emerges from Sec.s \ref{sec:bkg_summary_review_GE_ASD} and \ref{brainASD-s-o-a}, several studies suggest that transcriptome abnormalities in ASD are region-specific, other three important reasons why this approach is worth pursuing are listed below.
\begin{enumerate}
    \item \textbf{The existence of genes with highly conserved patterns across brain regions of healthy subjects.}\\
    The stereotyped structural and functional architecture of the human brain suggests that it should be the result of a highly consistent transcriptomic pattern across brain anatomical structures. With this hypothesis, a milestone study \cite{hawrylycz2015canonical} found a list of genes with high DS, i.e. genes with reproducible differential expression relationships across brain regions, and showed that the distribution of genes associated with psychiatric disorders, among which ASD, was shifted toward high DS genes as compared to the others. This suggests that altered spatial transcriptomic patterns may play a role in the pathogenesis of ASD.
    To corroborate this hypothesis, it has been noted that the number of DE genes between frontal and temporal cortex, typically high in TDCs, is significantly reduced in ASDs \cite{voineagu2011transcriptomic}. However a more exhaustive comparison on regional transcriptomic differences seems missing.
    \item \textbf{Gene expression spatial variation of MeCP2, whose mutation causes Rett's syndrome.}\\
    Rett's syndrome is a monogenic disorder that has a significant phenotypic overlap with ASD (despite appearing more severe) and for this reason, prior to the discovery of its genetic cause, the two disorders were both classified as Pervasive Developmental Disorders \cite{american1994diagnostic}. The mutated gene causing the Rett's syndorme, MeCP2, is an essential epigenetic regulator in human brain development. It has been proved that some MeCP2 target genes are implicated in ASD \cite{marshall2008structural,weiss2009genome,sebat2007strong}, explaining the symptomatological similarities between the two conditions. The MeCP2 expression varies across brain regions \cite{sanfeliu2019transcriptomic,ross2016exclusive}, as well as the expression of the genes that it regulates \cite{ben2009mouse}. This aspect suggests that gene expression regional differences kept stable in healthy subjects by a regulator may be altered in subjects with a regulator-related disorder. 
    Thus, despite it is well known that ASD is not a monogenic disorder, if a gene expression regulator is involved in the ASD etiology, studying regional differences might reveal hitherto undiscovered molecular correlates.
    \item \textbf{Gene expression regional differences within the same subjects may be less subject-dependant and more robustly associated to  disorder's traits.}\\
    A practical reason why it could be interesting to study regional differences is that in this way subject-related variability could be eliminated or at least attenuated.
    To understand under which conditions comparing individual regional GE differences between ASDs and TDCs helps to reveal gene dependency from the disorder, a simple simulation-based study has been conducted.\\ 
    Supposing that the normalization procedures have eliminated the sample dependence from the data, there are mainly other four sources of variability: 
    \begin{itemize}
        \item subject-dependence, which summarizes the contribution of various features like: sex, age, cause of death, PMI, brain bank, etc;
        \item disease-dependence, modeled as separated from subject-dependence for simulation purposes;
        \item tissue-dependence;
        \item noise due to random effects, measurement errors, correction errors, etc.
    \end{itemize}
    How these factors influence gene expression is of course unknown, but in a linear approximation (which is at the basis of many algorithms for GE analysis) the gene expression level of a gene $g$ in a tissue $x$ of the subject $y$ with disease $z$ can be modeled as:
    \begin{equation}
        g_{x,y,z}= g_0+n+t+s+d
    \end{equation}
    where $n, t, s$ and $d$ represent respectively the influence of noise, type of tissue, subject and disease to the value of $g_0$ which is an hypothetical baseline level of expression of gene $g$.
    To simplify the simulations, $g_0$ has been set equal to 0 and the others elements have been modeled as real numbers taken from the following normal distributions ($\mathcal{N}$):
    \begin{align*}
        n &\in \mathcal{N}(0,\sigma_n)\\
        t &\in \mathcal{N}(\mu_x,\sigma_x)\\
        s &\in \mathcal{N}(0,\sigma_y)\\
        d &\in \mathcal{N}(\mu_{z,x},\sigma_{z,x})
    \end{align*}
    With these choices only the influence of tissue and disease can shift the value of $g$ from $g_0 = 0$, while the dependence from subject is modeled as an increased variance in the value of the data. 
    As a consequence of what has been previously discussed, it is assumed that the dependence of GE from disease is different for each type of tissue $x$, but for the sake of simplicity the variance $\sigma_{z,x}$ has been set equal for each tissue and thus renamed $\sigma_{z}$.\\
    Under these assumptions, several simulations have been conducted, varying all the mentioned parameters, to verify in which conditions it can be considered convenient to analyze regional GE differences.
    It has been observed that when the following conditions are true, studying regional GE differences allow to better discriminate between ASDs and TDCs (see Fig. \ref{fig:sim_3}):
    \begin{equation}
         I)\;\;\; \sigma_{y}>>\sigma_{x} \;\;\;\;\;\;\;\;\;\; II)\;\;\; \mu_{z,x=1} \neq \mu_{z,x=2}  
    \label{cond}
    \end{equation}
    These conditions basically mean that the variability due to subject identity is much higher than the variability within a tissue (which sounds intuitively correct) and that the disease affects in different ways the two tissues (which is the main hypothesis behind this approach). 
    Within reasonable bounds, this phenomenon holds almost regardless of the value of the other parameters and even when the various factors influencing GE are sampled from distributions different than normal.\\
    Fig. \ref{fig:heat_1} shows two heatmaps representing the variation of the effect size, i.e. the normalized distance between the ASD and TDC distributions, measured analyzing only the GE in one tissue (Fig. a) or the GE difference between two tissues (Fig. b). It clearly emerges that the GE difference is insensitive to subject-identity variability $\sigma_y$, which has been removed through the subtraction operation.
    Another interesting result is illustrated in Fig. \ref{fig:heat_2} that shows that even when it is difficult to separate ASDs from TDCs from single tissue data, if the disease affects differently the two tissues, analyzing regional GE difference becomes much more discriminant.\\
    This simulation has no claim to be an accurate reproduction of GE dependencies, it simply provides an intuitive demonstration that regional GE difference is a useful measure because it can reduce the dependence of subject-related factors and may allow to capture disease effects that would be otherwise impossible to discover.
\end{enumerate}

\begin{figure} 
\centering
\begin{tabular}{cc}
\subfloat[]{\includegraphics[width = 0.45\textwidth]{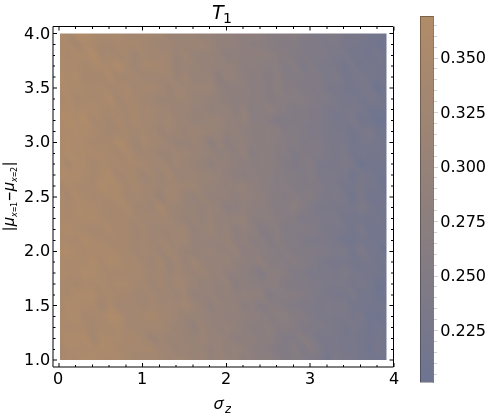}} &
\subfloat[]{\includegraphics[width = 0.45\textwidth]{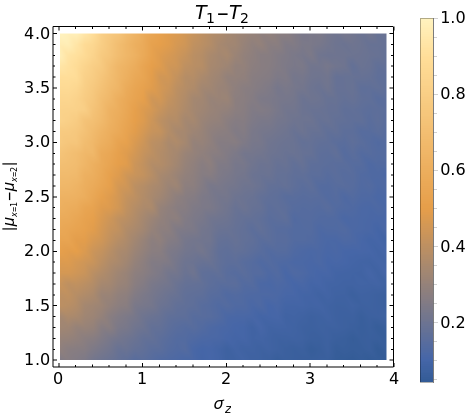}} 
\end{tabular}
\caption{Heatmaps showing how the effect size varies with respect to the difference between the mean variation induced by the type of tissue $|\mu_{x=1}-\mu_{x=2}|$ and the variability caused by the disease $\sigma_{z}$. The results obtained analyzing GE data from one sample are illustrated in Fig. (a), while the ones from regional GE difference are reported in Fig. (b).}\label{fig:heat_2}
\end{figure}
\chapter{Results of the confounder analysis}
\label{app:confounders}
\begin{figure}
\begin{subfigure}{.5\textwidth}
  \centering
  \includegraphics[width=.8\linewidth]{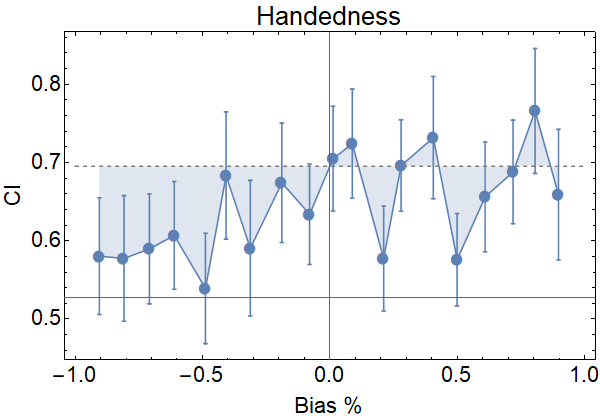}
  \caption{1a}
  \label{fig:ci_hand}
\end{subfigure}%
\begin{subfigure}{.5\textwidth}
  \centering
  \includegraphics[width=.8\linewidth]{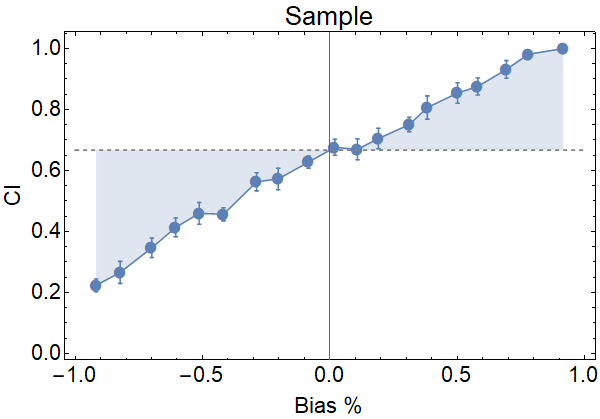}
  \caption{1b}
  \label{fig:ci_sample}
\end{subfigure}
\begin{subfigure}{.5\textwidth}
  \centering
  \includegraphics[width=.8\linewidth]{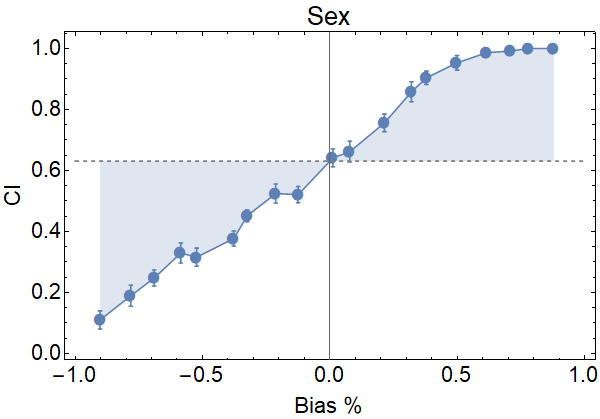}
  \caption{1a}
  \label{fig:ci_sex}
\end{subfigure}%
\begin{subfigure}{.5\textwidth}
  \centering
  \includegraphics[width=.8\linewidth]{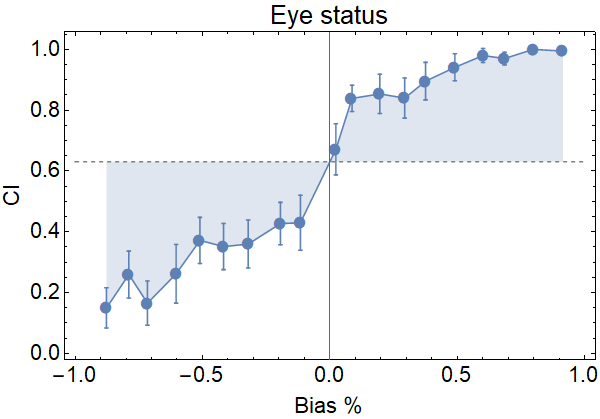}
  \caption{1b}
  \label{fig:ci_eye_status}
\end{subfigure}
\begin{subfigure}{.5\textwidth}
  \centering
  \includegraphics[width=.8\linewidth]{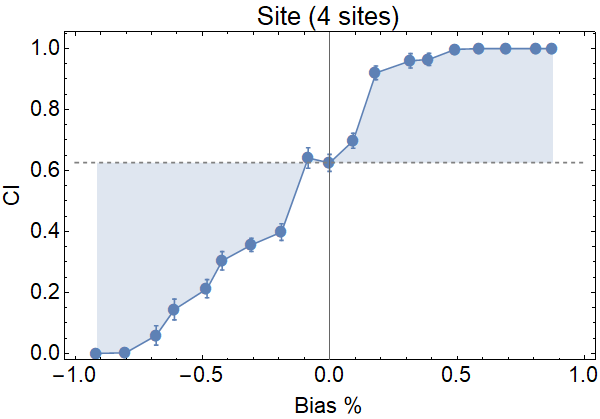}
  \caption{1a}
  \label{fig:ci_site_4_sites}
\end{subfigure}%
\begin{subfigure}{.5\textwidth}
  \centering
  \includegraphics[width=.8\linewidth]{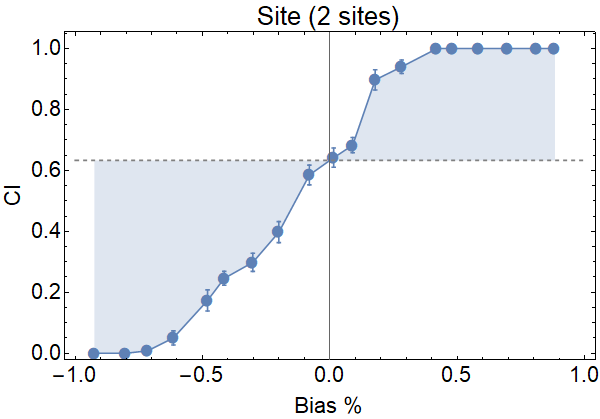}
  \caption{1b}
  \label{fig:ci_site_2_sites}
\end{subfigure}
\caption{Plots of the CI calculation. The dashed line in each plot shows the AUC obtained for $bias = 0$. The (signed) filling represents the CI value.}
\label{fig:ci_discrete}
\end{figure}

This appendix contains the plots of the CI calculation for the different confounder variables. Categorical confounders are reported in Fig. \ref{fig:ci_discrete}. All plots with the exception of handedness show a well-defined growing trend, monotonous with the bias, characteristic of strong confounders.

\begin{figure}
\begin{subfigure}{.5\textwidth}
  \centering
  \includegraphics[width=.8\linewidth]{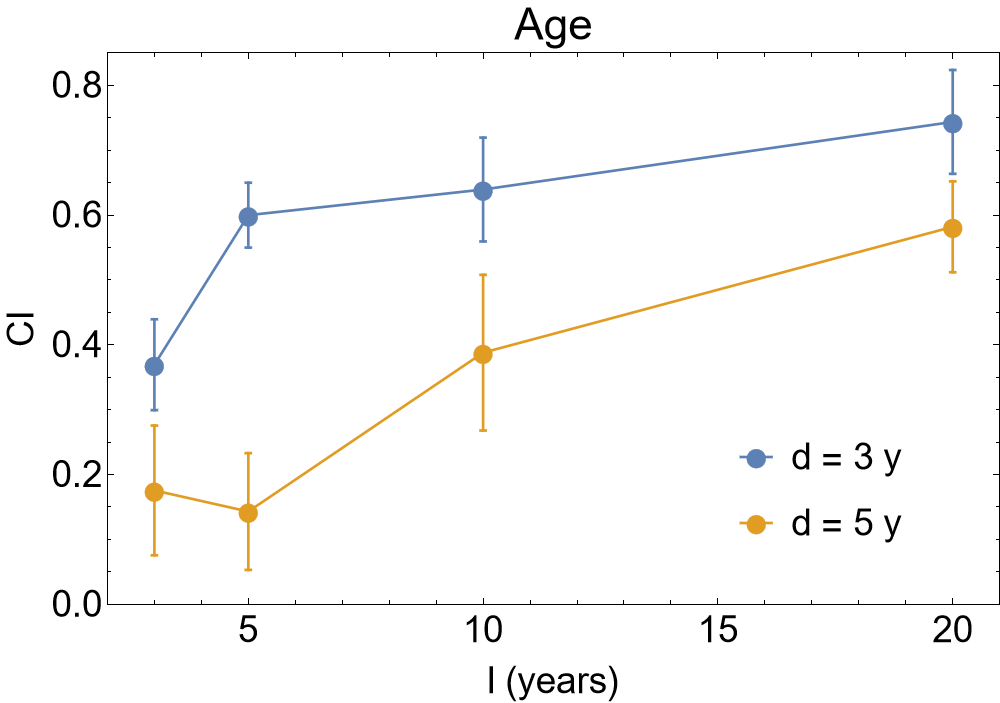}
  \caption{1a}
  \label{fig:ci_age}
\end{subfigure}%
\begin{subfigure}{.5\textwidth}
  \centering
  \includegraphics[width=.8\linewidth]{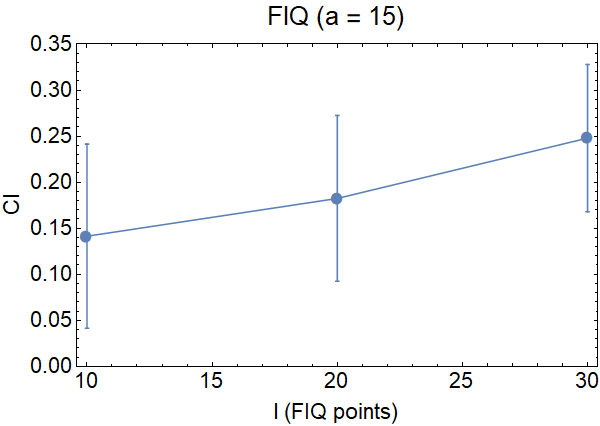}
  \caption{1b}
  \label{fig:ci_fiq}
\end{subfigure}
\caption{CI of the continuous variables as a function of $l$ and $d$. Age (left), FIQ (right)}
\label{fig:ci_continuous}
\end{figure}

For continuous variables instead, the CI value as a function of $l$ for different values of $d$ is shown in Fig. \ref{fig:ci_continuous}. For both age and FIQ, an increasing $l$, i.e., an increase in the difference between the two groups that are compared, produces an increase in CI. 

\chapter{Resilience of the neural network to confounders}
\label{app:resilience}

\begin{figure}[hb!]
    \centering
    \includegraphics[width=0.8\textwidth]{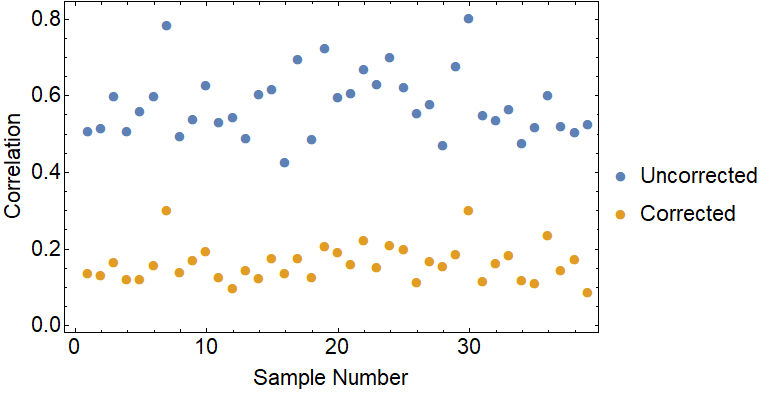}
    \caption{Correlation between true sample and sample prediction for the uncorrected (blue) and corrected (orange) networks.}
    \label{fig:correlations_corr_uncorr}
\end{figure}

Fig. \ref{fig:correlations_corr_uncorr} compares the sample-by-sample average correlation between the prediction of the $\mathbb{C}$ component of the neural network and the true sample on the training, validation and test datasets together. It can be observed that the use of the confounder correction substantially reduces the correlation. It must also be noted that none of the sites shows abnormally high correlation values, meaning that the mean correlation across all sites, reported in the main body of the thesis, is a good representation of the network performance.

\begin{figure}
\begin{subfigure}{.5\textwidth}
  \centering
  \includegraphics[width=.8\linewidth]{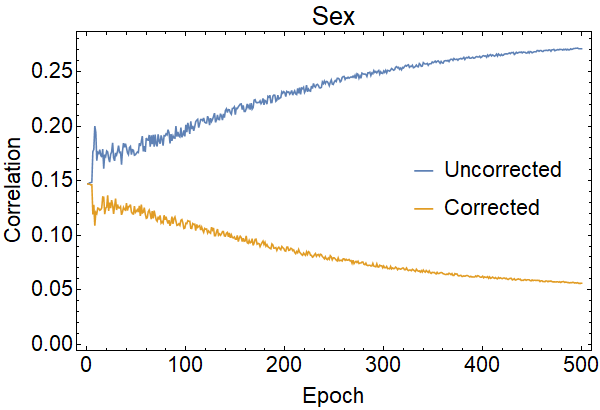}
  \label{fig:sex_corr}
\end{subfigure}%
\begin{subfigure}{.5\textwidth}
  \centering
  \includegraphics[width=.8\linewidth]{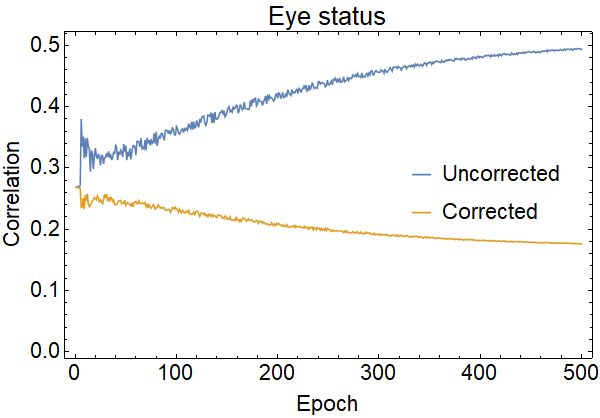}
  \label{fig:eyes_corr}
\end{subfigure}
\begin{subfigure}{.5\textwidth}
  \centering
  \includegraphics[width=.8\linewidth]{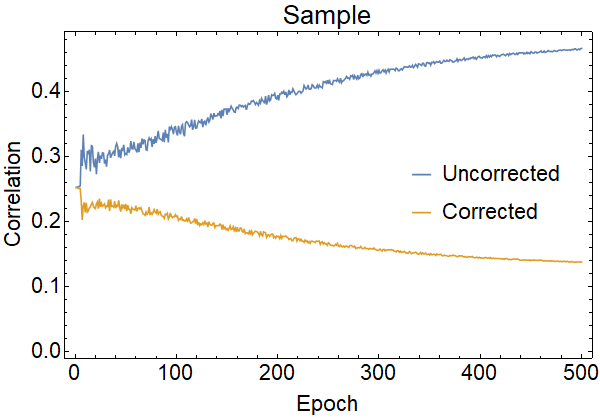}
  \label{fig:sample_corr}
\end{subfigure}%
\begin{subfigure}{.5\textwidth}
  \centering
  \includegraphics[width=.8\linewidth]{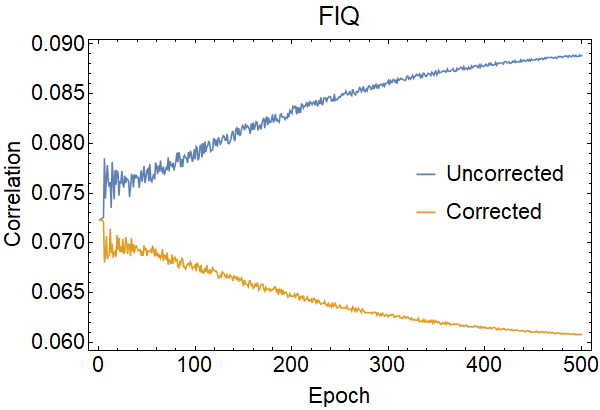}
  \label{fig:fiq_corr}
\end{subfigure}
\centering
\begin{subfigure}{.5\textwidth}
  \centering
  \includegraphics[width=.8\linewidth]{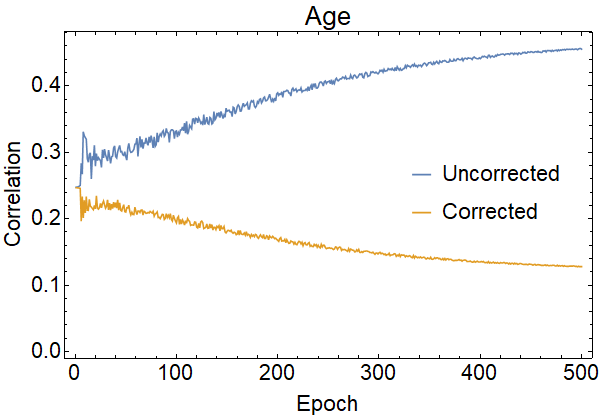}
  \label{fig:age_corr}
\end{subfigure}%
\caption{Plots of correlations between the prediction of $\mathbb{C}$ for the value of a confounder and its real value. The orange and blue lines show the prediction of the network with and without the confounder correction applied, respectively.}
\label{fig:correlations_trend_epochs}
\end{figure}

Fig. \ref{fig:correlations_trend_epochs} shows the trend of the correlation between the predictions of $\mathbb{C}$ and the true value for all confounder variables, as a function of the training epoch, for both the corrected and uncorrected networks. It can be seen that, for all variables, both networks start from the same value but in the uncorrected case the correlation steadily increases, while it decreases in the corrected one.

\chapter{Genetics preprocessing}
\label{app:genetics_preprocessing}

\begin{figure}[b!]
\centering
\begin{tabular}{ccc}
\subfloat[]{\includegraphics[width =.31\textwidth]{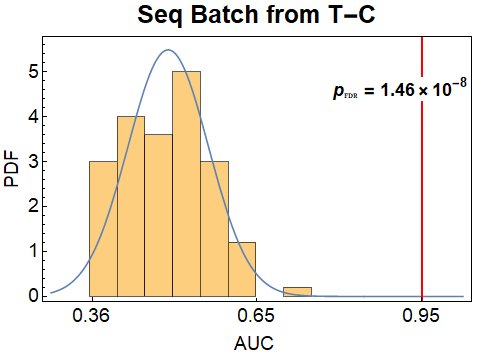}} &
\subfloat[]{\includegraphics[width =.31\textwidth]{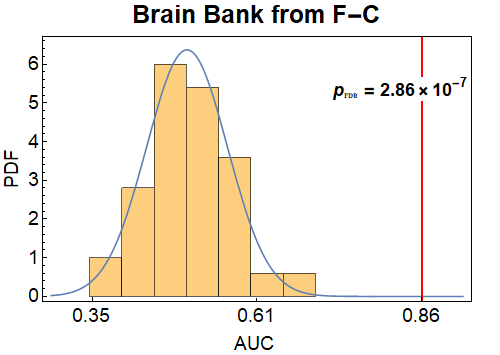}}
&
\subfloat[]{\includegraphics[width =.31\textwidth]{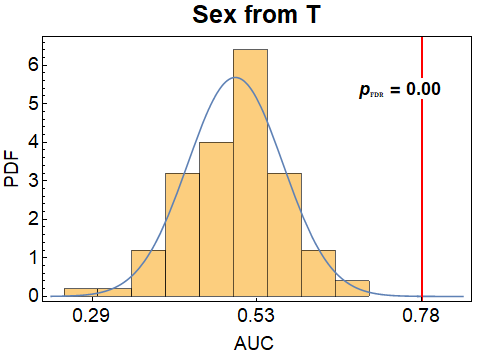}}
\end{tabular}
\begin{tabular}{cc}
\subfloat[]{\includegraphics[width =.31\textwidth]{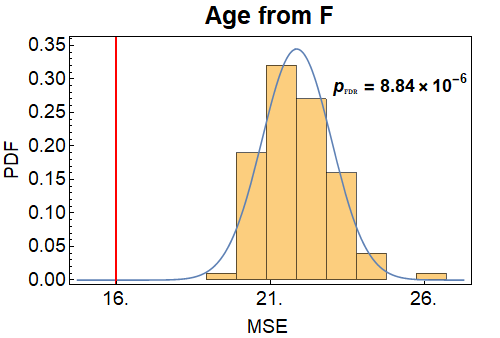}}
&
\subfloat[]{\includegraphics[width =.31\textwidth]{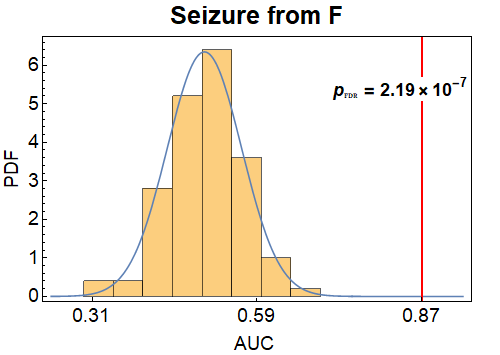}}
\end{tabular}
\caption{Examples of permutation test to understand which features can be imputed from gene expression data alone. Histograms show the results of the permutation, while the red line shows the performance of a classifier trained to predict the variable from GE or $\Delta$GE data.}
\label{fig:relevQ_hist}
\end{figure}

Fig. \ref{fig:relevQ_hist} shows some examples of the results of a permutation test aimed at identifying which variables strongly affect GE or $\Delta$GE (summarized in Tab. \ref{tab:relevantQ_feat}). In the pictures shown, it can be seen that the AUC obtained randomly permuting labels (the histograms) are much worse than the ones obtained with true labels (the red lines). The p-values have been calculated with a gaussian fit of the histograms and FDR-corrected with the BH method.

\chapter{Age dependency of GE}
\label{app:genetics_age_dependency}

\begin{figure}[b!]
\centering
\includegraphics[width =.75\textwidth]{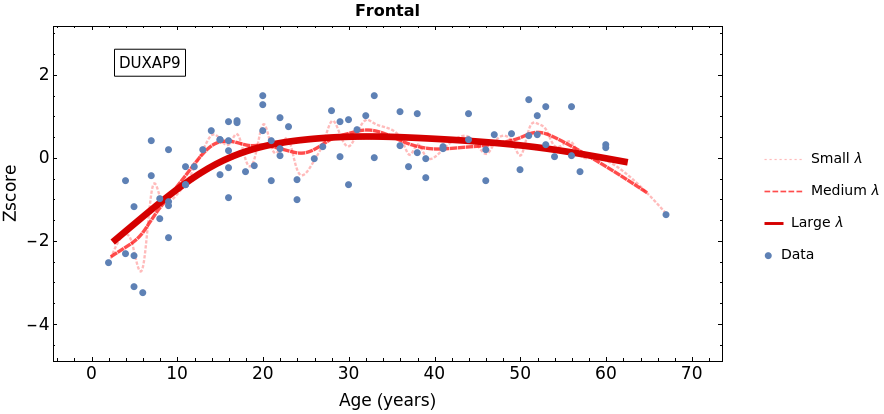} 
\caption{Example of spline estimation from data.}
\label{fig:splineExample}
\end{figure}

This appendix reports an in-depth analysis on the type of dependency that GE and $\Delta$GE values have on the age of the subject. This analysis was performed using splines, that are piecewise polinomials often used in interpolating data. 
Often, splines are regularized so that they do not overfit the data, according to the following formula:
\begin{equation}
    \sum_i{(x_i-f(x_i))^2} - \lambda \int_a^b{ (f''(x))^2 \,dx}
\end{equation}
Where the second term, weighed by the hyperparameter $\lambda$, is a penalty on the total squared value of the second derivative of the estimated spline. Increasing $\lambda$ generates a smoother function, while decreasing it improves the faithfulness to the data, as shown in Fig. \ref{fig:splineExample}.\\

\begin{figure}[h!]
\centering
\begin{tabular}{cc}
\subfloat[]{
\includegraphics[width =.35\textwidth]{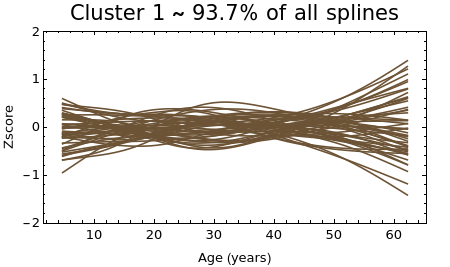}} 
&
\subfloat[]{\includegraphics[width =.35\textwidth]{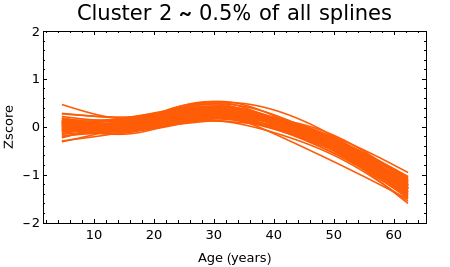}}
\\
\subfloat[]{\includegraphics[width =.35\textwidth]{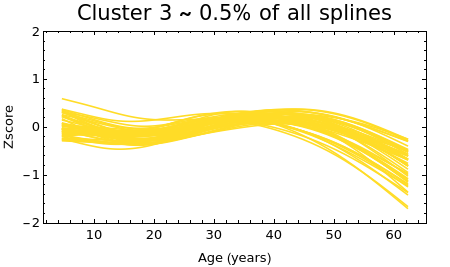}}
&
\subfloat[]{\includegraphics[width =.35\textwidth]{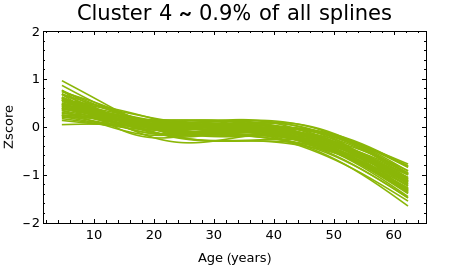}}
\\
\subfloat[]{\includegraphics[width =.35\textwidth]{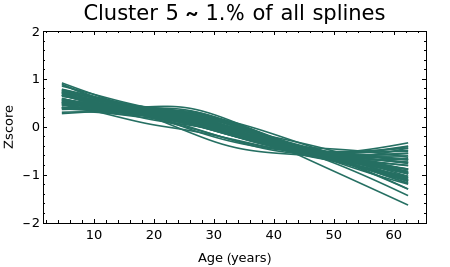}}
&
\subfloat[]{\includegraphics[width
=.35\textwidth]{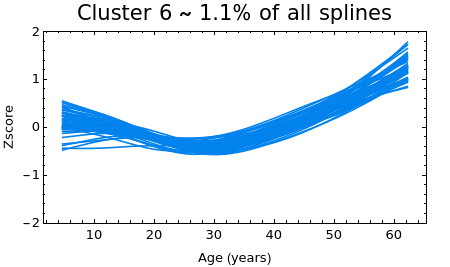}} 
\\
\subfloat[]{\includegraphics[width =.35\textwidth]{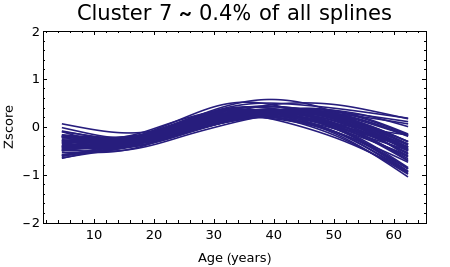}}
&
\subfloat[]{\includegraphics[width =.35\textwidth]{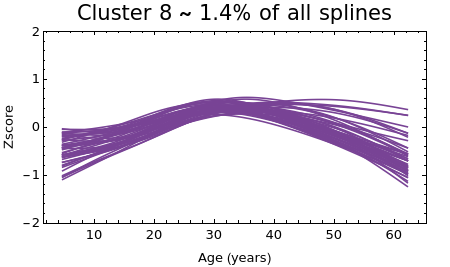}}
\\
\subfloat[]{\includegraphics[width =.35\textwidth]{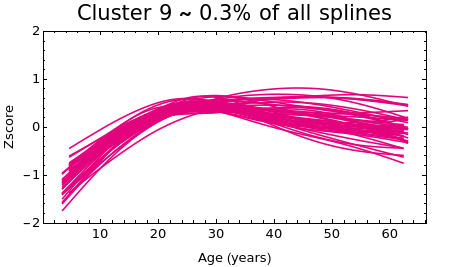}}
&
\subfloat[]{\includegraphics[width =.35\textwidth]{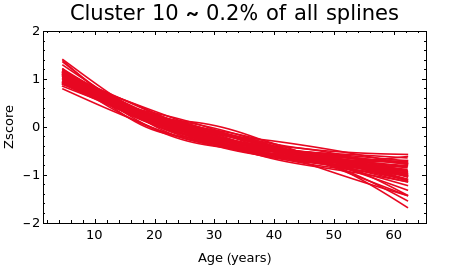}}

\end{tabular}
\caption{Spline clusters of all the genes in both the three tissues and their differences. Most of the splines do not exhibit a common trend and are grouped in the very heterogeneous cluster 1. However all the clusters, except for clusters 5 and 10, show a clear non-linear trend with respect to age.}
\label{fig:splines_cluster}
\end{figure}

Smoothed splines have been applied to approximate age dependency of the logarithm of GE and $\Delta$GE values, and clustered together to see if there was a large number of genes that had very similar splines. Fig. \ref{fig:splines_cluster} shows the results of such clustering analysis. The first cluster, the one with the least defined trend, contains 93.7\% of all the splines, i.e., of all the genes. All the other clusters contain a very small percentage of splines and have varying shapes, however, with the exception of cluster 5 and 10, they all show a clear non-linear trend with respect to age. This behaviour suggests that age dependency, when it is present, is mostly non-linear.

\begin{figure} 
\centering
\begin{tabular}{cc}
\subfloat[]{
\includegraphics[width =.45\textwidth]{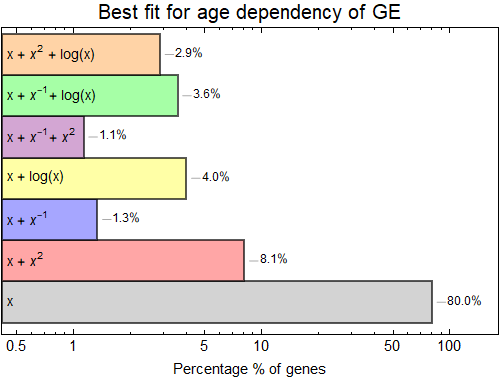}} 
&
\subfloat[]{\includegraphics[width =.45\textwidth]{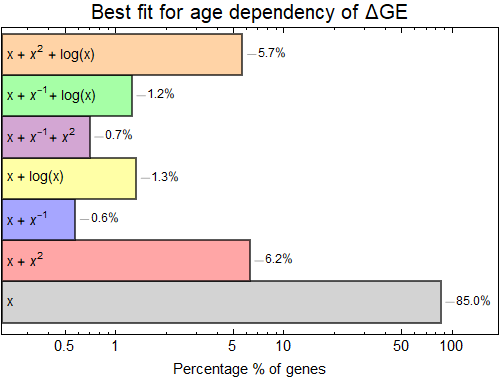}}

\end{tabular}
\caption{Results of the R-adjusted analysis on GE (a) and $\Delta$GE (b) dependency from age.}
\label{fig:Age_dependance}
\end{figure}

\begin{figure} 
\centering
\includegraphics[width =.75\textwidth]{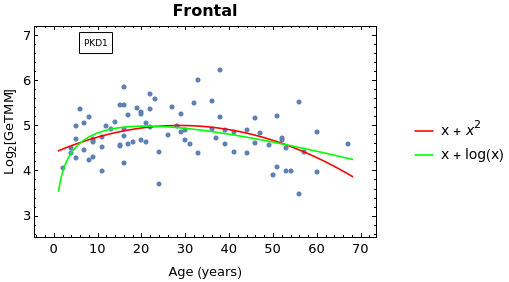}
\caption{Example showing that a logarithmic fit is not significantly better than a quadratic one.}
\label{fig:age_dep_quad_vs_log}
\end{figure}

To understand the best functional form to approximate age dependency, several fits with different functions have been performed and ranked by their R-adjusted value in Fig. \ref{fig:Age_dependance}. These forms include quadratic and logarithmic dependency on age.
Note that the R-adjusted contains a correction for the number of degrees of freedom of the fit, thus fits with different numbers of parameters can be compared.
The plots show that, in the majority of cases, approximately 80\%, a linear fit is the best approximation. This is expected as these cases obviously contain most of the genes without a clear dependency on age, because a simple constant is the best fit for them. 
The other form that is most often the best fit is the linear+quadratic form. Such form is the best among the tested ones for approximately 8\% of the genes, for both GE and $\Delta$GE values. Furthermore, if one observes some examples of the remaining genes, that are best fitted with logarithmic functions, it can be observed that they are not significantly better than the quadratic function (see Fig. \ref{fig:age_dep_quad_vs_log}).
For these reasons, a linear+quadratic form was used in the LMM analysis for regressing out age dependency.

\chapter{GO enrichment analysis of the ASD dysregulated genes}
\label{app:lmm_enrichment}
This appendix contains a comprehensive summary of the results of the GO enrichment analyses performed on the lists of genes found to be significantly differentially expressed between ASDs and TDCs.\\
The lists analyzed are the ones obtained from each single tissue and differences between each pair, the union of the lists obtained from the ST datasets and the DT ones and finally the union of all the lists.
For each of these lists the following elements are provided:
\begin{itemize}
    \item A scatter plot, made with Revigo, that shows the most significant (and non redundant) GO biological process terms associated to the list under examination. The relative position of the terms indicates their similarity based on semantic properties. The colours and the marker sizes indicate the enrichment significance of the term and the occurrence of the term in the GO annotation database, respectively.
    \item A table listing the top ten GO terms (including both biological processes, molecular functions and cellular components) ranked according to their FDR corrected p-values.
\end{itemize}

\begin{figure}
     \vspace{-2cm}
     \centering
     \makebox[\textwidth][c]{\begin{subfigure}[b]{1.05\textwidth}
         \centering
         \caption{Cerebellum (biological process)}
         \includegraphics[width=\textwidth]{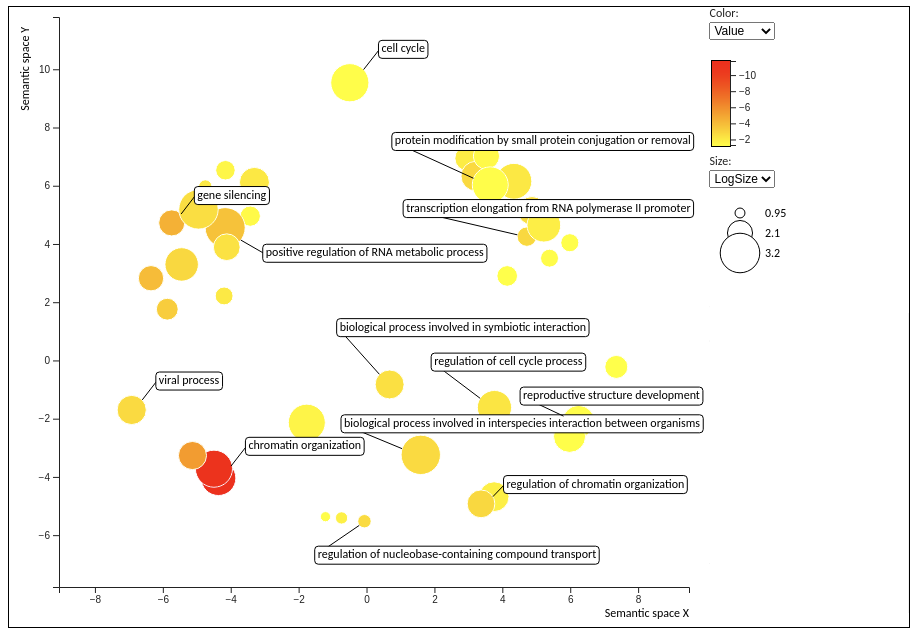}
         \label{fig:lmm_enrichment_c_picture}
     \end{subfigure}}\\
     \vspace{3mm}
     \begin{subfigure}[b]{0.9\textwidth}
         \centering
         \caption{Cerebellum (Top 10 GO terms)}
         \includegraphics[width=\textwidth]{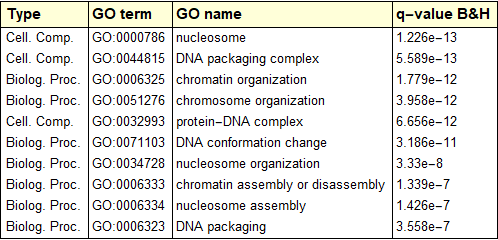}
         \label{fig:lmm_enrichment_c_tab}
     \end{subfigure}\\     
    \caption{Fig. (a) shows a selection of the most informative GO terms related to the biological process of the gene list found to be disregulated between ASDs and TDCs when analyzing the gene expression of the Cerebellum. The images have been obtained with the data visualization tool of Revigo using a MDS based on the semantic similarity of the GO terms.
    Fig. (b), instead, shows the top ten of most relevant GO terms on according to the GO enrichment analysis performed using WebGestalt. \revigopictures}
    \label{fig:lmm_enrichment_c}
\end{figure}

\begin{figure}
     \vspace{-2cm}
     \centering
     \makebox[\textwidth][c]{\begin{subfigure}[b]{1.05\textwidth}
         \centering
         \caption{Frontal (biological process)}
         \includegraphics[width=\textwidth]{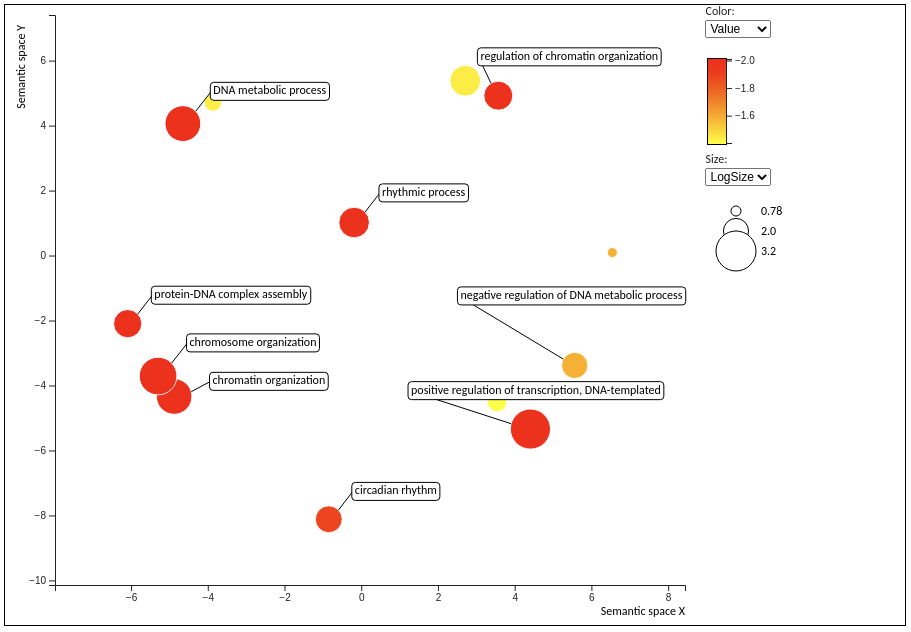}
         \label{fig:lmm_enrichment_f_picture}
     \end{subfigure}}\\
     \vspace{3mm}
     \begin{subfigure}[b]{0.9\textwidth}
         \centering
         \caption{Frontal (Top 10 GO terms)}
         \includegraphics[width=\textwidth]{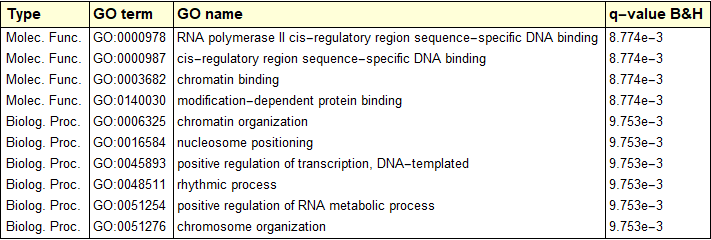}
         \label{fig:lmm_enrichment_f_tab}
     \end{subfigure}\\     
    \caption{Fig. (a) shows a selection of the most informative GO terms related to the biological process of the gene list found to be disregulated between ASDs and TDCs when analyzing the gene expression of the Frontal cortex. The images have been obtained with the data visualization tool of Revigo using a MDS based on the semantic similarity of the GO terms.
    Fig. (b), instead, shows the top ten of most relevant GO terms on according to the GO enrichment analysis performed using WebGestalt. \revigopictures}
    \label{fig:lmm_enrichment_f}
\end{figure}

\begin{figure}
     \vspace{-2cm}
     \centering
     \makebox[\textwidth][c]{\begin{subfigure}[b]{1.05\textwidth}
         \centering
         \caption{Temporal (biological process)}
         \includegraphics[width=\textwidth]{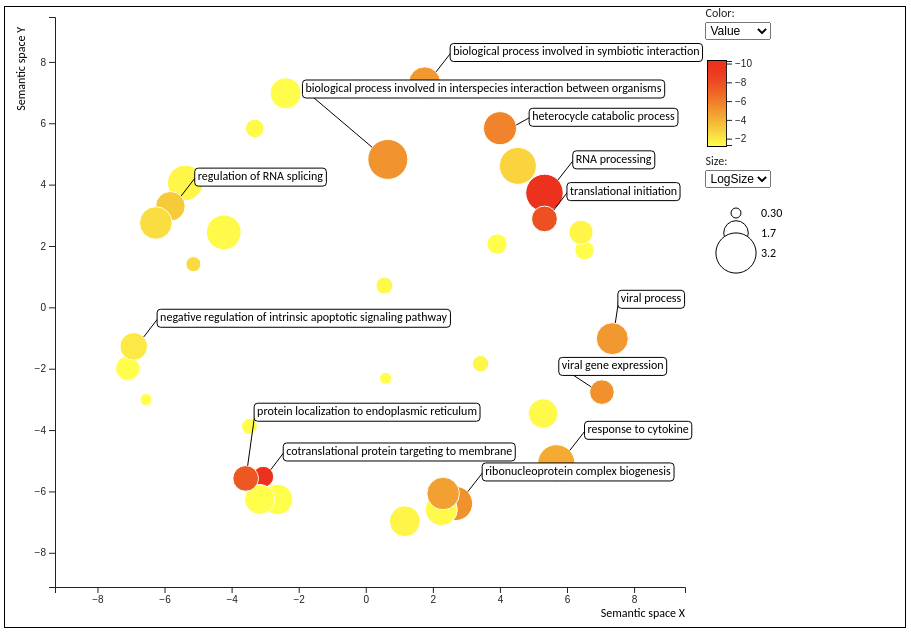}
         \label{fig:lmm_enrichment_t_picture}
     \end{subfigure}}\\
     \vspace{3mm}
     \begin{subfigure}[b]{0.9\textwidth}
         \centering
         \caption{Temporal (Top 10 GO terms)}
         \includegraphics[width=\textwidth]{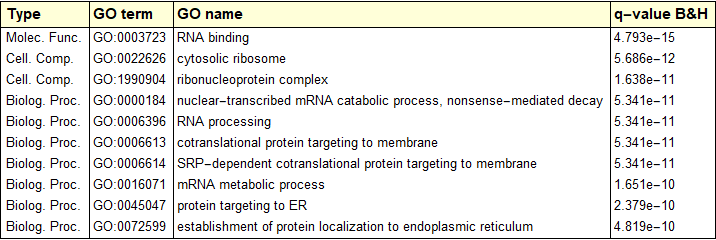}
         \label{fig:lmm_enrichment_t_tab}
     \end{subfigure}\\     
    \caption{Fig. (a) shows a selection of the most informative GO terms related to the biological process of the gene list found to be disregulated between ASDs and TDCs when analyzing the gene expression of the Temporal cortex. The images have been obtained with the data visualization tool of Revigo using a MDS based on the semantic similarity of the GO terms.
    Fig. (b), instead, shows the top ten of most relevant GO terms on according to the GO enrichment analysis performed using WebGestalt. \revigopictures}
    \label{fig:lmm_enrichment_t}
\end{figure}

\begin{figure}
     \vspace{-2cm}
     \centering
     \makebox[\textwidth][c]{\begin{subfigure}[b]{1.05\textwidth}
         \centering
         \caption{Temporal-Cerebellum (biological process)}
         \includegraphics[width=\textwidth]{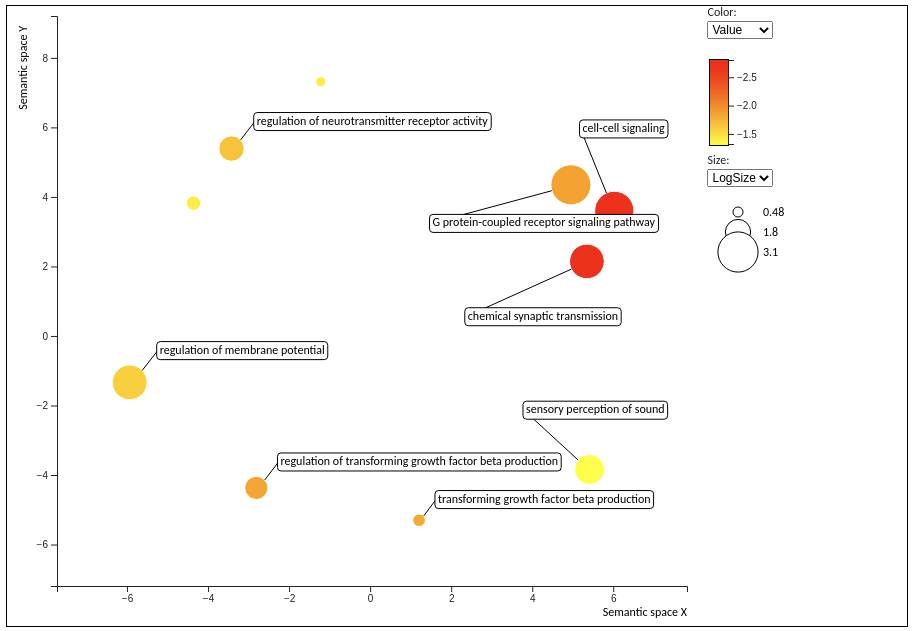}
         \label{fig:lmm_enrichment_ct_picture}
     \end{subfigure}}\\
     \vspace{3mm}
     \begin{subfigure}[b]{0.9\textwidth}
         \centering
         \caption{Temporal-Cerebellum (Top 10 GO terms)}
         \includegraphics[width=\textwidth]{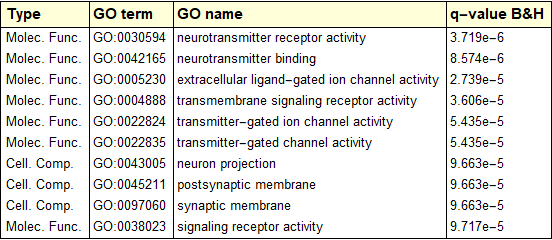}
         \label{fig:lmm_enrichment_ct_tab}
     \end{subfigure}\\     
    \caption{Fig. (a) shows a selection of the most informative GO terms related to the biological process of the gene list found to be disregulated between ASDs and TDCs when analyzing the gene expression differences between Temporal cortex and Cerebellum. The images have been obtained with the data visualization tool of Revigo using a MDS based on the semantic similarity of the GO terms.
    Fig. (b), instead, shows the top ten of most relevant GO terms on according to the GO enrichment analysis performed using WebGestalt. \revigopictures}
    \label{fig:lmm_enrichment_ct}
\end{figure}

\begin{figure}
     \vspace{-2cm}
     \centering
     \makebox[\textwidth][c]{\begin{subfigure}[b]{1.05\textwidth}
         \centering
         \caption{Temporal-Frontal (biological process)}
         \includegraphics[width=\textwidth]{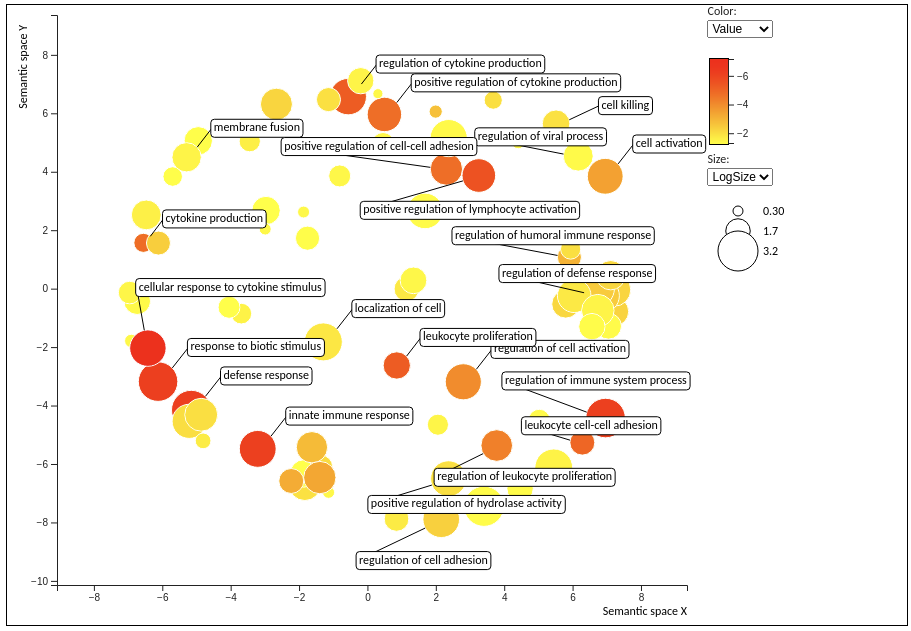}
         \label{fig:lmm_enrichment_ft_picture}
     \end{subfigure}}\\
     \vspace{3mm}
     \begin{subfigure}[b]{0.9\textwidth}
         \centering
         \caption{Temporal-Frontal (Top 10 GO terms)}
         \includegraphics[width=\textwidth]{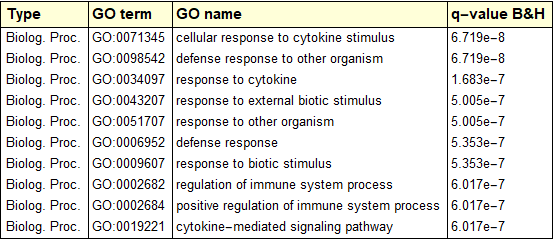}
         \label{fig:lmm_enrichment_ft_tab}
     \end{subfigure}\\     
    \caption{Fig. (a) shows a selection of the most informative GO terms related to the biological process of the gene list found to be disregulated between ASDs and TDCs when analyzing the gene expression differences between Temporal and Frontal cortices. The images have been obtained with the data visualization tool of Revigo using a MDS based on the semantic similarity of the GO terms.
    Fig. (b), instead, shows the top ten of most relevant GO terms on according to the GO enrichment analysis performed using WebGestalt. \revigopictures}
    \label{fig:lmm_enrichment_ft}
\end{figure}

\begin{figure}
     \vspace{-2cm}
     \centering
     \makebox[\textwidth][c]{\begin{subfigure}[b]{1.05\textwidth}
         \centering
         \caption{Frontal-Cerebellum (biological process)}
         \includegraphics[width=\textwidth]{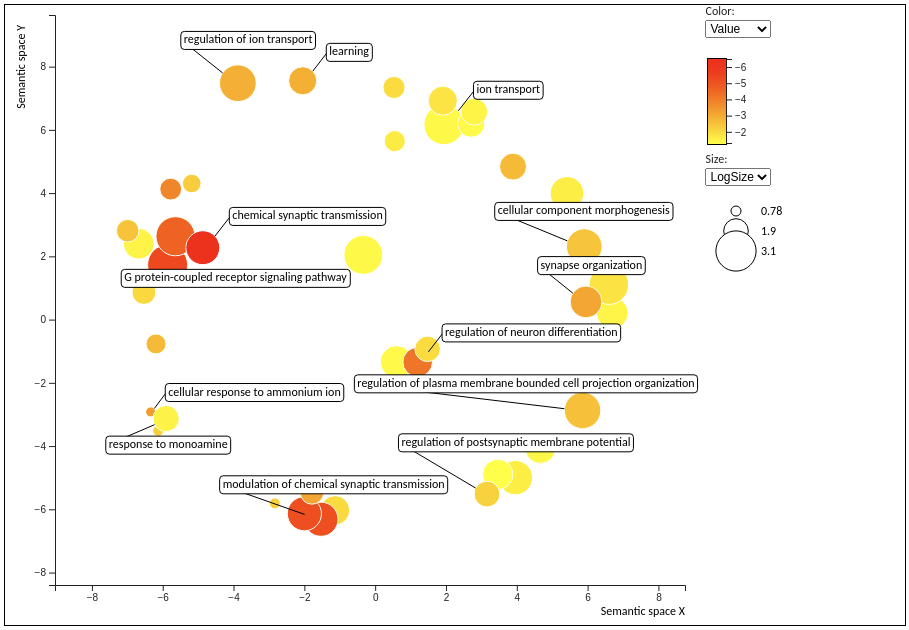}
         \label{fig:lmm_enrichment_cf_picture}
     \end{subfigure}}\\
     \vspace{3mm}
     \begin{subfigure}[b]{0.9\textwidth}
         \centering
         \caption{Frontal-Cerebellum (Top 10 GO terms)}
         \includegraphics[width=\textwidth]{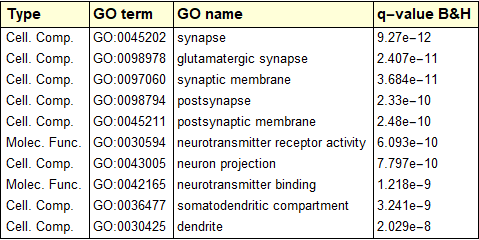}
         \label{fig:lmm_enrichment_cf_tab}
     \end{subfigure}\\     
    \caption{Fig. (a) shows a selection of the most informative GO terms related to the biological process of the gene list found to be disregulated between ASDs and TDCs when analyzing the gene expression differences between Frontal cortex and Cerebellum. The images have been obtained with the data visualization tool of Revigo using a MDS based on the semantic similarity of the GO terms.
    Fig. (b), instead, shows the top ten of most relevant GO terms on according to the GO enrichment analysis performed using WebGestalt. \revigopictures}
    \label{fig:lmm_enrichment_cf}
\end{figure}

\begin{figure}
     \vspace{-2cm}
     \centering
     \makebox[\textwidth][c]{\begin{subfigure}[b]{1.05\textwidth}
         \centering
         \caption{All Singles (biological process)}
         \includegraphics[width=\textwidth]{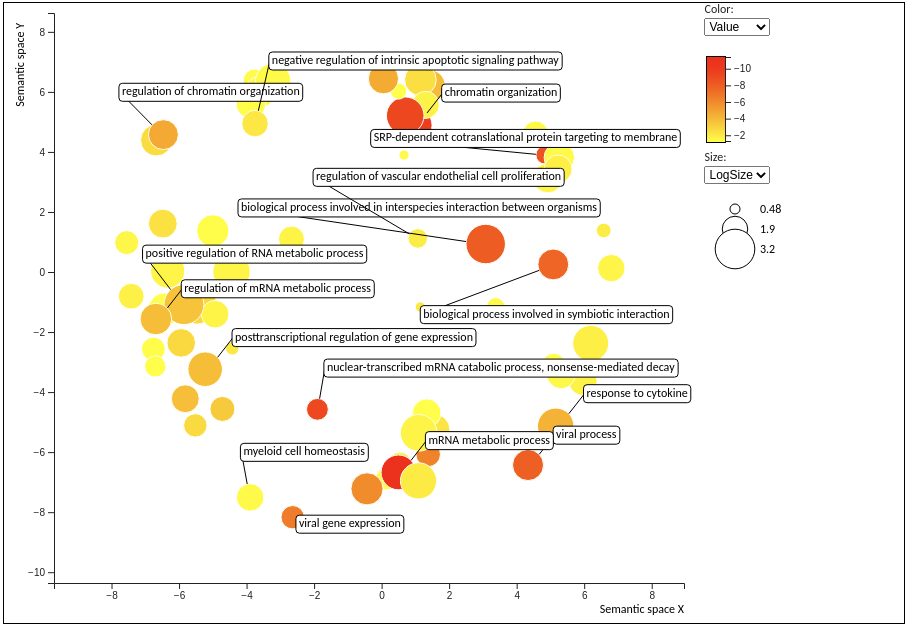}
         \label{fig:lmm_enrichment_cft_picture}
     \end{subfigure}}\\
     \vspace{3mm}
     \begin{subfigure}[b]{0.9\textwidth}
         \centering
         \caption{All Singles (Top 10 GO terms)}
         \includegraphics[width=\textwidth]{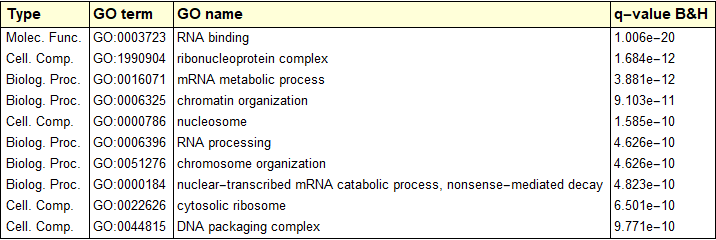}
         \label{fig:lmm_enrichment_cft_tab}
     \end{subfigure}\\     
    \caption{Fig. (a) shows a selection of the most informative GO terms related to the biological process of the gene list found to be disregulated between ASDs and TDCs when analyzing the gene expression of Cerebellum, Frontal cortex and Temporal cortex together. The images have been obtained with the data visualization tool of Revigo using a MDS based on the semantic similarity of the GO terms.
    Fig. (b), instead, shows the top ten of most relevant GO terms on according to the GO enrichment analysis performed using WebGestalt. \revigopictures}
    \label{fig:lmm_enrichment_cft}
\end{figure}

\begin{figure}
     \vspace{-2cm}
     \centering
     \makebox[\textwidth][c]{\begin{subfigure}[b]{1.05\textwidth}
         \centering
         \caption{All Diff. (biological process)}
         \includegraphics[width=\textwidth]{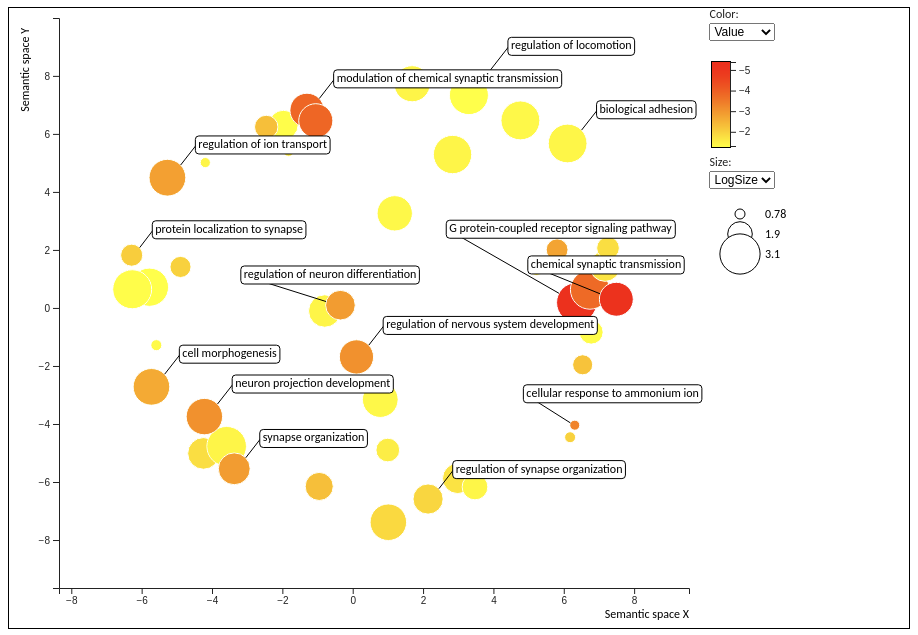}
         \label{fig:lmm_enrichment_ct_cf_ft_picture}
     \end{subfigure}}\\
     \vspace{3mm}
     \begin{subfigure}[b]{0.9\textwidth}
         \centering
         \caption{All Diff. (Top 10 GO terms)}
         \includegraphics[width=\textwidth]{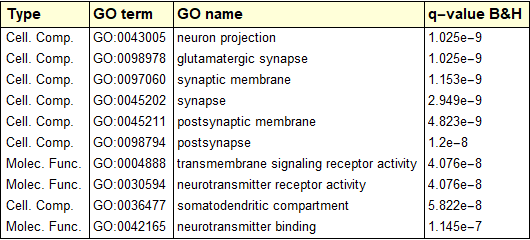}
         \label{fig:lmm_enrichment_ct_cf_ft_tab}
     \end{subfigure}\\     
    \caption{Fig. (a) shows a selection of the most informative GO terms related to the biological process of the gene list found to be disregulated between ASDs and TDCs when analyzing the gene expression differences of all the DT datasets. The images have been obtained with the data visualization tool of Revigo using a MDS based on the semantic similarity of the GO terms.
    Fig. (b), instead, shows the top ten of most relevant GO terms on according to the GO enrichment analysis performed using WebGestalt. \revigopictures}
    \label{fig:lmm_enrichment_ct_cf_ft}
\end{figure}

\begin{figure}
     \vspace{-2cm}
     \centering
     \makebox[\textwidth][c]{\begin{subfigure}[b]{1.05\textwidth}
         \centering
         \caption{All (biological process)}
         \includegraphics[width=\textwidth]{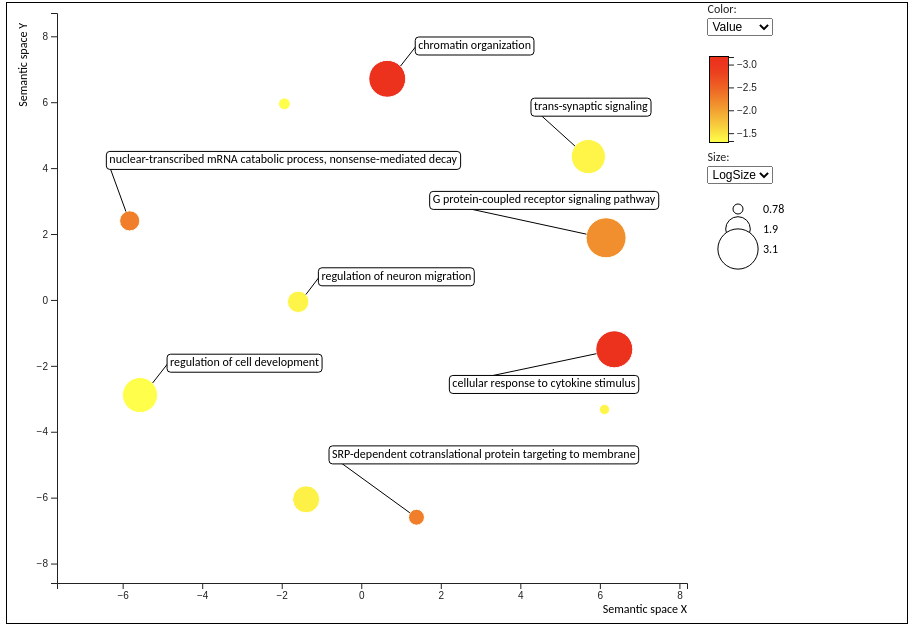}
         \label{fig:lmm_enrichment_cft-cf-ft-fc_picture}
     \end{subfigure}}\\
     \vspace{3mm}
     \begin{subfigure}[b]{0.9\textwidth}
         \centering
         \caption{All (Top 10 GO terms)}
         \includegraphics[width=\textwidth]{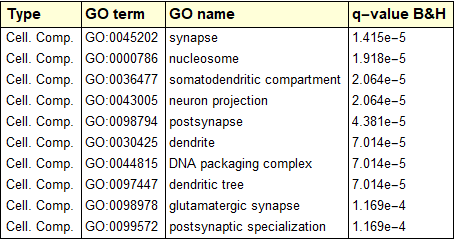}
         \label{fig:lmm_enrichment_cft_cf_ft_fc_tab}
     \end{subfigure}\\     
    \caption{Fig. (a) shows a selection of the most informative GO terms related to the biological process of the gene list found to be disregulated between ASDs and TDCs when analyzing the gene expression of all the ST and DT datasets. The images have been obtained with the data visualization tool of Revigo using a MDS based on the semantic similarity of the GO terms.
    Fig. (b), instead, shows the top ten of most relevant GO terms on according to the GO enrichment analysis performed using WebGestalt. \revigopictures}
    \label{fig:lmm_enrichment_cft-cf-ft-fc}
\end{figure}
\chapter{Gene co-expression network construction and module selection}
\label{app:network_construction}
This appendix explains the logic with which the power selection for the network construction was made. Since the adjacency matrix contains values between 0 and 1, elevating it to a power greater than 1 can be thought of as a soft threshold, that pushes small values closer to 0. Higher powers perform a stronger thresholding, while lower powers have smaller effects.
Typically, the correct power is chosen by looking for a scale-free topology, as described in Sec. \ref{sec:bkg_network}. It has been observed empirically that this topology is more easily reached at high power values, for which however the mean connectivity of the network also decreases, because the thresholding effect is stronger. 
A compromise must thus be made between the correct topology and the conservation of enough information. In fact, when the mean connectivity reaches 0, it can be said that the network does not contain any information.

\begin{figure}[h!]
\centering
\begin{tabular}{cc}
\subfloat[]{\includegraphics[trim={0 0 0 0},clip,width  =.45\textwidth]{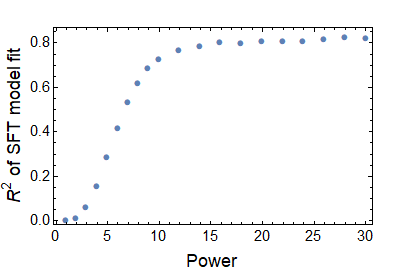}} 
&
\subfloat[]{\includegraphics[trim={0 0 0 0},clip,width
 =.45\textwidth]{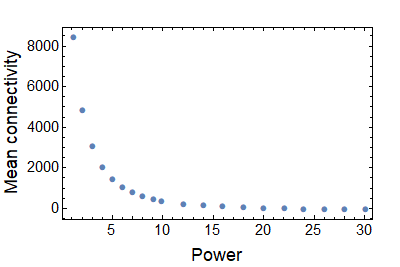}} 
\end{tabular}

\caption{Plots used for the choice of the power for building the TOM matrix. Fig. a shows the trend of the $R^2$ of the scale-free-topology model fit while Fig. b shows the one of the mean connectivity.}
\label{fig:power_choice}
\end{figure}

Fig. \ref{fig:power_choice} contains two plots, the one on the left showing the $R^2$ of the scale-free topology fit (i.e., how close the network is to a scale-free topology network, with a value of 1 indicating perfect topology) and the one on the right showing the mean connectivity.
According to these plots, a power of 16 has been chosen, which is the first power for which $R^2>0.8$, and before the last dive of the mean connectivity that happens around a power of 20.

\chapter{Gene co-expression network modules and GO enrichment analyses}
\label{app:network_enrichment}
In this appendix some visual representations of the co-expression network, its module and of the GO enrichment analyses performed on the latters are provided.
\section{Network modules visualization}
\begin{figure}[h!]
\centering
\includegraphics[width =0.7\textwidth]{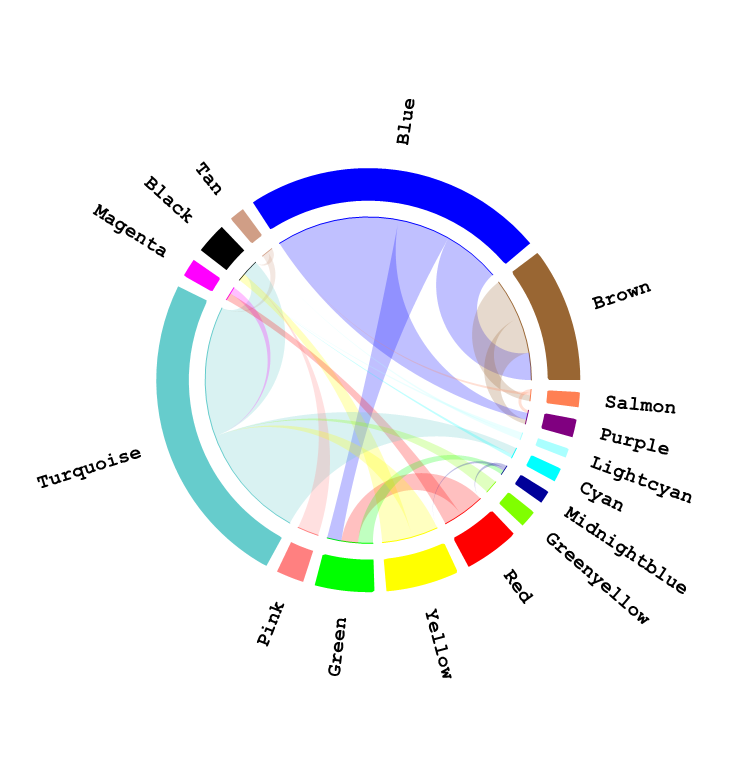}
\caption{Circular graph representation of the network}\label{fig:circulargraph}
\end{figure}

\begin{figure} 
\centering
\begin{tabular}{cc}
\subfloat{\includegraphics[trim={0 2mm 0mm 2mm},clip,width  =.282\textwidth]{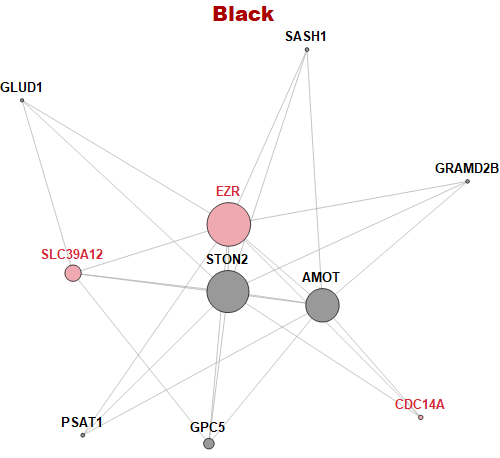}} 
&
\subfloat{\includegraphics[trim={0 0 4cm 0},clip,width
 =.08\textwidth]{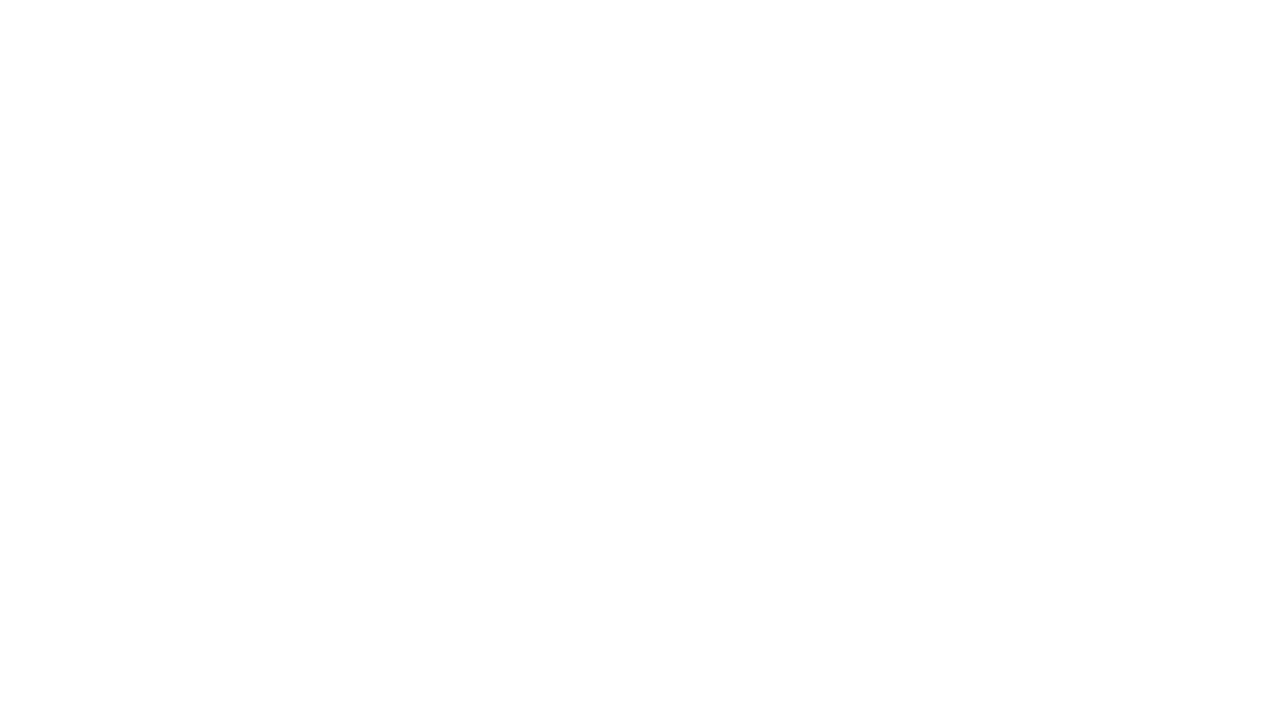}} 
\\
\subfloat{\includegraphics[trim={0 2mm 0mm 2mm},clip,width  =.282\textwidth]{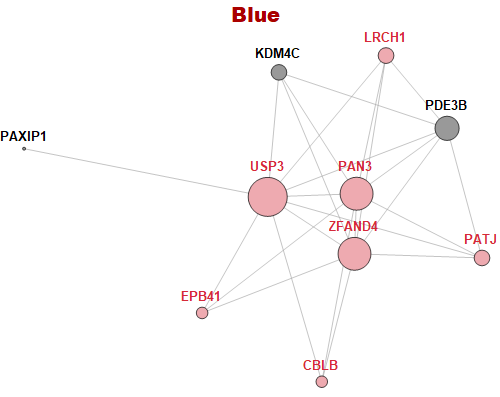}} 
&
\subfloat{\includegraphics[trim={0 0 4cm 0},clip,width
 =.08\textwidth]{Figures/Fig_Results/Fig_Report/white.png}}
\end{tabular}
\begin{tabular}{c}
\subfloat{\includegraphics[trim={0 2mm 0mm 2mm},clip,width  =.282\textwidth]{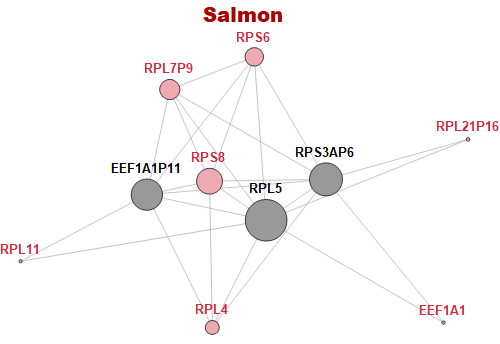}}
\\
\subfloat{\includegraphics[trim={0 2mm 0mm 2mm},clip,width  =.282\textwidth]{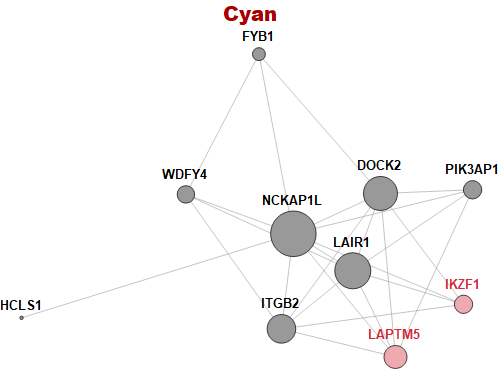}} 
\\
\subfloat{\includegraphics[trim={0 2mm 0mm 2mm},clip,width  =.282\textwidth]{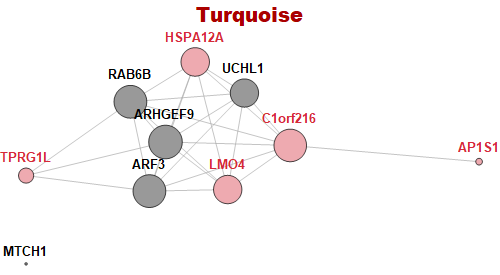}} 
\\
\end{tabular}
\caption{Visual representation of the top 15 more connected hubs and top 50 stronger connections for each module. The modules here represented are enriched by the genes found to be relevant for ASD.}\label{fig:modules_enriched}
\end{figure}

\begin{figure} 
\centering
\begin{tabular}{cc}
\subfloat{\includegraphics[trim={0 2mm 0mm 2mm},clip,width  =.2825\textwidth]{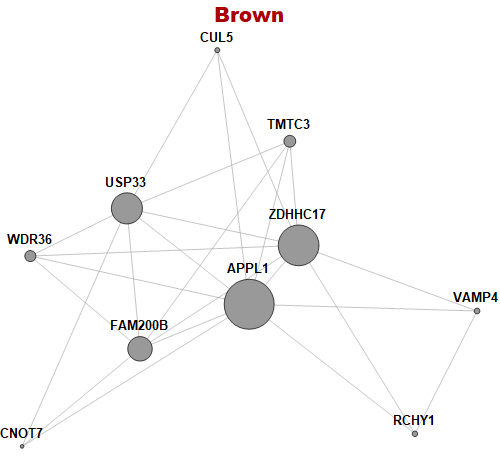}} 
&
\subfloat{\includegraphics[trim={0 0 4cm 0},clip,width
 =.08\textwidth]{Figures/Fig_Results/Fig_Report/white.png}} 
\\
\subfloat{\includegraphics[trim={0 2mm 0mm 2mm},clip,width  =.282\textwidth]{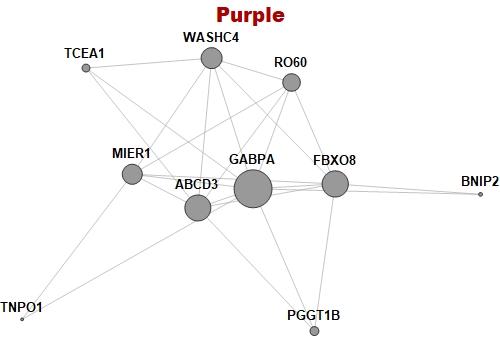}} 
&
\subfloat{\includegraphics[trim={0 0 4cm 0},clip,width
 =.08\textwidth]{Figures/Fig_Results/Fig_Report/white.png}}
\\
\subfloat{\includegraphics[trim={0 2mm 0mm 2mm},clip,width  =.282\textwidth]{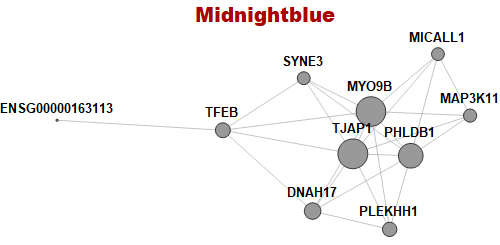}} 
&
\subfloat{\includegraphics[trim={0 0 4cm 0},clip,width
 =.08\textwidth]{Figures/Fig_Results/Fig_Report/white.png}}
\\
\subfloat{\includegraphics[trim={0 2mm 0mm 2mm},clip,width  =.282\textwidth]{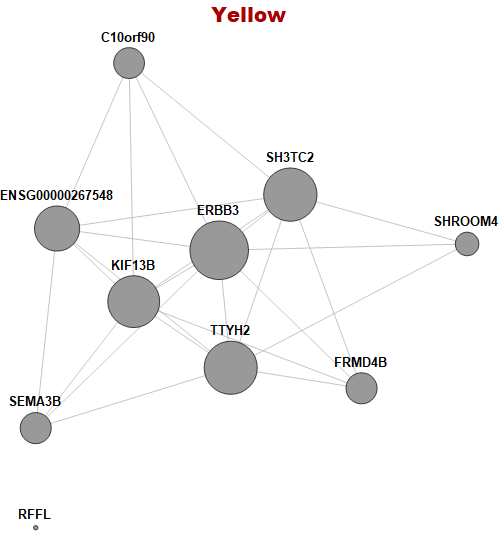}} 
&
\subfloat{\includegraphics[trim={0 0 4cm 0},clip,width
 =.08\textwidth]{Figures/Fig_Results/Fig_Report/white.png}}
\\
\subfloat{\includegraphics[trim={0 2mm 0mm 2mm},clip,width  =.282\textwidth]{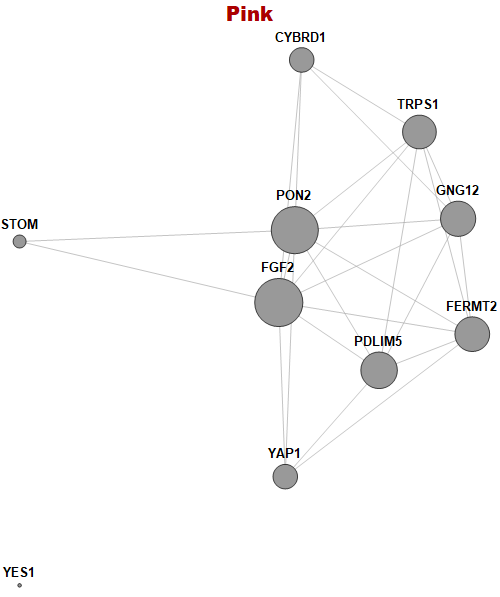}} 
&
\subfloat{\includegraphics[trim={0 0 4cm 0},clip,width
 =.08\textwidth]{Figures/Fig_Results/Fig_Report/white.png}}
\end{tabular}
\begin{tabular}{c}
\subfloat{\includegraphics[trim={0 2mm 0mm 2mm},clip,width  =.282\textwidth]{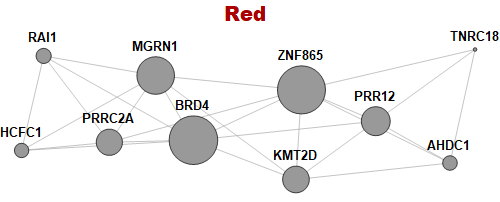}}
\\
\subfloat{\includegraphics[trim={0 2mm 0mm 2mm},clip,width  =.282\textwidth]{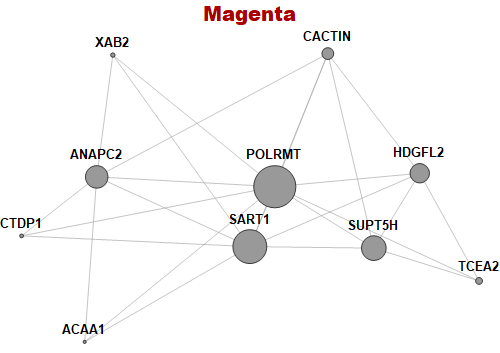}} 
\\
\subfloat{\includegraphics[trim={0 2mm 0mm 2mm},clip,width  =.282\textwidth]{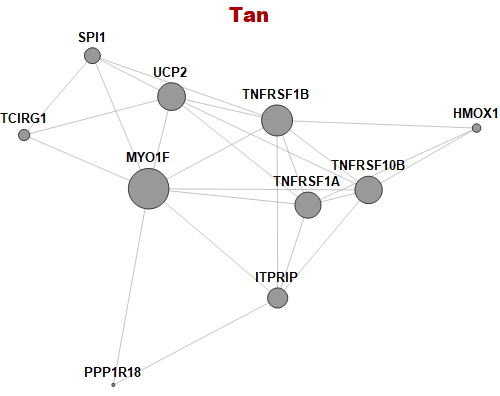}} 
\\
\subfloat{\includegraphics[trim={0 2mm 0mm 2mm},clip,width  =.282\textwidth]{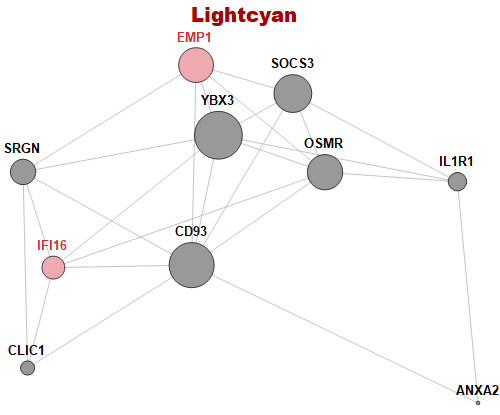}} 
\\
\subfloat{\includegraphics[trim={0 2mm 0mm 2mm},clip,width  =.282\textwidth]{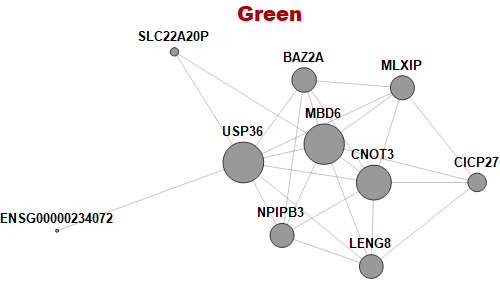}}
\\
\subfloat{\includegraphics[trim={0 2mm 0mm 2mm},clip,width  =.282\textwidth]{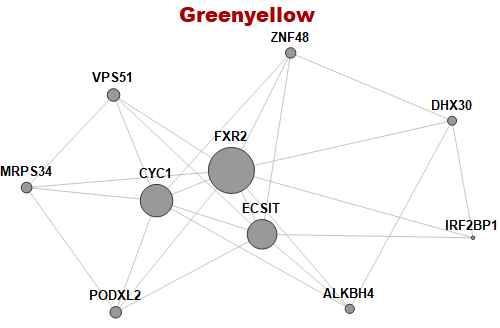}}
\end{tabular}
\caption{Visual representation of the top 15 more connected hubs and top 50 stronger connections for each module. The modules here represented are not enriched by the genes found to be relevant for ASD.}
\label{fig:modules_not_enriched}
\end{figure}

Fig. \ref{fig:circulargraph} depicts a circular graph of the network, while Fig. \ref{fig:modules_enriched} shows the top 10 more connected hubs for each of the 5 modules of interest, enriched by the lists of genes found with the LMM analysis. In these representations, the size of a hub reflects its connectivity within the module and the markers are pink coloured when the gene is included in the lists of genes found with LMM, and gray otherwise. Using the same representation, Fig. \ref{fig:modules_not_enriched} shows all the the other modules of the network. 

\section{GO enrichment analysis on network's modules of interest}
In this section a comprehensive summary of the results of the GO enrichment analyses performed on the five modules of interest is reported. 
For each of them, the following elements are provided:
\begin{itemize}
    \item Three scatter plots, made with Revigo, that show the most significant (and non redundant) GO terms associated to the module under examination, considering the biological processes, molecular functions and cellular components, respectively. The relative position of the terms indicates their similarity based on semantic properties. The colours and the marker sizes indicate the enrichment significance of the term and the occurrence of the term in the GO annotation database, respectively.
    \item A table listing the top ten GO terms ranked according to their FDR corrected p-values.
\end{itemize}

\begin{figure}
     \vspace{-2cm}
     \centering
     \makebox[\textwidth][c]{\begin{subfigure}[b]{1.05\textwidth}
         \centering
        \caption{Black module (biological process)}
         \includegraphics[width=\textwidth]{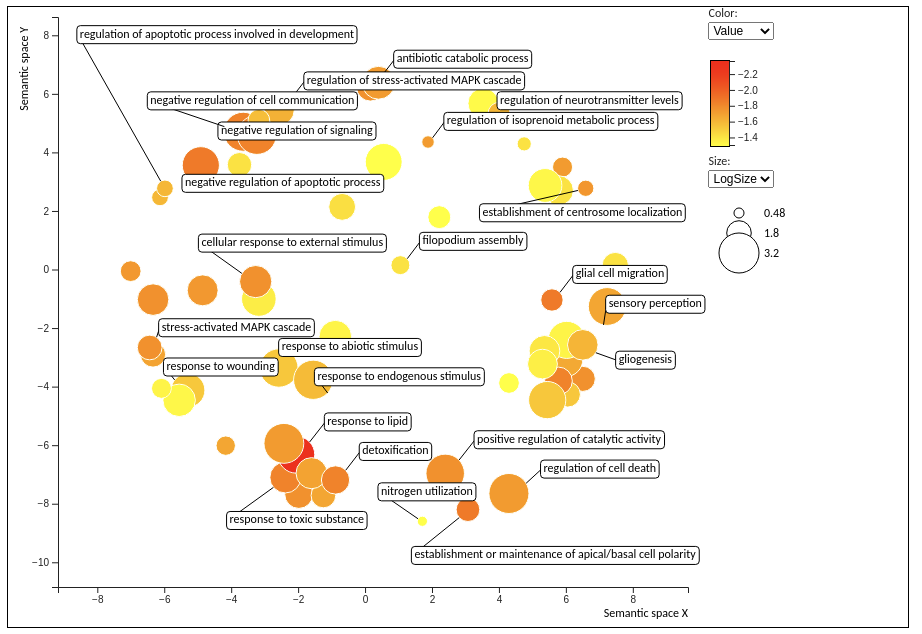}
         \label{fig:black_bio}
     \end{subfigure}}\\
     \makebox[\textwidth][c]{\begin{subfigure}[b]{1.05\textwidth}
         \centering
         \caption{Black module (cellular component)}
         \includegraphics[width=\textwidth]{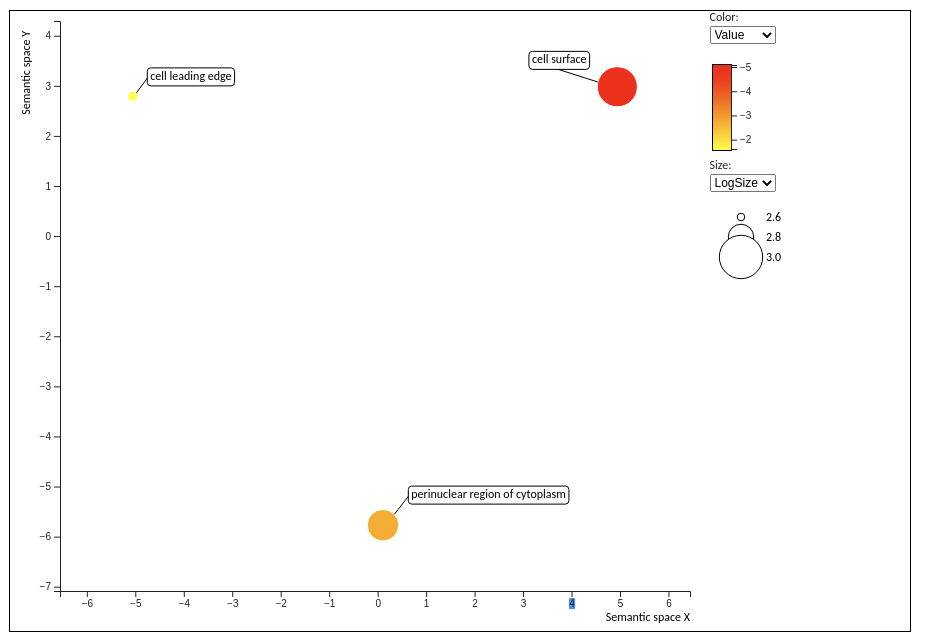}
         \label{fig:black_cell}
     \end{subfigure}}\\
    \caption{}
    \label{fig:black_network_enrichment_1}
\end{figure}

\begin{figure}
    \ContinuedFloat
     \vspace{-2cm}
     \centering
     \makebox[\textwidth][c]{\begin{subfigure}[b]{1.05\textwidth}
         \centering
         \caption{Black module (molecular function)}
         \includegraphics[width=\textwidth]{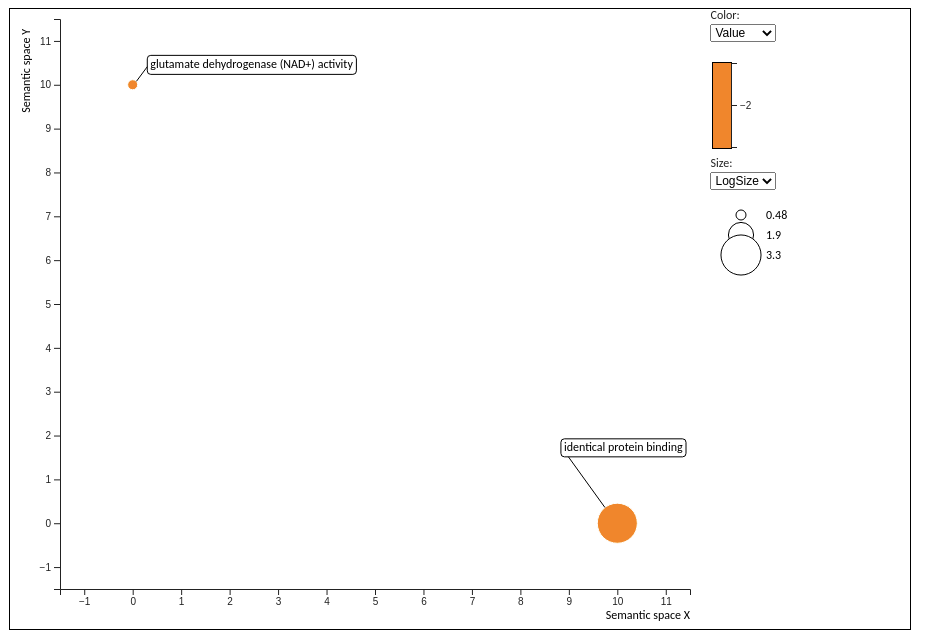}
         \label{fig:black_mol}
     \end{subfigure}}\\
     \vspace{3mm}
     \begin{subfigure}[b]{0.9\textwidth}
         \centering
         \caption{Black module (Top 10 GO terms)}
         \includegraphics[width=\textwidth]{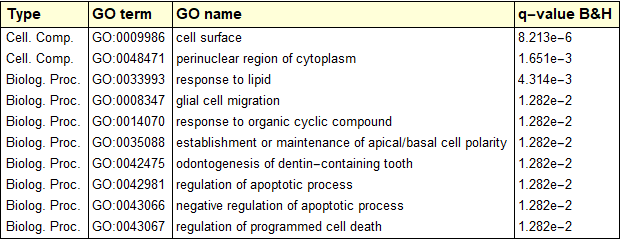}
         \label{fig:black_tab}
     \end{subfigure}\\     
    \caption{Fig.s (a), (b) and (c) show a selection of most informative GO terms related to the biological process, cellular component and molecular function of the Black module. The images have been obtained with the data visualization tool of Revigo using a MDS based on the semantic similarity of the GO terms.
    Fig. (d), instead, shows the top ten of most relevant GO terms on according to the GO enrichment anaysis performed using WebGestalt. \revigopictures}
    \label{fig:black_network_enrichment_2}
\end{figure}

\begin{figure}
     \vspace{-2cm}
     \centering
     \makebox[\textwidth][c]{\begin{subfigure}[b]{1.05\textwidth}
         \centering
        \caption{Blue module (biological process)}
         \includegraphics[width=\textwidth]{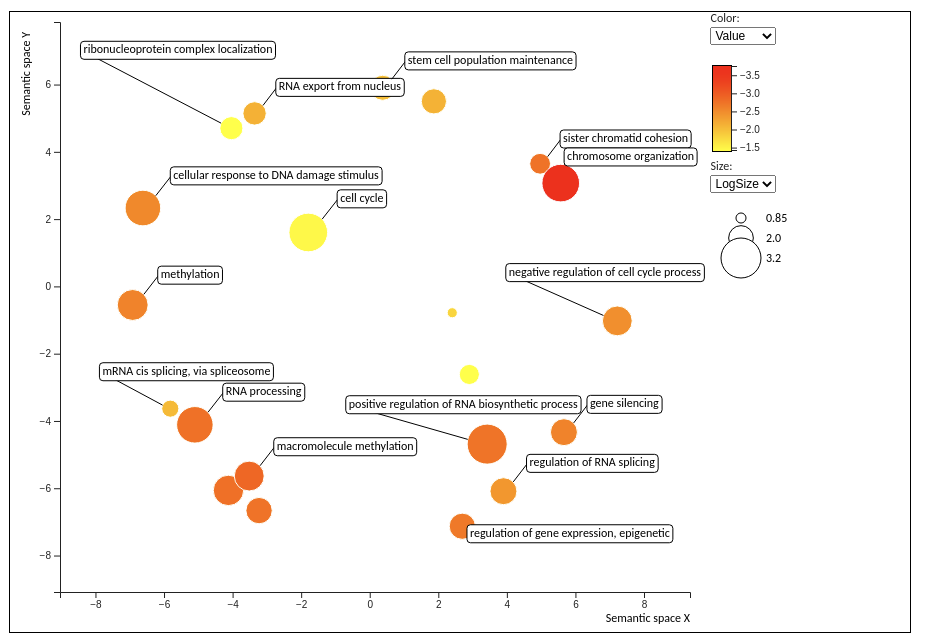}
         \label{fig:blue_bio}
     \end{subfigure}}\\
     \makebox[\textwidth][c]{\begin{subfigure}[b]{1.05\textwidth}
         \centering
         \caption{Blue module (cellular component)}
         \includegraphics[width=\textwidth]{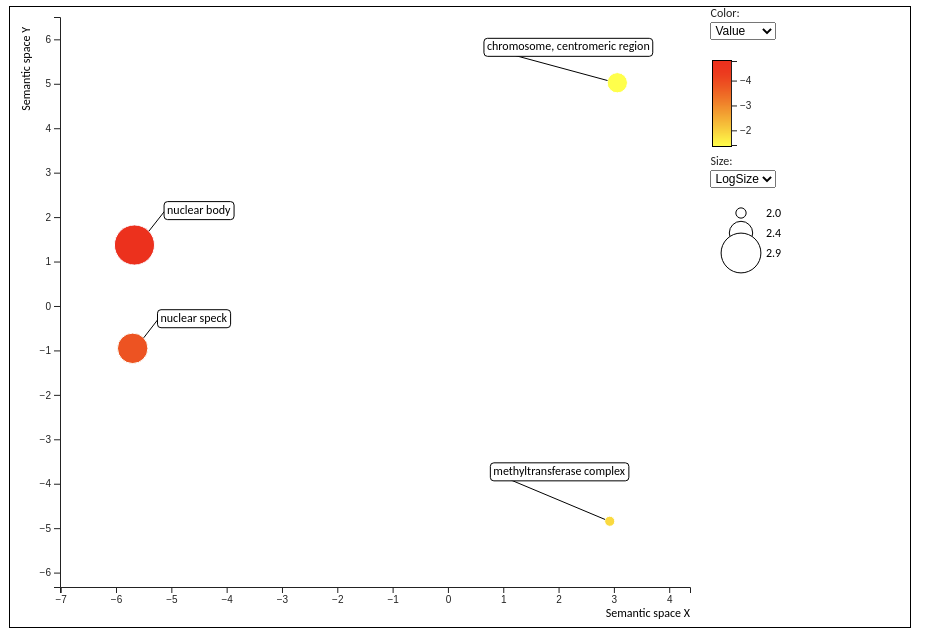}
         \label{fig:blue_cell}
     \end{subfigure}}\\
    \caption{}
    \label{fig:blue_network_enrichment_1}
\end{figure}

\begin{figure}
    \ContinuedFloat
     \vspace{-2cm}
     \centering
     \makebox[\textwidth][c]{\begin{subfigure}[b]{1.05\textwidth}
         \centering
         \caption{Blue module (molecular function)}
         \includegraphics[width=\textwidth]{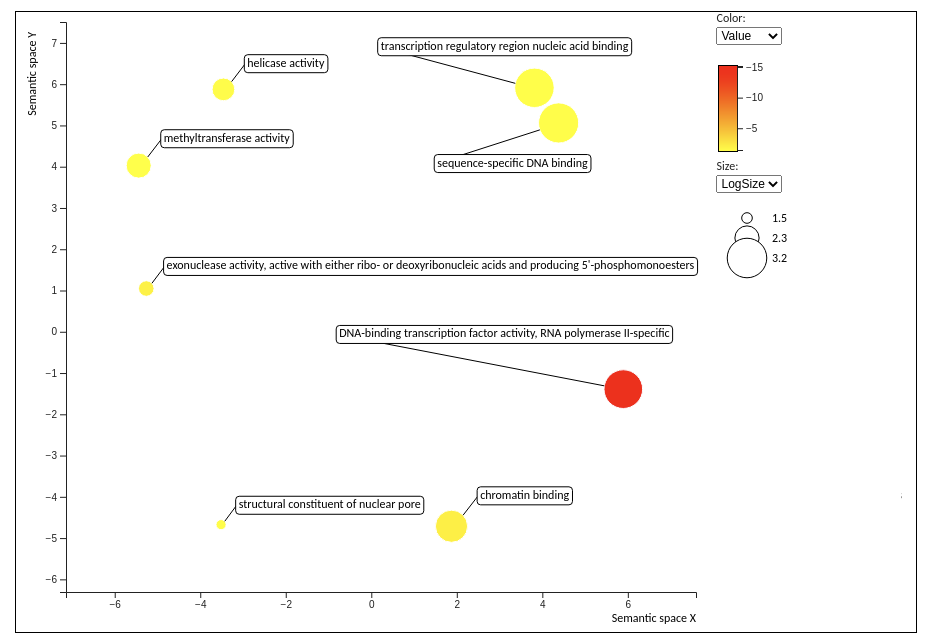}
         \label{fig:blue_mol}
     \end{subfigure}}\\
     \vspace{3mm}
     \begin{subfigure}[b]{0.9\textwidth}
         \centering
         \caption{Blue module (Top 10 GO terms)}
         \includegraphics[width=\textwidth]{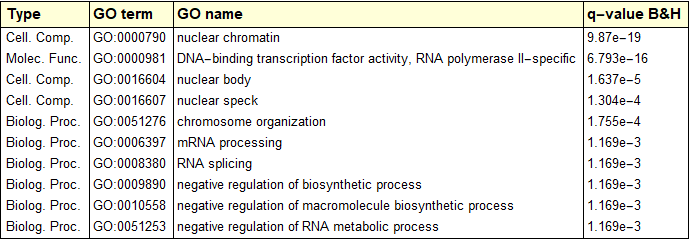}
         \label{fig:blue_tab}
     \end{subfigure}\\     
    \caption{Fig.s (a), (b) and (c) show a selection of most informative GO terms related to the biological process, cellular component and molecular function of the Blue module. The images have been obtained with the data visualization tool of Revigo using a MDS based on the semantic similarity of the GO terms.
    Fig. (d), instead, shows the top ten of most relevant GO terms on according to the GO enrichment anaysis performed using WebGestalt. \revigopictures}
    \label{fig:blue_network_enrichment_2}
\end{figure}

\begin{figure}
     \vspace{-2cm}
     \centering
     \makebox[\textwidth][c]{\begin{subfigure}[b]{1.05\textwidth}
         \centering
        \caption{Cyan module (biological process)}
         \includegraphics[width=\textwidth]{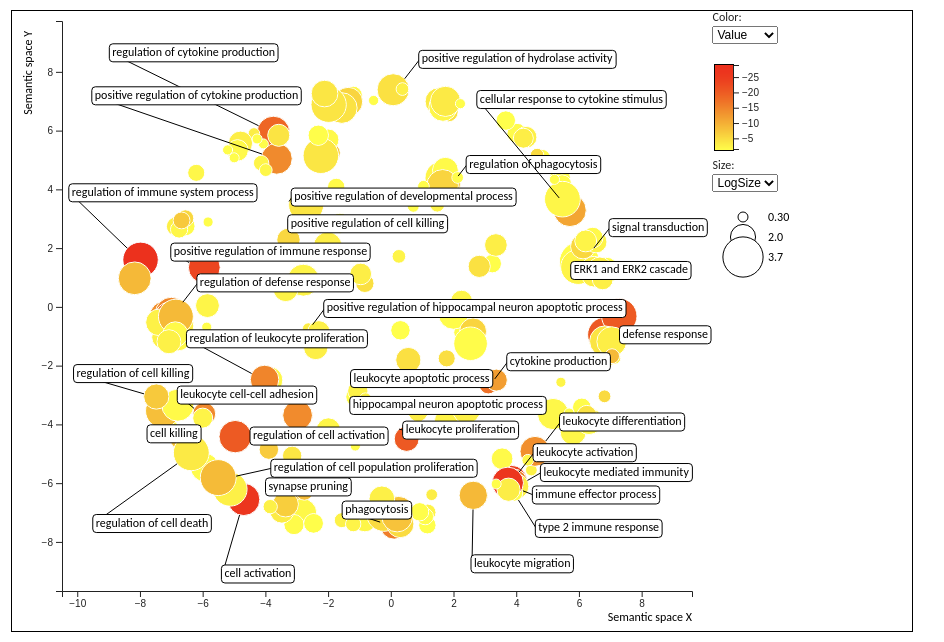}
         \label{fig:cyan_bio}
     \end{subfigure}}\\
     \makebox[\textwidth][c]{\begin{subfigure}[b]{1.05\textwidth}
         \centering
         \caption{Cyan module (cellular component)}
         \includegraphics[width=\textwidth]{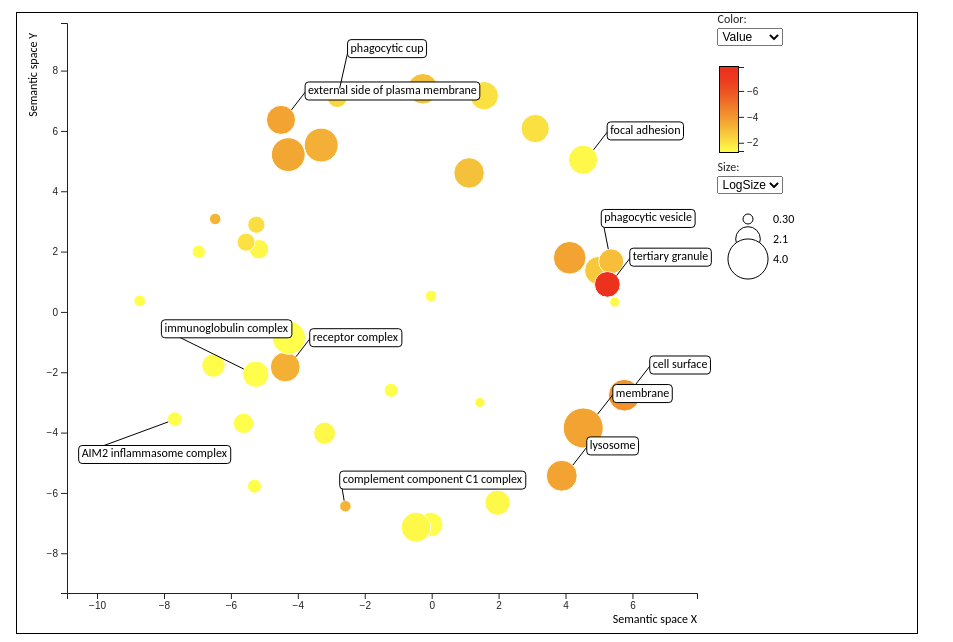}
         \label{fig:cyan_cell}
     \end{subfigure}}\\
    \caption{}
    \label{fig:cyan_network_enrichment_1}
\end{figure}

\begin{figure}
    \ContinuedFloat
     \vspace{-2cm}
     \centering
     \makebox[\textwidth][c]{\begin{subfigure}[b]{1.05\textwidth}
         \centering
         \caption{Cyan module (molecular function)}
         \includegraphics[width=\textwidth]{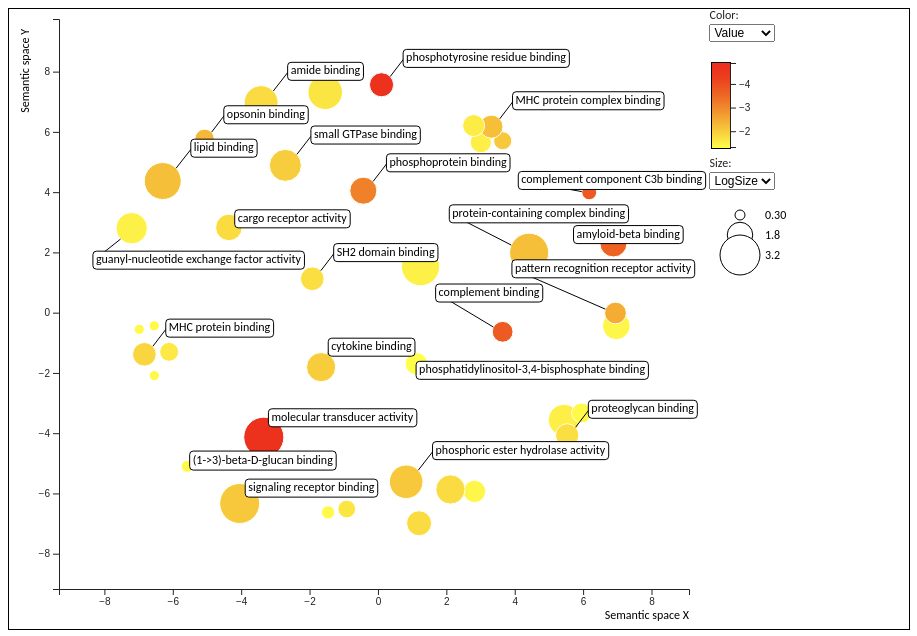}
         \label{fig:cyan_mol}
     \end{subfigure}}\\
     \vspace{3mm}
     \begin{subfigure}[b]{0.9\textwidth}
         \centering
         \caption{Cyan module (Top 10 GO terms)}
         \includegraphics[width=\textwidth]{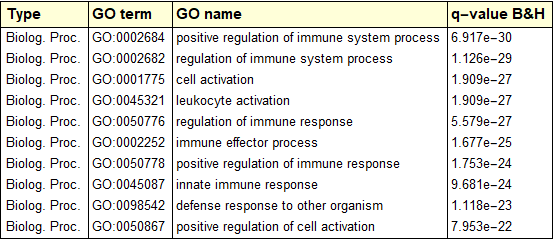}
         \label{fig:cyan_tab}
     \end{subfigure}\\     
    \caption{Fig.s (a), (b) and (c) show a selection of most informative GO terms related to the biological process, cellular component and molecular function of the Cyan module. The images have been obtained with the data visualization tool of Revigo using a MDS based on the semantic similarity of the GO terms.
    Fig. (d), instead, shows the top ten of most relevant GO terms on according to the GO enrichment anaysis performed using WebGestalt. \revigopictures}
    \label{fig:cyan_network_enrichment_2}
\end{figure}

\begin{figure}
     \vspace{-2cm}
     \centering
     \makebox[\textwidth][c]{\begin{subfigure}[b]{1.05\textwidth}
         \centering
        \caption{Salmon module (biological process)}
         \includegraphics[width=\textwidth]{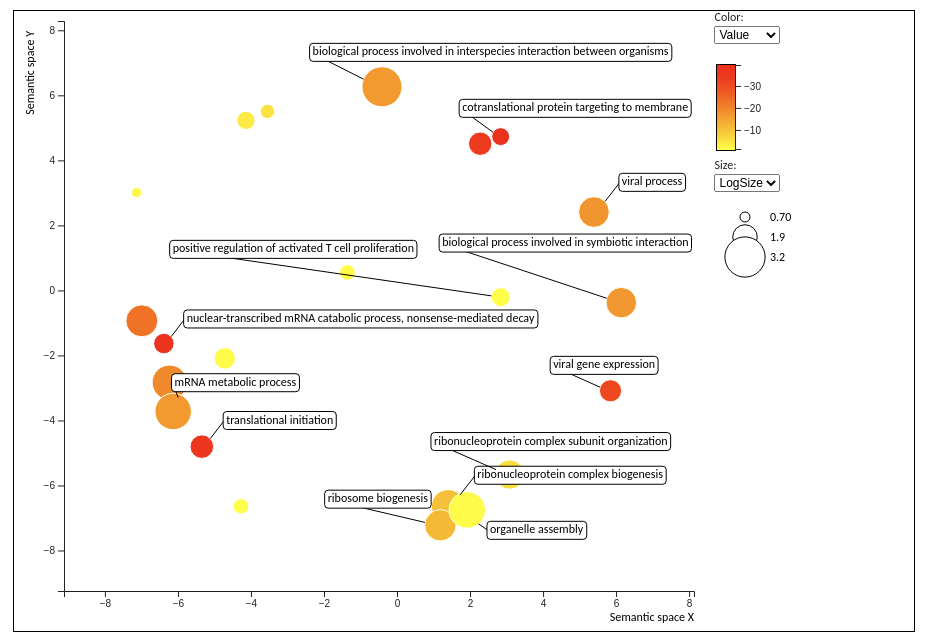}
         \label{fig:salmon_bio}
     \end{subfigure}}\\
     \makebox[\textwidth][c]{\begin{subfigure}[b]{1.05\textwidth}
         \centering
         \caption{Salmon module (cellular component)}
         \includegraphics[width=\textwidth]{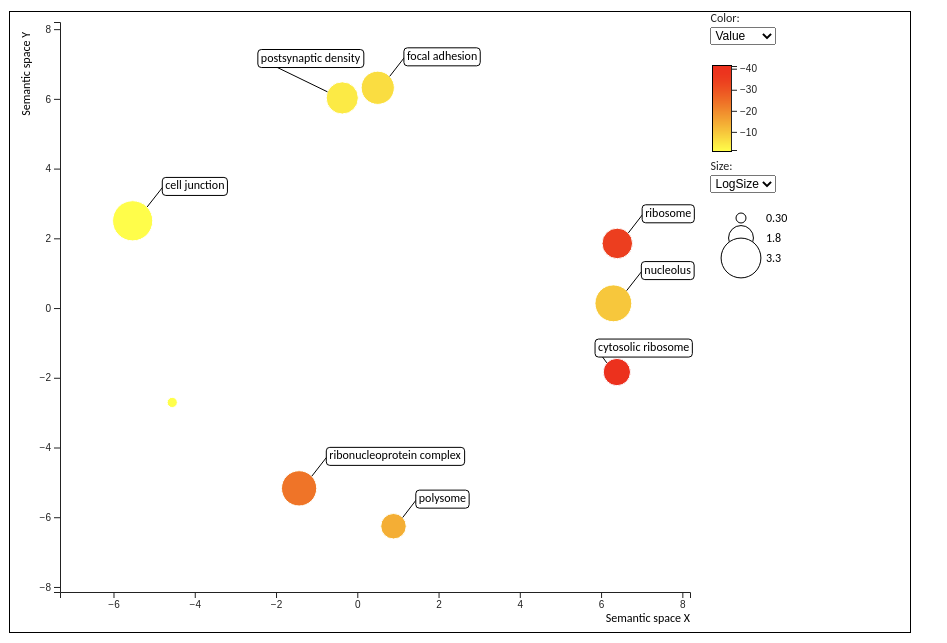}
         \label{fig:salmon_cell}
     \end{subfigure}}\\
    \caption{}
    \label{fig:salmon_network_enrichment_1}
\end{figure}

\begin{figure}
    \ContinuedFloat
     \vspace{-2cm}
     \centering
     \makebox[\textwidth][c]{\begin{subfigure}[b]{1.05\textwidth}
         \centering
         \caption{Salmon module (molecular function)}
         \includegraphics[width=\textwidth]{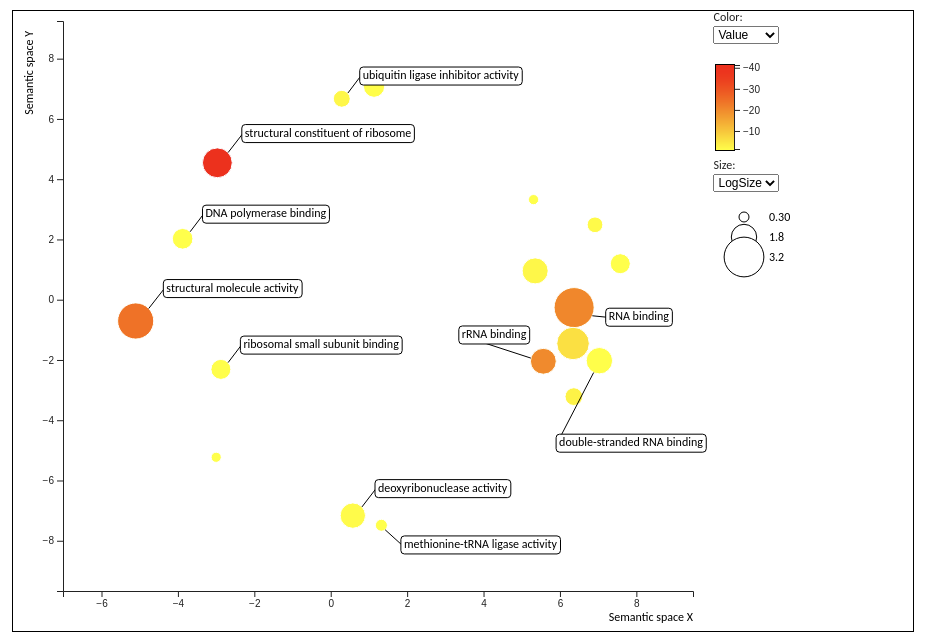}
         \label{fig:salmon_mol}
     \end{subfigure}}\\
     \vspace{3mm}
     \begin{subfigure}[b]{0.9\textwidth}
         \centering
         \caption{Salmon module (Top 10 GO terms)}
         \includegraphics[width=\textwidth]{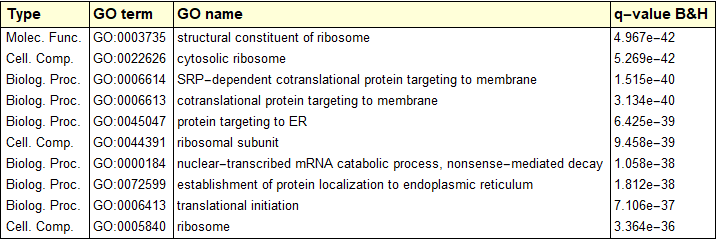}
         \label{fig:salmon_tab}
     \end{subfigure}\\     
    \caption{Fig.s (a), (b) and (c) show a selection of most informative GO terms related to the biological process, cellular component and molecular function of the Salmon module. The images have been obtained with the data visualization tool of Revigo using a MDS based on the semantic similarity of the GO terms.
    Fig. (d), instead, shows the top ten of most relevant GO terms on according to the GO enrichment anaysis performed using WebGestalt. \revigopictures}
    \label{fig:salmon_network_enrichment_2}
\end{figure}

\begin{figure}
     \vspace{-2cm}
     \centering
     \makebox[\textwidth][c]{\begin{subfigure}[b]{1.05\textwidth}
         \centering
        \caption{Turquoise module (biological process)}
         \includegraphics[width=\textwidth]{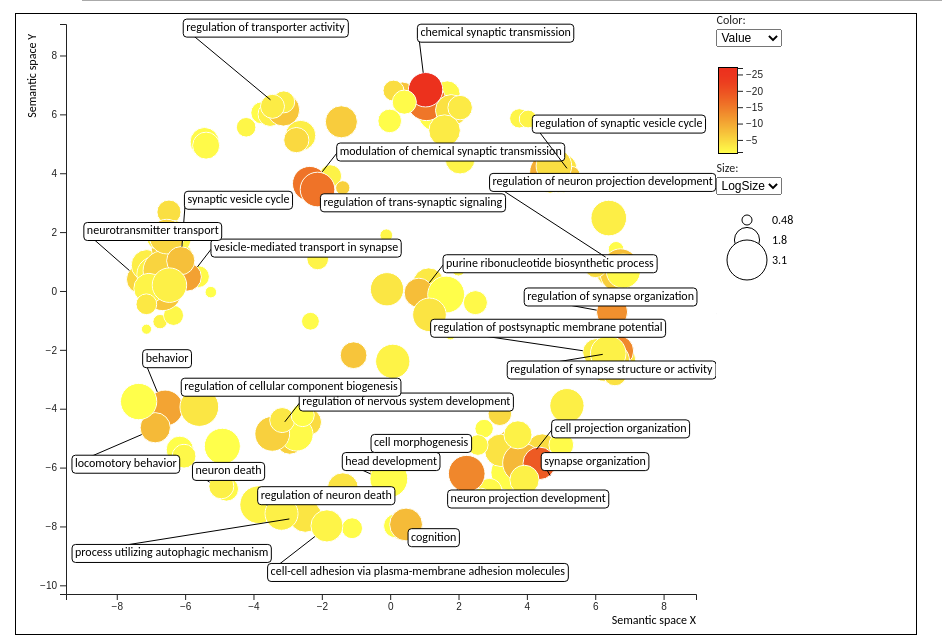}
         \label{fig:turquoise_bio}
     \end{subfigure}}\\
     \makebox[\textwidth][c]{\begin{subfigure}[b]{1.05\textwidth}
         \centering
         \caption{Turquoise module (cellular component)}
         \includegraphics[width=\textwidth]{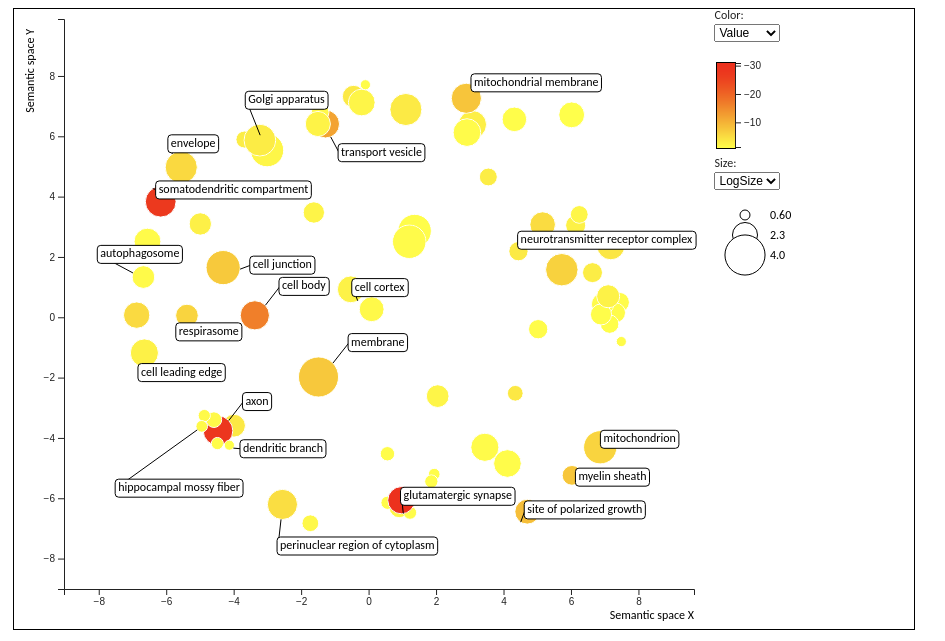}
         \label{fig:turquoise_cell}
     \end{subfigure}}\\
    \caption{}
    \label{fig:turquoise_network_enrichment_1}
\end{figure}

\begin{figure}
    \ContinuedFloat
     \vspace{-2cm}
     \centering
     \makebox[\textwidth][c]{\begin{subfigure}[b]{1.05\textwidth}
         \centering
         \caption{Turquoise module (molecular function)}
         \includegraphics[width=\textwidth]{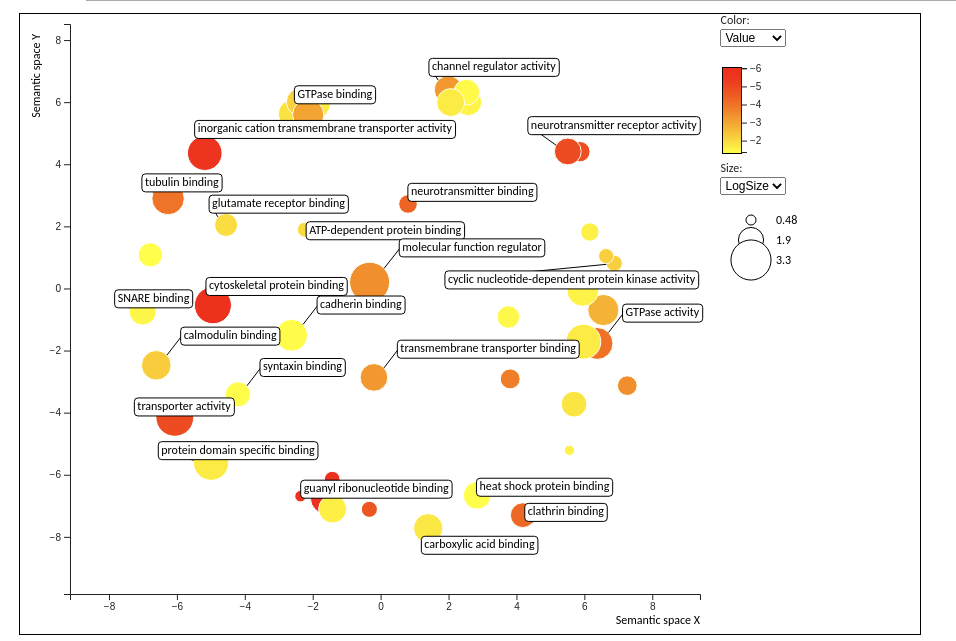}
         \label{fig:turquoise_mol}
     \end{subfigure}}\\
     \vspace{3mm}
     \begin{subfigure}[b]{0.9\textwidth}
         \centering
         \caption{Turquoise module (Top 10 GO terms)}
         \includegraphics[width=\textwidth]{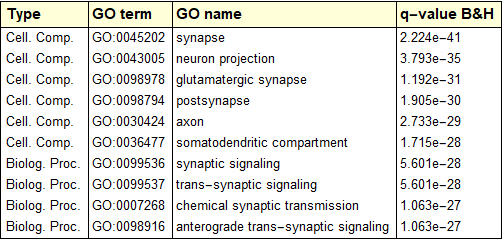}
         \label{fig:turquoise_tab}
     \end{subfigure}\\     
    \caption{Fig.s (a), (b) and (c) show a selection of most informative GO terms related to the biological process, cellular component and molecular function of the Turquoise module. The images have been obtained with the data visualization tool of Revigo using a MDS based on the semantic similarity of the GO terms.
    Fig. (d), instead, shows the top ten of most relevant GO terms on according to the GO enrichment anaysis performed using WebGestalt. \revigopictures}
    \label{fig:turquoise_network_enrichment_2}
\end{figure}
\chapter{Comparison between neuroimaging and transcriptomics results}
\label{app:imaging_genetics}

\begin{figure}[h!]
    \centering
    \begin{subfigure}{0.45\textwidth}
        \centering
        \includegraphics[width=\textwidth]{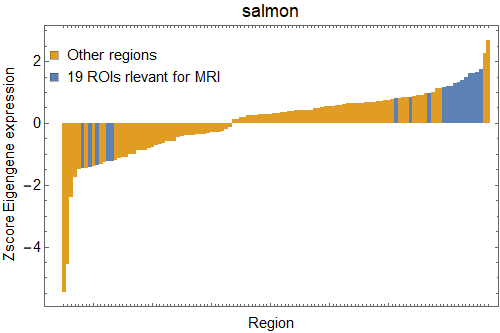}
        \caption{}
        \label{fig:app_modules_rel_1}
    \end{subfigure}
    \hfill
    \begin{subfigure}{0.45\textwidth}
        \centering
        \includegraphics[width=\textwidth]{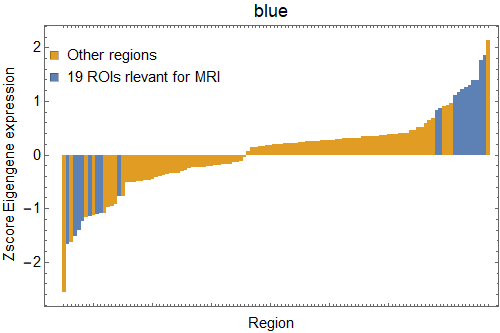}
        \caption{}
        \label{fig:app_modules_rel_2}
    \end{subfigure}\\
    \begin{subfigure}{0.45\textwidth}
        \centering
        \includegraphics[width=\textwidth]{Figures/Fig_Results/link/black.png}
        \caption{}
        \label{fig:app_modules_rel_3}
    \end{subfigure}
    \hfill
    \begin{subfigure}{0.45\textwidth}
        \centering
        \includegraphics[width=\textwidth]{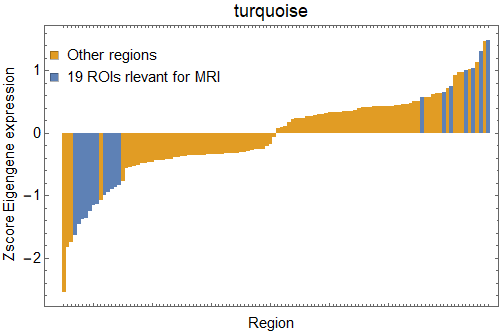}
        \caption{}
        \label{fig:app_modules_rel_4}
    \end{subfigure}\\   
    \begin{subfigure}{0.45\textwidth}
        \centering
        \includegraphics[width=\textwidth]{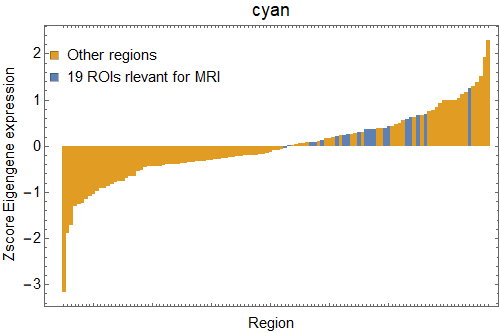}
        \caption{}
        \label{fig:app_modules_rel_5}
    \end{subfigure}\\   
    \caption{Plots used to identify gene modules abnormally expressed in the regions found relevant by neuroimaging. The modules shown are all enriched for the gene lists found in the LMM analysis.}
    \label{fig:app_modules_rel}    
\end{figure}

\begin{figure}[h!]
    \centering
    \begin{subfigure}{0.45\textwidth}
        \centering
        \includegraphics[width=\textwidth]{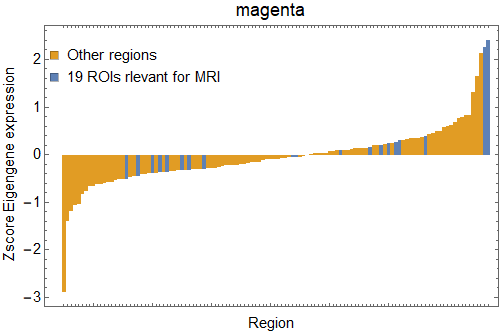}
        \caption{}
        \label{fig:app_modules_nonrel_1}
    \end{subfigure}
    \hfill
    \begin{subfigure}{0.45\textwidth}
        \centering
        \includegraphics[width=\textwidth]{Figures/Fig_Results/link/pink.png}
        \caption{}
        \label{fig:app_modules_nonrel_2}
    \end{subfigure}\\
    \begin{subfigure}{0.45\textwidth}
        \centering
        \includegraphics[width=\textwidth]{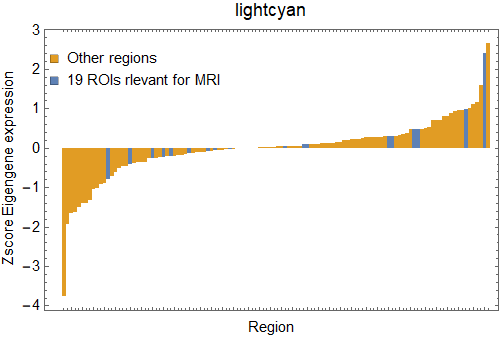}
        \caption{}
        \label{fig:app_modules_nonrel_3}
    \end{subfigure}
    \hfill
    \begin{subfigure}{0.45\textwidth}
        \centering
        \includegraphics[width=\textwidth]{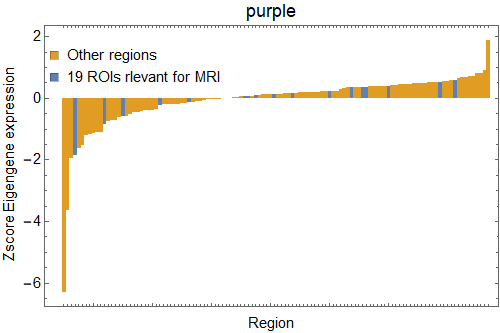}
        \caption{}
        \label{fig:app_modules_nonrel_4}
    \end{subfigure}\\   
    \begin{subfigure}{0.45\textwidth}
        \centering
        \includegraphics[width=\textwidth]{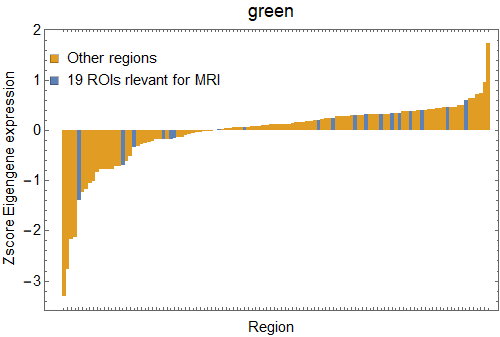}
        \caption{}
        \label{fig:app_modules_nonrel_5}
    \end{subfigure}
    \hfill
    \begin{subfigure}{0.45\textwidth}
        \centering
        \includegraphics[width=\textwidth]{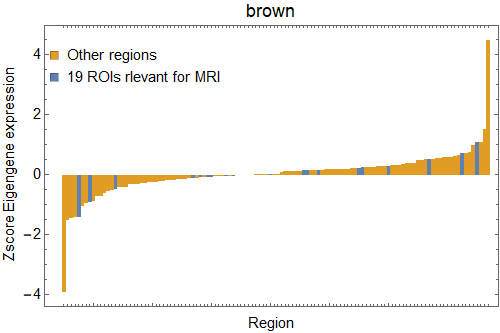}
        \caption{}
        \label{fig:app_modules_nonrel_6}
    \end{subfigure}\\   
    \caption{Plots used to identify gene modules abnormally expressed in the regions found relevant by neuroimaging. The modules shown are not enriched for the gene lists found in the LMM analysis (part A).}
    \label{fig:app_modules_nonrel_a}    
\end{figure}

\begin{figure}[h!]
    \centering
    \begin{subfigure}{0.45\textwidth}
        \centering
        \includegraphics[width=\textwidth]{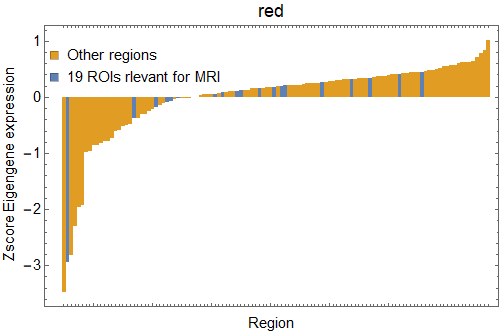}
        \caption{}
        \label{fig:app_modules_nonrel_7}
    \end{subfigure}
    \hfill
    \begin{subfigure}{0.45\textwidth}
        \centering
        \includegraphics[width=\textwidth]{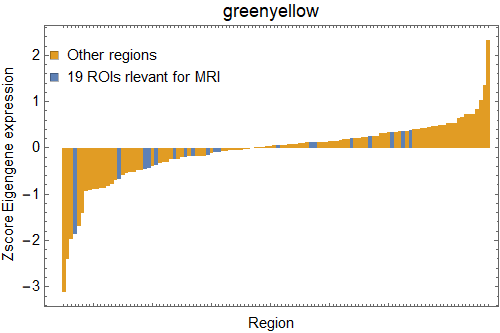}
        \caption{}
        \label{fig:app_modules_nonrel_8}
    \end{subfigure}\\
    \begin{subfigure}{0.45\textwidth}
        \centering
        \includegraphics[width=\textwidth]{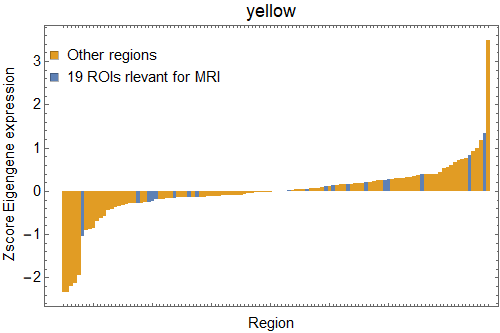}
        \caption{}
        \label{fig:app_modules_nonrel_9}
    \end{subfigure}
    \hfill
    \begin{subfigure}{0.45\textwidth}
        \centering
        \includegraphics[width=\textwidth]{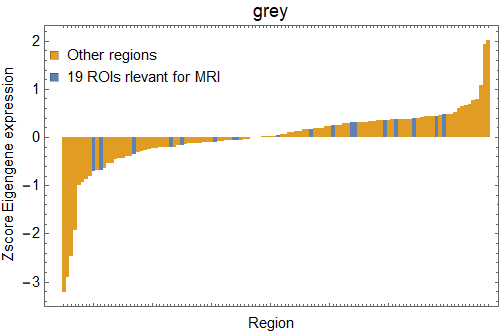}
        \caption{}
        \label{fig:app_modules_nonrel_10}
    \end{subfigure}\\   
    \begin{subfigure}{0.45\textwidth}
        \centering
        \includegraphics[width=\textwidth]{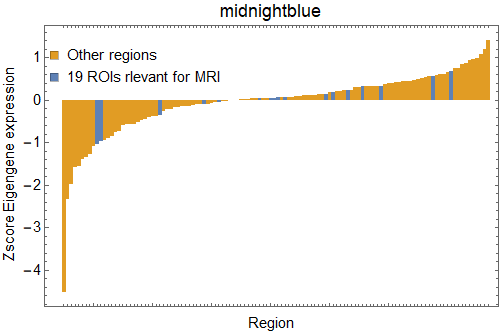}
        \caption{}
        \label{fig:app_modules_nonrel_11}
    \end{subfigure}
    \hfill
    \begin{subfigure}{0.45\textwidth}
        \centering
        \includegraphics[width=\textwidth]{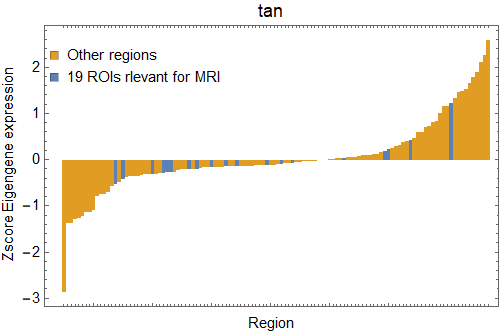}
        \caption{}
        \label{fig:app_modules_nonrel_12}
    \end{subfigure}\\   
    \caption{Plots used to identify gene modules abnormally expressed in the regions found relevant by neuroimaging. The modules shown are not enriched for the gene lists found in the LMM analysis (part B).}
    \label{fig:app_modules_nonrel_b}    
\end{figure}

This appendix contains the plots of the analysis described in Sec. \ref{sec:mm_enrichment_analysis}, aimed at discovering whether the modules enriched for the gene lists found with the LMM analysis, are abnormally expressed in the regions highlighted by the \textit{Smoothgrad} framework in the neuroimaging part.
The first group of plots (Fig. \ref{fig:app_modules_rel}), shows the modules of interest, while the remaining two groups (Fig.s \ref{fig:app_modules_nonrel_a} and \ref{fig:app_modules_nonrel_b}) display the other modules.
As underlined in chapter \ref{chap:res_link}, the modules of interest, with the exception of cyan, show a clustering towards the right-most and/or left-most part of the graphs of the regions found important in the neuroimaging analysis (the blue bars), while the other modules do not. This is a clue of the fact that the results of the neuroimaging analysis and those of the transcriptomic analysis are linked.

\chapter{Computational details}\label{Appendix_computational}
This appendix contains details regarding the computational resources used throughout this thesis, highlighting the most important technical difficulties that have been encountered.

\section{Neuroimaging part}
Uncompressed fMRI data at the resolution that has been used in this thesis require a lot of memory (approximately 60 million voxels per image, occupying 240 MB of memory).
For this reason, data is loaded and decompressed lazily in memory only when required for training the neural network. Furthermore, due to the large amount of memory required by each image, the batch size was limited to very small values, lengthening the duration of the training procedure.

The neural network has been trained on an NVIDIA 1080 Ti GPU on a system with a 20-core Xeon E5 and 64 GB of RAM. The main bottleneck was the GPU, with the other resources only requiring minor usage.

The grid search for hyperparameter optimization has explored approximately 1,000 different configurations, each trained for 100 epochs, for a total training time of 2 months. The final network has been trained for 500 epochs.

The most time consuming part of the study was the calculation of the CI, since it requires training the same model multiple times, with different degrees of dataset imbalance. In total, the combination of the hyperparameter optimization, the CI study and the training of the final model, required approximately 4 months of training time.

\section{Genetics part}
The genetics part of the study was conducted on a dataset containing approximately 20,000 genes for 250 samples, requiring about 40 MB of RAM. As the amount of memory required was very small, this analysis could have been conducted on a standard laptop, with the only exception being the network construction.

The adjacency matrix at the base of the network has in fact a dimensionality proportional to the number of genes squared, requiring several GB of memory for storage alone. A common strategy to reduce the required memory, is to calculate the matrix piece-wise, considering groups of genes. However, this reduces the reliability of the calculation and may introduce errors in the matrix. The use of a machine with 64 GB of RAM allowed to perform the calculation of the adjacency matrix in a single step.

Another point of note is the calculation of the mixed models. While each model is straightforward to calculate and does not require specialized computational resources, the sheer scale of the problem (one model for each gene had to be calculated for each tissue or pair of tissues with customized pre-processing) required a specialized implementation.
The Julia programming language was used, which supports just-in-time compilation, significantly speeding up the computation of mixed models with respect to interpreted languages such as python or R.

\section{Imaging genetics part}
The imaging genetics part of the study was conducted on the AHBA dataset. The scale of this dataset is impressive, with more than 80 million data points organized in a complex hierarchical structure, requiring several GBs of memory for each of the 6 subjects.
The large dimension of this dataset meant that the initial operations could be performed only loading the required parts in memory, while keeping the rest on disk. As the dimensionality of the data was reduced to get to the subject-wise average of eigengenes across regions, it was possible to perform the final parts of the analysis on a standard laptop without particular precautions.

\end{appendices}

\chapter*{Copyright information}
\addcontentsline{toc}{chapter}{Copyright information}
The images included in this thesis are of three types:
\begin{itemize}
    \item Images created by the author not published elsewhere. \\
    In this case no symbols are added in the figure caption.
    \item Images created by the author and already published in a peer-reviewed manuscript, allowed by the copyright owners to appear in this document. \\
    The caption of these images is marked with the symbol $\dagger$.
    \item Images taken from other published sources for which permission of reproduction in this document was obtained or was allowed explicitly by the license under which the source was published.\\ 
    The caption of these images is marked with the symbol $\ddagger$.
\end{itemize}

\end{document}